\documentclass[usenatbib,useAMS,twocolumn]{mn2e}

\usepackage{subfigure}
\usepackage{graphicx}
\usepackage[latin1]{inputenc}

\usepackage{rotating}

\usepackage{afterpage}

\usepackage{color}

\usepackage{verbatim} 
\usepackage{rotating} 

\usepackage{morefloats}
\usepackage{subfigure}

\bibpunct{(}{)}{;}{a}{}{,}

\def\arcsec{$^{\prime\prime}$}

\def\simgt{\mathrel{\raise0.35ex\hbox{$\scriptstyle >$}\kern-0.6em
\lower0.40ex\hbox{{$\scriptstyle \sim$}}}}
\def\lta{\mathrel{\raise0.35ex\hbox{$\scriptstyle <$}\kern-0.6em
\lower0.40ex\hbox{{$\scriptstyle \sim$}}}}

\def\gs{\mathrel{\raise0.35ex\hbox{$\scriptstyle >$}\kern-0.6em
\lower0.40ex\hbox{{$\scriptstyle \sim$}}}}
\def\ls{\mathrel{\raise0.35ex\hbox{$\scriptstyle <$}\kern-0.6em
\lower0.40ex\hbox{{$\scriptstyle \sim$}}}}

\def\factor{${1.0}$}

\def\jone{$^{12}$CO\,(1--0)}
\def\jtwo{$^{12}$CO\,(2--1)}
\def\jthree{$^{12}$CO\,(3--2)}
\def\jfour{$^{12}$CO\,(4--3)}

\def\jseven{$^{12}$CO(7--6)}

\title[A survey of CO in submm galaxies]{A survey of molecular gas in luminous sub-millimetre galaxies}
\author[M.\,S.\ Bothwell et al. ]
{M.\,S.\ Bothwell$^{1, 2, }$\thanks{E-mail:
matthew.bothwell@gmail.com}, 
Ian Smail,$^{\! 3}$
S.\,C.\ Chapman,$^{\! 2}$
R.\ Genzel,$^{\! 4}$
R.\,J.\ Ivison,$^{\! 5, 6}$ \newauthor
L.\,J.\ Tacconi,$^{\! 4}$
S.\ Alaghband-Zadeh,$^{\! 2}$
F.\ Bertoldi,$^{\! 7}$
A.\ W. Blain,$^{\! 8}$      
C.\,M.\ Casey,$^{\! 9}$ \newauthor
P.\ Cox,$^{\! 10}$
T.\,R.\ Greve,$^{\! 11}$
D.\ Lutz,$^{\! 4}$ 
R.\ Neri,$^{\! 10}$ 
A.\ Omont$^{12}$ \& 
A.\,M.\ Swinbank$^{3}$ \\
 \\ 
$^{1}$Steward Observatory, University of Arizona, Tucson, AZ 85721, USA\\
$^{2}$Institute of Astronomy, University of Cambridge, Cambridge, CB3 0HA, UK\\
$^{3}$Institute for Computational Cosmology, Durham University, Durham, DH1 3LE, UK \\
$^{4}$Max-Planck-Institut f\"ur extraterrestrische Physik, Postfach 1312, 85741 Garching, Germany \\
$^{5}$UK Astronomy Technology Centre, Royal Observatory, Blackford Hill, Edinburgh EH9 3HJ, UK \\
$^{6}$Institute for Astronomy, University of Edinburgh, Blackford Hill, Edinburgh EH9 3HJ, UK \\
$^{7}$Argelander-Institut f\"ur Astronomie, Auf dem H\"ugel 71, 53121 Bonn, Germany \\
$^{8}$Department of Physics \& Astronomy, University of Leicester, University Road, Leicester, LE1 7RH, UK\\
$^{9}$Institute for Astronomy, University of Hawaii, 2680 Woodlawn Dr, Honolulu, HI 96822, USA \\
$^{10}$Institut de Radio Astronomie Millimetrique (IRAM), St. Martin d'Heres, France \\
$^{11}$Dark Cosmology Centre, Niels Bohr Institute, University of Copenhagen, Juliane Maries Vej 30, DK-2100 Copenhagen, Denmark\\
$^{12}$CNRS and Institut d'Astrophysique de Paris, 98 bis boulevard Arago, 75014 Paris, France
}
\begin{document}
\date{Accepted ----. Received ---- in original form ----}

\pagerange{\pageref{firstpage}--\pageref{lastpage}} \pubyear{2012}

\maketitle
\begin{abstract}
We present the results from a survey for $^{12}$CO emission in 40 luminous sub-millimetre galaxies (SMGs), with 850-$\mu$m fluxes of S$_{850 \mu m} = 4-20$ mJy, conducted with the Plateau de Bure Interferometer. We detect $^{12}$CO emission in 32 SMGs at $z\sim 1.2$ -- 4.1, including 16 SMGs not previously published. Using multiple $^{12}$CO line ($J_{\rm up}=$\,2--7) observations, we derive a median spectral line energy distribution for luminous SMGs and use this to estimate a mean gas mass of $(5.3\pm 1.0)\times 10^{10}$\,M$_\odot$. We report the discovery of a fundamental relationship between $^{12}$CO FWHM and $^{12}$CO line luminosity in high-redshift starbursts, which we interpret as a natural consequence of the baryon-dominated dynamics within the regions probed by our observations. We use far-infrared luminosities to assess the star-formation efficiency in our SMGs, finding a steepening of the L$'_{\rm CO}$--L$_{\rm FIR}$ relation as a function of increasing $^{12}$CO $J_{\rm up}$ transition. We derive dynamical masses and molecular gas masses, and use these to determine the redshift evolution of the gas content of SMGs, finding that they do not appear to be significantly more gas rich than less vigorously star-forming galaxies at high redshifts. Finally, we collate X-ray observations, and study the interdependence of gas and dynamical properties of SMGs with their AGN activity and supermassive black hole masses (M$_{\mathrm{BH}}$), finding that SMGs lie significantly below the local M$_{\mathrm{BH}}$--$\sigma$ relation. We conclude that SMGs represent a class of massive, gas-rich ultraluminous galaxies with somewhat heterogeneous properties, ranging from starbursting disc-like systems with L$\sim10^{12}{\rm L}_{\sun}$, to the most highly star-forming mergers in the Universe. 

%
%We present the results from a survey for $^{12}$CO emission in luminous sub-millimetre galaxies (SMGs), conducted with the Plateau de Bure Interferometer. We observed a total of 40 SMGs at $z\sim $\,1--4 with 850-$\mu$m fluxes of S$_{850 \mu m} = $\,4--20 mJy.  We detect $^{12}$CO emission in 32 SMGs,  including 16 SMGs not previously observed in $^{12}$CO. Using multiple $^{12}$CO line ($J_{\rm up}=$\,2--7) observations, we derive a median spectral line energy distribution for luminous SMGs. We report the discovery of a fundamental relationship between $^{12}$CO FWHM and $^{12}$CO line luminosity  of high-redshift starbursts, which we interpret as a natural consequence of the baryon-dominated dynamics within the regions probed by our observations. We use far-infrared luminosities to assess the star-formation efficiency in our SMGs, finding an approximately linear slope of the L$'_{\mathrm{CO(1-0)}} $--L$_{\mathrm{FIR}}$ relation for SMGs. We derive dynamical masses and molecular gas masses, and go on to discuss the redshift evolution of the gas content of SMGs, finding that SMGs do not appear to have significantly higher baryonic gas fractions than less vigorously star-forming galaxies at comparable redshifts. Finally, we collate X-ray observations, and study the interdependence of gas properties with AGN activity and supermassive black hole mass (M$_{\mathrm{BH}}$), finding that SMGs lie significantly below the local M$_{\mathrm{BH}}$--$\sigma$ relation. 
%
\end{abstract}
\begin{keywords}
cosmology: observations --
galaxies: evolution --
galaxies: formation --
galaxies: ISM
\end{keywords}

%%%%%%%%%%%%%%%%%%%%%%%%%%

\nocite{2011MNRAS.415.2723C}

\section{Introduction}

The discovery of a population of sub-millimetre (sub-mm) bright, highly star-forming galaxies at high redshift has provided a
critical challenge for hierarchical galaxy evolution models (\citealt{Blain:1999aa}; \citealt{Baugh:2005aa}). These luminous ``sub-mm galaxies'' (SMGs) are young, highly dust obscured galaxies, at a median $<\! z\! > \sim 2.5$ (e.g.\
\citealt{Chapman:2003aa}; \citealt{2005ApJ...622..772C}, hereafter C05; \citealt{Wardlow:2011aa}), with extreme far-infrared (far-IR) luminosities ($\mathrm{L}_{\mathrm{FIR}} = 10^{12} $--$ 10^{13} \; \mathrm{L}_{\sun}$) implying prodigious star formation rates of $\gs 10^3 \; \mathrm{M}_{\sun}$\,yr$^{-1}$ (\citealt{2010MNRAS.409L..13C}; Magnelli et al.\ 2012). As they are selected at 850$\mu$m (corresponding to restframe $\lambda\sim 200$--400$\mu$m) these sources tend to
have significant masses of cold dust  (C05, see also Magnelli et al.\ 2012), as well as correspondingly substantial reservoirs of molecular gas ($\sim 10^{10} \; \mathrm{M}_{\sun}$; \citealt{2005MNRAS.359.1165G}). Together with their high stellar masses ($\sim 10^{11} \; \mathrm{M}_{\sun}$; e.g.\ \citealt{Hainline:2011lr}) this suggests this population reside in some of most massive dark matter haloes in the high-redshift Universe. 

\nocite{2010A&A...518L..28M} \nocite{2012A&A...539A.155M}
%Holland, W., MacIntosh, M., Fairley, A., et al. 2006, in SPIE Conf. Ser., 6275 
%add Hughes'98/Barger'98 references?
%Ivison et al. 1998
%add Dunlop's claims about disk-like SMGs?

SMGs have been studied extensively since their discovery  in the 850-$\mu$m atmospheric window using the Sub-mm Common User Bolometer Array (SCUBA) on the JCMT \citep{1997ApJ...490L...5S}, and it is clear that they represent a population of cosmological importance.   Recent advances in IR-detector technology have opened up the possibility of much wider field surveys at these wavelengths, including {\it Herschel} SPIRE at 500$\mu$m; (Chapman et al.\ 2010; Magnelli et al.\ 2010, 2012), in the millimetre-waveband with the South Pole Telescope  \citep{2010ApJ...719..763V} and surveys with the new SCUBA-2 camera (Holland et al.\ 2006) on JCMT.  These surveys are providing large numbers of dusty galaxies with properties similar to the original SCUBA population. In this work, we use the term ``SMG'' to refer to 850-$\mu$m-selected galaxies which comprise the bulk of our sample.

SMGs lie above the ``main sequence'' of star formation (defined as SFR/M$_* \sim {\rm M}_*^{\alpha} (1+z)^{\beta}$, \citealt{2007ApJS..173..315S}; \citealt{2007ApJ...660L..47N}; \citealt{2007ApJ...670..156D}; \citealt{Rodighiero:2010aa}). SMGs typically lie $\sim 1$ dex above this ``main sequence''; significantly more than the  scatter around the sequence, which is typically $\sim \pm 0.3$ dex. 

As a result of their location in the SFR--M$_*$ plane (along with corroborating kinematic and morphological evidence; Tacconi et al.\ 2006, 2008; Engel et al.\ 2010), many authors have argued that SMGs can be understood as ``scaled-up'' analogues of ultra-luminous IR galaxies (ULIRGs) in the local Universe, which typically lie an order of magnitude (or more) above the SFR--M$_*$ ``main sequence''. In this picture, luminous SMGs are best understood as major-merger events, with the extreme star formation being driven by the merger. 

There has also been claims, however, for a secular origin for some SMGs. Theoretical simulations that model SMGs as massive, star-forming discs  reproduce some of the physical properties of the SMG population, while avoiding some of the possible difficulties that arise from ascribing a merger-origin to all SMGs (such as the observed number counts). While there has been some corroborating evidence for this picture (e.g.\ Bothwell et al.\ 2010 presented evidence for a luminous SMG with disc-like properties), simulated disc-like SMGs generally fail to reproduce the extreme star formation rates characteristic of the bright end of the SMG population. 

In addition to these (apparently opposing) pictures, ``main-sequence'' galaxies can also fall within the standard SMG selection criterion (a simple flux cut at $850 \mu$m, S$_{850\mu \rm m}\gs 4$\,mJy). High-resolution imaging has resolved a few sub-mm sources into two (or more) less luminous star forming galaxies, which happen to fall within the same sub-mm detection beam, pushing the combined system above the SMG survey detection limits (see \citealt{Wang:2011aa}). In addition, Chapman et al.\ (2005) confirmed the presence of a population of lower luminosity galaxies at lower redshifts, which are selected at sub-mm wavelengths due to unusually cold dust temperatures.   An overarching theme of the results over the last few years has therefore been an appreciation of the diversity of the SMG population. The transition between the brightest starburst sources and the more secular systems is a undoubtably gradual one, and it is clear that a mix of models is required if we are to fully understand the entirety of the SMG population.

Observations of the cold molecular phase of the ISM have the power to resolve many of these issues, as well as providing unique insight into the physics and behaviour of these systems. Molecular line observations provide the most direct insight into galaxy dynamics (e.g. \citealt{Walter:2011aa}) with turbulent merging systems having very different kinematic profiles from ordered rotating discs (\citealt{2003ApJ...597L.113N}; \citealt{2006ApJ...640..228T}, 2008, 2010). In addition, molecular gas linewidths provide excellent estimates of the dynamical mass of galaxies, free from the uncertainties (extinction effects and non-gravitational motions, such as winds) that can potentially affect estimates derived from optical and near-IR observations.

\nocite{2008ApJ...680..246T} 
\nocite{Tacconi:2010aa}

From the intensity of the observed $^{12}$CO emission lines, it is possible to calculate the mass of the molecular gas
reservoir. As molecular hydrogen lacks a permanent electric dipole, it is all but invisible for even nearby
galaxies. Instead, $^{12}$CO is used as a tracer molecule, as it is produced in similar conditions to molecular hydrogen and is
collisionally thermalised by H$_2$ even at low densities (though it is typically optically thick). Once the relative ratio of $^{12}$CO/H$_2$ molecules is known (the
conversion factor is usually parameterised as $\alpha$), it is possible to calculate the mass of the underlying H$_2$
reservoir. When combined with far-IR luminosities (calculated directly, or inferred from radio continuum flux observations via the
far-IR/radio correlation), gas masses provide estimates of the star formation efficiencies in SMGs (e.g.\ Greve et al.\ 2005; \citealt{Genzel:2010aa}; \citealt{Daddi:2010aa}). Whether such extreme starburst galaxies have similar star formation efficiencies to more ``normal'' star forming galaxies at similar epochs, or whether modification is needed in order to explain their prodigious star-formation rates, has important implications for galaxy evolution models.

Historically, a number of instrumental limitations caused molecular gas in high-redshift galaxies to be poorly studied. Observing molecular gas is very time consuming with current instrumentation, with on-source integration times of  several hours being typical for even luminous $^{12}$CO sources. In addition, several steps are required in order to observe a $^{12}$CO line once an SMG has been detected at sub-mm wavelengths. After the galaxy has been detected with the typically low-resolution sub-mm or far-IR beam, an accurate position is required using radio continuum emission (Ivison et al.\ 2001; \citealt{Biggs:2011qy}). Only then can UV spectroscopy obtain an accurate redshift (C05), and only at that point could the relatively narrow-bandwidth receivers observe $^{12}$CO transitions. Of course, with the higher-resolution imaging in the far-IR/sub-mm made available by the advent of ALMA, the positioning of SMG-bright sources is becoming easier, and with increased bandwidths available at mm wavelengths the required redshift accuracy is now becoming less of an issue).  

One significant weakness of the studies of the gas content of high-redshift starbursts published to date is their poor redshift coverage, and subsequent limited ability to address evolutionary effects in the molecular gas properties of SMGs. The redshift evolution of molecular gas fractions and dynamical masses encodes important information relating to the assembly of massive galaxies, providing valuable constraints on models of structure formation.

In view of the  importance of this population, and of the paucity of existing datasets, we undertook a large survey for $^{12}$CO emission in a large sample of SMGs using the IRAM Plateau de Bure Interferometer (PdBI), aiming to draw significant conclusions about the SMG population as a whole across a wide range of redshifts. This makes it possible,  for the first time, to address evolutionary effects in the gas properties of luminous SMGs.

Our IRAM PdBI SMG survey was started in 2002. Prior to the start of the survey, only two SMGs had been detected in $^{12}$CO; SMM\,J02399$-$0136 ($z=2.81$: Ivison et al.\ 1998; \citealt{Frayer:1998aa}; \citealt{2003ApJ...584..633G}) and SMM\,J14011+0252 ($z=2.56$: \citealt{Frayer:1999aa}; \citealt{Ivison:2001aa}). The first results from the PdBI SMG survey were published by \cite{2003ApJ...597L.113N} and \cite{2005MNRAS.359.1165G}, who added $^{12}$CO detections of three and five new SMGs respectively. In addition, spatially resolved (sub-arcsecond) imaging for a total of 12 SMGs from the survey has been reported by Tacconi et al.\ (2006, 2008), \cite{Bothwell:2010aa}, and \cite{2010ApJ...724..233E}. The IRAM PdBI SMGs survey has now reached its conclusion with a total of 40 SMGs at $z\sim $\,1--4  observed in a range of $^{12}$CO transitions. 

\nocite{2006ApJ...640..228T}\nocite{2008ApJ...680..246T}

In \S\ref{sec:sample} we give details of the survey, the observing strategy used, and the reduction and analysis of the data. \S\ref{sec:ladder} presents a median spectral line energy distribution for the sample, from which luminosity ratios are estimated. In \S\ref{sec:properties}, we describe the physical properties of the SMG sample as revealed by the $^{12}$CO observations, including the kinematic properties, dynamical masses, star formation efficiencies and molecular gas properties. In \S\ref{sec:masscomp} we also compare baryonic and dynamical mass measurements, and discuss the implications for deriving physically-motivated parameters for SMGs. \S\ref{sec:agn} then discusses the effects of supermassive black hole (SMBH) activity on the SMG population, and \S\ref{sec:conc} presents our conclusions.
  
Throughout this work we adopt a cosmological model with ($\Omega_m, \; \Omega_{\Lambda},\; \mathrm{H}_0) = (0.27,\;
0.73,\; 71$ km\,s$^{-1}$\,Mpc$^{-1}$), and a \cite{2003PASP..115..763C} IMF.

\section{Observations, Reduction and Sample Properties}
\label{sec:sample}

\subsection{Sample Selection}
\label{sec:obs}

%
% Figure 1
%
\begin{figure*}
\centering
{\includegraphics[width=15cm]{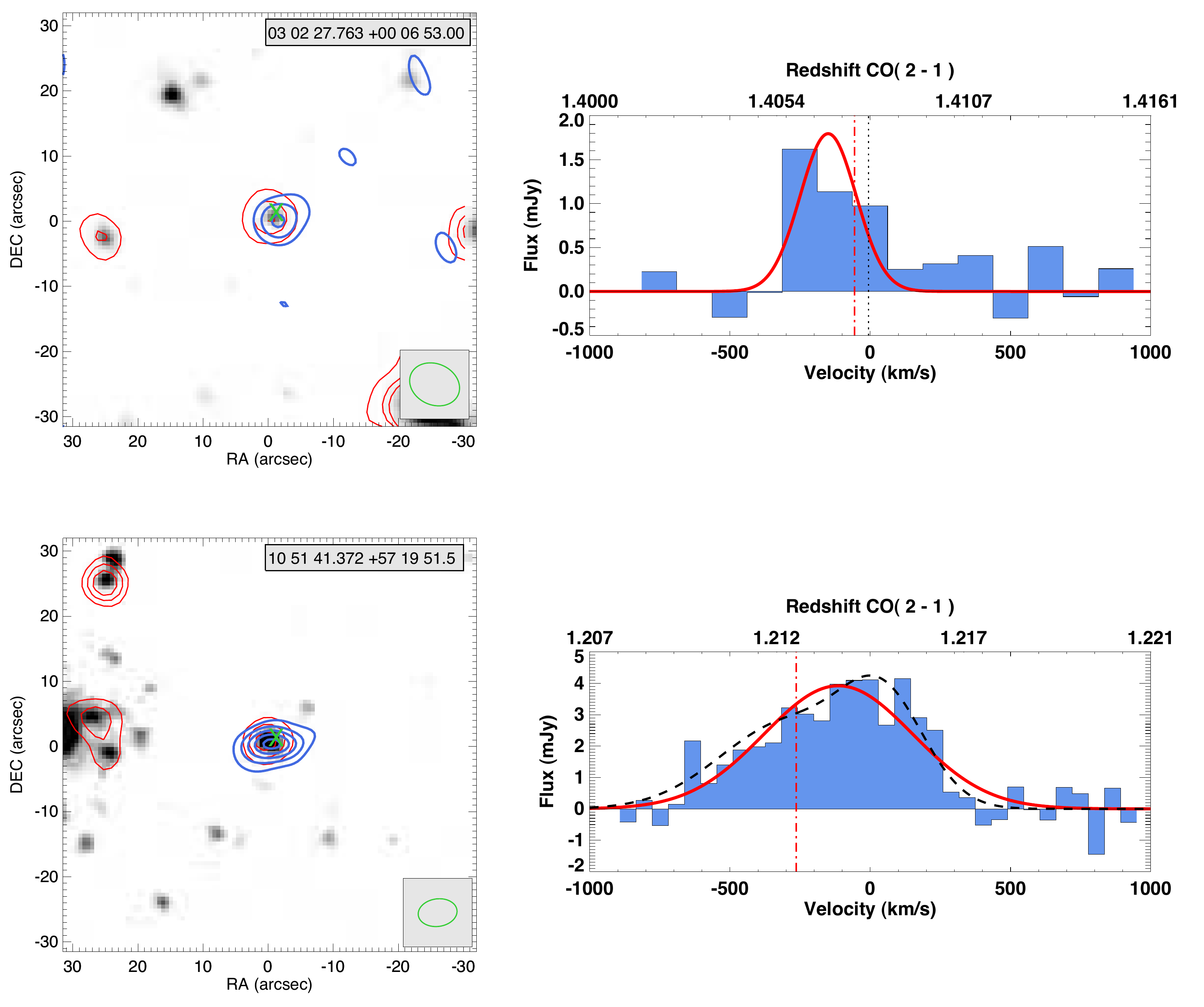}}
\caption{Example maps and $^{12}$CO spectra for two of the SMGs in our sample (the full sample is shown in Appendix A). {\it Top:} SMM\,J030227.73. {\it Bottom:} SMM\,J105141.31. The maps show $^{12}$CO (blue contours) and 24-$\mu$m (red contours) overlaid on near-IR images made from the combined {\it Spitzer} IRAC images (3.6, 4.5, 5.8, and 8$\mu$m). $^{12}$CO contours start at 2-$\sigma$ significance, and are  spaced in steps of 1$\sigma$ (top) and 2$\sigma$ (bottom). The  cross shows the radio position, and the ellipse in the bottom right shows the synthesised PdBI beam. The spectra are taken at the point of maximum $^{12}$CO intensity. The best-fitting single Gaussian profile is overlaid on each. Vertical dashed lines indicate redshift estimates from the literature: red shows redshifts derived from nebular recombination lines (H$\alpha$, or O[{\sc ii}]), black is from UV emission lines (C05).}
\label{maps}
\end{figure*}

%Smail et al. 2002
%Coppin et al. 2006

Our full sample of SMGs observed with the PdBI low-resolution programme consists of 40 galaxies\footnote{Four SMGs in our sample are members of close pairs (SMM\,J123711.98/SMM\,J123712.12, and SMM\,J123711.86/SMM\,J123708.80), and in each case a single pointing was sufficient to observe both SMGs.} at $z\sim 1$--4.   The initial target selection was drawn from the optical-spectroscopic survey of SMGs by \cite{2005ApJ...622..772C}, as well as the SCUBA Cluster Lens Survey (Smail et al.\ 2002) and the SHADES survey of the Subaru/XMM-Newton Deep Field (Coppin et al.\ 2006). Given the duration of the survey, our observing strategy naturally evolved with time, taking advantage of the rapid advance of other multi-wavelength projects studying this population.  Therefore as the survey progressed we also included a few of the first spectroscopically-identified millimetre-selected galaxies (from surveys with the MAMBO camera on the IRAM 30m; i.e.\ \citealt{Bertoldi:2000lr}; \citealt{2004MNRAS.354..779G}), as well as SMGs with more precise rest-frame optical redshifts derived from near- or mid-IR spectroscopy.  As a result the final target sample includes SMGs selected to have precise spectroscopic redshifts, derived using one or more of the following techniques: (i) Lyman-$\alpha$ emission (C05); (ii) H$\alpha$ emission (\citealt{2004ApJ...617...64S}), and (iii) PAH emission from mid-IR spectroscopy with {\it Spitzer} (\citealt{2008ApJ...675.1171P}; \citealt{2009ApJ...699..667M}).   

For those SMGs with only optical (restframe UV) spectroscopy at the time of observation, we used the median velocity offset between optical and near-IR redshifts for SMGs to estimate the likely velocity offsets of their systemic redshift from the optically-derived value.  The resulting redshift uncertainty is $\Delta z = 0.005$ and this should guarantee that the $^{12}$CO emission from an SMG at $z \geq 1$ would fall within the 580-MHz  bandwidth available  at 3\,mm at the time.

As we have radio fluxes for our sources, we can derive the far-IR luminosity for our SMGs from the 1.4-GHz continuum, via the far-IR-radio correlation in \cite{2001ApJ...554..803Y}:

\begin{equation}
\mathrm{L}_{\mathrm{FIR}} = 4\pi \mathrm{D}_{\mathrm{L}}^2 \;(8.4 \times 10^{14})\; \mathrm{S}_{1.4} \;(1+z)^{(\alpha - 1)},
\end{equation}
where D$_{\mathrm{L}}$ is the luminosity distance in metres, S$_{1.4}$ is the radio flux density at 1.4 GHz (in W\,m$^{-2}$\,Hz$^{-1}$), and $\alpha$ is the synchrotron slope used to K-correct the 1.4 GHz observations to the appropriate source redshift ($\alpha$ is taken here to be 0.8). This equation assumes $q_{\rm FIR} = 2.34$. Some  recent results from {\it Herschel} have suggested that a lower value ($q_{\rm FIR} \sim 2.0$) may be appropriate for luminous SMGs (Ivison et al.\ 2010; Magnelli et al.\ 2012), which would lower our derived far-IR luminosities by a factor of $\sim 2$. However, no clear conclusion has yet been reached on this issue, so following Magnelli et al.\ (2012) we assume the (local) value of $q_{\rm FIR} = 2.34$, and note the possibility of our far-IR luminosities being over-estimated. 

\subsection{Data Acquisition, Reduction and Analysis}

The observations of SMGs described here were obtained in observing campaigns on the Plateau de Bure Interferometer between 2002 and 2011, in good to excellent weather. The observations were undertaken with the lowest-resolution ``D'' configuration of the interferometer, in order to maximise the sensitivity, and used five of the six available antennae (giving a total of 10 baselines). The targets were observed in the 2- or 3-mm windows, depending upon the redshift of the source.  The typical resulting resolution of our 3-mm maps is $\sim4$\arcsec. Table~1 lists our observation log. 

For data reduction, the IRAM \textsc{gildas} software was used. Data were carefully monitored throughout the tracks, and any bad or high phase noise visibilities were flagged. One or more bright quasars were typically used for passband calibration, and flux calibration was determined using observations of the main calibrators, 3C\,454.3, 3C\,345, 3C\,273, and MWC\,349. Phase and amplitude variations within each track were calibrated out by interleaving reference observations of nearby compact calibrators every twenty minutes (see Table~1). For the subsequent analysis, the  routines \textsc{clic} and \textsc{mapping} were used, producing naturally-weighted datacubes which were then outputted for analysis with our own {\sc idl} routines. 

In a few cases, a line was detected close to the edge of the band pass in the initial track. In these situations the frequency  setting was then adjusted to centre the line in the bandpass, and  the source was re-observed. On average during the survey, a source was typically observed for 2--3 tracks (a total on-source integration time of 10--18 hours). If no signal had been detected after this, the source was classed as a non-detection.  Of the non-detections, subsequent observations (either in $^{12}$CO, or from spectroscopy -- either optical or near-/mid-IR) have shown that some of the originally targeted redshifts were incorrect and that the $^{12}$CO lines likely fall outside of our band (Table \ref{tab:wrong}).  We therefore conclude we have only eight true non-detections. Table~4 lists the observed $^{12}$CO properties of our SMGs. Figure~\ref{maps} shows example maps and spectra for two of the SMGs in our sample -- the full sample of detections and non-detections are shown in Appendices A and B respectively. 

\begin{table*}
\begin{center}
\begin{minipage}{140mm}
\caption{Observation log}
\begin{tabular}{@{}lcccccc@{}}
\hline
\hline

ID &
$\nu_{obs}$    &
RMS$_{\rm Band}$\footnote{RMS noise averaged over the bandwidth of observation.}  &
RMS$_{\rm Channel}$\footnote{RMS noise per channel.} &
On-source time &
Calibrator &
Reference \\

 &
(GHz) &
(mJy) &
(mJy) &
(hours) &
 &
  \\

\hline

SMM\,J021738.71$-$050339.7    &      151.9 &     0.15 &      0.80 & 4.1     &   3C454.3 & This work \\
SMM\,J021738.91$-$050528.4    &      130.6 &     0.15 &      0.80 & 5.1     &   3C454.3 & This work \\
SMM\,J021725.16$-$045934.7    &      139.5 &     0.15 &      0.80 & 5.3   &   3C454.3 & This work \\
SMM\,J030227.73+000653.3    &      95.73 &     0.15 &      0.80 & 13.1     &   0336-019 & This work \\
SMM\,J044315.00+021002.0	  &    150 	  &	 0.17       &	 0.7 &     ...  & ... &Neri et al.\ (2003)\\
SMM\,J094304.08+470016.2	 &   	150    &   	0.17		 &    	0.7 &  ...         & ... & Neri et al.\ (2003) \\ 
SMM\,J105141.31+571952.0   &      104.12  &    0.12 &      0.68 &   19.3      & 0923+392 &Engel et al.\ (2010)\\ 
SMM\,J105151.69+572636.0   &      87.92  &    0.05 &      0.51 &    8.7     & 0954+658 &This work \\ 
SMM\,J105227.58+572512.4    &      99.94 &     0.09 &      0.84 &   7.1     & 0954+658 &This work \\ 
SMM\,J105230.73+572209.5   &     96.02  &     0.12 &      0.65 &    21.9   & 1044+719 &This work \\ 
SMM\,J105307.25+572430.9    &      91.33  &     0.05 &      0.38 &   16.4    &0954+658 &Bothwell et al.\ (2010)\\ 
SMM\,J123549.44+621536.8    &      107.95  &    0.12 &      0.65 &   11.4    & 1150+497 & Tacconi et al.\ (2006) \\ 
SMM\,J123555.14+620901.7    &     80.46  &     0.19 &       1.0 &    11.0     & 1044+719 &This work \\ 
SMM\,J123600.10+620253.5  &     115.49 &    0.42 &       2.2 &     9.4    & 1044+719 & This work \\
SMM\,J123606.85+621047.2    &       98.65 &     0.14 &      0.78 &    18.0    & 1150+497 &This work\\ 
SMM\,J123618.87+621007.5   &      107.72 &    0.06 &      0.57 &      13.0   & 1418+546 &This work\\ 
SMM\,J123618.33+621550.5    &      153.73   &   0.09 &      0.64 &     18.6    & 1044+719 &Bothwell et al.\ (2010)\\ 
SMM\,J123621.27+621708.4  &     154.19 &    0.15 &       1.0 &    7.3     & 0954+658 &This work \\
SMM\,J123629.13+621045.8  &     114.52 &    0.34 &       1.8 &      6.1   & 1150+497 &This work \\
SMM\,J123632.61+620800.1    &     115.48    &   0.54 &       2.9 &     20.1    & 1044+719 &This work\\ 
SMM\,J123634.51+621241.0    &      103.71   &   0.13 &      0.79 &     15.1   & 1044+719 &Engel et al.\ (2010)\\ 
SMM\,J123707.21+621408.1    &      99.08  &     0.16 &      0.85 &     9.6  & 1150+497 &Tacconi et al.\ (2006)\\ 
SMM\,J123708.80+622202.0  &      159.86 &   0.21 &       1.0&      17.5  & 1044+719 &This work \\
SMM\,J123711.98+621325.7    &     115.41 &     0.33 &       1.8 &     12.3   & 1150+497 &Bothwell et al.\ (2010)\\ 
SMM\,J123711.86+622212.6    &     159.86  &     0.21 &       1.0 &     17.5   & 1044+719 &Casey et al.\ (2009)\\ 
SMM\,J123712.05+621212.3  &      88.37  &    0.08 &      0.55 &     17.8   & 1044+719 &This work \\
SMM\,J123712.12+621322.2    &     115.41  &     0.33 &       1.8 &   12.3     & 1150+497 &Bothwell et al.\ (2010)\\ 
SMM\,J131201.20+424208.8 &      99.60    &    0.39  &       0.70  &   10.2           &       1308+326        & Greve et al.\ (2005) \\
SMM\,J131208.82+424129.1   &       90.62  &    0.10 &      0.60 &    18.8    & 1308+326 & This work \\
SMM\,J131232.31+423949.5   &     103.90  &    0.39 &       2.1 &      0.9    & 1308+326 &This work\\ 
SMM\,J163631.47+405546.9  &      105.53 &   0.16 &      0.83 &    9.7     & 3C345 &This work \\
SMM\,J163639.01+405635.9   &     92.65   &    0.28 &       1.5 &       5.5  & 3C345 & This work \\ 
SMM\,J163650.43+405734.5   &      102.17 &     0.07 &      0.38 &     40.4   & 3C345 &Neri et al.\ (2003)\\
SMM\,J163655.79+405909.5  &     95.52  &     0.15 &       1.1 &      7.0   & 3C345 &This work \\
SMM\,J163658.78+405728.1   &     105.24  &    0.20 &       1.0 &      8.3    & 3C345 &This work \\ 
SMM\,J163658.19+410523.8   &      100.11 &    0.12 &      0.73 &      14.9  & 3C345 &Greve et al.\ (2005) \\ 
SMM\,J163706.51+405313.8   &      102.51 &     0.13 &      0.76 &      13.9  & 3C345 &Greve et al.\ (2005)\\ 
SMM\,J221804.42+002154.4   &      98.32  &    0.13 &      0.63 &     24.8    & 2223-052 &This work \\ 
SMM\,J221735.15+001537.2   &      84.44  &    0.10 &      0.52 &     16.5   & 2230+114 &Greve et al.\ (2005)\\ 
SMM\,J221737.39+001025.1   &      95.52  &    0.15 &      0.78 &     14.4    & 2223-052 &This work\\

\hline
\hline
\end{tabular}
\end{minipage}
\end{center}
\end{table*}

Of the  32 SMGs with detected $^{12}$CO emission, it is important that we identify any ``marginal'' cases, for which there is a chance that we have a false positive detection. For each of our detections, we identify any source with a peak flux $>5$ times the RMS map noise as a true detection. Sources with a  peak flux $<5\sigma$ were individually inspected for large spatial and velocity offsets (which we define as $>4 ''$ from expected phase centre or $>400$\,km\,s$^{-1}$ from expected velocity zero point respectively), either of which result in the source being classified as a ``candidate'' detection. We identify four such marginal cases; these are listed in Tables~4 and 5. In addition, we classify SMM\,J131208+4241 as a `candidate' detection -- despite being just $\sim 1''$ from phase centre and $\sim 300 $ km/s from the expected velocity zero point, the source is only weakly detected ($3.6 \sigma$), and as such we cannot say with confidence that it is a robust detection. These five candidate sources are hereafter identified separately in all Figures.

%%%%%%

Where we had detectable $^{12}$CO emission, we determined the redshift ($z$) of the $^{12}$CO emission by calculating the intensity-weighted peak redshift,

\begin{equation}
{z}_{\rm CO} = \frac{\sum(I(z) \, z)}{\sum I(z)}
\end{equation} 

A linewidth can be estimated by measuring the FWHM of a Gaussian profile fit to the spectrum. However, if the line in question is non-Gaussian (for any number of reasons, such as showing an asymmetry, or being double-peaked), then a Gaussian profile will be a poor fit. To avoid this complication, we use the intensity-weighted second moment ($s_\nu$) of the $^{12}$CO spectrum as a non-parametric estimator of the line width:

\begin{equation}
s_\nu = \frac{\int(\nu-\bar{\nu})^2\; I_\nu \, d\nu}{\int I_\nu \, d\nu}, 
\end{equation} 
where $\bar{\nu}$ is the intensity-weighted frequency centroid of the line, and $I_\nu$ is the flux as a function of frequency. In most cases derived widths are similar (to within the errors) to those achieved with traditional Gaussian-fitting. Variations occur only in low signal-to-noise galaxies, where Gaussian fits struggle to achieve sensible results. 

This non-parametric estimator does have the weakness of being sensitive to noise spikes in the spectrum -- a large spike at the edge of the receiver bandwidth can significantly broaden the effective width derived. To reduce this potential source of error, we used a Monte Carlo technique to generate multiple copies of each spectrum, with Gaussian noise added to each (see Table~1). We took as our ``true'' second moment the median value of the resultant distribution of intensity-weighted second moments.

The equivalent Gaussian FWHM of the line can then be calculated from the second moment,  using the relation 

\begin{equation}
\mathrm{FWHM} = 2 \sqrt{2 \ln 2} \; s_\nu \; \;  \simeq 2.35 s_\nu 
\end{equation}

The flux density of each observed $^{12}$CO line was found by velocity-integrating the $^{12}$CO spectrum,

\begin{equation}
I_{\rm CO} = \int_{z_{CO} - 2s_{\nu}}^{z_{CO} + 2s_{\nu}} I(v) dv
\end{equation}

(for a Gaussian emission line, this is equivalent to integrating the flux from $-2\sigma$ to $+2\sigma$ around the centre). The $^{12}$CO luminosities of the SMGs in our sample were then calculated using the standard relation given by \cite{Solomon:2005aa}:

\begin{equation}
 \mathrm{L}'_{\mathrm {CO}} = 3.25 \times 10^7 \; S_{\mathrm {CO}} \Delta v  \; \nu_{\mathrm {obs}}^{-2} \; D_{\rm L}^2 \; (1+z)^{-3},
\end{equation}
where  $S \Delta v$ is the velocity-integrated line flux, L$'_{\mathrm {CO}}$ is the line luminosity in K\,km\,s$^{-1}$\,pc$^{2}$, $\nu_{obs}$ is the observed central frequency of the
line, and $D_L$ is the luminosity distance in Mpc in our adopted cosmology. 

%We typically do not detect continuum emission in any of our sources. This is somewhat unsurprising, due to the sensitivity of our observations (most of our data were taken at 3mm, where the continuum level is typically <0.1 mJy). We identify two sources in our sample that exhibit measurable continuum emission (significant positive emission throughout the bandpass): SMM021738-0503 (which was observed with the new-generation receivers, and has a 2mm continuum flux of $0.33 \pm 0.12$ mJy), and SMM123618+6215 (which has a continuum flux of $0.28 \pm 0.10$ mJy, also at 2mm).

The fits included the option for a positive uniform continuum, but only 2 sources appear to have detected continuum (see Appendix A and B).  The derived limits on the continuum emission in the remaining sources are consistent with those expected from the observed dust spectrum, and exclude significant synchrotron contributions to the far-infrared/sub-millimetre continuum in these sources.

For sources which are not detected in our $^{12}$CO observations, we calculate 3-$\sigma$ line intensity limits based on the RMS channel noise:

\begin{equation}
\mathrm{I}_{\rm CO} < 3 \; \mathrm{RMS}_{\mathrm{channel}} \; \sqrt{\Delta V_{\rm CO}\;  dv}\;, 
\end{equation}
where $\mathrm{RMS}_{\mathrm{channel}}$ is the channel noise given in Table~1, $\Delta V_{\rm CO}$ is the mean linewidth of the detected sample, and $dv$ is the bin size in km\,s$^{-1}$.

%
%
%
%\section{Results and Discussion}
%\label{sec:results}

\begin{table*}
\centering
\caption{List of non-detection observations discarded due to a redshift measurement from the literature lying outside the band. ($^{\dagger}$  SMG followed up later at the correct redshift.) }
%\begin{tabular}{cccl}

\begin{tabular}{@{}lccc@{}}
\hline
ID & $z$   & $z$ & Redshift reference\\
   & (targeted) & (literature) & \\
 \hline
 \hline
SMM\,J105238.36+572435.3	&  3.045			&  2.99 & Men\'{e}ndez-Delmestre et al.\ (2009) \\
SMM\,J131232.34+423949.5$^{\dagger}$		&  2.321		&2.332 & Chapman et al.\ (2005)\\
SMM\,J123616.22+621513.2	& 2.55			&  2.578 & Chapman et al.\ (2005)\\
SMM\,J123553.33+621337.2	& 2.098			&  2.17 & Men\'{e}ndez-Delmestre et al.\ (2009)  \\
SMM\,J123635.67+621423.5	& 2.005			&  2.015 & Swinbank et al.\ (2004) \\
\hline
\end{tabular}
\label{tab:wrong}
\end{table*}

Importantly for the interpretation of the evolutionary properties of molecular gas, there is no apparent redshift bias in the detection probability: a Kolmogorov-Smirnov (K-S) analysis of the redshift distributions, sub-mm flux, or 1.4-GHz continuum flux suggests that the non-detections are  consistent with being a random selection drawn from the parent sample. 

\subsection{Comparing our sample to the wider SMG population}

\begin{figure*}
\centering
{\includegraphics[width=15cm, clip=true, trim = 1cm 12.8cm 5cm 3.2cm]{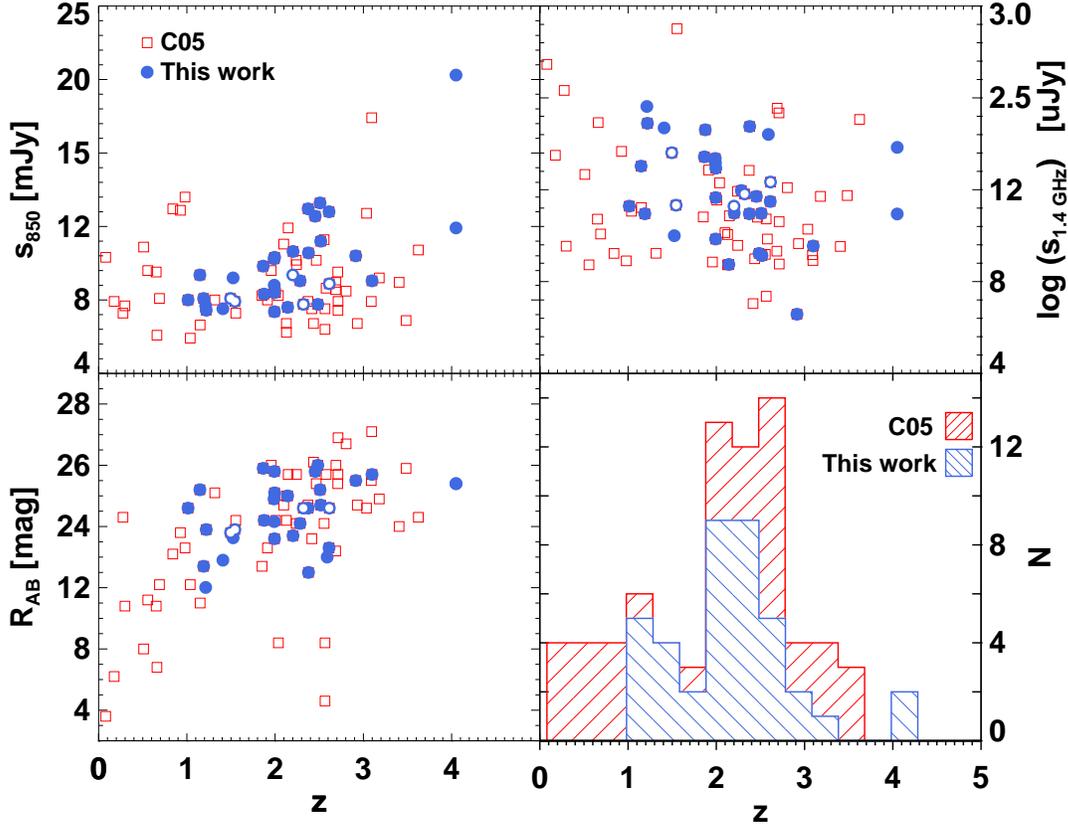}}
\caption{Plots comparing the observed properties of the $^{12}$CO sample presented in this work, along with the ``parent'' sample of C05. Here (and in plots hereafter), open circles denote the SMGs we identify as ``candidate'' detections, while filled circles are our secure detections.  All four panels have redshift as the abscissa. The panels:  \textit{Top left:} 850-$\mu$m flux. \textit{Top right:} 1.4-GHz continuum flux. \textit{Bottom left:} $R$-band magnitude. \textit{Bottom right:} Redshift distributions. Sources comprising our sample do not differ significantly from those in Chapman et al.\ (2005), over a similar redshift range.} 
\label{parent}
\end{figure*}

If we wish to use our sample to draw conclusions about the SMG population as a whole, it is important to investigate whether our $^{12}$CO-detected sample is statistically representative of the wider (i.e.\ 850-$\mu$m bright, radio-detected) SMG population. For the purpose of comparison, we take as our ``representative'' sample the large spectroscopic database of SMGs presented by C05, which functions as an approximate parent sample to the sub-mm-selected, radio-detected, and optical spectroscopy-confirmed SMGs presented in this work. Note that the selection of the C05 sample does imbue it with certain unavoidable biases -- being sub-mm selected, it is  biased towards colder sources (see \citealt{2004ApJ...614..671C}; \citealt{2009MNRAS.399..121C}), and the radio-selection will tend to bias against higher-redshift SMGs with weak radio emission. In addition, the necessity of a spectroscopic redshift confirmation will tend to bias the sample against sources which are UV-faint, or have weak emission or absorption features. Fig.~\ref{parent} shows a comparison of three observable parameters (1.4-GHz continuum flux, 850-$\mu$m  flux, and $R$-band magnitude), as well as redshift distribution histograms, for the detected subset and the C05 parent sample. 

Looking at the distribution of redshifts, our detected subset appears to be a representative sample of SMGs (with the aforementioned selection effects) at $z\sim 1$--3. We do not match the low-redshift coverage, however: low redshift ($z<1$) systems were not included in our programme, and as a result the lowest redshift SMG that was targeted in $^{12}$CO is SMM\,J123629.13 at $z=1.013$. Our observations also under-sample the higher redshift range $z>3$ when compared to C05. 

As the remaining three panels show, the observable properties of our sample show no significant differences when compared to the parent, aside from what results from the redshift restrictions. The brightest $R$-band sources, for example, lie at $z<1$ and do not appear in our sample. The  850-$\mu$m and 1.4-GHz flux distributions of sources appearing in our sample are also consistent with the equivalent distributions for the C05 sample sources, with no apparent bias towards more luminous systems. The mean 850$\mu$m flux for our sample is $6.5 \pm 0.6$\,mJy, compared to $5.9 \pm 0.4$\,mJy for the C05 sample (cut to match the redshift coverage -- i.e.\ $z>1$ -- of our sample); the respective radio flux densities are $114 \pm 11 \mu$Jy and $100 \pm 12 \mu$Jy. KS tests of the 850-$\mu$m and and 1.4-GHz fluxes distributions suggest that our sample is statistically consistent with being drawn at random from that of C05. As a result, our sample of $^{12}$CO-observed SMGs  can be seen as representative of the brighter-end of the population of $z\sim $\,1--3 SMGs as a whole, at the era where the activity in this population peaks.

%
% Figure 2
%

\section{CO excitation modelling}
\label{sec:ladder}

In order to derive the gas properties of our SMG sample, it is important to fully understand the gas excitation as shown by the $^{12}$CO spectral line energy distribution (SLED). The excitation of a molecular gas reservoir is controlled by the physical conditions within the host galaxy: star-forming ULIRGs (and SMGs) are expected to have dense, excited gas that may be thermalised up to $J_{\mathrm{\rm up}}=3$ or beyond (e.g.\ \citealt{2007ASPC..375...25W}; Danielson et al.\ 2011), in contrast to the low density, low-excitation gas which dominates the $^{12}$CO emission of more quiescent galaxies (e.g.\ \citealt{Crosthwaite:2007aa}; \citealt{2009ApJ...698L.178D}). 

Even in luminous SMGs, however, the assumption that higher-$J_{\rm up}$ transitions always trace fully thermalised gas is a poor one -- as $J_{\mathrm{\rm up}}$ increases, the observed gas becomes warmer and denser, and any underlying cold component can be missed (see \citealt{Harris:2010aa}; \citealt{2010MNRAS.404..198I}; \citealt{Ivison:2011aa}). If we wish to fully investigate the physical conditions within our multi-line survey (our data cover $J_{\mathrm{\rm up}}=2$--7), it is important that we fully understand the typical SLED of SMGs from low- to high-$J_{\rm up}$ transitions. This SLED can then be used to derive a self-consistent excitation model.

Throughout this work, we use $r_{J, J-1/10}$ to denote the ratio of $^{12}$CO luminosities\footnote{In units of K\,km\,s$^{-1}$\,pc$^{2}$} L$'_{\mathrm{CO}}(J, \; J-1)/\mathrm{L}'_{\mathrm{CO}}(1-0) \equiv \mathrm{T}_b(J, \; J-1)/\mathrm{T}_b(1-0) $, where $\mathrm{T}_b$ is the equivalent Rayleigh-Jeans brightness temperature in excess of that of the Cosmic Microwave Background.

In order to directly compare the SLEDs for multiple SMGs, it is important to remove the effect of the strong dependence of the $^{12}$CO line flux on far-IR luminosity -- without this step, the variation of L$'_{\mathrm{CO}}$ with L$_{\mathrm{FIR}}$  would introduce extraneous scatter, and a systematic offset between
transitions (due to  the gradient of the L$'_{\mathrm{CO}}$--L$_{\mathrm{FIR}}$ relation, combined with the different  luminosities sampled by the different transitions at different redshifts).  We therefore calculate a ``normalising'' factor, which will scale the line fluxes to the mean L$_{\mathrm{FIR}}$ of our sample ($= 6.0 \pm 0.6 \times 10^{12}$ L$_{\sun}$):

\begin{equation}
\Delta \mathrm{L}_{\mathrm{CO}} =  \frac{d \, \mathrm{L}'_{\mathrm{CO}}}{d \, \mathrm{L}_{\mathrm{FIR}}} \; ( \mathrm{L}_{\mathrm{FIR}} - <\! {\mathrm{L}_{\mathrm{FIR}}} \! >) 
\end{equation}

The gradient of the L$'_{\mathrm{CO}} $--L$_{\mathrm{FIR}}$ relation has been debated in recent years.  Efforts at high redshift with small samples found a sub-linear slope: $0.62 \pm 0.08$, as given in \cite{2005MNRAS.359.1165G}, for example. Recent analyses have suggested a somewhat steeper slope, with  \cite{Genzel:2010aa} reporting a near-linear slope for both SMGs and ``normal'' star forming galaxies. Below (\S4.2), we discuss the slope of the L$'_{\mathrm{CO}} $--L$_{\mathrm{FIR}}$ relation for the extrapolated \jone\ luminosity, a physical quantity that has important implications for the star formation efficiency of our SMGs. For the purposes of removing L$_{\mathrm{FIR}}$-induced scatter, however, it is necessary to examine the slope of the relation for the raw (un-extrapolated) luminosities of the $J_{\rm up}\geq2$ lines. With increasing $J_{\rm up}$, the molecular gas being observed traces progressively warmer, denser gas, which is more and more associated with star formation. The slope of the L$'_{\mathrm{CO}} $--L$_{\mathrm{FIR}}$ relation for the observed (i.e.\ $J_{\rm up}\geq2$) $^{12}$CO luminosities will therefore be dependent on $J_{\rm up}$.

\begin{figure}
\centering
{\includegraphics[width=8.5cm, clip=true, trim = 1.5cm 13cm 5.5cm 3.2cm]{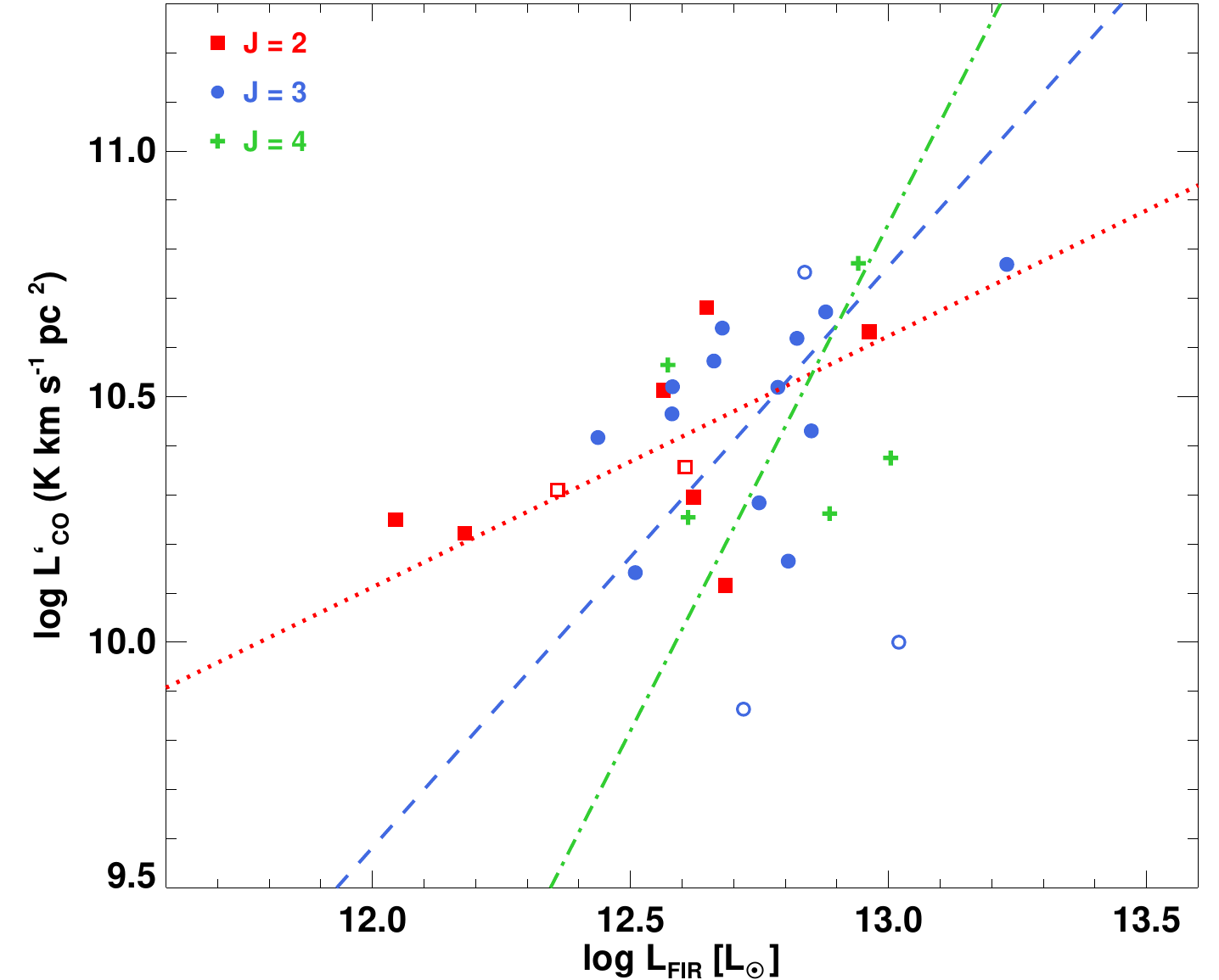}}
\caption{The L$'_{\mathrm{CO}} $--L$_{\mathrm{FIR}}$ relation for our sample of SMGs, colour-coded by the observed transition. Slopes are fitted to the respective samples. We find that a fit to the $J_{\rm up}=2$ points has a slope of $0.71\pm0.19$, the $J_{\rm up}=3$ points have a slope of $1.2 \pm 0.3$, and the $J_{\rm up}=4$ points to have a slope of $2.1 \pm 0.4$. Hence the  L$'_{\mathrm{CO}} $--L$_{\mathrm{FIR}}$ relation is progressively steeper for increasing values of $J_{\rm up}$, as the observed gas becomes a more efficient tracer of active star formation.}
\label{fig:slopes}
\end{figure}

Figure~\ref{fig:slopes} shows the L$'_{\mathrm{CO}} $--L$_{\mathrm{FIR}}$ correlation for our sample of SMGs, colour-coded by the observed transition. We find that the slope does indeed steepen with increasing $J_{\rm up}$, finding slopes of $d \, \mathrm{L}'_{\mathrm{CO(2-1)}} / d \, \mathrm{L}_{\mathrm{FIR}} = 0.71\pm0.19$, $d \, \mathrm{L}'_{\mathrm{CO(3-2)}} / d \, \mathrm{L}_{\mathrm{FIR}} = 1.2\pm0.3$, and $d \, \mathrm{L}'_{\mathrm{CO(4-3)}} / d \, \mathrm{L}_{\mathrm{FIR}} = 2.1 \pm 0.4$. Of course there is the potential for a redshift-induced bias in this result -- the $J_{\rm up}=2$ lines are observed to lower redshifts than the more excited transitions, which could result in the artificial ``flattening'' of the relation. We do note, however, that the resulting line luminosity ratios reached are relatively insensitive to the choice of slope -- adopting a shallower slope has the effect of lowering the line luminosity ratios, but for a reasonable range of slopes (i.e.\ $>0.6$) the variation is within the bootstrap errors. 

After the removal of this L$_{\mathrm{FIR}}$ dependence, SMGs with $\mathrm{L}_{\mathrm{FIR}} = \, <\! {\mathrm{L}_{\mathrm{FIR}}}\! > $ will have their line fluxes unchanged, and SMGs which are less (more) luminous will have their line fluxes proportionally increased (decreased). 

\begin{figure}
\centering
{\includegraphics[width=8.1cm, clip=true, trim = 0cm 0cm 0cm 0cm]{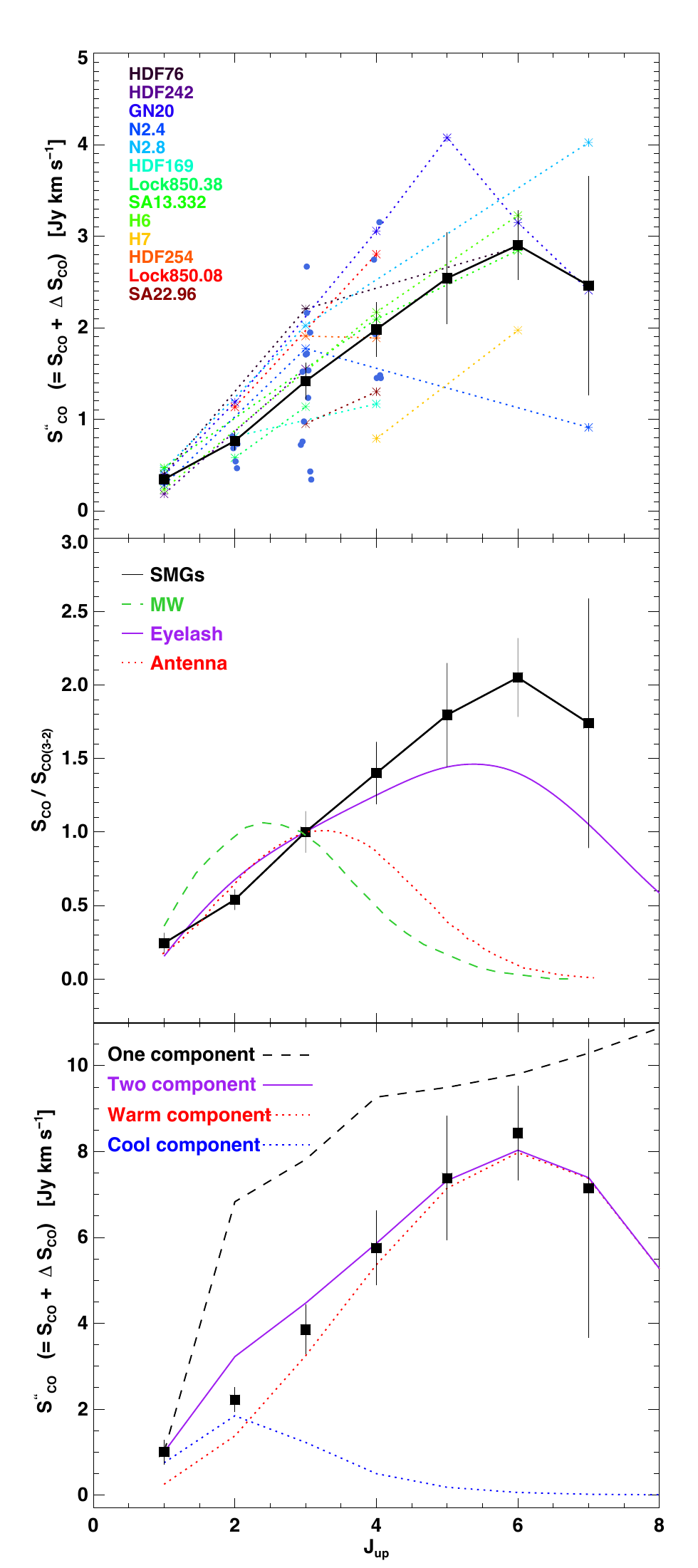}}
\caption{The $^{12}$CO SLED for the SMGs in our sample. All SMGs have been normalised to the mean far-IR luminosity of the sample as detailed in the text, and fluxes have been adjusted to the median redshift of the sample ($z=2.2$). {\it Top:} The $^{12}$CO SLED for our SMGs, in (normalised) flux units. The SMGs in our sample with multiple-$J_{\rm up}$ observations available in the literature are shown by connecting dotted lines. The colour-coding is shown in the inset legend. SMGs with only a single observed transition are shown as blue circles. The median SLED of the sample, along with the bootstrapped error, is overlaid. {\it Middle:} The median SLED compared to other well-studied galaxies, normalised to the flux in the \jthree\ transition. {\it Bottom:} The median SLED, fit with PDR models as discussed in the text. The median SLED shows a moderate excitation, similar to that of SMM\,J2135-0102 and is best fit by a two-component model of the ISM.}
\label{ladder}
\end{figure}

The resulting $^{12}$CO-line SLED for our sample of SMGs, adjusted to a mean far-IR luminosity, is shown in Fig.~\ref{ladder}. The fluxes have also been adjusted to a common source redshift of $z=2.2$.  In addition to the $^{12}$CO line observations presented in this work, we have compiled all the observed transitions available in the literature for SMGs in our sample (compiled from \citealt{2010ApJ...714.1407C}; \cite{Riechers:2010aa}; Ivison et al.\ 2011). SMGs with multiple-$J_{\rm up}$ transitions are highlighted in Fig.~\ref{ladder} and we also show the trend of the median $^{12}$CO luminosity/flux as a function of $J_\mathrm{\rm up}$, which represents the SLED of ``typical'' SMGs.

As can be seen from the median line flux in the top panel of Fig.~\ref{ladder}, the $^{12}$CO SLED shows a moderate excitation. There is a monotonic increase in S$_{\mathrm{CO}}$  up to $J_\mathrm{\rm up}\sim 5$, with a flattening or turn-over above this. This is similar to local starburst galaxies (see, e.g.\ \citealt{2007ASPC..375...25W}). The behaviour of the best sampled SMG in our sample (SMM\,J123711.86 at $z=4.04$ has five observed transitions) is not dissimilar to that of the median, showing an increase up to the $J_{\rm up}=$\,5 transition, and a turnover at higher excitations. We do caution, however, that the behaviour of the median SLED above the apparent turnover is poorly sampled, and dominated by low number statistics; hence constraints on the higher excitation components must be taken as tentative. 

The middle panel of Fig. \ref{ladder} shows our derived $^{12}$CO SLED, compared to three other well-studied systems, normalised to the \jthree\ flux. The moderate excitation found for our SMGs is certainly more excited than less-active, local galaxies. It is very similar, however, to the well-studied, strongly lensed SMG SMM\,J2135$-$0102 (``The Cosmic Eyelash''; \citealt{Danielson:2011lr}), showing a similar peak and falloff towards higher-$J_{\rm up}$.

The bottom panel of Fig.~\ref{ladder} shows the results of fitting our SLED with the Photon Domimated Region (PDR) model of \cite{Meijerink:2007fk}. As can be seen, a single-component model provides a very poor fit to the data. Hence we are driven
to a two-component model, consisting of separate ``warm'' and ``cool'' components, which provides a much better fit to the data. The best fit is achieved with a combination is with 75\% of a cool component plus 25\% of the warm component. The cool component has a density $\log_{10}(n)=2.0$\,cm$^{-3}$ and $\log_{10}(G)=1.5$ (where $n$ is the Hydrogen volume density density, and $G$ is the far-UV field strength in Habing units), and the warm component has a density $\log_{10}(n)=5.5$\,cm$^{-3}$ and a radiation field of $\log_{10}(G)=2.0$.   Thus the volume-averaged radiation field experienced by both components is between 30--100\,$\times$ that of the Milky Way. Given the high star-formation rates for our SMG sample ($\sim 1000\times$ the Milky Way), these relatively modest radiation fields suggest their star formation must be spatially extended.  Similarly, the characteristic densities we derive range between that expected for the ISM in typical star-forming galaxies and that in dense starburst systems (see Fig.~4 in Danielson et al.\ 2011).  As the bulk of the mass in these galaxies is in the cool component, we can use the best-fit density for this component and  assume the
gas is in a 1-kpc thick disk with a radius of $R\sim 3$\,kpc (see \S4.4) to predict a typical gas mass for an SMG of the order of $\sim 10^{11}$\,M$_\odot$ comparable to the masses we estimate later in \S4.4.  However, we caution that acceptable model fits to the SLED span a significant range in parameters:  the cool component can have a density of $2.0 < \log_{10}{(n)} < 3.0$ and a range of acceptable radiation fields of $1.0 < \log_{10} {(G)} < 2.5$, while the warm component parameters have ranges $4.5 < \log_{10}{(n)} < 6.5$ and $2.0 < \log_{10}{(G)} < 3.0$.

%The bottom panel of Fig.  \ref{ladder} shows the results of fitting our SLED with the Photon Domimated Region (PDR) model of \cite{Meijerink:2007fk}. As can be seen, a single-component model provides a very poor fit to the data. The best fitting single-component model has a density of $\log_{10}(n)=2.5$\,cm$^{-3}$, and $\log_{10}(G)=2.5$ (where $n$ is the Hydrogen volume density density, and $G$ is the far-UV field strength in Habing units).

%A two-component model, consisting of separate ``warm'' and ``cool'' components, provides a much better fit to the data. The best fit is achieved with a combination with 75\% of the total mass in a ``cool'' component plus 25\% in the ``warm'' component. The cool component has density $\log_{10}(n)=2.5$\,cm$^{-3}$ and $\log_{10}(G)=1.5$, and the warm component has density $\log_{10}(n)=5.5$\,cm$^{-3}$ and $\log_{10}(G)=2.0$. It must be noted that the errors on these values are significant. The cool component has a range of $2.0 < \log_{10}{(n)} < 3.0$ and $1.0 < \log_{10} {(G)} < 2.5$, while the warm component parameters have ranges $4.5 < \log_{10}{(n)} < 6.5$ and $2.0 < \log_{10}{(G)} < 3.0$. (As a result, the contribution from the warm component varies from $\sim$\,10--35\%). Within the errors, these parameters for the ``cool'' and ``warm'' components are roughly comparable to those measured for normal star-forming galaxies and ULIRGs respectively (see Danielson et al. 2010), though the far-UV field strength required to fit our data, $G$, seems to be towards the lower end of the distribution measured for other galaxies.

Using the SLED shown in Fig.~\ref{ladder}, we calculate median brightness temperature ratios (equivalent to line luminosity ratios) which we use to convert our  $J_\mathrm{\rm up}\geq 2$ observations into an equivalent $^{12}$CO\,$J=$\,1--0 flux. These are given in Table~\ref{tab:ladder}, and we use them throughout this work. These values agree well with values reported elsewhere in the literature (e.g.\  Ivison et al.\ 2011).

\begin{table}
\centering
\caption{Median brightness temperature ratios for the SMGs in our sample. All errors estimated using bootstrap resampling. }
\begin{tabular}{@{}cc@{}}
\hline
Transition & Value\\
 \hline
 \hline
$r_{21/10}$ & $0.84 \pm 0.13$ \\
$r_{32/10}$ & $0.52\pm0.09$ \\
$r_{43/10}$ & $0.41 \pm 0.07$ \\
$r_{54/10}$ & $0.32 \pm 0.05$ \\
$r_{65/10}$ & $0.21\pm 0.04$ \\
$r_{76/10}$ & $0.18 \pm 0.04$\\
\hline
\end{tabular}
\label{tab:ladder}
\end{table}

%
% Figure 3
%

\section{The physical properties of SMGs}
In this section, we describe the modes and methods used to derive physical parameters for our sample of SMGs -- these are given in Table 5.
\label{sec:properties}

\subsection{$^{12}$CO Kinematics}
\label{sec:kinematics}

% Histograms%%%%%%%%%%%%%%%%%%%%%%%%%%%%

%
% Figure 5
%
\begin{figure}
\centering
\mbox
{
\includegraphics[width=8.5cm, clip=true, trim = 1.5cm 12cm 5.5cm 3cm]{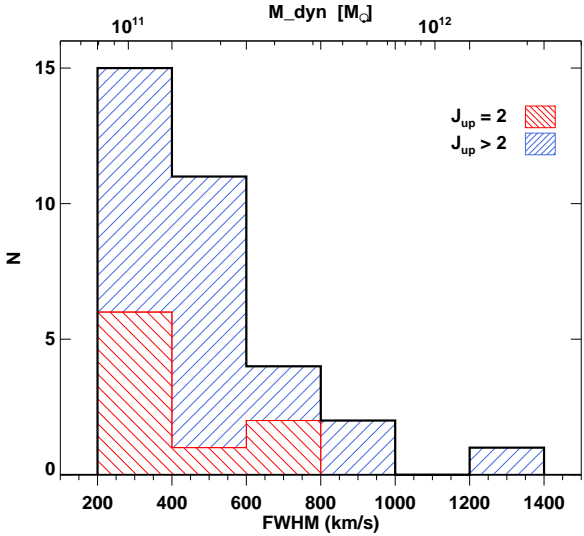}
}
\caption{Distribution of Gaussian-equivalent FWHMs for the sourced detected in $^{12}$CO. FWHMs are derived from the intensity-weighted second moment, as described in \S\ref{sec:mdyn}. SMGs detected in ``high-$J_{\rm up}$'' $(J_\mathrm{\rm up}\geq3)$ and ``low-$J_{\rm up}$'' ($J_\mathrm{\rm up}=2$) are plotted.  As we discuss in the main text, we believe that the lower FWHM displayed by the lower-$J_{\rm up}$ $^{12}$CO lines are likely to be due to their lower mean redshifts and hence lower luminosities.}
\label{hist1}
\end{figure}
%%%%%%%%%%%%%%%%%%%%%%%%%%%%%%%%%%%
%%%%%%%%%%%%%%%%%%%%%%%%%%%%%%%%%%%

The $^{12}$CO line emission from SMGs traces the kinematics of the potential well in which the molecular gas lies.  Previous studies of this emission have
uncovered very broad lines (\citealt{2005MNRAS.359.1165G}; Tacconi et al.\ 2006, 2008):  \cite{2005MNRAS.359.1165G} found a mean FWHM linewidth of 780\,km\,s$^{-1}$ for their SMG sample, suggestive of deep gravitational potential wells.

\begin{figure*}
\centering
\mbox
{
  \subfigure{\includegraphics[width=8.2cm, clip=true, trim = 1.5cm 12cm 5.4cm 3cm]{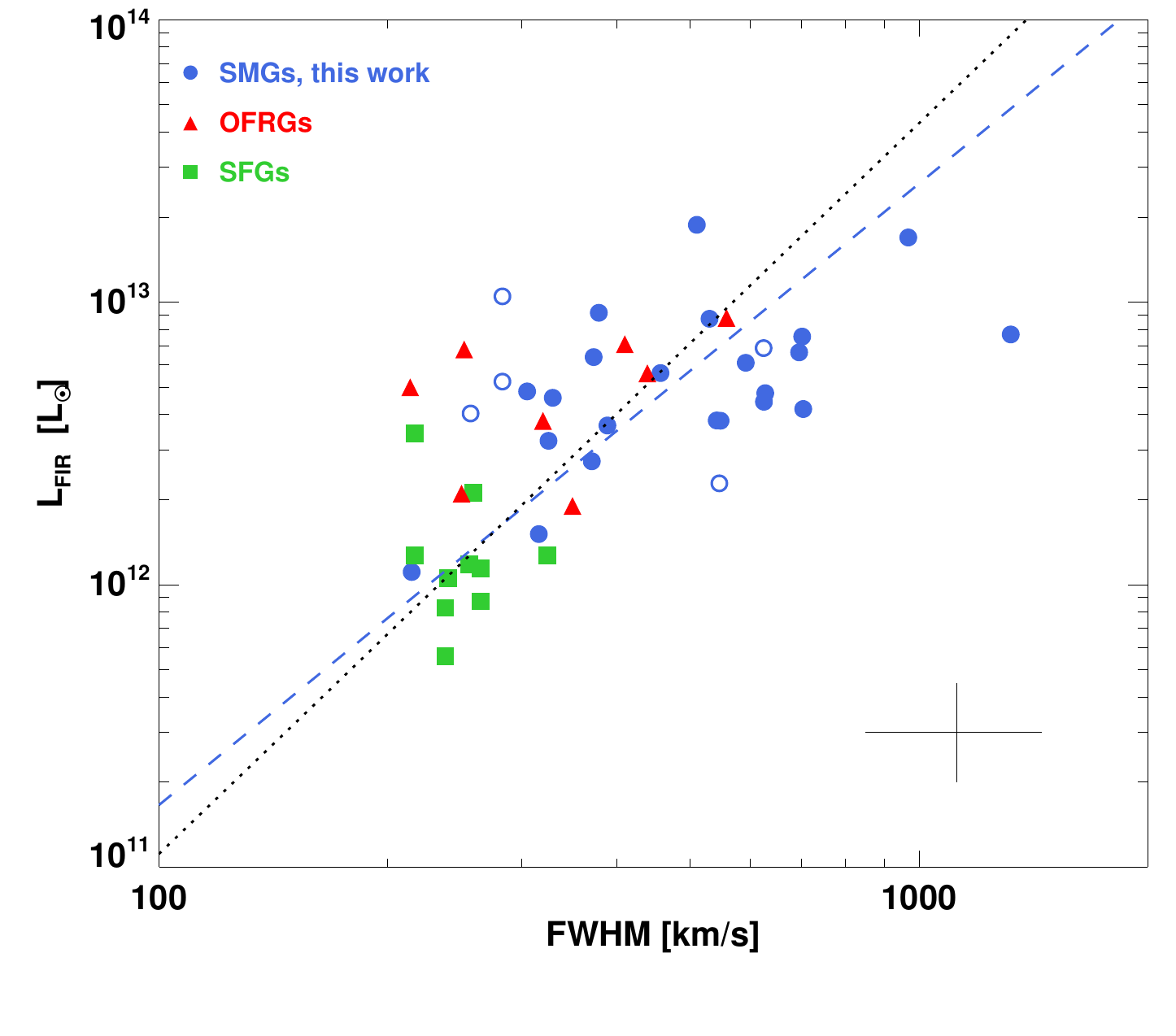}} \quad
  \subfigure{\includegraphics[width=8.2cm, clip=true, trim = 1.5cm 12cm 5.4cm 3cm]{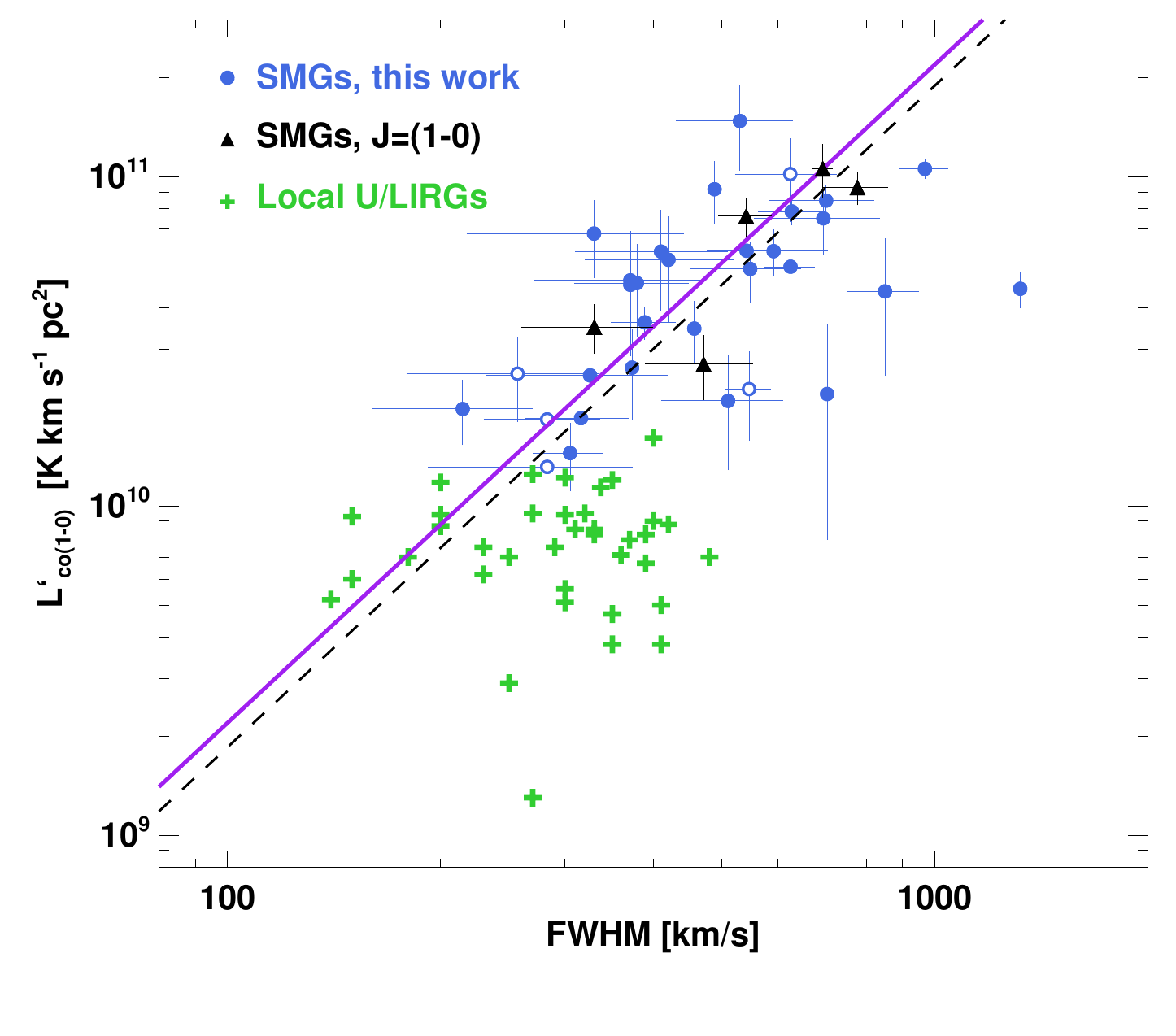}}   
}
\caption{\textit{Left panel:} $\mathrm{L}_{\mathrm{FIR}}$ plotted against FWHM for the sample of $^{12}$CO-detected SMGs. Also plotted are SFRGs from Chapman et al.\ (2008) and Casey et al.\ (2011), and $z\sim2$ SFGs from Tacconi et al.\ (2010). The SFGs have IR luminosities several times lower than SMGs and SFRGs with compatible linewidths. The blue dashed line shows a power-law fit to the SMGs alone (index = $2.2\pm0.2$), and the black dotted line shows a fit to all three samples of galaxies (index = $2.6\pm0.3$). The SMGs have broader typical linewidths than other populations, and have enhanced far-IR luminosities compared to SFGs (even at comparable linewidths). \textit{Right panel:} $\mathrm{L}'_{\mathrm{CO(1-0)}}$ plotted against FWHM for the same sample. Also included are values for local U/LIRGs from Downes \& Solomon (1998), and $J=$\,1--0 observations of SMGs from the literature,  but these are not included in the fits. The solid purple line shows the relation described in \S \ref{sec:kinematics}, while the dashed black line shows a simple fit to the points. The SMGs show a tight correlation between  $\mathrm{L}'_{\mathrm{CO(1-0)}}$ and FWHM. }
\label{fig:linewidth}
\end{figure*}

The mean FWHM of our sample (which includes the Greve et al.\ 2005 SMGs) is $510 \pm 80$\,km\,s$^{-1}$. This is somewhat lower than that found by \cite{2005MNRAS.359.1165G} and is most likely attributable to the bias towards IR-luminous sources in  \cite{2005MNRAS.359.1165G} -- the far-IR luminosities of their  sample are higher than our sample, by a factor of $\sim 2.5$ (they quote $<\mathrm{L}_{\mathrm{FIR}}>= (15\pm7)\times 10^{12}$\,L$_{\sun}$, whereas our sample has $<\! \mathrm{L}_{\mathrm{FIR}}\! >= (6.0\pm0.6)\times 10^{12}$\,L$_{\sun}$). As there is a strongly positive L$_{\mathrm{FIR}} $--$ \mathrm{M}_{\mathrm{dyn}}$ correlation (see Fig. \ref{fig:linewidth}), it is to be expected that the more IR-luminous SMGs presented in \cite{2005MNRAS.359.1165G} would have broader line widths.

The distribution of FWHM values for our sample is shown in Fig.~\ref{hist1}. The values derived from low-$J_{\rm up}$ (i.e.\  $J=$\,2--1) and higher-$J_{\rm up}$ (i.e.\  $J_\mathrm{\rm up}\geq 3$) lines have been highlighted separately. There is a slight bias towards lower line widths in the low-$J_{\rm up}$ subset: the $J = 2$--1 observations have a mean FWHM of $470 \pm 80$ km s$^{-1}$, compared to $550 \pm 90$ km s$^{-1}$ for the $J_{\rm up} \geq 3$ sample. This trend is the opposite of what might be expected (everything else being equal, the lower-$J_{\rm up}$ lines trace more extended gas than the higher-$J_{\rm up}$ lines, and therefore the linewidth should reflect a higher dynamical mass), and suggests a redshift-induced bias in the population, with low-$J_{\rm up}$ transitions being measured in lower-redshift, typically lower-luminosity SMGs.  

Figure \ref{fig:linewidth}, left panel, shows the FWHM of the observed emission line plotted against the (radio-derived) far-IR luminosity. Also shown on the plot are the ``submillimetre-faint radio galaxies'' (Chapman et al.\ 2008; \citealt{2011MNRAS.415.2723C}), warmer dust counterparts of SMGs, and $z\sim 2$ optically selected star-forming galaxies (Tacconi et al.\ 2010), less active star forming galaxies at these epochs. 

Of the three samples of galaxies shown in Fig.~\ref{fig:linewidth}, our SMGs clearly have the largest linewidths. There is a significant overlap between the SFRGs and the SMGs, suggesting that the ``hotter dust'' ULIRGs are kinematically similar to SMGs. The less-actively star-forming galaxies (SFGs), however, have only a slight overlap with the SMGs, having typical linewidths $<300$ km\,s$^{-1}$ -- lower than all but the most narrow-line SMGs. Interestingly, the SFGs have far-IR luminosities a factor of several lower than even those narrow-line SMGs/SFRGs with compatible $^{12}$CO FWHMs, suggesting that even SMGs which are kinematically comparable to optically-selected galaxies have enhanced star formation. Dissecting the mechanisms driving this strong emission in the narrow-line SMG population requires high resolution $^{12}$CO imaging, in order to spatially resolve the kinematics.

Having measured $^{12}$CO line luminosities and line widths for our sample, we now turn to the correlation between these two observables, which respectively trace the mass of the gas reservoir and the dynamics of the potential well in which it lies. Figure~\ref{fig:linewidth}, right panel, shows the FWHM of the $^{12}$CO emission line plotted against the derived \jone\ luminosity for our sample of SMGs. Also shown on the plot are the \jone\ observations for SMGs by Ivison et al.\ (2011). Included on the plot (but not included in any of the fits) are the $z\sim0$ U/LIRGs from Downes \& Solomon (1998). We fit a power-law fit to the SMG sample, deriving a power-law index of $1.98 \pm 0.07$.  The scatter around this power-law is low -- just  $\Delta \,\mathrm{L}'_{\mathrm{CO}} / \mathrm{L}'_{\mathrm{CO}} =0.38$.

As $\mathrm{L}'_{\mathrm{CO}}$ is sensitive to the total mass of molecular gas, while the FWHM is sensitive to both dynamical mass and any inclination effects,  some dispersion around the relation would be expected.   A population of thin, randomly orientated discs would introduce an inclination-based dispersion of $\sim 50\%$ -- the fact that our overall scatter is lower than this expected value strongly suggests that we are instead seeing gas in the form of thick discs (or turbulent ellipsoids), for which the line of sight velocity depends less strongly on inclination.

We also overlay a simple functional form for $\mathrm{L}'_{\mathrm{CO(1-0)}}$:
\begin{equation}
\mathrm{L}'_{\mathrm{CO(1-0)}} = \frac{C \, ({V/2.35})^{2} R} {1.36 \alpha G},
\end{equation}
where $V$ is the FWHM $^{12}$CO line width; $1.36 \alpha$ is the conversion factor from $^{12}$CO line luminosity to gas mass (we adopt $\alpha= 1$; see \S4.4 below), $C$ is a constant parameterising the kinematics of the galaxy (where we have taken $C = 2.1$ appropriate for a disc: see  \citealt{2006ApJ...646..107E}), $R$ is the radius of the $^{12}$CO $J=1$--0 emission region which, following Ivison et al.\ (2011) we have taken to be 7\,kpc (see \S\ref{sec:mdyn} below for discussion of the issue of $^{12}$CO sizes), and $G$ is the gravitational constant.  

This simple model provides a very good fit to the observed trend. Adopting $\mathrm{L}'_{\mathrm{CO(1-0)}} \propto V^{2}$ is clearly a  good match to luminous SMGs. The tight correlation seen in high-redshift ULIRGs is in stark contrast to $z\sim0$ U/LIRGs, which show very little  correlation between $V$ and $\mathrm{L}'_{\mathrm{CO(1-0)}}$.  We suggest that this is due to a combination of different geometry of the gas reservoirs; the thin nuclear gas discs and rings seen in local U/LIRGs mean the lack of any inclination correction produces significant scatter in our plot, combined with a  range in $f_{\rm gas}$ and a more significant stellar mass contribution to the dynamics of the potential well (there is little or no correlation between the stellar masses of the SMGs in our sample and their $^{12}$CO linewidths).  

The strength of the L$'_\mathrm{CO}$--FWHM correlation either indicates a very uniform ratio of gas-to-stellar contribution to the dynamics of the region probed by the $^{12}$CO emission, or that the gas mass dominates in this region (although that then requires the stellar mass  to be significantly extended beyond the gas radius probed with our observations). There is little evidence for this, however; \cite{Swinbank:2010ab} report {\it Hubble Space Telescope} NICMOS $H$-band radii of $2.8 \pm 0.4$ kpc for the SMGs, comparable to the extent of the high-$J_{\rm up}$ $^{12}$CO emission (\citealt{2010ApJ...724..233E}; \citealt{Bothwell:2010aa}).

\subsubsection{Double-peaked sources}
Looking in more detail at the spectra, some SMGs exhibit double-peaked $^{12}$CO spectra, a potential indication of kinematically distinct components within these systems. \cite{2005MNRAS.359.1165G} found that 4--6 of their 12 $^{12}$CO-detected SMGs displayed evidence of multiple kinematic components (as did Tacconi et al.\ 2006, 2008 and Engel et al.\ 2010), and also suggested these tended to be the lower-redshift systems. Our larger sample allows us to investigate these trends in more detail: We find that 7 of the 32 $^{12}$CO-detected SMGs in the full sample are clearly double-peaked, and a further 2 have spectra that are better-fit by a double than a single Gaussian profile (measured by the reduced $\chi^{2}$ of the fits). This is a lower fraction, 20--28\%, than previous findings. In addition we do not find any evidence that the double-peaked sources lie at systematically lower redshift than the sample as a whole, in contrast to the suggestion of \cite{2005MNRAS.359.1165G}. 

It must be noted that our observations can fail to detect kinematically-distinct components in two situations. Firstly, if the velocity separation between the components is too small they will appear as a single line. The typical peak separation of the double-peaked SMGs is $\sim$470\,km\,s$^{-1}$, comparable to the mean line width of the sample ($\sim 500$\,km\,s$^{-1}$) -- velocity separations much smaller than the line-width will be blended. Secondly, the  low spatial resolution of our maps cannot distinguish spatially-separated components with similar line-of-sight velocities; our source SMM\,J094303+4700, for example, has been observed with higher-resolution imaging, finding two distinct components (termed H6 and H7; Ledlow et al.\ 2002). This is also true of SMM\,J123707+6214 (Tacconi et al.\ 2008) and SMM\,J105307+5724 (Bothwell et al.\ 2010).

Thus our 20--28\% fraction of double-peaked profiles should be interpreted as a lower limit to the true rate in SMGs.  

Examining the physical properties of the double-peaked sources, we find that they do not seem to differ significantly from the population as a whole. The mean gas mass for double-peaked spectra is $(6.2 \pm 1.2) \times 10^{10}$ M$_{\sun}$, compared to $(5.3 \pm 1.0) \times 10^{10}$ M$_{\sun}$ for the complete sample. The mean far-IR luminosities for the respective classes differ by a comparable about, with the double-peaked SMGs having a mean L$_{\rm FIR} = (7.2 \pm 1.1) \times 10^{12} \; {\rm L}_{\sun}$, while the single-peaked galaxies have a mean L$_{\rm FIR} = (6.0 \pm 0.6) \times 10^{12} \;{\rm L}_{\sun}$. That is, double-peaked sources have molecular gas reservoirs and far-IR luminosities approximately 20\% times greater than those with single peaks, but given the small numbers of sources examined these results are certainly not significant. 

%The tendency of double-peaked sources to be more far-IR- and $^{12}$CO-luminous and gas rich than average suggests a higher merger fraction with increasing far-IR luminosity. Double-peaked emission profiles are a characteristic feature of gravitationally orbiting material, originating either from a rotating disc, or from a late-stage merger (where the velocity-separated components represent the remnants of the pre-merger galaxies). If this latter model is the better descriptor of bright, double-peaked SMGs (as high resolution observations suggest; \citealt{2010ApJ...724..233E}), then the trend towards a higher fraction of double-peaked sources with increasing far-IR luminosity is unsurprising. In this model, the brightest SMGs are best described by a major merger model, while the less IR-luminous SMGs could be better described as more homogeneous, disc-like systems\footnote{See Swinbank et al.\ 2011, for an analysis of a IR-faint, disc-like SMG.} (which could also be mergers, but at a post-inspiral phase).

\subsection{Dynamical Masses}
\label{sec:mdyn}

Our measurement of the kinematics of the $^{12}$CO emission from SMGs allows us to estimate masses for the galaxies themselves
from the width of the $^{12}$CO line, with an assumption about the dynamical structure and extent of the system. For example if the
$^{12}$CO emission arises in a virialised body of radius $R$, with one dimensional velocity dispersion $\sigma = s_{\nu}$, then the dynamical mass is given by:

\begin{equation}
\mathrm{M}_{\rm dyn}  \; [\mathrm{M}_{\sun}] = 1.56 \times 10^{6} \; \sigma^{2} \; R, 
\end{equation}

An alternative model for calculating dynamical masses is to assume that the line emission originates from a rotating disc. In this case, the dynamical mass is given by (i.e. \citealt{2003ApJ...597L.113N}):

\begin{equation}
\mathrm{M}_{\rm dyn} \sin^{2} i  \; [\mathrm{M}_{\sun}] = 4 \times 10^{4} \; V^{2} \; R, 
\end{equation}

where $V$ is the FWHM velocity, and $i$ is the inclination of the disc. For the purpose of calculating dynamical masses using this method, we adopt a mean inclination angle, appropriate for a population of randomly-orientated discs, of $< \sin i> = \pi/4 \simeq 0.79$ (see Appendix A of \citealt{2009ApJ...697.2057L}).

The choice of which estimator to use depends on the assumed form of the molecular ISM. If we assume that bright SMGs are predominantly major merger events (as argued by \citealt{2010ApJ...724..233E}) then the virial estimator may be the correct choice.  However, recently the dynamics of the $^{12}$CO emission from some SMGs (such as the lensed $z=2.3$ SMG, SMM\,J2135$-$0102, which has been mapped at 100-pc resolution; Swinbank et al.\ 2010, 2011), have revealed that the molecular gas appears to lie in a rotating disc. However, the intrinsic luminosity is at the fainter end of our sample (Ivison et al.\ 2010).  This is consistent with the suggestion made above that there are a mix of disc-like and virialised systems in the SMG population, with discs being more common at the lowest luminosities, while the brighter end of the SMG population has a higher frequency of merging systems. Of course, a rotating disc configuration does not preclude the conclusion that the galaxy is undergoing a merger, though it does suggest that any such merger is likely to be late-stage.

\nocite{2011ApJ...742...11S}

The final source of uncertainty in these calculations is the extent of the $^{12}$CO emission. While the compact-configuration PdBI observations in our programme are well-suited to a detection survey, they have the disadvantage of lacking the angular resolution necessary to resolve the emission in typical SMGs.  Higher-resolution observations from PdBI (typically in the higher-$J_{\rm up}$ transitions) have suggested that the warm gas reservoirs in SMGs have measured radii of 2--3 kpc (\citealt{2008ApJ...680..246T} \citealt{Bothwell:2010aa}; \citealt{2010ApJ...724..233E}). Ivison et al.\ (2011), using \jone\ observations of SMGs, suggested that the derived radius is also dependent on the transition observed; higher-$J_{\rm up}$ lines preferentially trace denser and more centrally concentrated star-forming gas, yielding smaller radii, than the more extended reservoirs traced by lower-$J_{\rm up}$ emission (which includes a more-extended component which may not be directly associated with vigorous star formation). 

Examining all SMGs observed in high-resolution $^{12}$CO (see the compilation presented by \citealt{2010ApJ...724..233E}), there seems to be something of a transition-based effect on the observed size of the $^{12}$CO reservoirs in SMGs; the galaxies observed in \jfour\ emission have a mean radius of $2.1 \pm 0.7$\,kpc, while those observed in \jthree\ have a mean radius of $3.1 \pm 1.4$ kpc (though these results should be taken as somewhat tentative, as sample sizes are small here). The redshift ranges covered by the respective samples are similar, suggesting that this is not an evolutionary effect. These high-$J_{\rm up}$ radii are smaller than the extents measured for \jone\ emission; Ivison et al.\ (2011) found typical radii of $\sim7$\,kpc for a smaller sample of five SMGs, observed with the JVLA.  Those authors also find some evidence that the FWHM of the \jone\ emission is $\sim 15$\% broader than the higher-$J_{\rm up}$ lines measured for the same SMGs, roughly consistent with the size variations.

With a likely variation in radius with transition, it could be argued that the correct radius to use for the dynamical calculation is the one appropriate for the transition used to derive the FWHM.  However, one of the reasons for deriving the dynamical masses is to compare them to the ``total'' gas and baryonic masses of these galaxies, we need to bear in mind that these ``total'' measurements are effectively measured in an aperture equal to the size of the \jone\ emission (as we are using galaxy-integrated  $r_{J, J-1/10}$ values). The JVLA results suggesting a larger $^{12}$CO radius are however, as yet, based on small sample sizes.

To avoid introducting more uncertainty due to the choice of assumed size (which our low-resolution survey is not optimised to study), we explicitly retain the size dependence in our estimates of the dynamical masses. (For ease of comparison with other studies, we shall also quote the final values assuming both ``extreme'' cases)

The median dynamical mass derived for our sample depends on the choice of mass estimator. Adopting the ``virial'' estimator given in Eq.~6, the median dynamical mass of the sample is $(7.1 \pm 1.0) \times 10^{10} \; R_{\rm kpc} \, {\rm M}_{\sun}$\footnote{The error here is the statistical error resulting from uncertainty on the FWHM measures.}. Removing the dependency on radius, this value corresponds to $(2.1 \pm 0.3) \times 10^{11} \; {\rm M}_{\sun}$ if we assume $R = 3$\,kpc, and $(5.0 \pm 0.7) \times 10^{11} \; {\rm M}_{\sun}$ if we assume $R = 7$\,kpc.

Alternately, adopting the ``rotational'' (i.e. disc-like) estimator, given in Eq. 7, results in a median dynamical mass of  $(1.6 \pm 0.3) \times 10^{10} \; R_{\rm kpc} \, {\rm M}_{\sun}$. Again removing the dependency on radius, this value corresponds to $(4.8 \pm 0.9) \times 10^{10} \; {\rm M}_{\sun}$ if we assume R = 3 kpc, and $(1.1 \pm 0.2) \times 10^{11} \; {\rm M}_{\sun}$ if we assume R = 7 kpc.

Each of these results are roughly in line with what might be expected from total halo masses; \cite{2012MNRAS.tmp.2285H} found, using a clustering analysis, that SMGs typically lie within Dark Matter haloes of masses M$_{\rm halo} = 9 \times 10^{12} \; {\rm M}_{\sun}.$

These dynamical masses are also reasonably consistent with recent estimates for SMGs derived using other indicators. Swinbank et al.\ (2004; 2006) used the kinematics of the spatially-resolved H$\alpha$ emission line to estimate the dynamical mass of a sample of bright SMGs, finding that a mass of $(5 \pm 3) \times 10^{11} \; {\rm M}_{\sun}$ represented the ensemble population well. The high dynamical masses found for SMGs are, in general far greater than those found in other, less extreme samples of galaxies: galaxies comprising the UV-selected SINS sample, for example have a typical dynamical mass of $8 \times 10^{10}$ M$_{\sun}$ (\citealt{2006ApJ...645.1062F}; \citealt{Shapiro:2009aa}). Due to uncertain model parameters (size, and choice of kinematic structure) inherent to a low-resolution study such as this, the range of possible dynamical masses is large. It is worth noting that a comparison to values derived from other studies favours the upper end of our estimates, suggesting that systems which are both compact and disc-like  perhaps do not comprise a significant proportion of the SMG population, and that extended and/or merging SMGs represent most of the SMGs in current studies. See \S\ref{sec:masscomp} for more discussion of this issue.

\subsection{The L$'_{\mathrm{CO}}$ -- L$_{\mathrm{FIR}}$ correlation}
\label{sec:sfe}

%
% Figure 9
%
\begin{figure}
\centering
\includegraphics[width=8.5cm, clip=true, trim = 1.6cm 13cm 5.5cm 3.2cm]{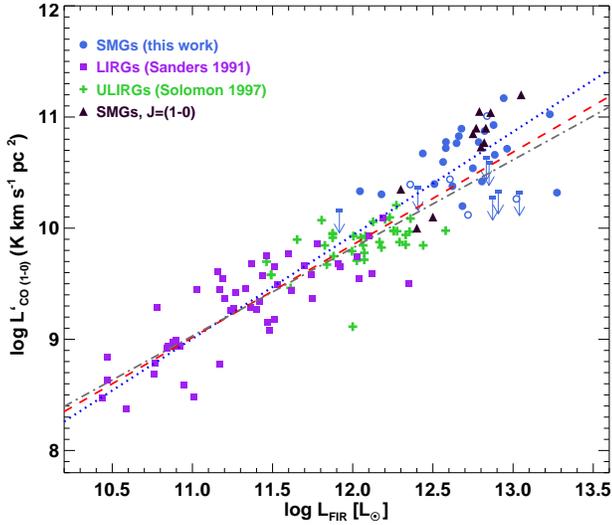}
\caption{The star formation efficiency, L$'_{\mathrm{CO}}$ vs. $\mathrm{L}_{\mathrm{FIR}}$. Included on the plot are two samples of local ULIRGs, and the J=($1-0$) SMG observations of Ivison et al.\ (2011). Best fitting slopes to the SMGs alone (blue dotted line), the local ULIRGs alone (grey dot-dashed line), and all three combined samples (red dashed line)  are overplotted. They have slopes of $0.93 \pm 0.14$, $0.79 \pm 0.08$, and $0.83 \pm 0.09$ respectively.}
\label{Lco_vs_Lir}
\end{figure}

A useful quantity to measure for our sample of SMGs is the efficiency with which their molecular gas is being converted into stars. The star formation efficiency (SFE) is sometimes defined as SFR/M(H$_{2}$) -- the inverse of the gas depletion time -- but here we take the approach of parameterising this as a ratio of observable quantities:  L$_{\mathrm{FIR}}$ and L$'_{\mathrm{CO(1-0)}}$.

There has been debate, in recent years, as to the value of the slope of the L$'_{\mathrm{CO(1-0)}} - $\,L$_{\mathrm{FIR}}$ relation. Whereas earlier we discussed the difference in slopes between the various transitions observed (in order to derive the median SLED), here we present the relation between the derived $^{12}$CO\,$J=$\,1--0 luminosity and L$_{\mathrm{FIR}}$ -- a relation which describes, in observable terms, the relationship between the luminosity due to star formation, and the {\it total} gas content.

In their study of 12 SMGs, \cite{2005MNRAS.359.1165G} found a slope of the relation between L$_{\mathrm{FIR}}$ and L$'_{\mathrm{CO(1-0)}}$ of $0.62 \pm 0.08$ fit a combined sample of lower-redshift LIRGs, ULIRGs, and SMGs -- identical to the slope derived for the local LIRGs/ULIRGs alone. The small number of galaxies in \cite{2005MNRAS.359.1165G}, however, prevented a full investigation of the SFE slopes within the SMG population itself. Some recent authors, however, have found the slope to be closer to linear --  \cite{Genzel:2010aa} found a slope of $0.87 \pm 0.09$ fit a combined sample of star-forming galaxies across a wide range of redshifts. 

Figure~\ref{Lco_vs_Lir} shows the L$'_{\mathrm{CO}} - \mathrm{L}_{\mathrm{FIR}}$ relation for our sample of SMGs. Included on the plot are datapoints for local (U)LIRGs, as measured by \cite{1991ApJ...370..158S} and \cite{1997ApJ...478..144S}. We also show three power-law fits; to the local (U)LIRGs alone, the SMGs alone, and the combined sample. We find the SMGs to lie slightly above the best fitting (sub-linear) line for local (U)LIRGs, necessitating a steeper slope. The power-law fit to the local (U)LIRGs alone has a slope of $0.79 \pm 0.08$, while the fit to the combined sample of SMGs and (U)LIRGs has a slope of $0.83 \pm 0.09$ -- very close to linear. It can also be seen that a fit to the SMG sample alone has an even steeper slope of $0.93 \pm 0.14$ -- although, within the uncertainty, this is consistent with the slope for the combined samples. These results are in good agreement with most previous findings; the near-linear slopes (which agree well with those found by Genzel et al.\ 2010) would imply a roughly constant gas depletion timescale across the entire range of far-IR luminosities shown here.

However, we caution that our analysis requires extrapolating from high-$J_{\rm up}$ $^{12}$CO transitions to \jone.  As we showed in Fig.~3, plotting L$'_{\rm CO}$--L$_{\rm FIR}$ for different transitions shows increasingly steep slopes with higher $J_{\rm up}$ and hence using a single brightness temperature ratio to correct each of these correlations to a ``common'' transition will introduce significant scatter and uncertainty into the resulting slope.  Indeed, Ivison et al.\ (2010) have undertaken  a similar analysis based solely on \jone\ observations and homogeneously-derived far-IR luminosities and conclude that the L$'_{\rm CO(1-0)}$--L$_{\rm FIR}$ has a slope substantially below unity (consistent with the trend of slope with $J_{\rm up}$ seen in Fig.~3).  We therefore suggest that it is difficult to draw any strong conclusions from high-$J_{\rm up}$  observations about the gas depletion timescales of the different populations or the form of the Kennicutt-Schmidt relation in these galaxies, as these are too uncertain without brightness temperature ratio measurements for individual sources.

\subsection{Molecular gas masses}
\label{sec:gas}

Part of the power of observations of $^{12}$CO emission from high-redshift galaxies is that they provide a powerful tool to derive the mass of the reservoir of molecular gas in these systems, which is mostly in the form of H$_2$. This is of critical importance because this reservoir is the raw material from which the future stellar mass in these systems is formed. Along with the existing stellar population, it therefore gives some indication of the potential stellar mass of the resulting galaxy at the end of the starburst phase (subject, of course, to the unknown contribution from in-falling and out-flowing material).

Estimating the mass of H$_2$ from the measured L$'_{\rm CO}$ requires two  steps. Firstly, luminosities originating from higher transitions ($J_{\rm up}\geq2$) must be transformed to an equivalent $^{12}$CO $J=$\,1--0 luminosity, using a brightness ratio. We have derived the necessary brightness ratios using our composite SLED as discussed in \S\ref{sec:ladder} above. Once a L$'_{\rm CO(1-0)}$ has been determined, it must be converted into a H$_2$ mass by adopting a conversion factor $\alpha$: $\mathrm{M (H}_2) = \alpha L'_{\rm CO}$, where $\alpha$ is in units of M$_{\sun}$ (K\,km\,s$^{-1}$\,pc$^{2})^{-1}$ (when discussing $\alpha$ hereafter, we omit these units for the sake of brevity).   This can then be converted to a total gas mass, including He, $\mathrm{M}_{\rm gas} = 1.36 \mathrm{M} (\mathrm{H}_2)$.

There is a large body of work, both observational and theoretical, dedicated to determining the value -- and ascertaining the metallicity or environmental dependence -- of $\alpha$ (e.g.\ \citealt{1991ARA&A..29..581Y}; \citealt{Solomon:2005aa}; \citealt{Liszt:2010aa}; \citealt{2011MNRAS.418..664N}; \citealt{Genzel2011aa}; \citealt{2012arXiv1202.1803P}).  While secular discs such as the Milky Way have a relatively ``high'' value of $\alpha \sim 3$--5, using this value for the gas in nuclear discs/rings within merging systems and starbursts at $z\sim 0$ leads to the molecular gas mass sometimes exceeding their dynamical masses. As such, a lower value -- motivated by a radiative transfer model of the $^{12}$CO kinematics -- is typically used for the intense nuclear starbursts in the most IR-luminous local systems:  $\alpha \sim 0.8$, with a range of 0.3--1.3 (Downes \& Solomon 1998).  However, some recent results have suggested that this value might, in fact, {\it underestimate} the true value in high-redshift SMGs. Bothwell et al.\ (2010) found that applying the canonical ULIRG value to two $z\sim 2$ SMGs resulted in gas fractions of $<10\%$, incongruous with their extreme star formation rates. Similarly, a dynamical analysis has been undertaken on the high-redshift SMG, SMM\,J2135$-$0102, by Swinbank et al.\ (2011) yielding a higher value, $\alpha\sim 2$ (supported by LVG modelling; Danielson et al.\ 2011).

Here we adopt a value of $\alpha = $\factor, and caution that all gas masses derived are dependent on this uncertain parameter. Using this value, the resulting mean H$_{2}$ mass of our sample SMGs (including limits from the non-detections) is $(5.3 \pm 1.0) \times 10^{10}$ M$_{\sun}$. This is comparable to the findings of \cite{2005MNRAS.359.1165G}, once the conversion factor and excitation model (detailed in \S\ref{sec:ladder}) is accounted for; adjusted for our model values, the SMGs presented in that work have a mean gas mass of $(6.5 \pm 2.2) \times 10^{10}$ M$_{\sun}$. 

\begin{figure}
\centering
\mbox
{
  \includegraphics[width=8.5cm, clip=true, trim = 1.5cm 12cm 5.5cm 3cm]{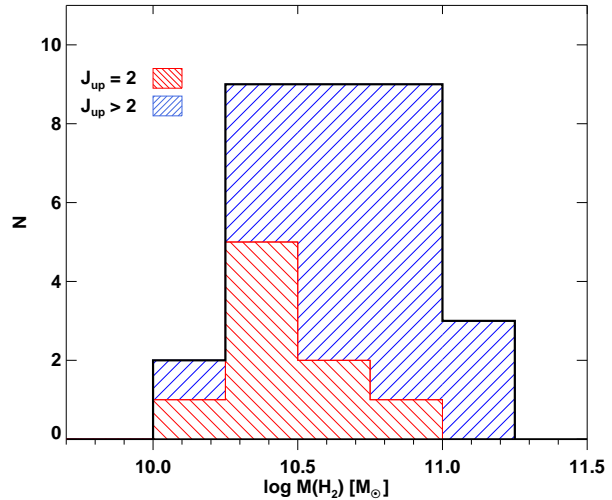} 
}
\caption{ Distribution of H$_{2}$ gas masses for the full sample. To calculate H$_{2}$ masses, a $^{12}$CO/H$_{2}$ conversion factor of $\alpha =$\factor\ has been assumed as detailed in \S\ref{sec:gas}. We identify both ``high-J$_{\rm up}$'' and ``low-J$_{\rm up}$'' derived masses; the offset between these most probably arises due to a redshift bias, whereas the ``low-J$_{\rm up}$'' sources are smaller galaxies lying at a lower mean redshift.}
\label{hist2}
\end{figure}

The distribution of H$_{2}$ masses for our sample is shown in  Fig.~\ref{hist2}. As previously, we have differentiated in the distribution between low-$J_{\rm up}$ ($J_\mathrm{\rm up}=2$) and higher-$J_{\rm up}$ ($J_\mathrm{\rm up}\geq 3$) lines. There is a notable tendency for the low-$J_{\rm up}$ lines to result in lower gas masses than the higher-$J_{\rm up}$ subsample. The mean gas mass for the $J_\mathrm{\rm up} = 2$ subsample is $(3.2 \pm 2.1) \times 10^{10}$ M$_{\sun}$, compared to a mean gas mass of $(6.2 \pm 0.2) \times 10^{10}$\,M$_{\sun}$ for the SMGs observed in higher $J_{\rm up}$ transitions. This factor of $\sim2$ difference is substantially more than could be attributed to random errors in the adopted brightness model, and it likely a result of the low-$J_{\rm up}$ observed SMGs lying at a lower mean redshift (and therefore having a lower mean far-IR luminosity; see Fig.~\ref{hist1}, where a similar effect is seen in the distribution of linewidths).  Splitting our SMG sample into low-redshift ($z < 2$) and high-redshift ($z > 2$) subsamples, we find H$_{2}$ masses of $(4.1 \pm 1.0) \times 10^{10}$ M$_{\sun}$ and $(6.1 \pm 0.2) \times 10^{10}$ M$_{\sun}$ respectively. 

The gas masses determined for our sample of SMGs are roughly comparible to those derived for other high-redshift star forming galaxies (though differences in $^{12}$CO/H$_2$ conversion factor make comparisons difficult). \cite{Tacconi:2010aa} find mean molecular hydrogen masses of $(7.9 \pm 2.9) \times 10^{10}$ M$_{\sun}$ and $(1.3 \pm 0.7) \times 10^{11}$ M$_{\sun}$ respectively for their low-redshift ($z\sim1.2$) and high-redshift ($z\sim 2.3$) samples of ``main sequence''  star-forming galaxies (selected from the AEGIS and BX/BM surveys).  These are similar to our estimates, but we note that the two samples were calculated with very different values of $\alpha$ -- \cite{Tacconi:2010aa} adopt a value of $\alpha=3.2$, more appropriate for secular disc galaxies and a factor of three higher than the $\alpha=1$ we use for our SMGs.   Using a similar choice of $\alpha$ for both samples would obviously lead to disagreements in their gas masses.

\subsection{The evolution of the gas fraction}
\label{sec:gasfrac}

%
% Figure 7
%
\begin{figure}
\centering
{\includegraphics[width=8.5cm, clip=true, trim = 1.8cm 13cm 5.5cm 3.2cm]{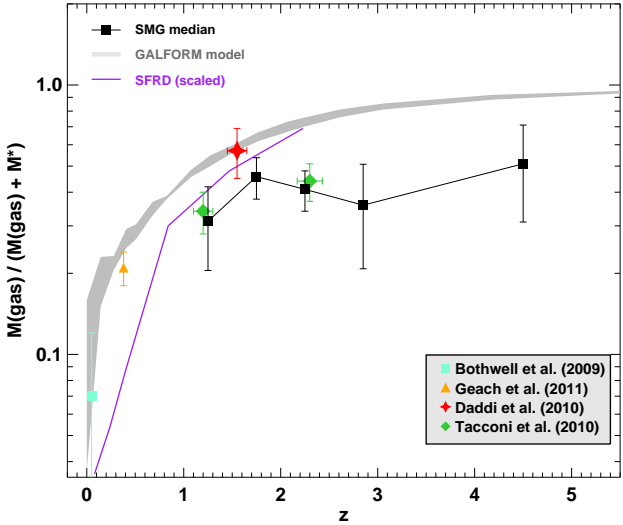}}
\caption{The redshift evolution of the baryonic gas fraction. The evolution of the mean baryonic gas fraction for our SMG sample, calculated in bins of width $\Delta z = 0.5$, is overlaid (due to the paucity of high-redshift SMGs the final bin is the mean for all $z > 3$ SMGs). We also plot star-forming galaxies from the literature as follows: at $z\sim 0$ (Bothwell et al.\ 2009); $z\sim 0.4$ \citep{2011ApJ...730L..19G}; $z\sim 1.5$ (Daddi et al.\ 2010); $z\sim 1.2$ and $z\sim 2.3$ (Tacconi et al.\ 2010).
We also plot, as a grey shaded area, the GALFORM predictions for the evolution of the baryonic gas fraction for LIRGs (L$_{\mathrm{FIR}} > 10^{11}$ L$_{\sun}$) with halo masses between $10^{12}$  and $10^{13}$\,M$_{\sun}$. There appears to be very little evolution in the mean SMG gas fraction over a wide interval of cosmic time. Furthermore, our SMGs exhibit similar gas fractions to lower luminosity, optically-selected populations at comparable redshifts. A schematic representation of the evolution of the star formation rate density (Sobral et al.\ 2012) is overlaid -- it can be seen that the behaviour of the SFRD broadly follows the global evolution of the gas fractions.}
\label{evo-fbar}
\end{figure}

If it is true that all star formation follows a universal scaling law dependent solely on  the available gas, then the peak in cosmic star-formation activity -- and subsequent decline -- is simply a reflection of the availability of molecular gas within galaxies. Recent results have strengthened the  expectation that normal, star-forming galaxies at high redshift are substantially more gas rich than their $z\sim0$ counterparts, by perhaps a factor of 3--10  (\citealt{Tacconi:2010aa}; \citealt{Daddi:2010aa}; Geach et al.\ 2011).  As pointed out by \cite{2011MNRAS.416.1354D}, the drop in gas fraction between high-redshift and the present day could be a straightforward result of the gas supply rate dropping faster than the gas consumption rate. Conversely, a gas fraction which rises with time represents a rapid accretion rate which swamps the consumption rate, leading to the accumulation of large gas reservoirs (thought to occur at $z \simgt 4$; \citealt{Papovich:2011aa}; \citealt{2011MNRAS.416.1354D}). The implication of this, of course, is that the enhanced star-formation rates seen in these normal star-forming galaxies at $z > 1$ do not necessarily represent a ``super-efficient'' mode of star formation, but could instead simply  reflect  the larger gas reservoirs available in the early Universe (potentially driven by correspondingly larger accretion rates from their surroundings).  

The gas content of galaxies is controlled by three main processes; gas accretion via inflow from the intergalactic medium (IGM), gas consumption (and associated outflows) driven by star formation (and perhaps AGN), and outflows. We can define the baryonic gas fraction, $f_\mathrm{gas} = \mathrm{M}_{\mathrm{gas}} / (\mathrm{M}_{\mathrm{gas}} + \mathrm{M}_{*})$, where M$_{\rm gas}$ includes Helium.  The gas fraction represents the product of these opposing forces, and the evolution of $f_\mathrm{gas}$ therefore encodes important information about their relative strength over time.
 
Figure~\ref{evo-fbar} shows the baryonic gas fraction  for our sample of SMGs. To extend the redshift range of the plot we include four SMGs at high redshift ($z>4$), as recently presented by \cite{Schinnerer:2008aa}, \cite{2009ApJ...695L.176D}, \cite{Coppin:2010aa}, and \cite{Riechers:2010aa}, in addition to GN20 in our sample at $z=4.05$. We see a substantial variation in $f_\mathrm{gas}$ across the population, with values ranging from 0.1--0.9. However, overall we find the SMGs to be very gas rich, as expected, with a median $f_{gas} =0.43\pm 0.05$.    The evolution of the mean SMG $f_\mathrm{gas}$ is overplotted on Fig.~\ref{evo-fbar}. We include on this plot the mean value for star-forming galaxies in the local Universe \citep{Bothwell:2009aa} as well measurements of UV-selected star-forming galaxies at $z\gs 1$ from \cite{Tacconi:2010aa} and  \cite{Daddi:2010aa}, and  mid-IR-selected $z\sim 0.4$ LIRGs from Geach et al.\ (2011).  Note that these samples of galaxies use differing values of $\alpha$ appropriate for their respective types; Tacconi et al.\ (2010) adopt $\alpha=3.2$, Daddi et al.\ (2010) use $\alpha=3.6$, while Geach et al.\ (2011) and Bothwell et al.\ (2009) use a local ``Galactic'' value of 4.6.

It is notable that while SMGs are clearly highly gas-rich systems: at every epoch in our sample molecular gas represents 40--60\% of their total baryonic content, they do \textit{not} appear to be more gas rich than ``normal'' star-forming galaxies at comparable redshifts, despite being selected to be a strongly star-forming population. Indeed, although they have comparable baryonic gas fractions, the Tacconi et al.\ (2010) galaxies have mean SFRs of  100\,M$_{\sun}$\,yr$^{-1}$ at $z\sim 1.2$, and 150\,M$_{\sun}$\,yr$^{-1}$ at $z\sim 2.3$, lower than the SFRs of our SMG sample at the same redshifts (Table~5) by factors of $\sim 4$ and $\sim 8$ respectively. The comparable gas fractions indicate that the SMGs are no less evolved -- in terms of their gas properties -- than less-active galaxies at similar epochs.

It is also interesting to note the redshift evolution of the baryonic gas fraction. The median gas fraction of the sample remains approximately constant across the redshift range of our sample, down to $z\sim1.5$. There is a steep decrease in the mean gas fraction of galaxies below $z\sim1$, however, with median gas fractions dropping to $\sim 20\%$ by $z\sim 0.4$ (Geach et al.\ 2011), and $<10\%$ for typical star-forming galaxies in the local Universe (Bothwell et al.\ 2009). This drop below $z\sim 1$ coincides with both the observationally observed drop in global star-formation rate density (e.g.\ \citealt{Hopkins:2006aa}; \citealt{Sobral:2011uq} -- shown on Fig. \ref{evo-fbar}), and the theoretically-predicted fall in specific accretion rates \citep{Krumholz:2011aa}. It seems that with globally falling accretion rates below $z\sim2$, the mean size of the gas reservoirs available to fuel starbursts decreases. 

In order to compare to simulations, we also include in Fig.~\ref{evo-fbar} the predictions from the semi-analytic galaxy formation model {\sc galform} (\citealt{2011MNRAS.418.1649L}; see also Swinbank et al.\ 2008). For this analysis, we applied a ``LIRG'' selection criteria, so that only galaxies with L$_{\mathrm{FIR}} > 10^{11}$\,L$_{\sun}$ were considered. In order to have a reasonable comparison to the SMG population, we also only considered galaxies inhabiting the most  massive haloes, with masses of 10$^{12}$--10$^{13}$\,M$_{\sun}$ \citep{2012MNRAS.tmp.2285H}. It is clear that the real SMGs are somewhat gas deficient compared to those in the semi-analytical model, which predicts very high median gas fractions of $>70\%$ above $z\sim 2$.  Swinbank et al.\ (2008) drew a similar conclusion from their comparison of {\sc galform}'s predictions for SMGs with observations:  that the model had too many of the baryons in $z\sim 2$ sub-mm-selected ULIRGs in the form of gas, compared to their mass of stars.  Thus it appears that the model needs to increase the efficiency of star-formation in massive galaxies at high redshifts to be match the observations.

Other cosmological models fare somewhat better when compared to our observational data. The cosmological hydrodynamic simulations of Dav\'e et al (2011) predict smaller gas fractions for massive galaxies at high redshift -- typically $\sim 30\%$ for a galaxy with a stellar mass of $10^{11}$\,M$_{\sun}$ at $z\sim2$  closer to the current observations (the precise figure depends on  input parameters, such as the choice of wind model). However, a major weakness of this comparison is that these model galaxies do not have a ``high-SFR'' criterion applied to them (to attempt to approximate SMGs) and so the relevance of this model comparison to the SMG population is not clear.

\subsection{A comparison of mass measurements}
\label{sec:masscomp}
%
% Figure 6
%
\begin{figure*}
\centering
{\includegraphics[width=17cm]{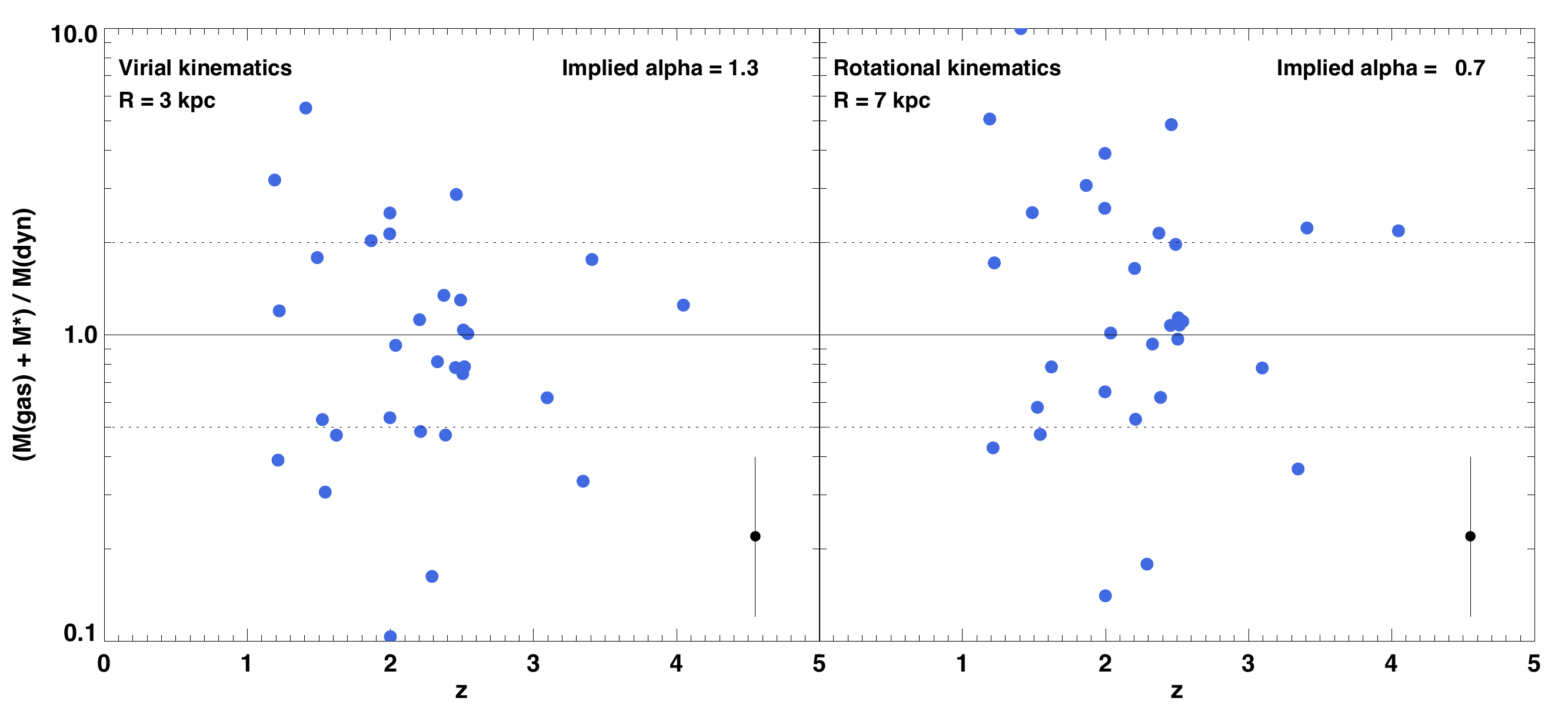}}
\caption{The ratio of total baryonic mass (the sum of the gas plus stellar masses) to dynamical mass, plotted against redshift, for both kinematical models of SMGs (rotational kinematics shown in the right panel, and virialised kinematics shown in the left panel). As the central regions of the SMGs -- traced by the molecular gas -- are expected to be baryon dominated we determine the value of the $^{12}$CO/H$_2$ conversion factor ($\alpha$) necessary to result of this ratio of unity (the values are reported in the respective panels). This figure also illustrate the scatter of the SMG population around this median ratio, which is larger than the 1-$\sigma$ error (shown in the bottom right), and may, in part, be driven by our assumption of a single model to describe to all SMGs. A median ratio of unity is indicated with a solid like, and dotted lines indicate a factor of 2 variation.}
\label{masscomp}
\end{figure*}

As explained above, there are several properties of the SMG population which a  detection survey such as ours is not ideally suited to study. These include the size of the $^{12}$CO reservoir, as well as the kinematic mode dominating the gas dynamics (which we present as either rotational or virialised). These quantities can be studied in tandem, however, in order to try to explore the range of physically-motivated parameters that describe the SMG population. As an example, it is unlikely that the SMG population as a whole is both rotation-dominated and compact\footnote{Which, as above, we take to mean a fiducial radius of 3 kpc; similarly, in this discussion we take ``extended'' to mean a fiducial radius of 7 kpc.}, as the total dynamical masses derived under these assumptions are typically lower than our adopted stellar masses (which provide only one component of the total gravitational potential). 

It is possible to explore the range of parameters which result in physically-plausible outcomes. The dynamical masses (derived above) and the total baryonic masses (our gas masses, plus stellar masses derived by \citealt{Hainline:2011lr}) of our SMGs effectively function as independent measurements of the total mass in the central regions of the galaxies (where the baryons are expected to dominate the kinematics; see \S \ref{sec:kinematics}). As a test of what constitutes a ``physically plausible'' set of parameters, we allow $\alpha$ to vary, calculating the value of $\alpha$ required in order to force agreement between the two independent measurements of the total mass. In other words, we assume

\begin{equation}
\frac{1.36 \alpha L'_{\rm CO} + M_{*} }{M_{\rm dyn} } = 1,
\end{equation}

(Where the factor 1.36 is to account for interstellar helium). Just as a combination of compact (3\,kpc) sizes and rotational kinematics results in a mean $M_{*} > M_{\rm dyn}$ (in our formalism above, implying a negative value of $\alpha$), a combination of extended (7\,kpc) sizes and virialised kinematics causes a similarly unphysical outcome. The dynamical masses derived in this circumstance are large enough to require a value of $\alpha = 4.9\pm0.8$ -- greater even than the value adopted for secular discs at $z\sim0$, and inappropriate for highly luminous starbursts, such as SMGs (see \citealt{Solomon:2005aa}). Alternately, these large dynamical masses imply a dominant dark matter component to the kinematics of the SMG, something which is again unlikely (see \S\ref{sec:kinematics}).

The two combinations that result in a physically motivated value of $\alpha$, compatible (to within a factor of 2) to values adopted typically adopted for ULIRGs/SMGs, are (1) compact sizes and virialised kinematics, and (2) extended sizes and rotational kinematics. Figure~\ref{masscomp} shows the ratio of the total baryonic mass to dynamical mass for both of these models of our $^{12}$CO-detected sample, plotted against redshift. 

There is a large scatter in each plot, indicative of the large uncertainties on each parameter (including the population averaged value of $\sin(i)$ adopted in the rotational estimator). However, the scatter in the population is, in each case, larger than the expected 1$\sigma$ error (indicated in the bottom right), implying that there is a significant  variation in the value of our derived mass ratio across the population. Given that the innate mass ratio is likely to be very close to unity (modulo a small dark matter contribution), this additional scatter, beyond the formal error, is likely a result of the globally-chosen parameters applying poorly to some individual SMGs -- an incorrect value of $\alpha$, for example, or an incorrect $^{12}$CO radius. This suggests that the models being discussed here almost certainly do not apply universally, across the entire SMG population -- even in an average sense. 

As discussed above, it seems likely that the SMG population is a heterogeneous mix of merger-like and disc-like galaxies, with the relative contribution of these to the total population varying as a function of luminosity. Considered within this framework, it could be argued that the SMG population contains both objects which are ``compact and virialised'', and ``extended and rotation-dominated'', with the former corresponding to the luminous ``merger'' systems, and the latter representing the more ``disc-like'' objects (such as the ``Eyelash'', SMM\,J2135$-$0102). Solidifying this picture, however, will require higher-resolution measurements of $^{12}$CO in a large number of SMGs, capable of resolving the kinematic structure and the extent of the $^{12}$CO reservoir. Using such measurements, it could also be possible to use this technique to place dynamical constraints on the value of the $^{12}$CO/H$_2$ factor $\alpha$ for the SMG population.

\section{The Evolution of SMBHs in SMGs}
\label{sec:agn}

The characterisation of super-massive black holes (SMBHs) in high-redshift SMG hosts has the potential to shed  light on the co-evolution of SMBHs and spheroids, an important facet of galaxy formation studies.  It is well known that there exists a correlation between the mass of a supermassive black hole (M$_{\mathrm{BH}}$), and the velocity dispersion ($\sigma$) of its host spheroid: M$_{\mathrm{BH}}$--$\sigma$  \citep{Magorrian:1998aa}, implying well-regulated feedback mechanisms in place which couple the two components. Indeed, the correlation is so good that the scatter is no more than would be expected from pure measurement error alone \citep{Ferrarese:2000aa}. This tight correlation suggests a symbiosis between galaxy formation and the formation of the central SMBHs, feedback from which is thought to play an integral part in the evolution of the most massive galaxies (e.g.\ \citealt{Benson:2003aa}).

\subsection{AGN-dominated SMGs}

As reported by \cite{Alexander:2005aa}, the  high AGN fraction seen in the SMG population indicates relatively continuous BH growth occurring throughout the intense star formation phase. This is in line with theoretical models of ULIRGs and SMGs (i.e. \citealt{Narayanan:2010aa}): major mergers efficiently transport gas into the dense central regions, which -- as well as fuelling a nuclear starburst -- can efficiently ``feed'' a BH.

While observational studies of SMGs have generally concluded that their prodigious bolometric luminosities are powered star-formation activity (e.g. \citealt{Frayer:1998aa};  \citealt{Frayer:1999aa}; Alexander et al.\ 2005b; but see \citealt{2011MNRAS.410..762H}), many studies have found that a significant minority of SMGs do host a luminous AGN (Alexander et al.\ 2005a, 2008; \citealt{Lutz:2010aa}; \citealt{Hainline:2011lr}; \citealt{Wardlow:2011aa}; Ivison et al. 2011). There are a number of methods that can be used to identify AGN activity, including X-ray emission \citep{Alexander:2003aa}, mid-IR colour selection \citep{Ivison:2004aa}, mid-IR spectral properties (\citealt{2007ApJ...660.1060V}; \citealt{2009ApJ...699..667M}), and optical spectral properties \citep{2004ApJ...617...64S}. \cite{Hainline:2011lr} analysed a sample of X-ray-observed SMGs, finding that the fractional contribution to the mid-IR from non-stellar (i.e.\ power-law) emission provides an excellent proxy for the hard X-ray luminosity and so provides a robust AGN diagnostic. 

We have compiled the AGN classifications for our sample: we find 13/40 SMGs in our sample have been observed to have AGN activity, based on one (or more) of the above diagnostics.  It must be noted, however, that the selection of the C05 sample could result in a bias towards SMGs containing AGN, as a result radio preselection, and/or the presence of strong emission lines in their restframe UV spectra (this is discussed in \citealt{Hainline:2011lr}; see also \citealt{Wardlow:2011aa}). If this is the case, then our sample -- being substantially drawn from the C05 sample -- could also be biased towards a higher percentage of AGN than the general SMG population.  
Nevertheless, we can still use our sample to investigate the correlation between the presence of an AGN and its luminosity, with other physical properties of the SMG hosts.

In fact, we find little or no correlation between the presence of an actively-fueled SMBH, shown by AGN activity, and the $^{12}$CO properties of the SMGs in our sample: gas masses, gas fractions, and FWHM-derived dynamical masses for the AGN-hosting-subsample are all consistent with being identical to the $^{12}$CO sample as a whole. This is in-line with the picture that SMGs are star-formation dominated, with activity from a central AGN playing little part in determining the properties of the system.  Indeed, it implies that the AGN activity occurs in all classes of SMGs, irrespective of dynamical or gas masses.  The lack of any difference in the gas fractions of the SMGs showing AGN activity and those without, may be indicating that any feedback effect from the AGN influences both the dust and gas reservoirs in these galaxies.  In this way those ``evolved'' systems where the AGN have had the most influence would no longer appear in our 850-$\mu$m-selected SMG sample due to an increase in dust temperature or a reduction in cold dust mass.

\subsection{The M$_{\mathrm{BH}}$--$\sigma$ relation}

%
% Figure 10
%
\begin{figure}
\centering
{\includegraphics[width=8.5cm, clip=true, trim = 1.5cm 13cm 5.5cm 3cm]{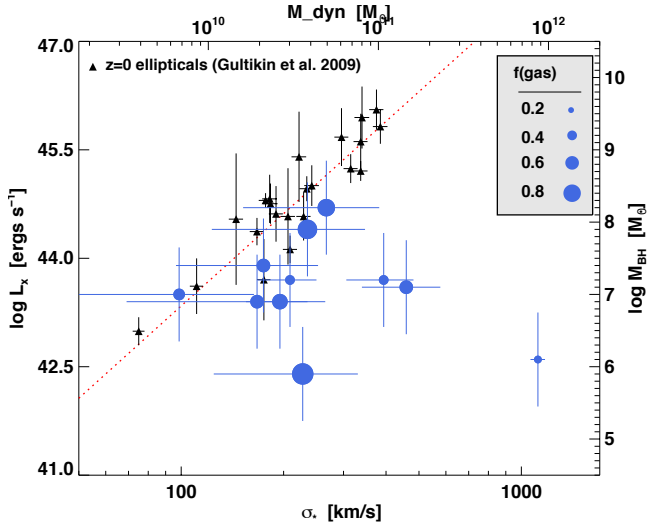}}
\caption{The M$_{\mathrm{BH}}$--$\sigma_*$ relation for the SMGs in our sample with measured X-ray luminosities, which have been converted into SMBH masses following Alexander et al.\ (2005). We also plot the same quantities for local elliptical galaxies, as presented by G\"{u}ltekin et al.\ (2009), and the best fitting relation from the same work (with a slope $\beta = 4.24$). The SMGs have SMBH masses almost an order of magnitude lower than their dynamical properties would suggest, suggesting that they represent a population yet to evolve on to the M$_{\mathrm{BH}}$--$\sigma$ relation. The size of the SMG points scales with the baryonic gas fraction, two examples of which are shown in the key. There is no correlation between gas fraction and position on the diagram, suggesting that the rapid SMBH ``feeding'' stage may occur after the starburst episode.}
\label{xray}
\end{figure}

SMGs have been proposed as the precursors of the massive ellipticals seen in the local Universe (e.g.\ Swinbank et al.\ 2006; Hickox et al.\ 2012). This population also exhibit the tightest M$_{\mathrm{BH}}$--$\sigma$ relation \citep{Gultekin:2009aa}, and as such examining their M$_{\mathrm{BH}}$--$\sigma$ connection during the time of their peak star-formation activity may be useful if we are to understand the role of feedback in the galaxy formation process. Unfortunately, one difficulty with addressing the evolution of the M$_{\mathrm{BH}}$--$\sigma$ relation in SMGs using direct kinematic measures is the fact that observations of the central stellar velocity dispersions at high redshift are not possible with the current generation of telescopes.  Similarly, use of optical and near-IR emission line gas as a tracer of host galaxy dynamics is complicated by the mixture of obscuration and potential outflow in these systems (e.g.\ Swinbank et al.\ 2006; Alexander et al.\ 2010).
Previous attempts to study the  M$_{\mathrm{BH}}$--$\sigma$ relation in SMGs have therefore been forced to use estimates of the stellar and gas masses as a proxy for $\sigma$ of the host galaxy (Borys et al.\ 2005; Alexander et al.\ 2005, 2008).  These studies uncovered an apparent offset between the relation for SMGs and that seen for spheroids at $z\sim 0$, in the sense that the SMBH masses were lower in the SMGs, relative to the host galaxy mass.  This behaviour is the reverse of that seen in studies of the SMBHs in QSOs \citep{2006ApJ...649..616P}, including those using $^{12}$CO dynamics \citep{2008MNRAS.389...45C}, suggesting that SMGs must grow their SMBHs to lie on the present day relation.

Our large sample of SMGs with $^{12}$CO observations (providing well-constrained individual baryonic masses and robust kinematic information) is also a useful tool for analysing the relationship of the mass of the SMBH to that of their galaxy hosts, and compare this to the present-day M$_{\mathrm{BH}}$--$\sigma$ relation to search for evolutionary changes.  We therefore start by estimating the SMBH masses for the SMGs  following \cite{Alexander:2005ab}. We derive SMBH masses from the X-ray luminosity integrated from 0.5--8\,keV \citep{Alexander:2003aa}, assuming that the X-ray emission from the AGN accounts for $6^{+12}_{-4}\%$ of its total bolometric luminosity \citep{1994ApJS...95....1E}. Next, an ``Eddington ratio'', $\eta$, can be adopted, which is the ratio of the AGN bolometric luminosity to the  Eddington luminosity. As the Eddington luminosity is simply a function of black hole mass, this allows us to estimate the mass of the central SMBH. Of course, the adoption of an assumed Eddington ratio is a critical step here, and a source of uncertainty. \cite{Alexander:2008aa} investigated a sample of SMGs taken from the parent sample of \cite{2005ApJ...622..772C}, and estimate a typical value for the Eddington ratio of $\eta = 0.2$--0.5\footnote{This range in estimations of $\eta$ is primarily driven by the unknown distribution of gas within the Broad Line Region (BLR) of the black hole.}. Here we adopt a value of $\eta = 0.4^{+0.1}_{-0.2}$, and include the uncertainty in $\eta$ in the resulting uncertainty on the derived SMBH masses. 

We then derive the equivalent velocity dispersion of the galaxy from our $^{12}$CO linewidths, using the prescription of  \cite{Ho:2007aa}:

\begin{equation}
\log_{10} \sigma = (1.26 \pm 0.05) \log\left( \frac{V_{20}}{2} \right) - (0.78 \pm 0.11), \\
\end{equation}
where $\sigma_{*}$ is the stellar velocity dispersion of the bulge, and $V_{20}$ is the velocity width at 20\% of the peak. This can be somewhat uncertain for lower luminosity galaxies (and, of course, varies for sources with non-Gaussian emission profiles), but for IR-bright galaxies (L$_{\rm IR} \sim 10^{12}$ L$_{\sun}$) such as our SMGs the relationship between $V_{20}$ and FWHM ($ = V_{50}$) approaches the analytic relation for a Gaussian profile, $V_{20} \simeq 1.5 \times {\rm FWHM}$ \citep{Ho:2007aa}.

Figure~\ref{xray} shows $\sigma$ versus L$_{\rm X}$ (and the resulting derived values of M$_{\mathrm{BH}}$ and M$_{\mathrm{dyn}}$) for our sample of SMGs, which have been coded according to their baryonic gas fraction. Included on the plot are local elliptical galaxies (thought to be the low-redshifts decedents of SMGs), as well as a fit to the local data (G\"{u}ltekin et al.\ 2009).  The SMGs in our sample lie below the local M$_{\mathrm{BH}}$--$\sigma$ relationship, with SMBH masses approximately an order of magnitude lower than would be predicted based on their dynamical properties.   This confirms the offset previously seen for SMGs using indirect estimates of their  masses (e.g.\ \citealt{2005ApJ...635..853B}; Alexander et al.\ 2008).

As discussed by  \cite{Alexander:2008aa}, if we wished to remove this offset from the local  M$_{\mathrm{BH}}$--$\sigma$  by adopting a lower Eddington Ratio, we would be forced to an extremely low value of $\eta \sim 0.05$. This would imply an unfeasibly slow SMBH growth rate during a phase where the necessary gas supply should be abundant. We conclude that SMGs represent a population which has yet to evolve onto the low-redshift M$_{\mathrm{BH}}$--$\sigma$ relation. While their high FWHMs (especially in the high-$J_{\rm up}$ lines, which trace the compact, central potentials) are indicative of massive galaxies, the central SMBHs are still relatively immature, having yet to grow substantially  due to rapid mass accretion.

An SMG's position in the M$_{\mathrm{BH}}$--$\sigma$ plane may evolve with age, as the central SMBH accretes mass and the gas is turned into stars. A low baryonic gas fraction could be interpreted as a sign that the SMG is a ``more evolved'' spheroid towards the end of the starburst episode, having already converted much of its gas reservoir into stars. As a result, if the SMBH growth occurs concomitantly with the starburst phase, we might expect gas fractions to correlate inversely with the ratio of SMBH to dynamical mass.  As we see in Figure~\ref{xray}, there is no apparent trend for lower $f_{\rm gas}$ SMGs to be closer to the present-day 
M$_{\mathrm{BH}}$--$\sigma$ relation, although the small number of SMGs and the significant scatter which may be present in $\sigma$ due to geometrical projection, weaken the strength of any conclusions.  Equally, the lack of any apparent correlation may suggest that the rapid BH ``feeding'' stage occurs separately, after the starburst episode which causes the galaxies to be sub-mm selected.   Indeed the timescales required to grow the SMBHs to the point that they resemble those of their $z \sim 0$ counterparts (which have M$_{\rm BH} \simgt 4 \times 10^{8} \; \mathrm{M}_{\sun}$; \citealt{Alexander:2008aa}) are signficantly longer than the expected sub-mm-bright lifetime (100--300 Myr; \citealt{Swinbank:2006aa}; Hickox et al.\ 2012), again indicating the need for a subsequent phase of predominantly SMBH-growth, most likely associated with strong AGN activity. Interferometric observations of far-IR-bright QSOs could be a way to approach this phase from the ``other side'', potentially shedding light on this interesting problem (Simpson et al.\ in prep).

\section{Conclusions}
\label{sec:conc}

We have presented results from a large PdBI survey for molecular gas in luminous SMGs. We observed 40 SMGs with well-defined redshifts, detecting $^{12}$CO emission in 32. For the remaining eight SMGs without detected $^{12}$CO emission, we constrain the upper limits on their $^{12}$CO flux.  We used gas masses and dynamical masses from our sample, combined with ancillary multi-wavelength data (including far-IR luminosities and stellar masses), to discuss the physical properties of the SMG population. Our main conclusions are as follows:

\begin{itemize}

\item{Analysing the median $^{12}$CO spectral line energy distribution for our SMG sample, after normalisation to a mean far-IR luminosity, we find the SLED rises up to $J_{\rm up} \sim 5$. Data are sparse at $J_{\rm up} > 5$, but within the errors we find evidence for a turnover. } \\

\item{Plotting $^{12}$CO luminosity against linewidth, we find that $\mathrm{L}'_{\mathrm{CO(1-0)}} \propto \sigma^{2}$ is a good fit to the trend seen in our sample. The scatter around this relation is lower than would be expected for a population of randomly orientated discs. This  suggests that the ISM in luminous SMGs typically exists as a thick disc or turbulent ellipsoid. }\\

\item{The $^{12}$CO linewidths are, in general, broad, having a mean  FWHM of $510 \pm 80$\,km\,s$^{-1}$.  Using these
line widths we derive dynamical masses and find mean values -- for virial and rotational mass estimators -- of $(7.1 \pm 1.0) \times 10^{10} \; R_{\rm kpc} \, {\rm M}_{\sun}$ and $(1.6 \pm 0.3) \times 10^{10} \; R_{\rm kpc} \, {\rm M}_{\sun}$ respectively. In these calculations we have kept the dependence on the size of the $^{12}$CO reservoir, which is not well constrained for the majority of our sources. We also find that 20--28\% of the sample exhibit double-peaked $^{12}$CO profiles, which we interpret as a signature of an on-going merger.} \\

%\item{We find an approximately linear slope to the L$'_{\mathrm{CO(1-0)}} $--$ \mathrm{L}_{\mathrm{FIR}}$ relation for our SMG sample, steeper than the fit for local ULIRGs. This is in line with some recent findings, and implies a relatively constant star formation efficiency across the population. }\\

\item{We use far-infrared luminosities to assess the star-formation efficiency in our SMGs, finding a steepening of the L$'_{\rm CO}$--L$_{\rm FIR}$ relation as a function of increasing $^{12}$CO $J_{\rm up}$ transition. While we find an approximately linear slope to the L$'_{\mathrm{CO(1-0)}} $--$ \mathrm{L}_{\mathrm{FIR}}$ relation, without brightness temperature ratio measurements for individual sources this result is uncertain, and we suggest that it is difficult to draw any strong conclusions about the gas depletion timescales of the different populations from high-$J_{\rm up}$  observations.} \\

\item{We see little evidence of  evolution in the baryonic gas fraction in SMGs with redshift. The median value of our sample 40--60\%  at all redshifts. By comparison to recently derived gas fraction for other high-redshift populations, we conclude that SMGs do not have significantly higher gas fractions than more modestly star-forming galaxies at similar redshifts.} \\

\item{We compile a variety of archival multi-wavelength data in order to analyse the AGN contribution to the SMG population as a whole. We find that 30\% of our sample have some indication of AGN activity, but the presence of an AGN does not seem to have a significant effect on the gas properties of the host SMG. We also make use of deep X-ray data to estimate SMBH masses for our SMGs. We find that the SMBHs in SMGs lie substantially below the $z\sim0$ M$_{\mathrm{BH}}$--$\sigma$ relation, and that there is no correlation between SMG  gas fraction and SMBH mass. We conclude that the SMBH growth phase occurs separately from, and after, the ``star formation'' phase, which our sample of SMGs is undergoing.} \\

\end{itemize}

\section*{Acknowledgments}
This study is based on observations made with the IRAM Plateau de Bure Interferometer. IRAM is supported by INSU/CNRS (France), MPG (Germany) and IGN (Spain). We acknowledge the use of \textsc{gildas} software (http://www.iram.fr/IRAMFR/GILDAS). We are grateful to the Great Observatories Origins Deep Survey (GOODS) team for use of their ACS data. MSB and IRS acknowledges the support of STFC and IRS also acknowledges support from a Leverhulme Senior Fellowship.

\nocite{Karim:2011aa}
\nocite{2002ApJ...577L..79L}
\nocite{2012arXiv1202.3436S}

\nocite{2009MNRAS.400..670C}
\nocite{2009ApJ...691..560C}

\begin{table*}
\label{tab:obsparam}
\begin{minipage}{140mm}
\centering
%\begin{center}
\small
\caption{Observational parameters of the programme SMGs. (Full table available from authors upon request).}
\begin{tabular}{@{}lccccccc@{}}
\hline
\hline

ID &
Transition & 
CO position &
$z$ &
I$_{\rm CO}$ &
FWHM&
S$_{850\mu m}$\footnote{From C05 (and references therein), unless otherwise stated.}   &
S$_{1.4 {\rm GHz}}^{a}$  \\

 &
 &
 (J2000) &
 &
 (Jy km s$^{-1}$) &
 (km s$^{-1}$) &
 (mJy) &
 ($\mu$Jy) \\

\hline

\textit{Detections} \\
SMM\,J105141+571952     & (2--1) &    	10 51 41.31  +57 19 52.0    &  1.2138 &           $2.50 \pm 0.19$   & $625 \pm 50  $ &	4.6	&	295	\\ 
\noalign{\smallskip}
\textit{Non-Detections}\footnote{CO upper limits derived as detailed in \S4.3}$^,$\footnote{Redshifts for non-detections are quoted at the centre of the receiver bandwidth.}
 \\ 
SMM\,J105230+572209    & (3--2) & --  &    2.6011 &     $ <$\,0.23 &  -- & 11.0 &  86 \\ 
\hline
\hline
\end{tabular}
\end{minipage}
%\end{center}
\end{table*}

\begin{table}
\centering
\begin{minipage}{140mm}
\small
\caption{Physical properties of the SMGs. (Full table available from authors upon request).}
\begin{tabular}{@{}lcccccccc@{}}
\hline
\hline

ID & L$'_{\mathrm{CO}}$ &  L$'_{\mathrm{CO}(1-0)}$\footnote{Converted using an excitation model as detailed in \S3} & $\log_{10}$ M(H$_{2}$)\footnote{Assuming $\alpha$ = 1.0 M$_{\sun}$ (K\,km\,s$^{-1}$\,pc$^{2})^{-1}$}  & $\log_{10}$ M$_{\rm dyn}$  & $\log_{10}$ M$_{\rm dyn}$  & $\log_{10}$ M$_{\ast}$ \footnote{From Hainline et al. (2011)} & $\log_{10}$ L$_{\rm FIR}$    & AGN?\footnote{Letters represent AGN activity diagnosed using the following estimators:\\
 A = X-ray (Alexander et al.\ 2005) \\ B = H$\alpha$  (Swinbank et al.\ 2004; Men\'{e}ndez-Delmestre et al.\ 2009) \\ C = Power-law NIR component fractions (Hainline et al.\ 2011) \\ D = IRS spectral properties (Men\'{e}ndez-Delmestre et al.\ 2009) \\ E = Colour selection (Ivison et al.\ 2004; Hainline et al.\ 2009)}
   \\
      &  (10$^{10}$   &  (10$^{10}$     &      & (rotational) & (virial) &  & & \\
      &   K\,km\,s$^{-1}$\,pc$^{2}$)  & K\,km\,s$^{-1}$\,pc$^{2}$)  &  (M$_{\sun}$) & ($R_{\rm kpc}$ M$_{\sun}$)    &  ($R_{\rm kpc}$ M$_{\sun}$)   & (M$_{\sun}$) & (L$_{\sun}$)\\                     
\hline

\textit{Detections} \\ 
SMM\,J105141+571952  &  $4.8\pm 0.4$ &  $5.8  \pm 0.5$      &  $ 10.76 \pm 0.03$ &     $  10.39 \pm 0.04$ &   $11.04  \pm0.04$ & ...         &  12.64     & ...\\             
\noalign{\smallskip}
\textit{Non-Detections} \\
SMM\,J105230+572209  & $<$\,0.8 &  $<$\,1.5 &  $<$\,10.17&  ... &  ... & 11.34 &  12.90  & ... \\ 
\hline
\hline
\end{tabular}
\end{minipage}
\end{table}

\bibliography{/Users/Matt/Documents/mybib}{}
\bibliographystyle{mn2e}

%\section{APPENDIX: Maps and spectra for sample}

\appendix

\section{Sample maps and spectra}

%%%%%%%%%%%%%%%%%%%%%%%%%%%%%%%%

\begin{figure*}
\centering
\mbox
{
  \subfigure{\includegraphics[width=8cm, clip=true, trim=50 350 70 0]{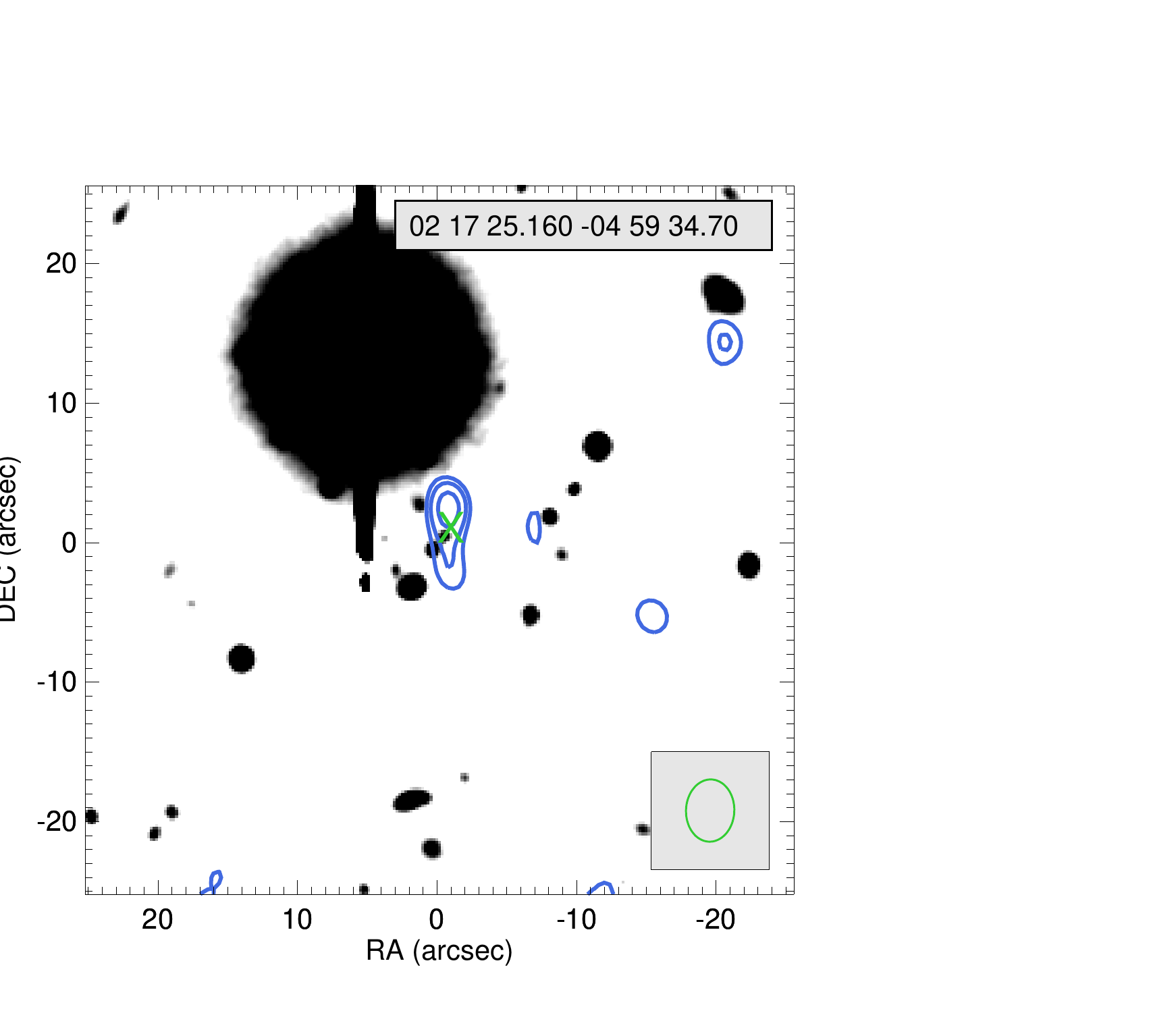}} \hspace{-2cm}
  \subfigure{\includegraphics[width=10cm, clip=true, trim=30 300 70 0]{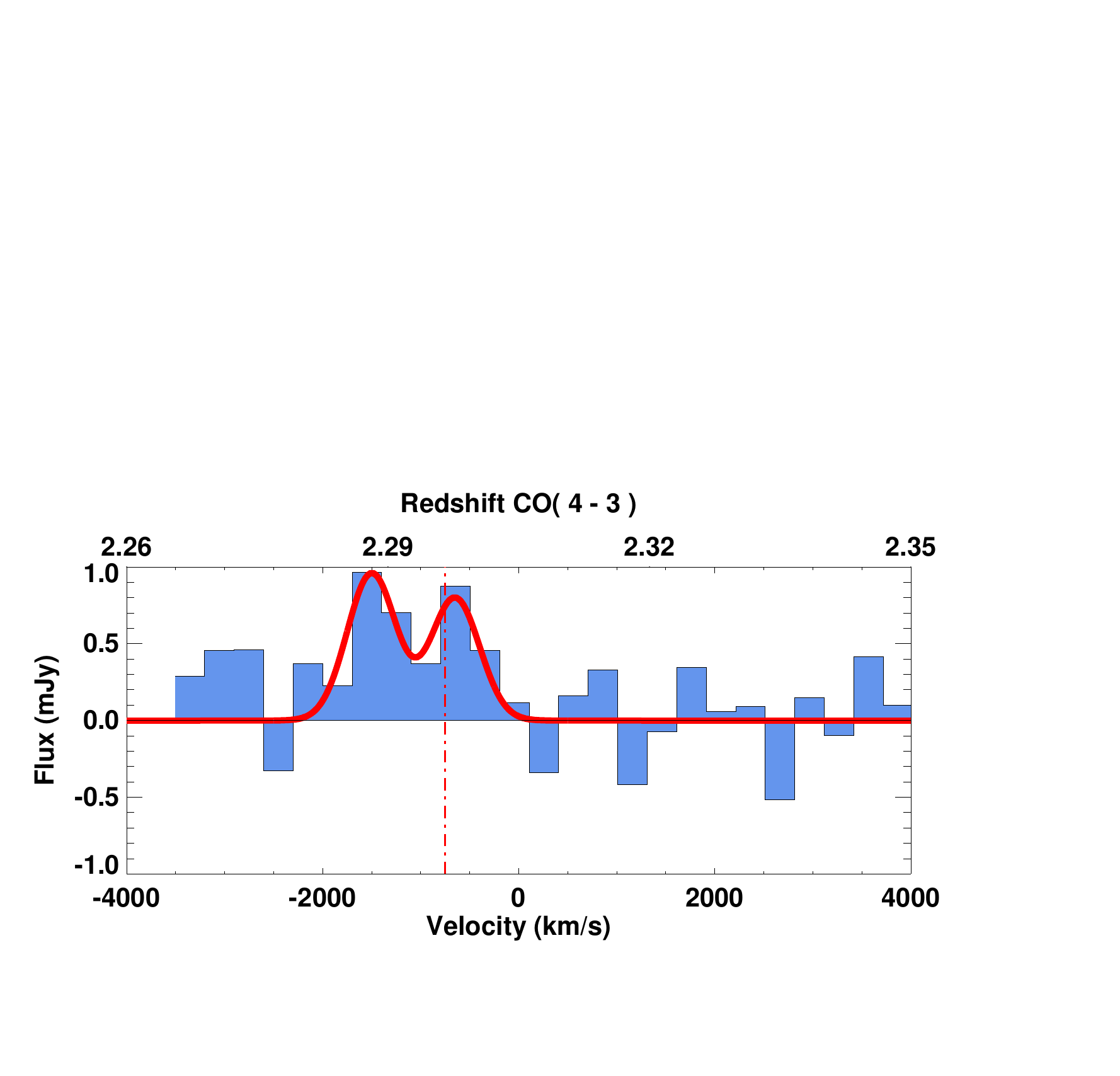}}
}
\caption[SMM021725-0459]{SMM021725-0459. 
Note: All maps and spectra in the appendices are displayed identically to this, unless otherwise stated in the respective captions. 
\textit{Left panel}: Multi-wavelength source map. Blue contours: integrated $^{12}$CO image in 1$\sigma$ steps, starting at 2$\sigma$. Red contours: 24$\mu$m (where available). The background is a stacked image of all 4 Spitzer IRAC bands (where available; otherwise we have used other Near-IR imaging;  $K$-band, is shown here). The FWHM of the synthesised beam is shown in the corner, and the green cross indicates the centre of the radio emission. \textit{Right panel}: $^{12}$CO spectrum, taken at the point of maximum $^{12}$CO flux, with the best fitting Gaussian profile overlaid in red. In the cases where a double-peaked profile provided a superior fit (as described in the text), the best fitting double Gaussian profile is shown. Vertical lines indicate previously obtained redshifts, derived as from the following wavelengths: black dashed line = from UV; red dot-dashed line = from H$\alpha$ or other nebular emission line; green solid line = PAH. For the sources not detected in CO, the spectrum has been taken at phase centre. The SMM021725-0459 image has been stretched to account for the bright nearby star. $^{12}$CO is detected here at $4.7 \sigma$.  The 850$\mu$m source,  SMM021725-0459 (Coppin et al.\ 2006), was followed up spectroscopically (Alaghband-Zadeh et al.\ 2012), detecting H$\alpha$ as indicated, and the CO(4-3) and [CI](1-0) spectra are published in (S.\ Alaghband-Zadeh et al.\ in prep).
}
\label{figure_sxdf11}
\end{figure*}

%%%%%%%%%%%%%%%%%%%%%%%%%%%%%%%%

\begin{figure*}
\centering
\mbox
{
  \subfigure{\includegraphics[width=8cm, clip=true, trim=50 350 70 0]{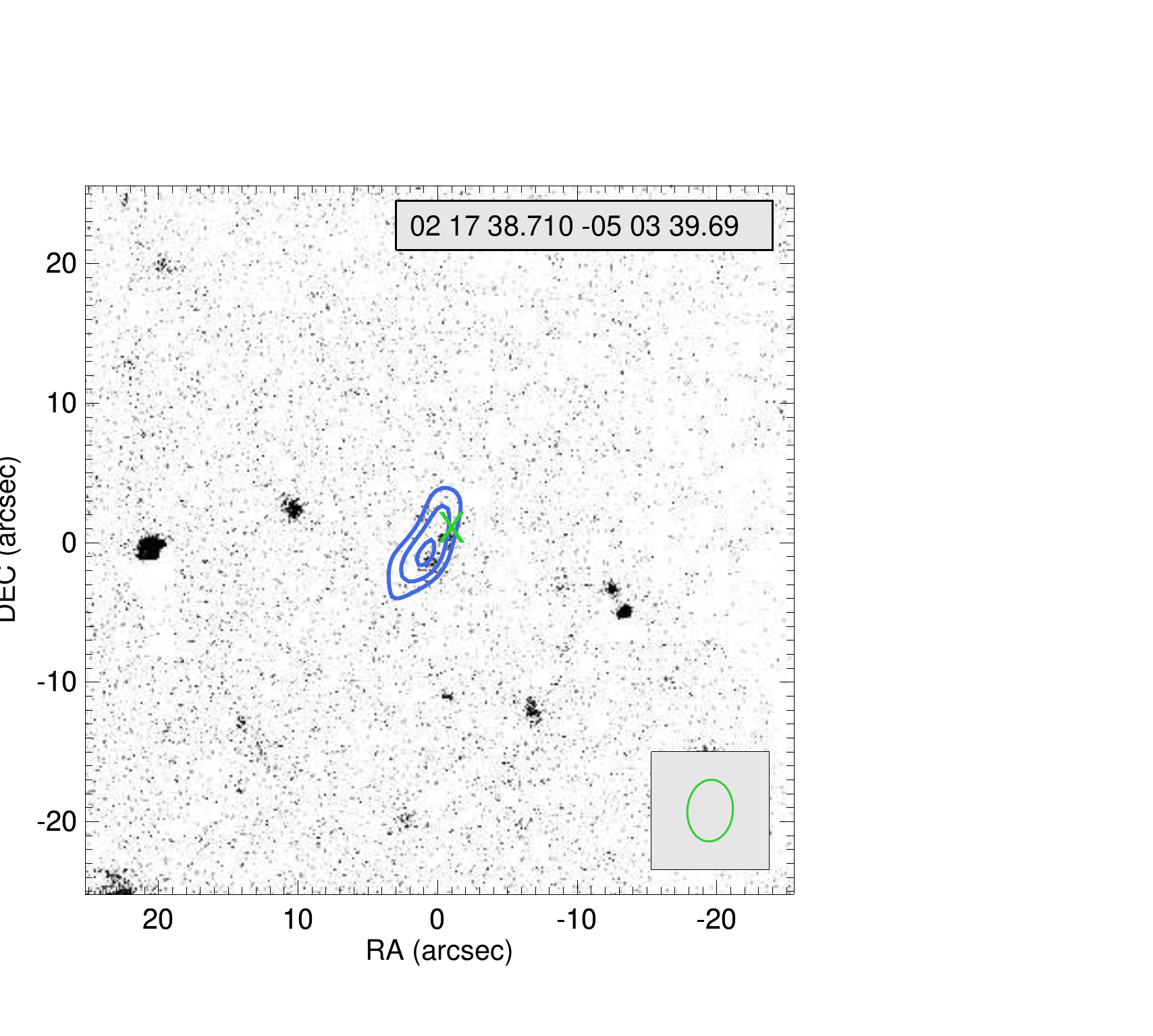}} \hspace{-2cm}
  \subfigure{\includegraphics[width=10cm, clip=true, trim=30 300 70 0]{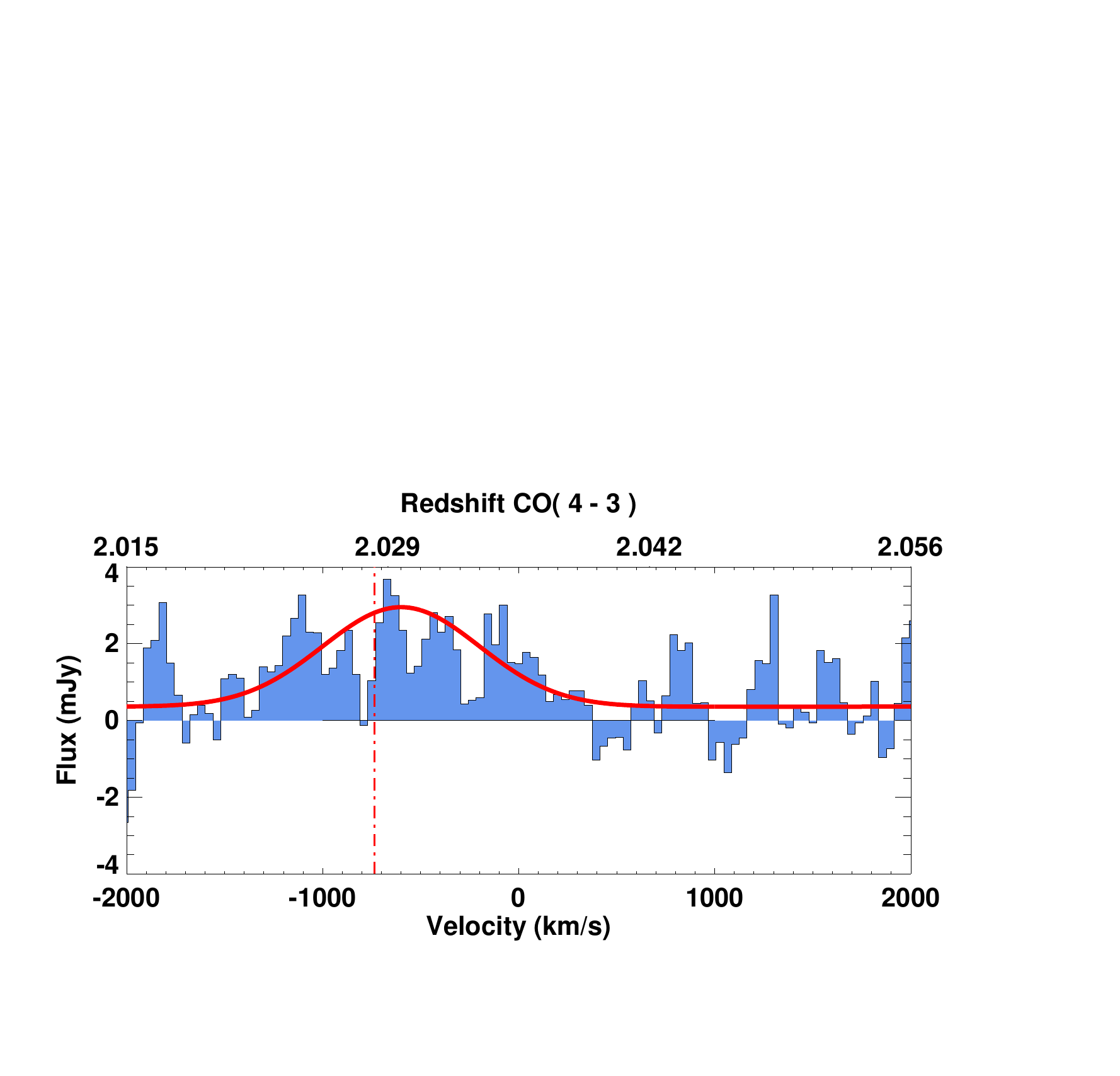}}
}
\caption[SMM021738-0503]{SMM021738-0503. $^{12}$CO is detected at 5.3~$\sigma$. The 850$\mu$m source,  SMM021738-0503 (Coppin et al.\ 2006), was followed up spectroscopically (Alaghband-Zadeh et al.\ 2012)
with H$\alpha$ derived  redshifts for both $K$-band sources lying under the CO contours. The redshift of the northern $K$-band source is indicated on the CO spectrum, the southern source lying  +700~km/s offset. 
The CO(4-3) and [CI](1-0) spectra are published in (S.\ Alaghband-Zadeh et al.\ in prep).
}
\label{figure_sxdf4}
\end{figure*}

%%%%%%%%%%%%%%%%%%%%%%%%%%%%%%%%

\begin{figure*}
\centering
\mbox
{
  \subfigure{\includegraphics[width=8cm, clip=true, trim=50 350 70 0]{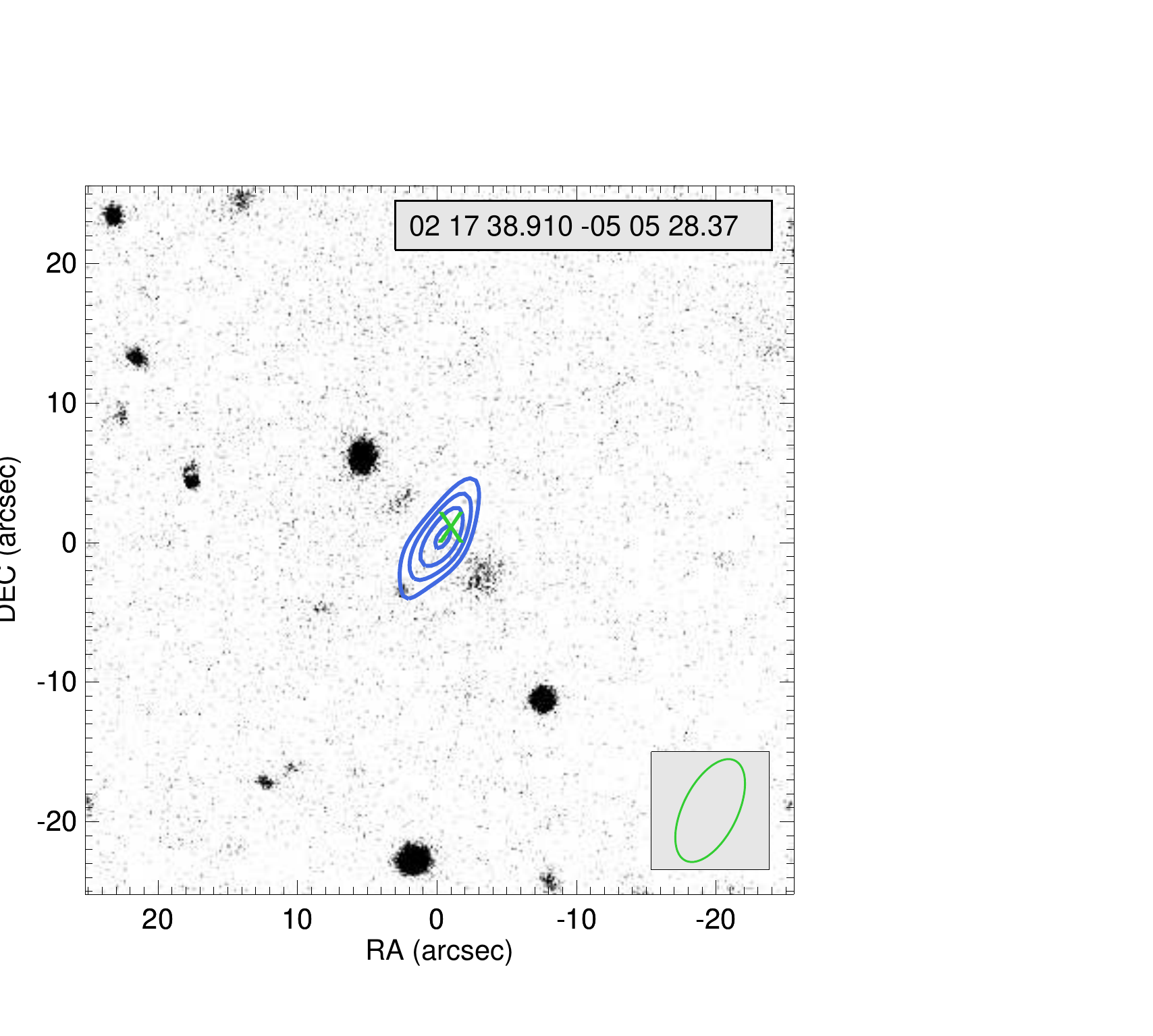}} \hspace{-2cm}
  \subfigure{\includegraphics[width=10cm, clip=true, trim=30 300 70 0]{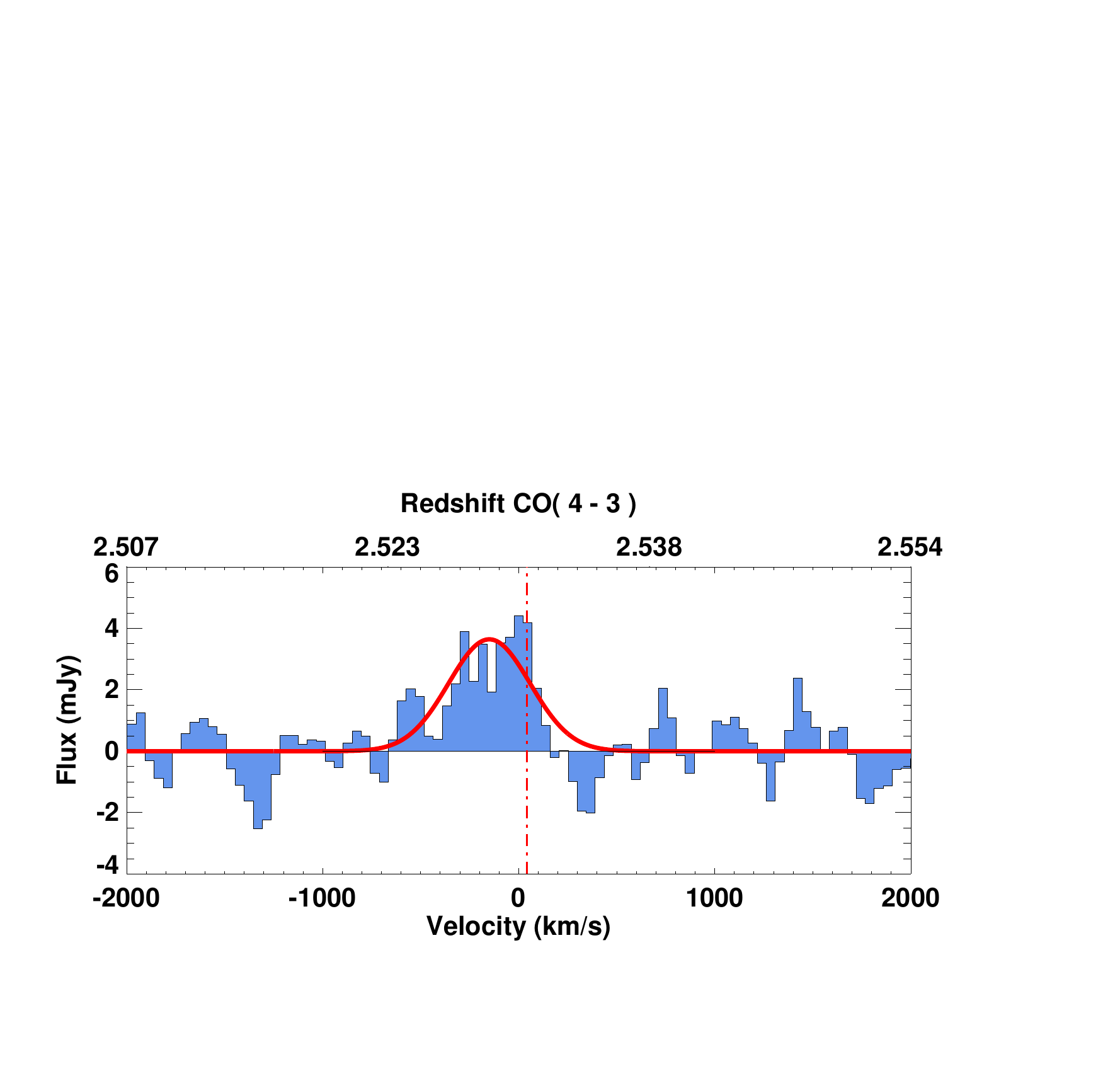}}
}
\caption[SMM021738-0505]{SMM021738-0505. $^{12}$CO is detected at 6.2 $\sigma$.  
The 850$\mu$m source,  SMM021738-0505 (Coppin et al.\ 2006), was followed up spectroscopically (Alaghband-Zadeh et al.\ 2012), detecting H$\alpha$ as indicated, and the CO(4-3) and [CI](1-0) spectra are published in (S.\ Alaghband-Zadeh et al.\ in prep).
}
\label{figure_sxdf7}
\end{figure*}

%%%%%%%%%%%%%%%%%%%%%%%%%%%%%%%%%%%%%%%%%%%%%%%%

\begin{figure*}
\centering
\mbox
{
  \subfigure{\includegraphics[width=8cm, clip=true, trim=50 350 70 0]{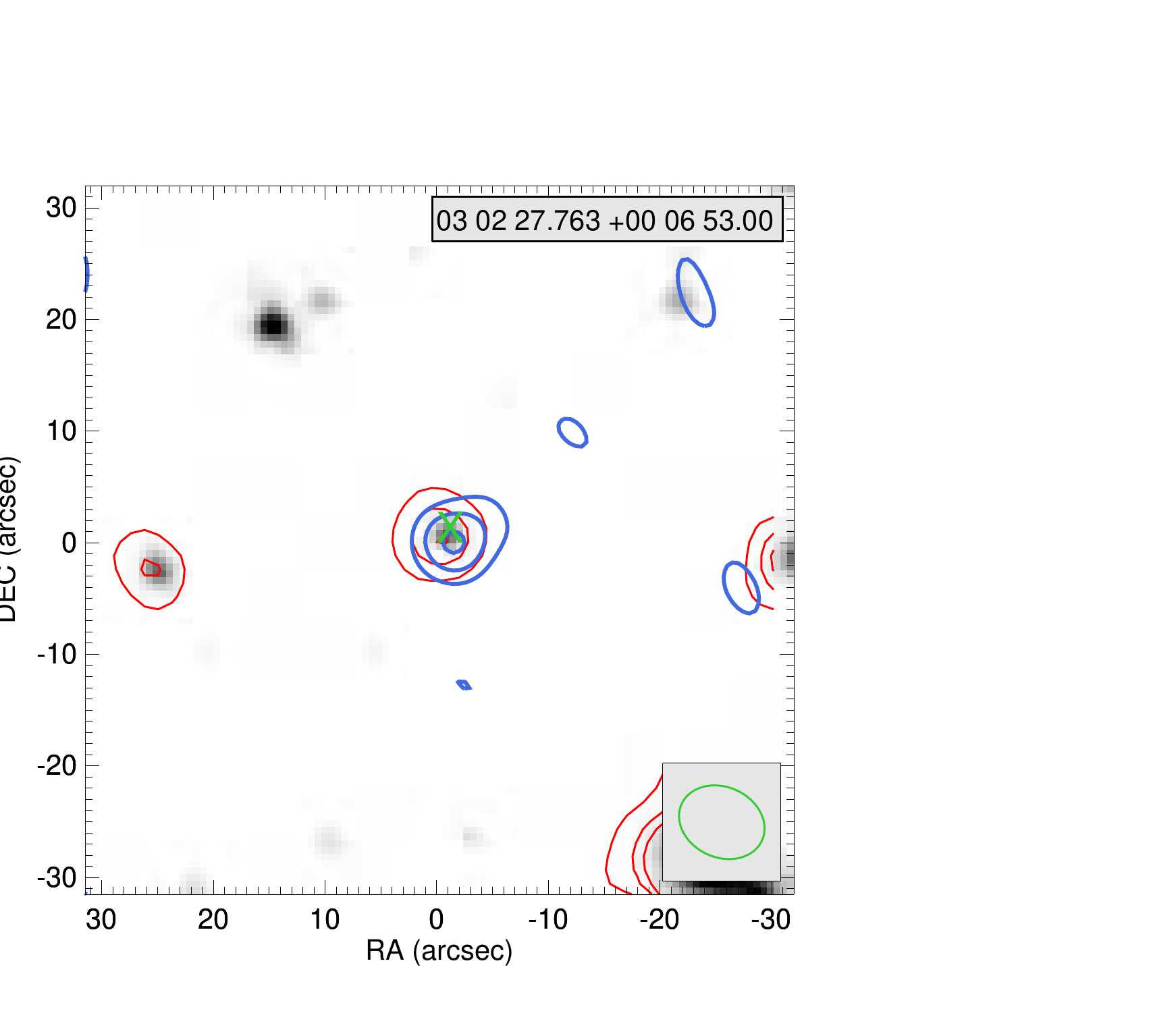}} \hspace{-2cm}
  \subfigure{\includegraphics[width=10cm, clip=true, trim=30 300 70 0]{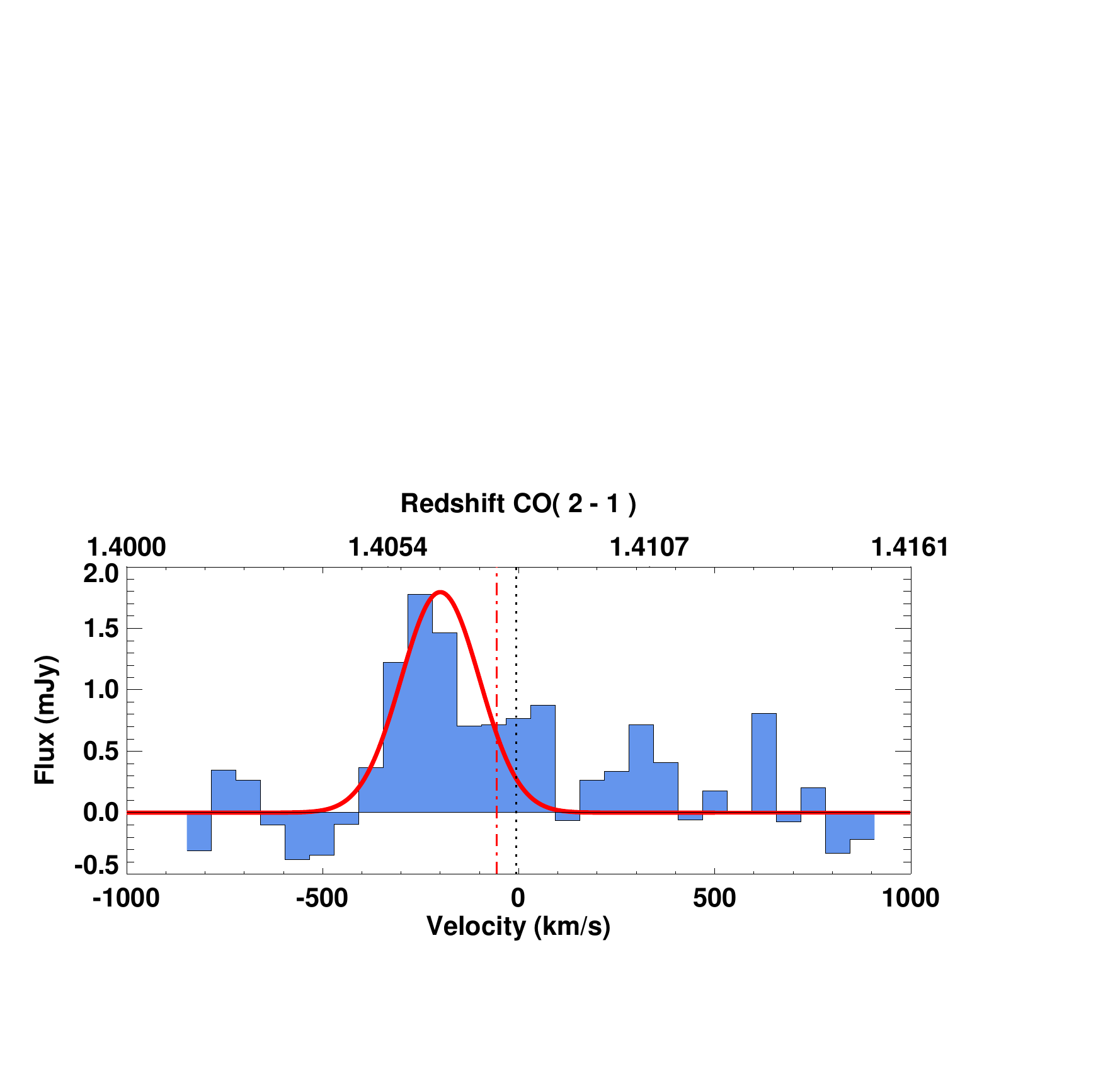}}
}
\caption[SMM030227+0006]{SMM030227+0006. $^{12}$CO is detected at 4.2 $\sigma$, and is lined up well with both IRAC and 24$\mu$m emission. The PAH redshift is slightly inconsistent, but consistent with $dz\sim0.02$ redshift fitting errors (Menendez-Delmestre et al.\ 2007, 2009) ($z=1.408$ from UV, $z=1.4076$ from H$\alpha$, $z=1.43$ from PAH).
}
\label{figure_ob2a}
\end{figure*}

%%%%%%%%%%%%%%%%%%%%%%%%%%%%%%%%

\begin{figure*}
\centering
\mbox
{
  \subfigure{\includegraphics[width=7.5cm, clip=true, trim=10 0 0 0]{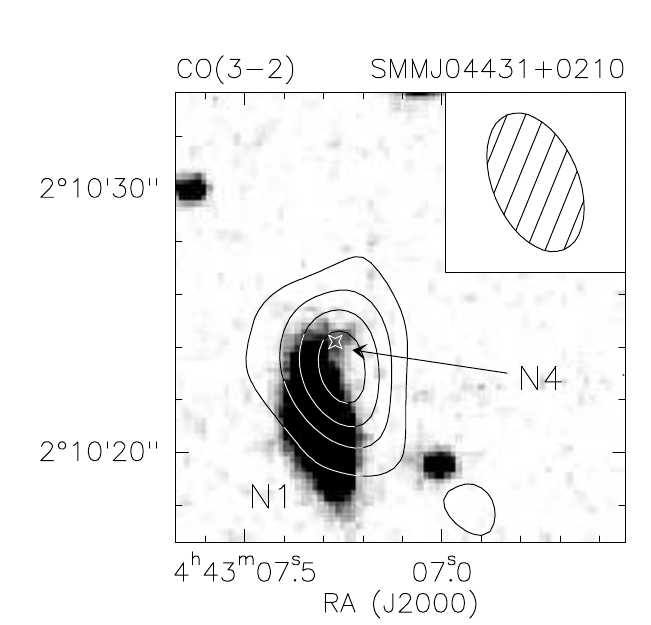}} \hspace{0cm}
  \subfigure{\includegraphics[width=10cm, clip=true, trim=30 300 70 0]{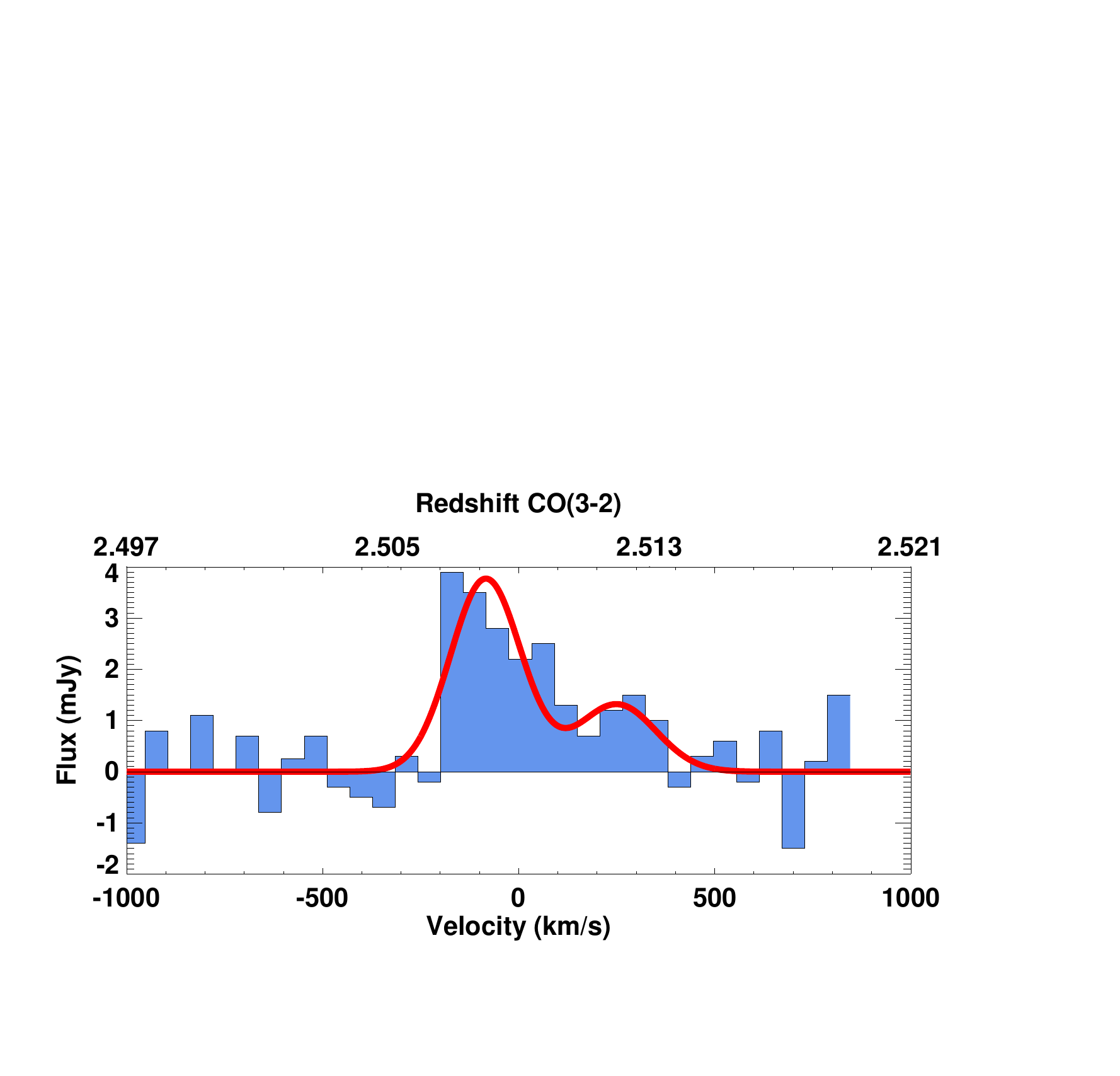}}
}
\caption[SMM044315+0210]{SMM044315+0210. $^{12}$CO is detected at 7.0 $\sigma$. 
This SMG was previously published by Neri et al. (2003); Greve et al. (2005); Tacconi et al. (2006).}
\label{figure_044}
\end{figure*}

%%%%%%%%%%%%%%%%%%%%%%%%%%%%%%%%

\begin{figure*}
\centering
\mbox
{
  \subfigure{\includegraphics[width=7.5cm, clip=true, trim=10 0 0 0]{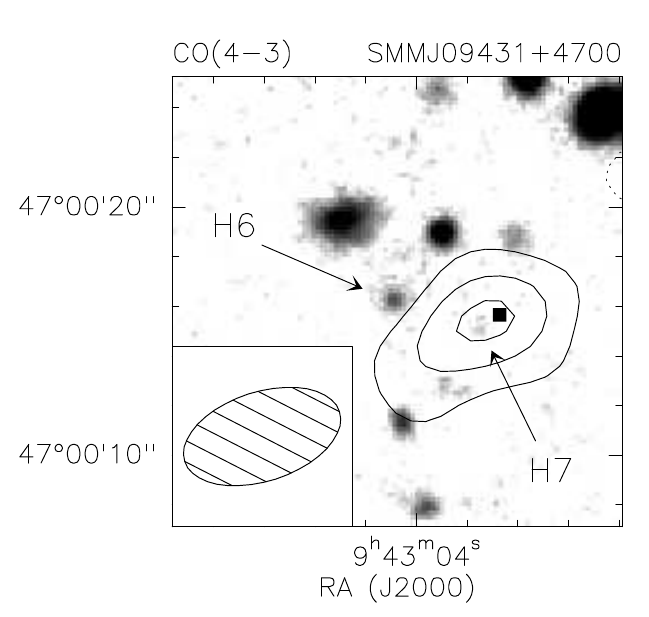}} \hspace{0cm}
  \subfigure{\includegraphics[width=10cm, clip=true, trim=30 300 70 0]{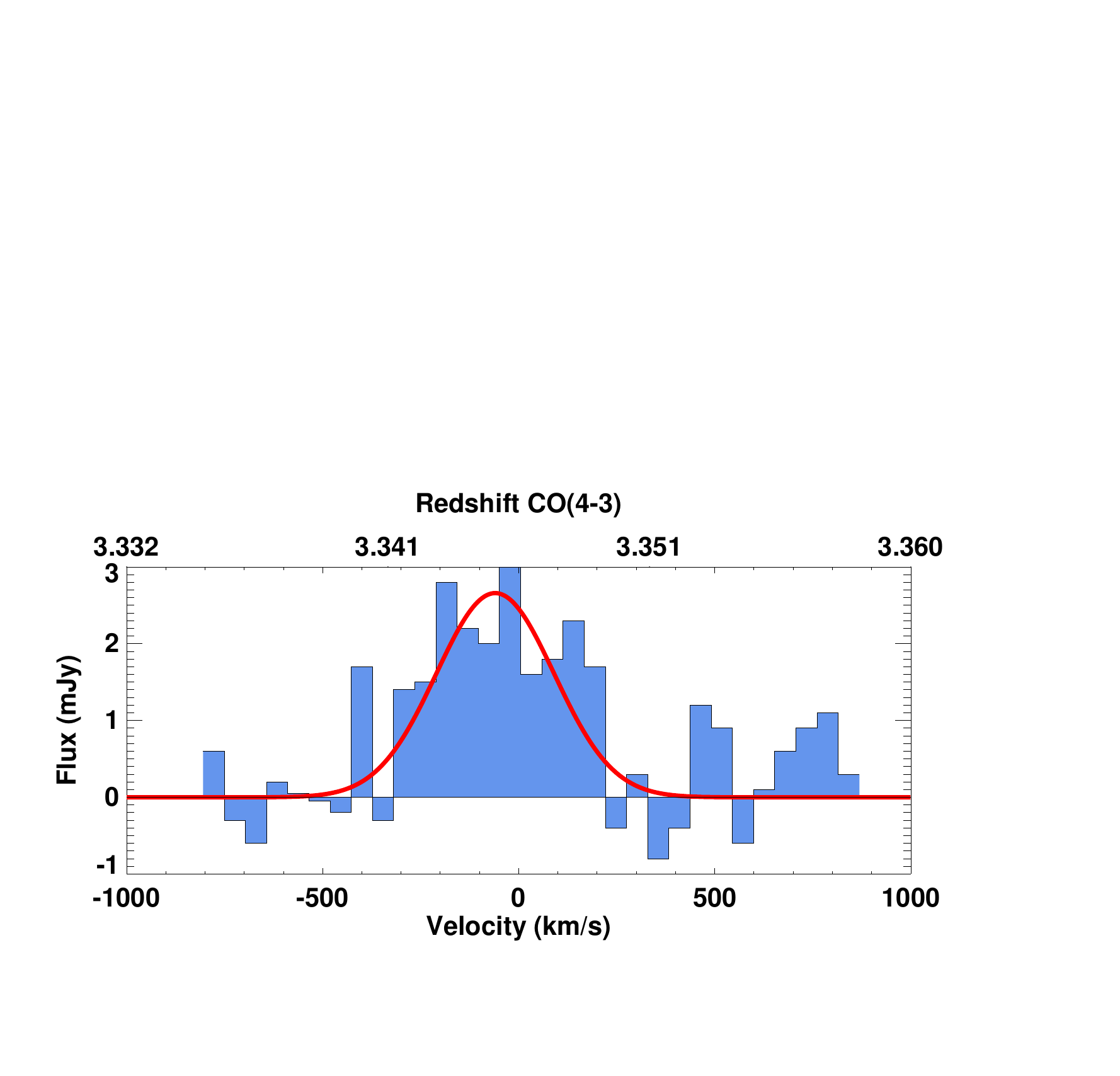}}
}
\caption[SMM094304+4700]{SMM094304+4700. $^{12}$CO is detected at 11.0 $\sigma$. This SMG was previously published by Neri et al. (2003); Greve et al. (2005); Tacconi et al. (2006).
In Engel et al.\ (2010), the H6 component of this 30~kpc separated system
is also  detected  at 5$\sigma$ in CO(6-5).
}
\label{figure_094}
\end{figure*}

%%%%%%%%%%%%%%%%%%%%%%%%%%%%%%%%

\begin{figure*}
\centering
\mbox
{
  \subfigure{\includegraphics[width=8cm, clip=true, trim=50 350 70 0]{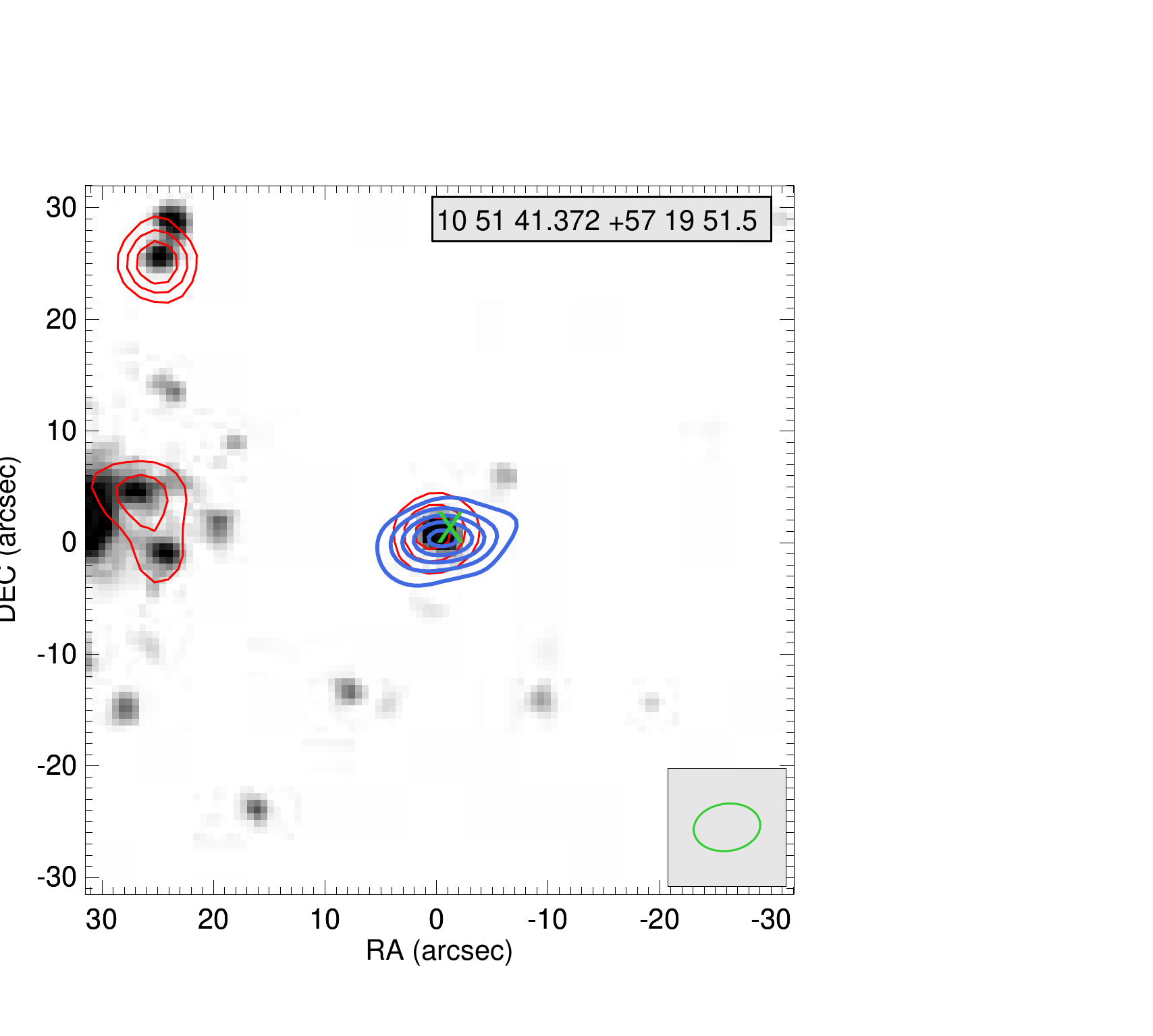}} \hspace{-2cm}
  \subfigure{\includegraphics[width=10cm, clip=true, trim=30 300 70 0]{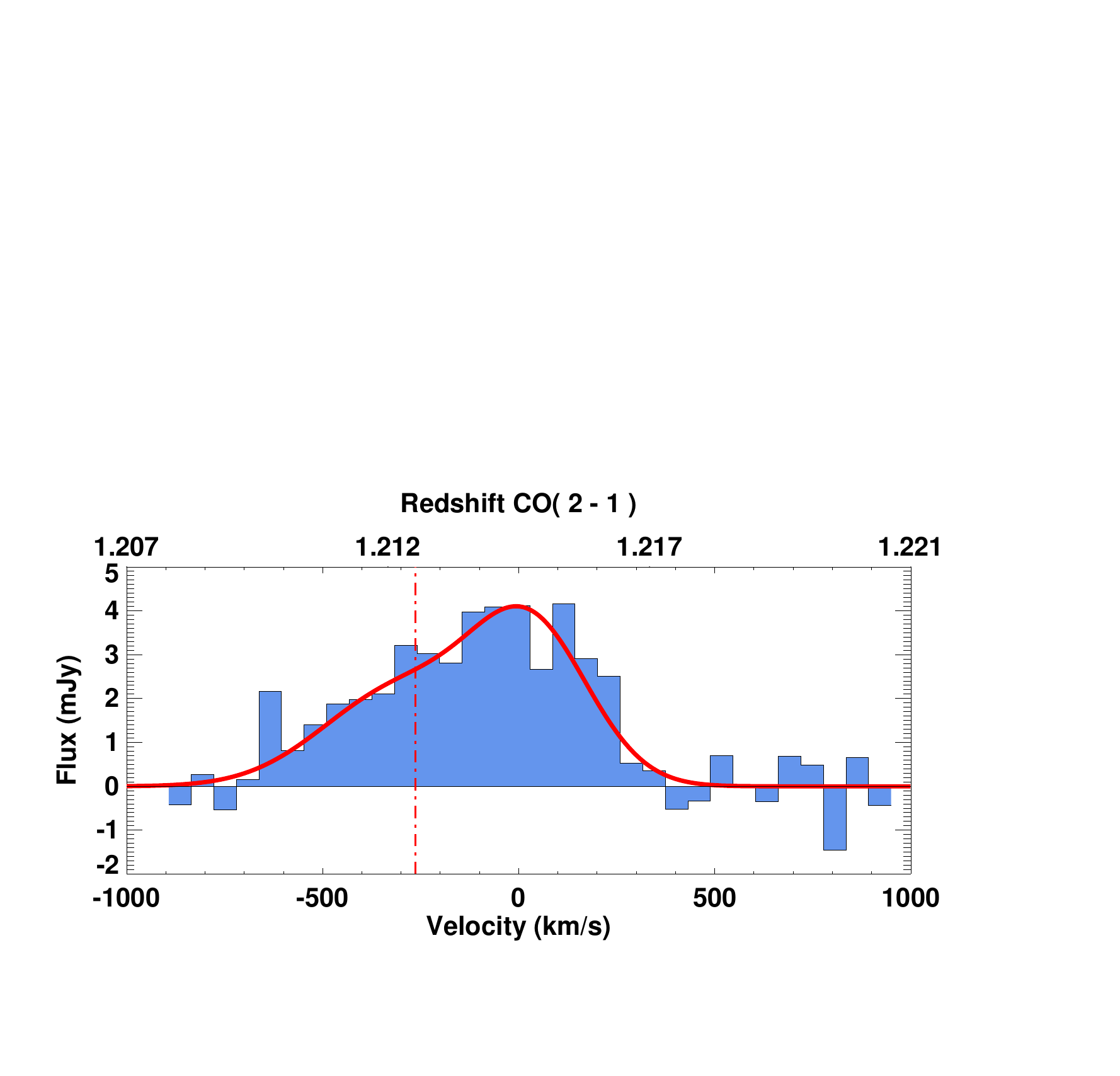}}
}
\caption[SMM105141+5719]{SMM105141+5719.  $^{12}$CO is detected at 10.7$\sigma$, and CO contours are shown double spaced (2$\sigma$, 4$\sigma$, 6$\sigma$, etc). The 24$\mu$m emission is well aligned with the CO. In addition, high spatial resolution (A-config) CO(4-3) data for this SMG is published in Engel et al. (2010), resolving
the CO into two components.}
\label{figure_oa67}
\end{figure*}

%%%%%%%%%%%%%%%%%%%%%%%%%%%%%%%%

\begin{figure*}
\centering
\mbox
{
  \subfigure{\includegraphics[width=8cm, clip=true, trim=50 350 70 0]{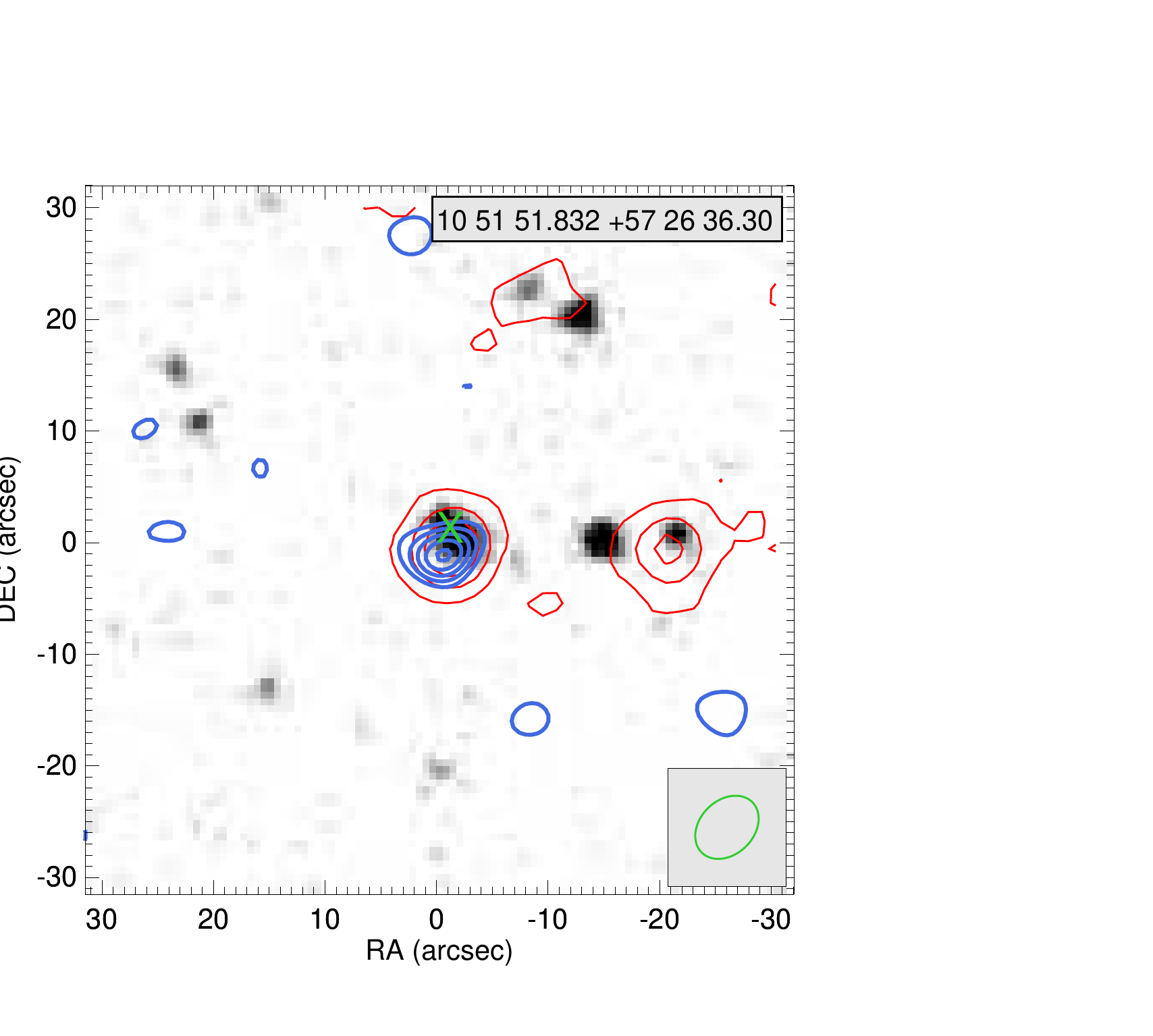}} \hspace{-2cm}
  \subfigure{\includegraphics[width=10cm, clip=true, trim=30 300 70 0]{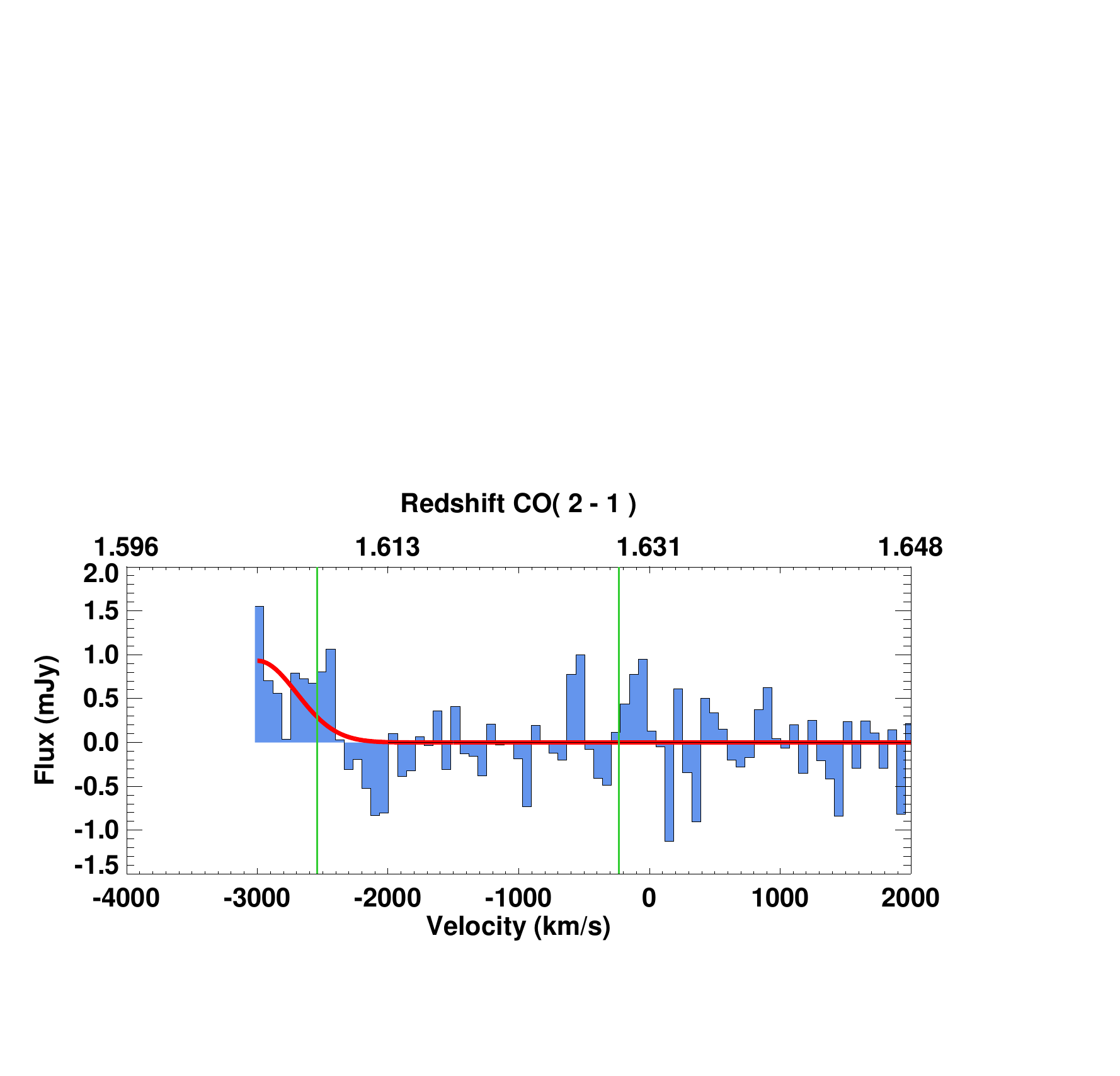}}
}
\caption{SMM105151+5726. $^{12}$CO is detected at $6.2 \sigma$, but near the edge of the bandpass. All emission components are aligned to the radio centre within errors. The published UV-derived redshift, $z=1.147$ in Ivison et al.\ (2005) results from a foreground UV-bright galaxy. The PAH redshift can be difficult to constrain in this redshift range, and we show the best two fits -- the higher of the two estimates was the initial estimate from Menendez-Delmestre et al.\ (2009), while the lower estimate is in agreement with the CO redshift.}
\label{figure_sj3b}
\end{figure*}

%%%%%%%%%%%%%%%%%%%%%%%%%%%%%%%%

\begin{figure*}
\centering
\mbox
{
  \subfigure{\includegraphics[width=8cm, clip=true, trim=50 350 70 0]{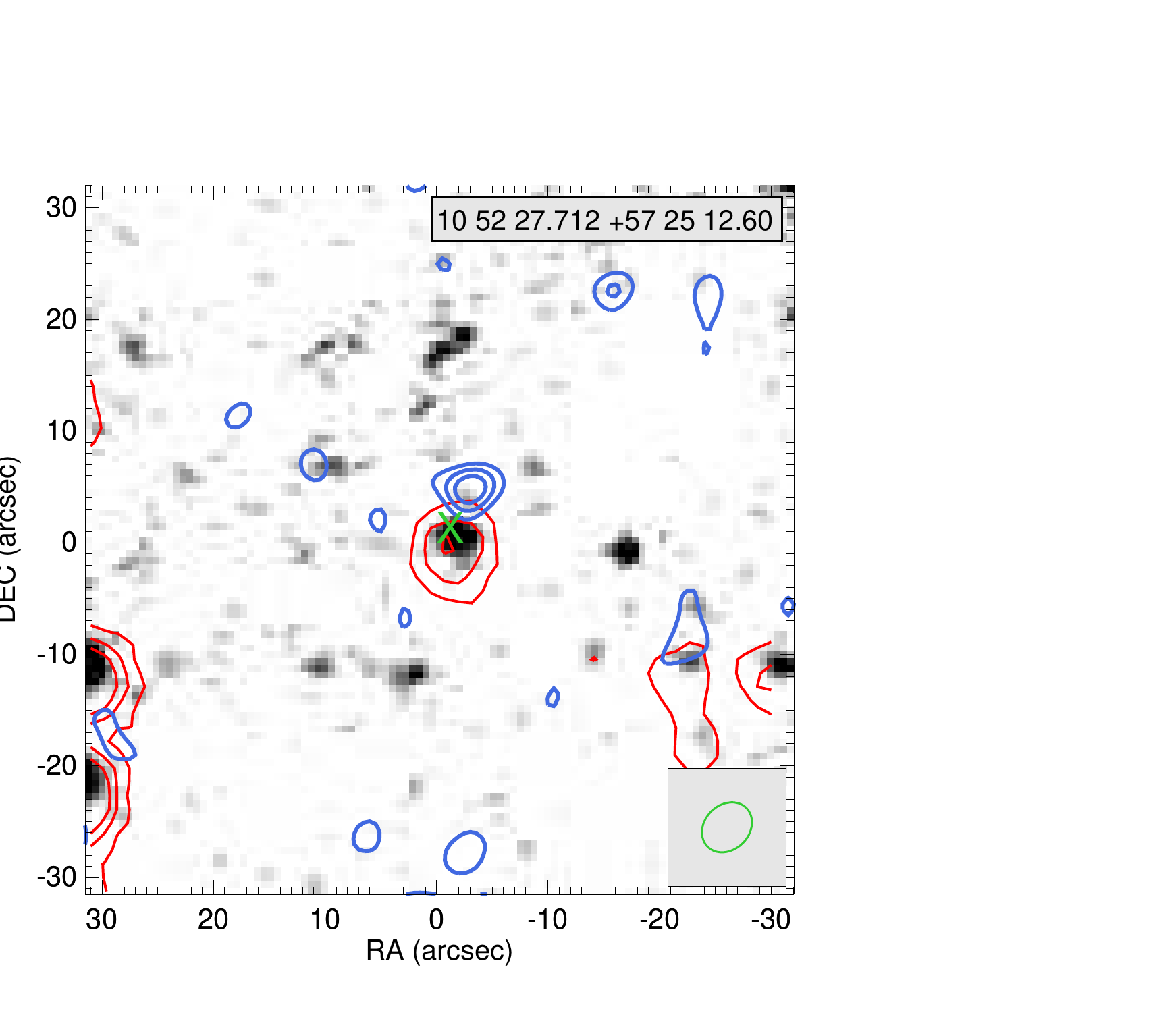}} \hspace{-2cm}
  \subfigure{\includegraphics[width=10cm, clip=true, trim=30 300 70 0]{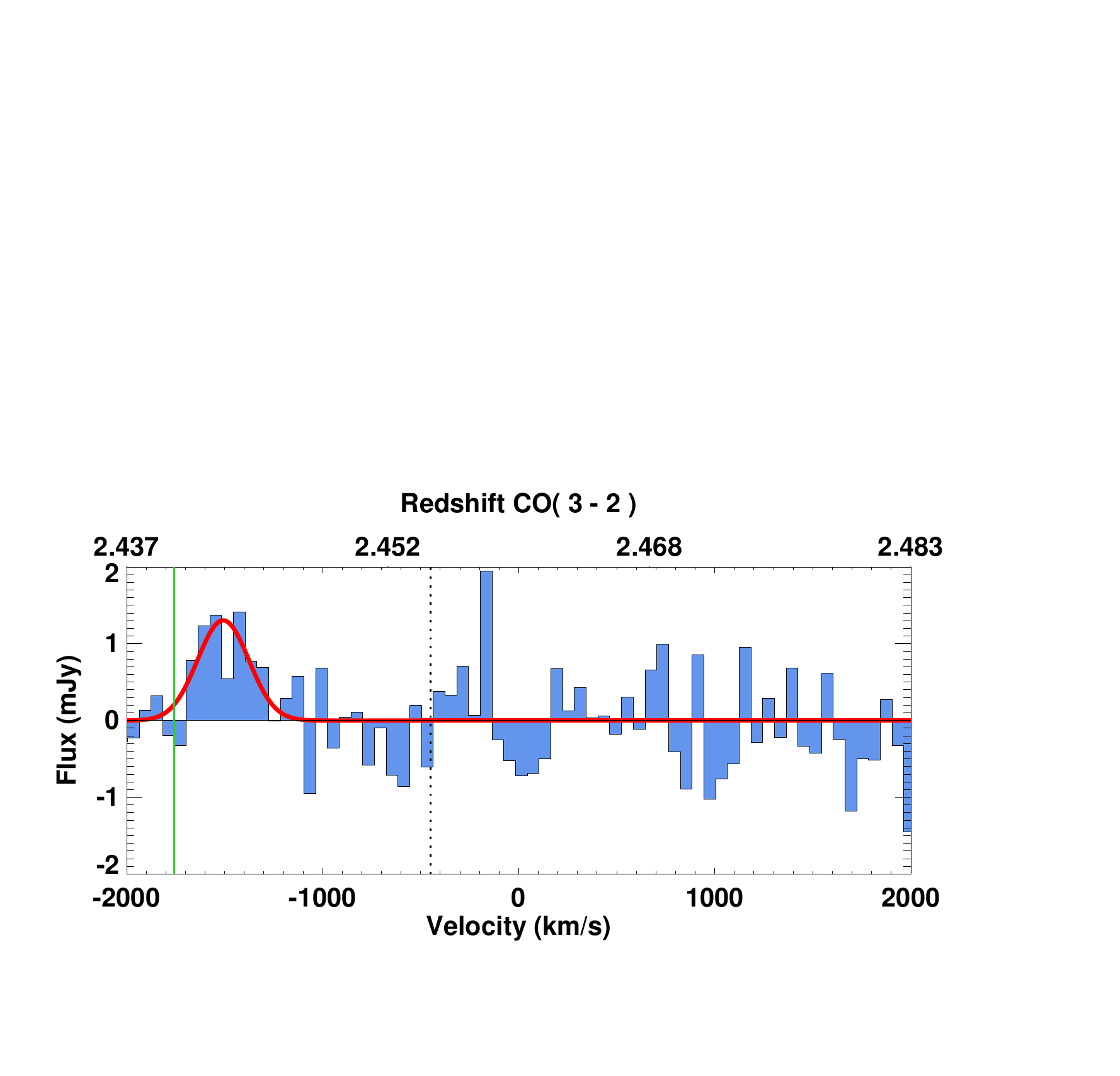}}
}
\caption{SMM105227+5725. $^{12}$CO is detected at $5.0\sigma$, though it is $\sim$5\arcsec\ north of other emission components.}
\label{figure_sf3b}
\end{figure*}

%%%%%%%%%%%%%%%%%%%%%%%%%%%%%%%%%%%%%%%%%%%%%%%%

\begin{figure*}
\centering
\mbox
{
  \subfigure{\includegraphics[width=8cm, clip=true, trim=50 350 70 0]{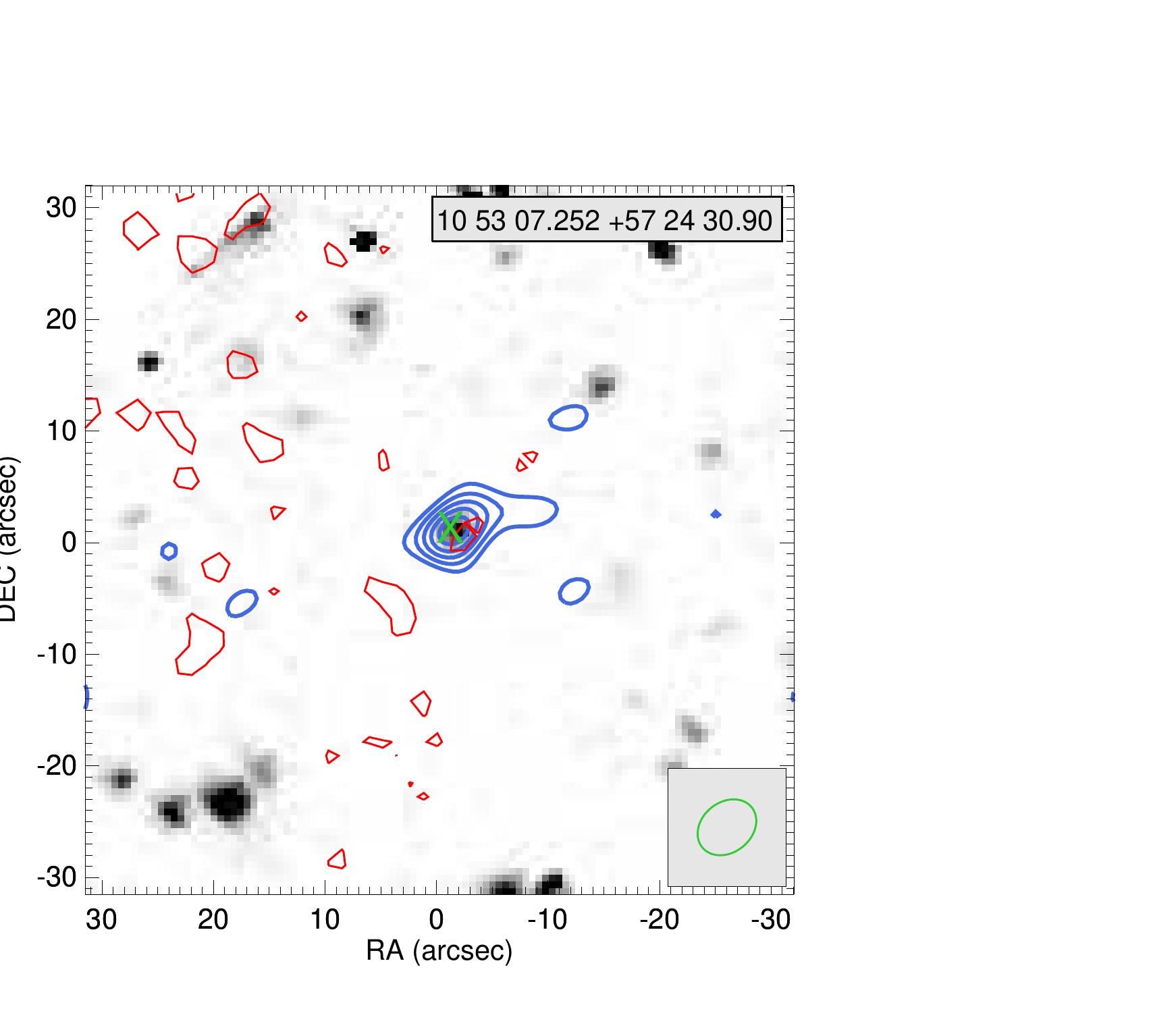}} \hspace{-2cm}
  \subfigure{\includegraphics[width=10cm, clip=true, trim=30 300 70 0]{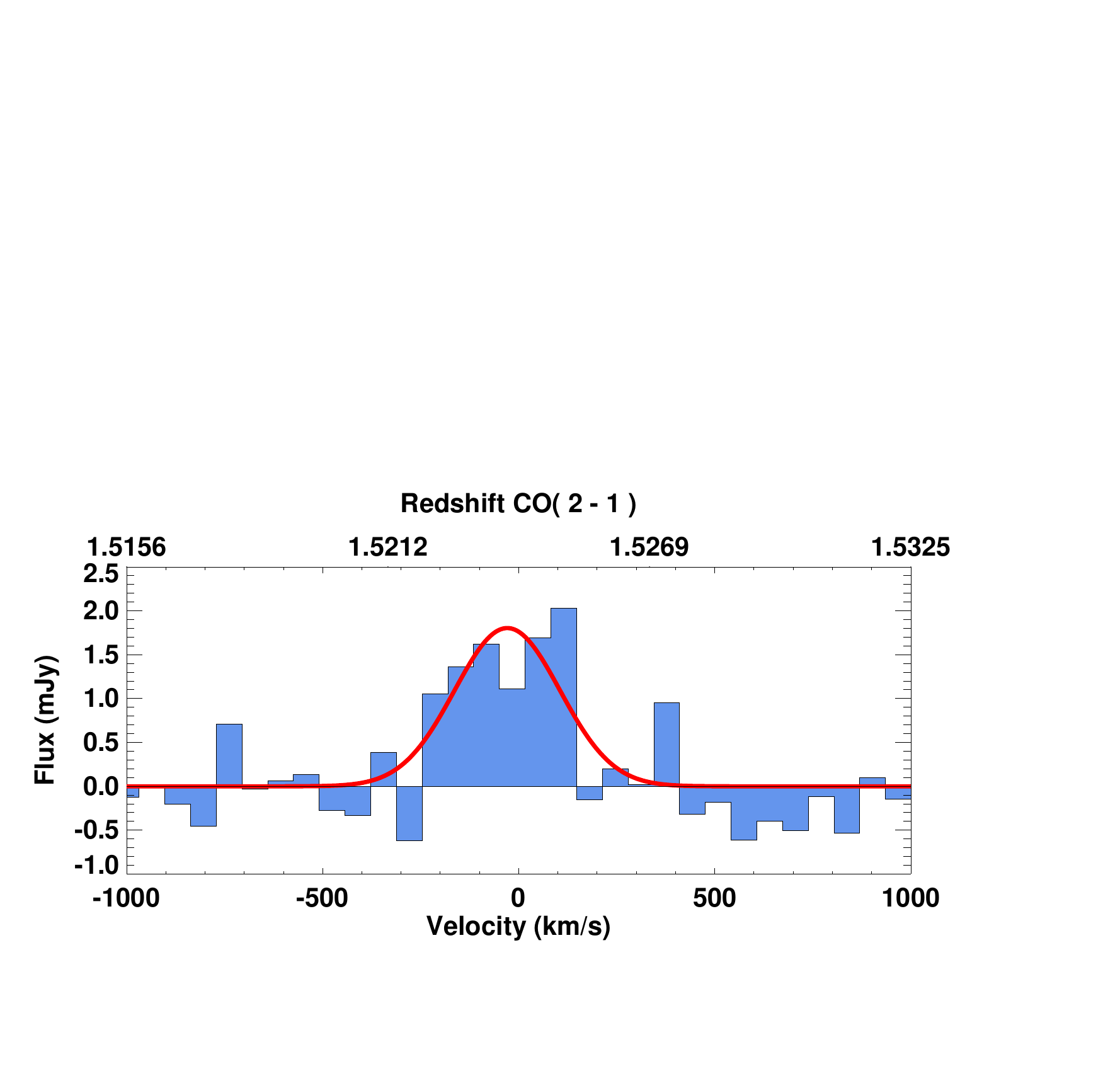}}
}
\caption{SMM105307+5724. $^{12}$CO is well detected at $6.9\sigma$, and all emission components are close to radio centre. In addition, high spatial resolution (A-config) CO(3-2) data for this SMG is published in Bothwell et al. (2010), resolving the CO into a north-south, 1.5$''$ extended source. }
\label{figure_sn3b}
\end{figure*}

%%%%%%%%%%%%%%%%%%%%%%%%%%%%%%%%%%%%%%%%%%%%%%%%

\begin{figure*}
\centering
\mbox
{
  \subfigure{\includegraphics[width=8cm, clip=true, trim=50 350 70 0]{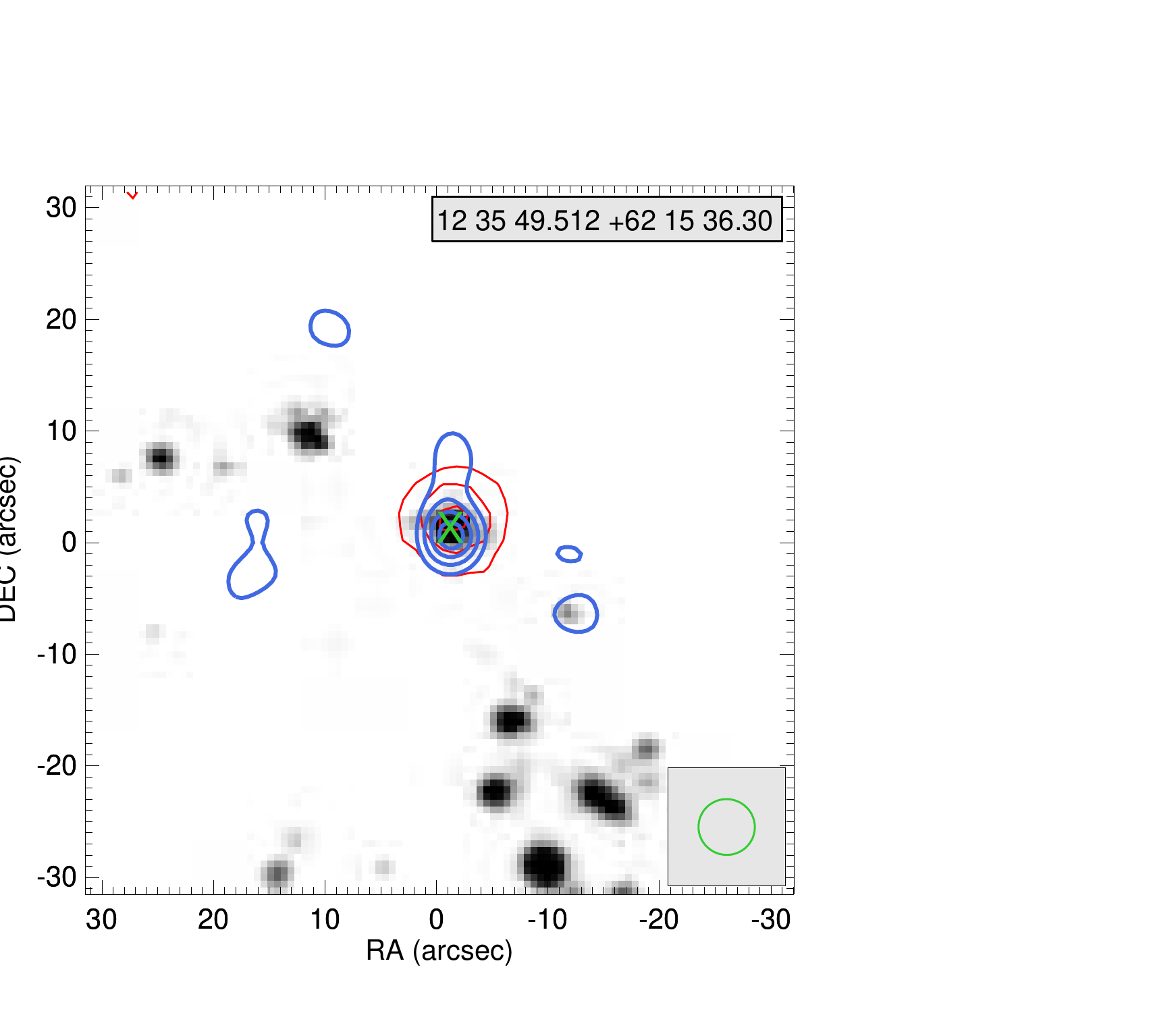}} \hspace{-2cm}
  \subfigure{\includegraphics[width=10cm, clip=true, trim=30 300 70 0]{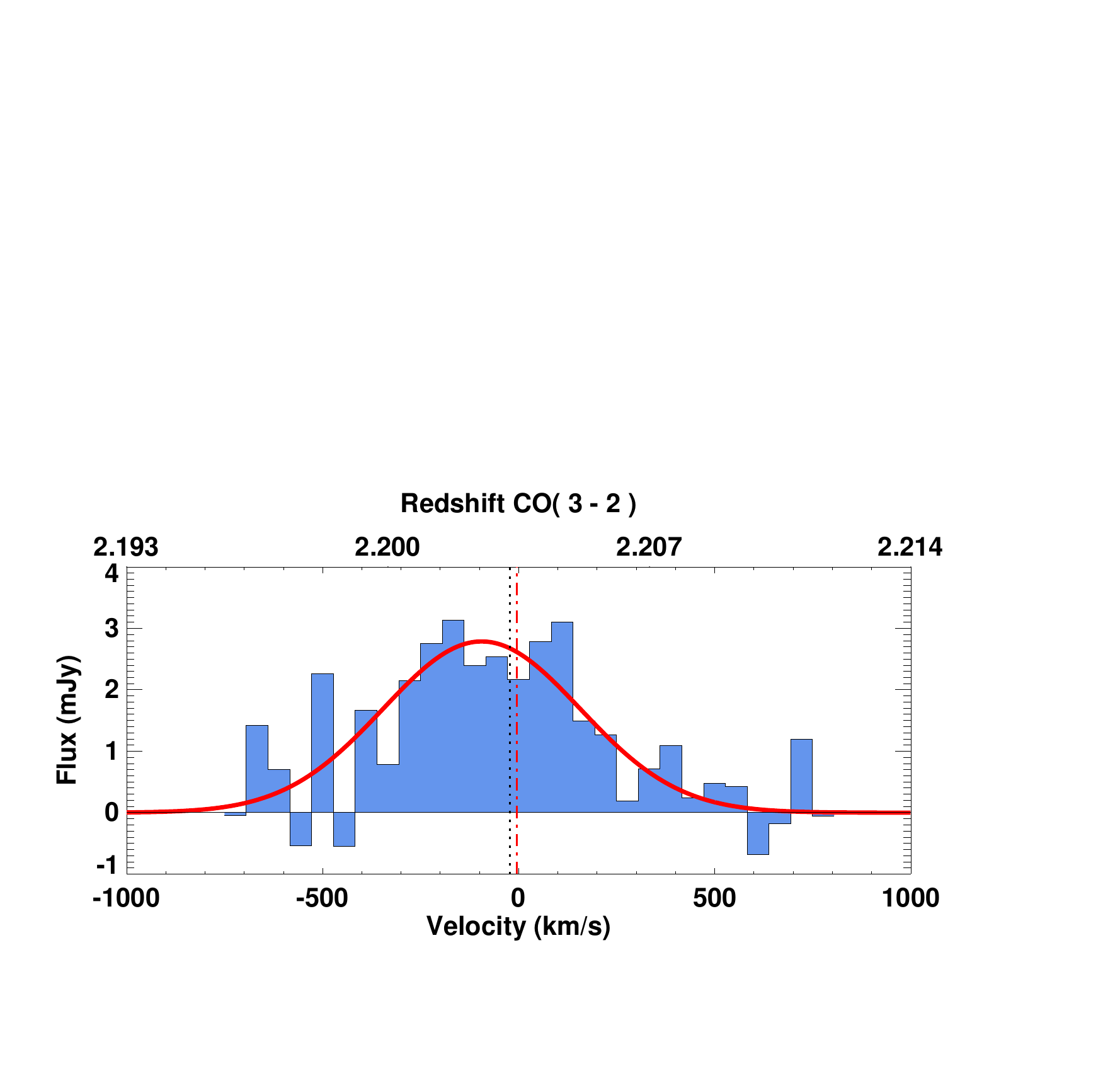}}
}
\caption[SMM123549+6215]{SMM123549+6215. CO contours are double spaced (2$\sigma$, 4$\sigma$, 6$\sigma$, etc). $^{12}$CO is well detected at 9.3$\sigma$. The PAH redshift is inconsistent with other determinations ($z=2.203$ from UV, $z=2.2032$ from H$\alpha$; compare to $z=2.24$ from PAH), slightly larger than the $dz=0.02$ rms error on the PAH fit. This SMG was previously published by Tacconi et al.\ (2006).}
\label{figure_oc2a}
\end{figure*}

%%%%%%%%%%%%%%%%%%%%%%%%%%%%%%%%

\begin{figure*}
\centering
\mbox
{
  \subfigure{\includegraphics[width=8cm, clip=true, trim=50 350 70 0]{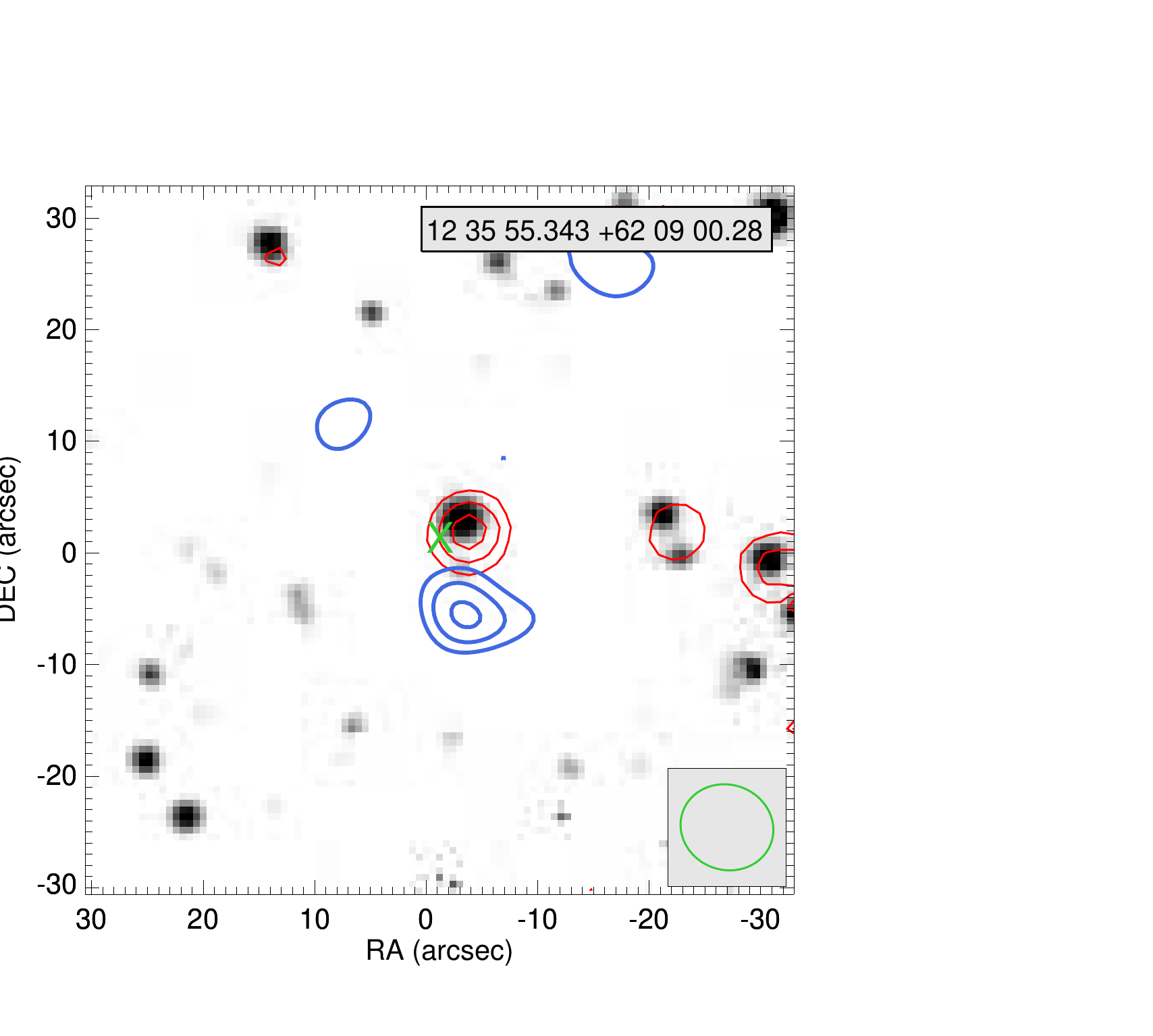}} \hspace{-2cm}
  \subfigure{\includegraphics[width=10cm, clip=true, trim=30 300 70 0]{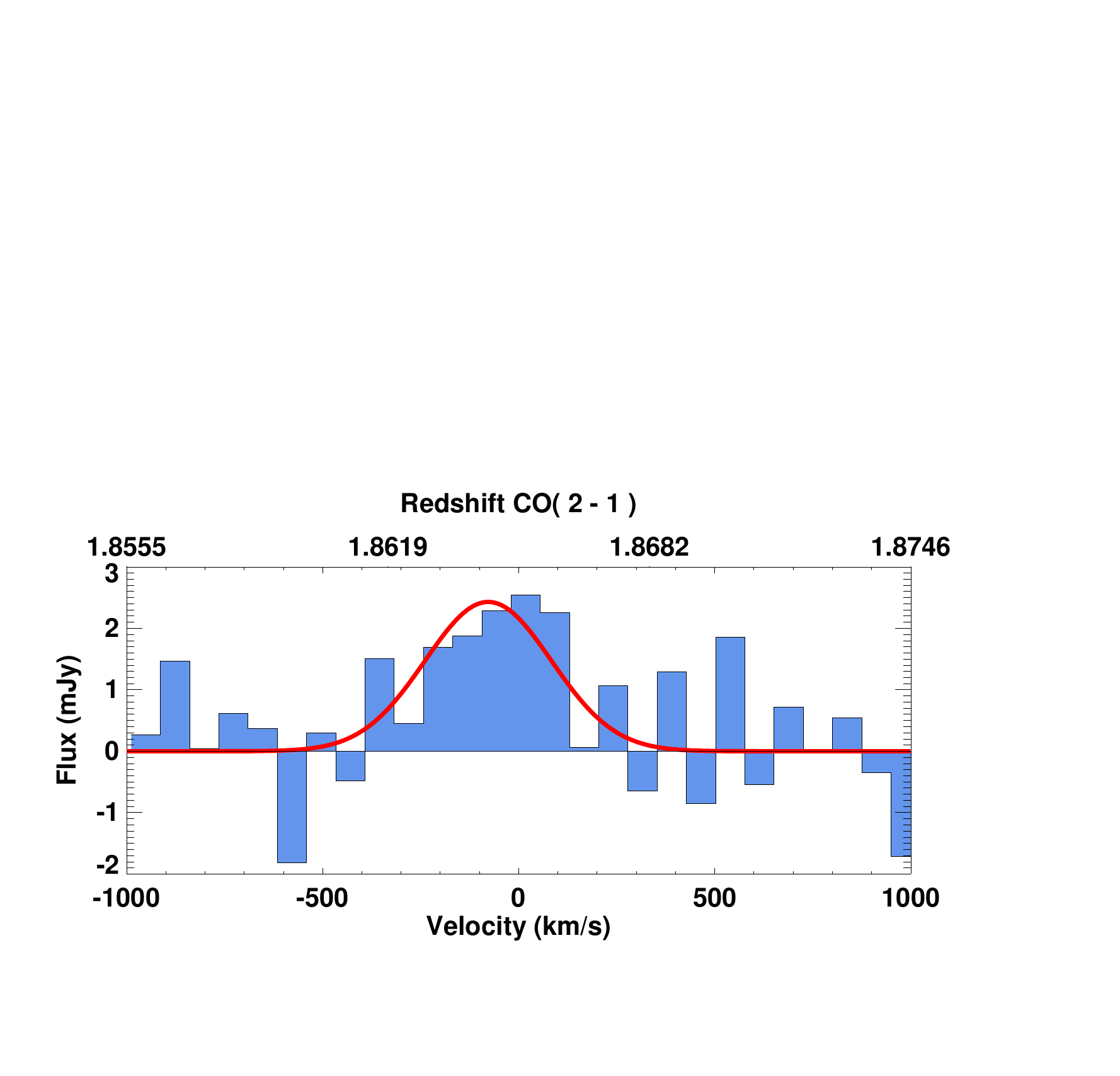}}
}
\caption{SMM123555+6209. $^{12}$CO is detected at 4.3$\sigma$. The peak of emission is between 6 to 8 arcsec to the south of the IRAC/radio/24$\mu$m positions. Both the published Ly$\alpha$-based UV redshift and the PAH-derived redshift are higher than the CO: $z=1.875$ from UV and $z=1.88$ from PAH, while the $^{12}$CO line is detected at $z=1.864$.
Followup spectroscopic analysis suggests a slightly lower systemic redshift $z=1.868$ based on weak inter-stellar absorption features blueshifted relative to the Ly$\alpha$ emission line. 
}
\label{figure_pa32}
\end{figure*}

%%%%%%%%%%%%%%%%%%%%%%%%%%%%%%%%

\begin{figure*}
\centering
\mbox
{
  \subfigure{\includegraphics[width=8cm, clip=true, trim=50 350 70 0]{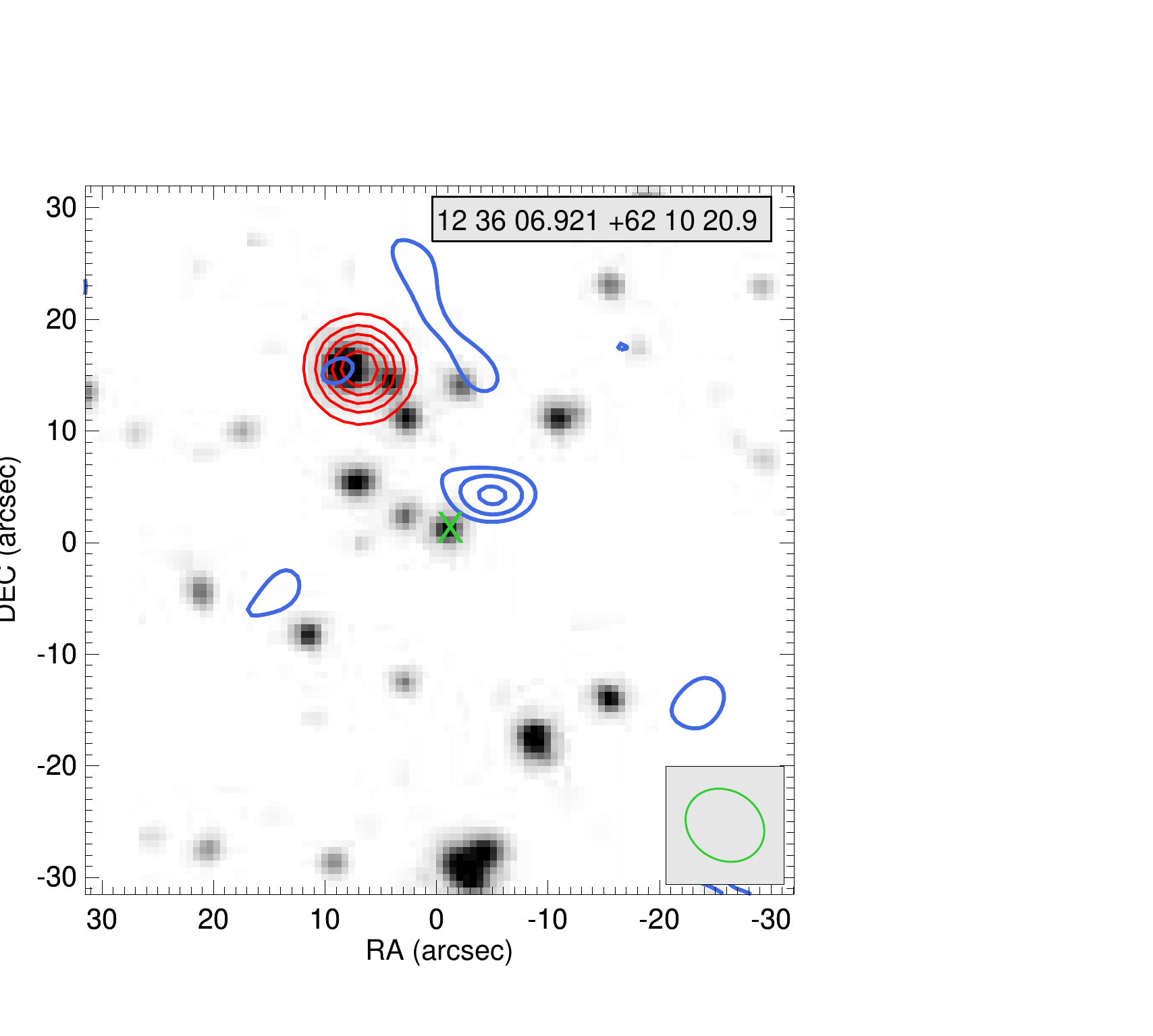}} \hspace{-2cm}
  \subfigure{\includegraphics[width=10cm, clip=true, trim=30 300 70 0]{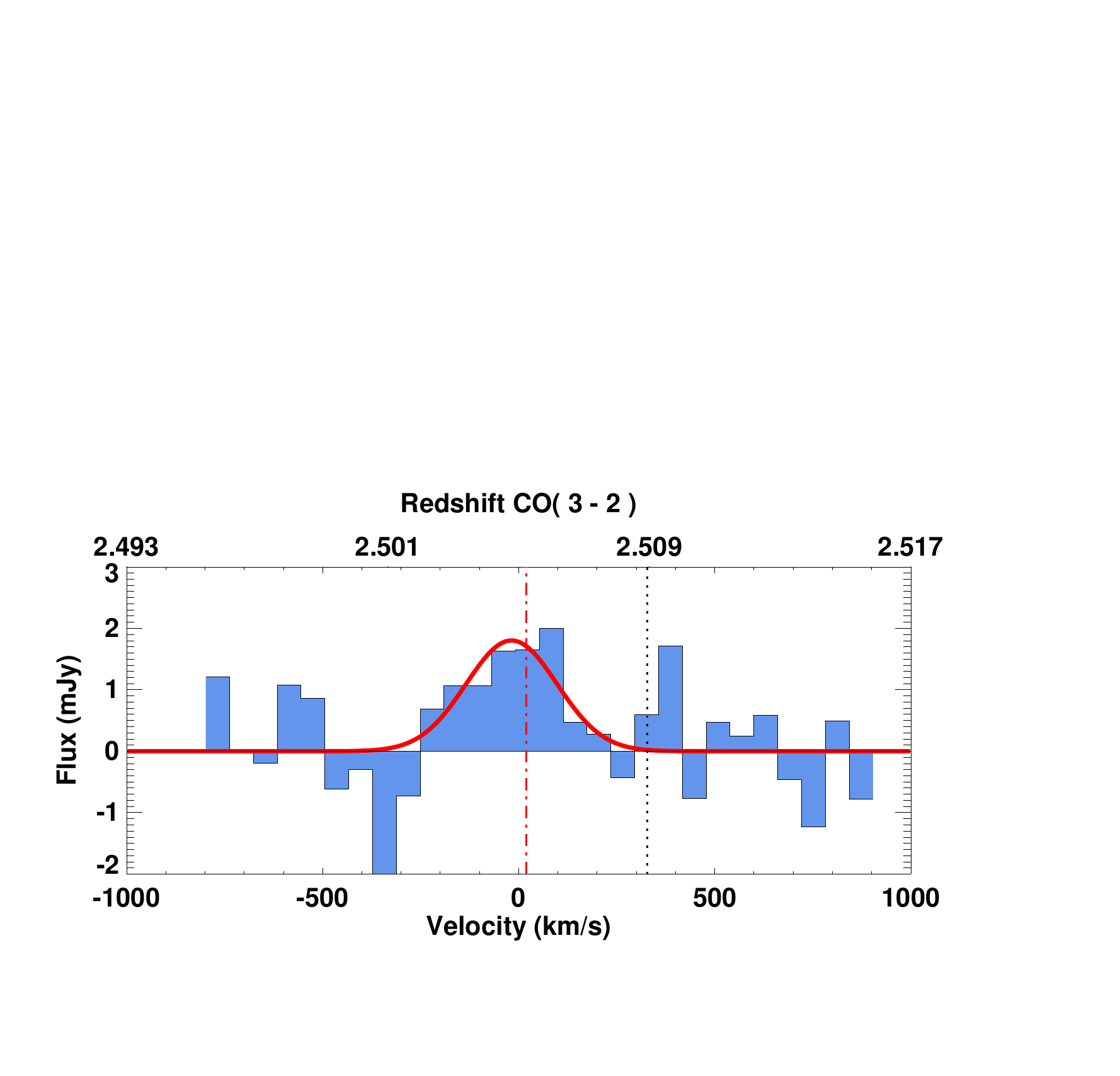}} 
}
\caption{SMM123606+6210. $^{12}$CO is reasonably well detected at $4.3\sigma$ (approximately 5'' to the NE of the radio centre) but there is no apparent near-IR counterpart. A radio source is detected by {\it Spitzer}-MIPS and IRAC, 20 arcseconds away to the NE, but is likely unconnected with the SMG.
}
\label{figure_nd61}
\end{figure*}

%%%%%%%%%%%%%%%%%%%%%%%%%%%%%%%%

\begin{figure*}
\centering
\mbox
{
  \subfigure{\includegraphics[width=8cm, clip=true, trim=50 350 70 0]{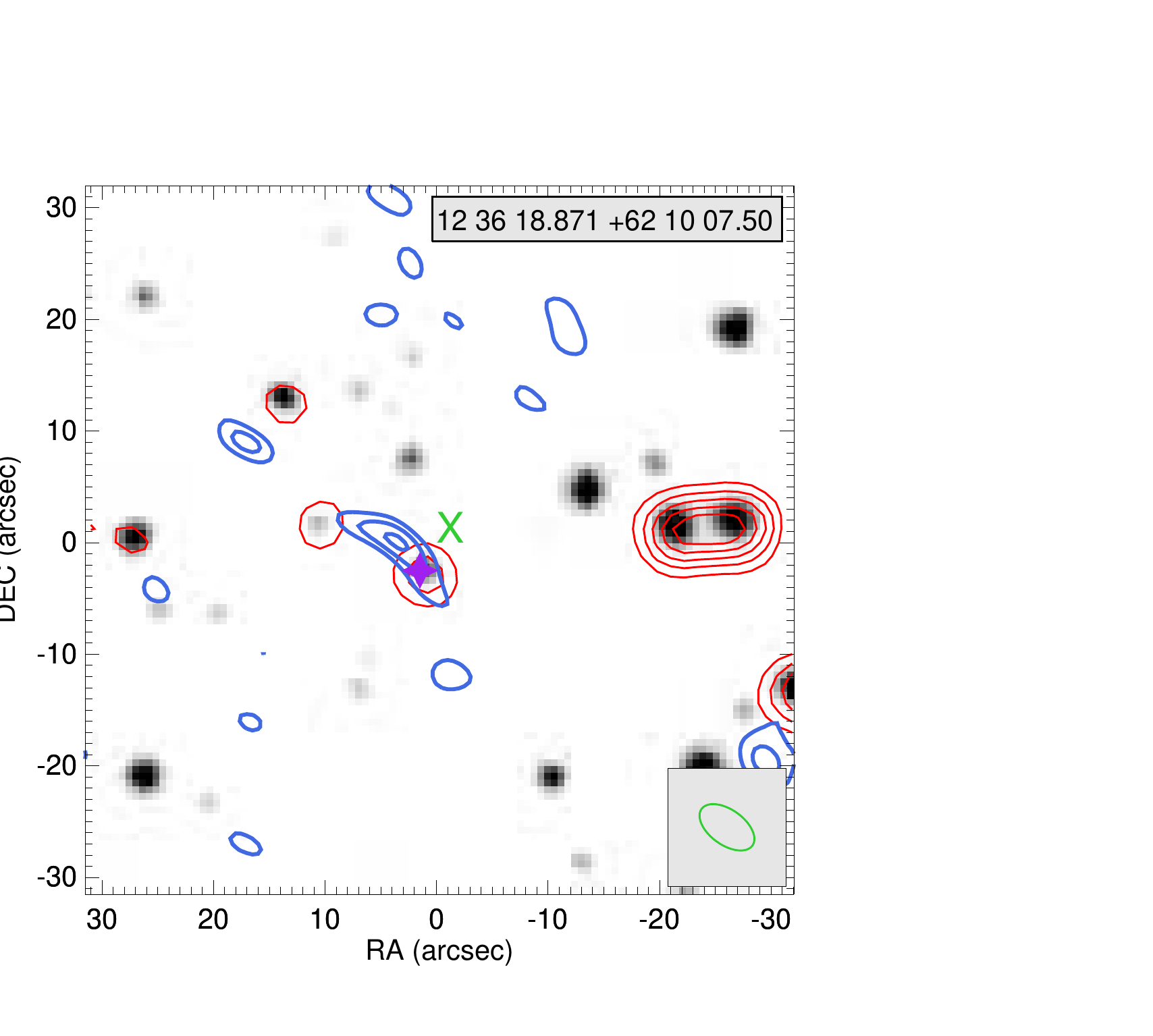}} \hspace{-2cm}
  \subfigure{\includegraphics[width=10cm, clip=true, trim=30 300 70 0]{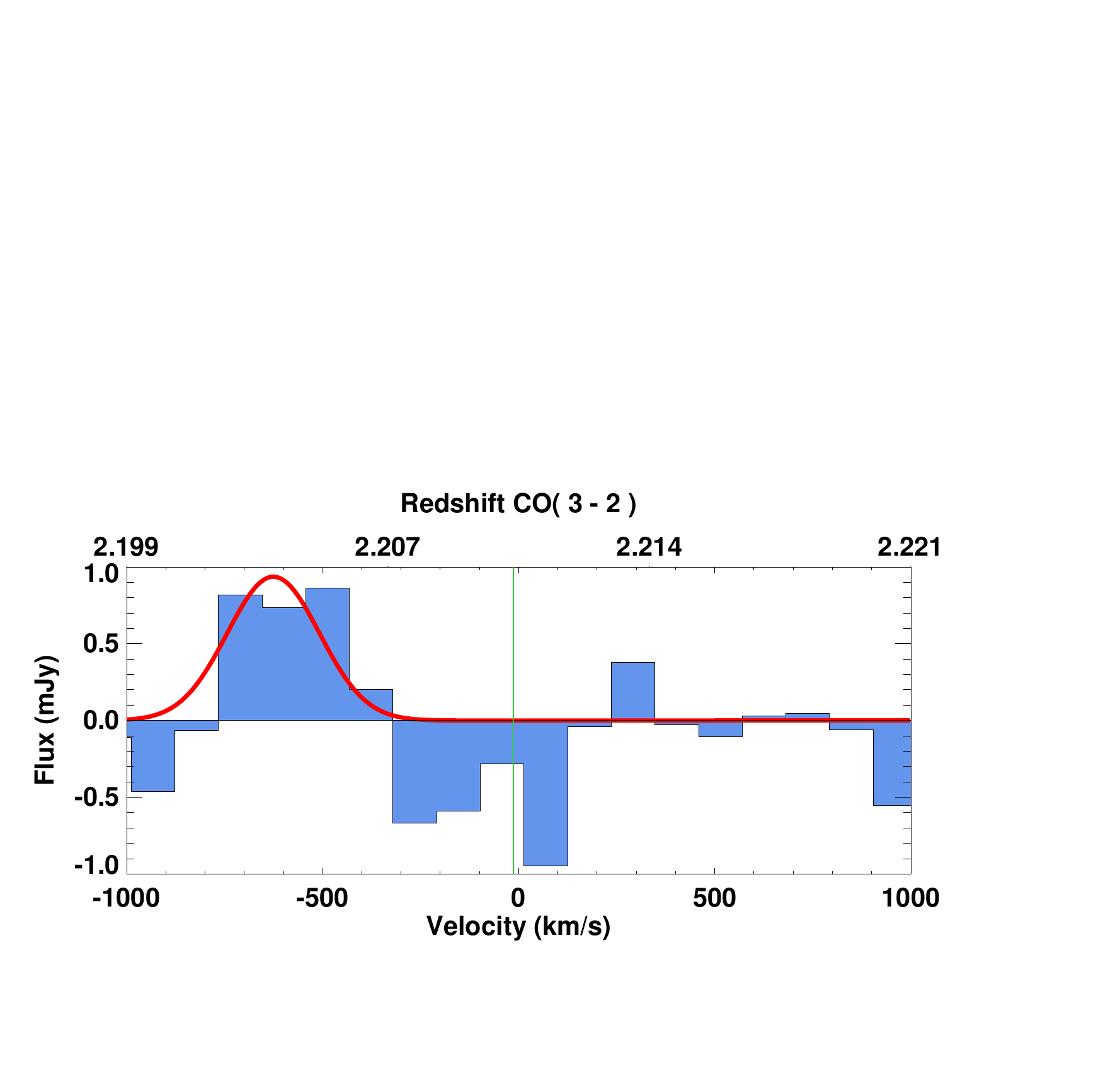}}
}
\caption{SMM123618+6210. $^{12}$CO is detected at $4.0\sigma$ close to the radio centre and within the elongated beam resulting from the poor UV-plane coverage, here marked marked with a purple star. The green cross indicates here the $850 \mu$m centre (which was chosen for pointing).
 We classify this as a candidate detection as in the text, as a result of the offset, CO significance, and lack of an optical
 redshift to verify the PAH.}
\label{figure_se3b}
\end{figure*}

%%%%%%%%%%%%%%%%%%%%%%%%%%%%%%%%

\begin{figure*}
\centering
\mbox
{
  \subfigure{\includegraphics[width=8cm, clip=true, trim=50 350 70 0]{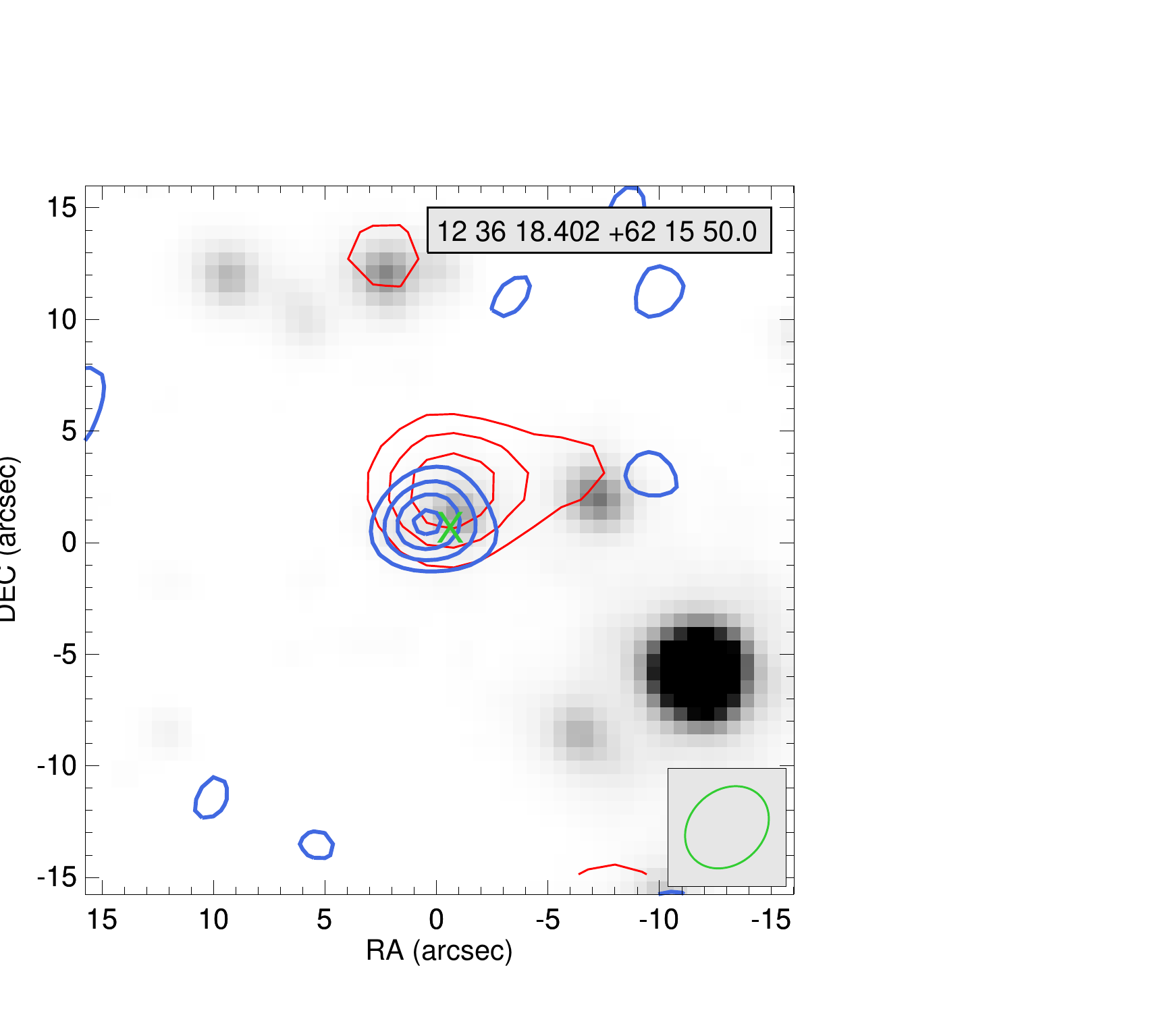}} \hspace{-2cm}
  \subfigure{\includegraphics[width=10cm, clip=true, trim=30 300 70 0]{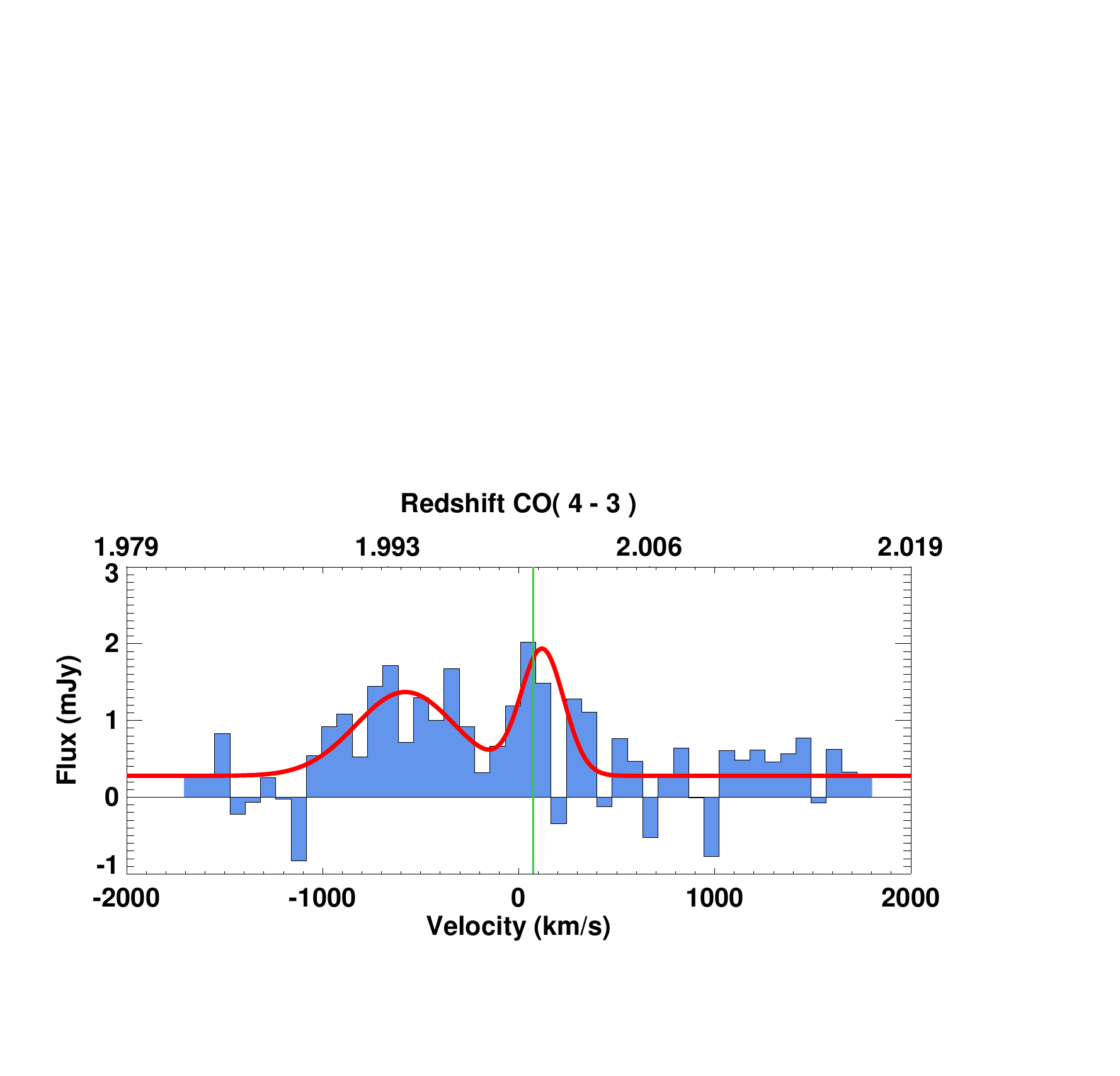}}
}
\caption[SMM123618+6215]{SMM123618+6215. CO contours are double spaced (2$\sigma$, 4$\sigma$, 6$\sigma$, etc). $^{12}$CO is strongly detected at $8.6 \sigma$ and aligned  with the 24$\mu$m emission. As discussed in Chapman et al.\ (2009) and Bothwell et al.\ (2010 -- where higher resolution CO(4-3) data is published), the redshift originally derived from UV is for the galaxy 0.8$''$ to the west,  which led to a CO non-detection in a separate observation.
}
\label{figure_sb3b}
\end{figure*}

%%%%%%%%%%%%%%%%%%%%%%%%%%%%%%%%

\begin{figure*}
\centering
\mbox
{
  \subfigure{\includegraphics[width=8cm, clip=true, trim=50 350 70 0]{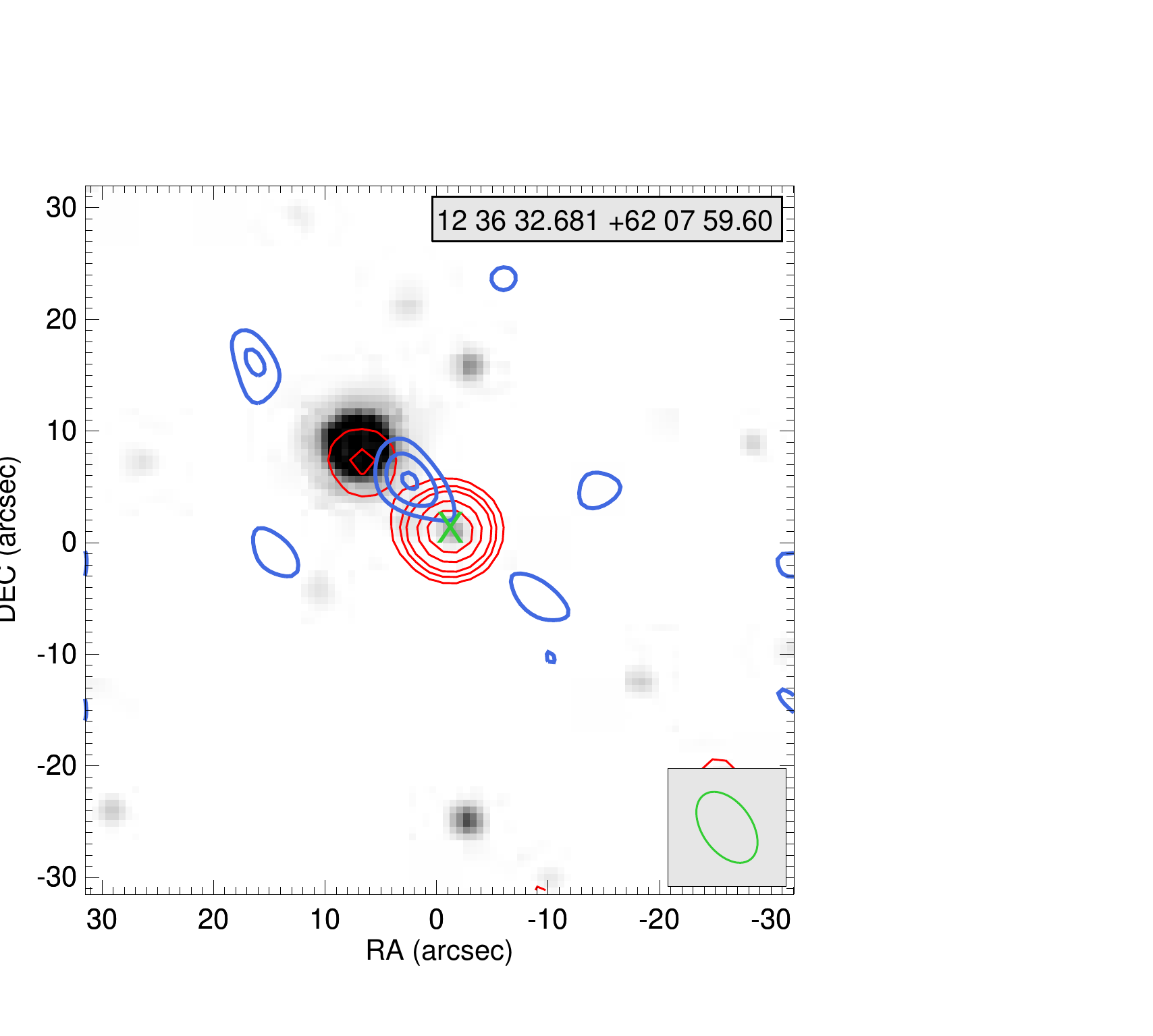}} \hspace{-2cm}
  \subfigure{\includegraphics[width=10cm, clip=true, trim=30 300 70 0]{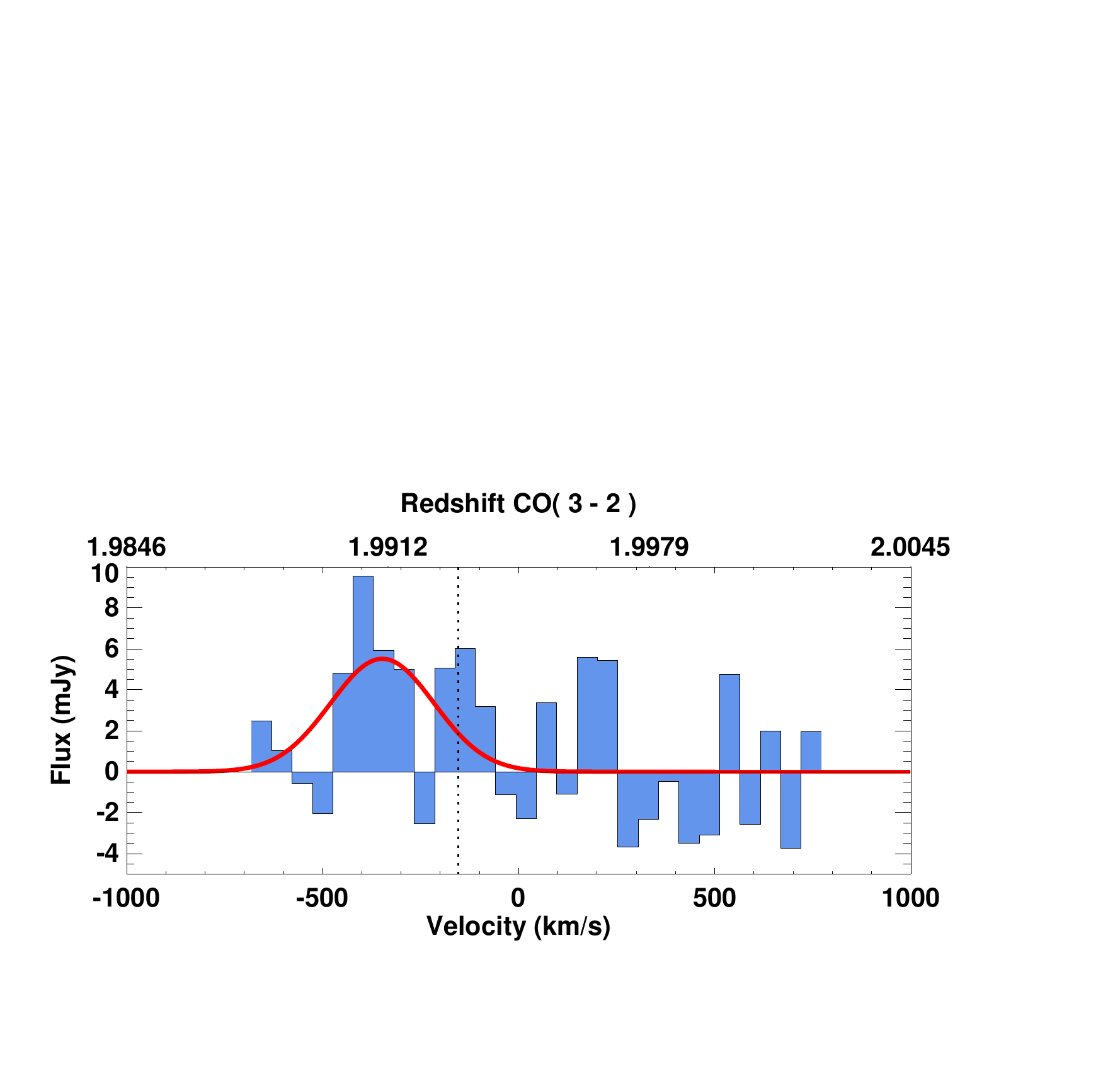}}
}
\caption{SMM123632+6207. $^{12}$CO emission is detected at $4.5\sigma$, offset by $\sim 7$'' from the radio centre, but still consistent at 2$\sigma$ within the elongated CO beam. }
\label{figure_oe67}l
\end{figure*}

%%%%%%%%%%%%%%%%%%%%%%%%%%%%%%%%

\begin{figure*}
\centering
\mbox
{
  \subfigure{\includegraphics[width=8cm, clip=true, trim=50 350 70 0]{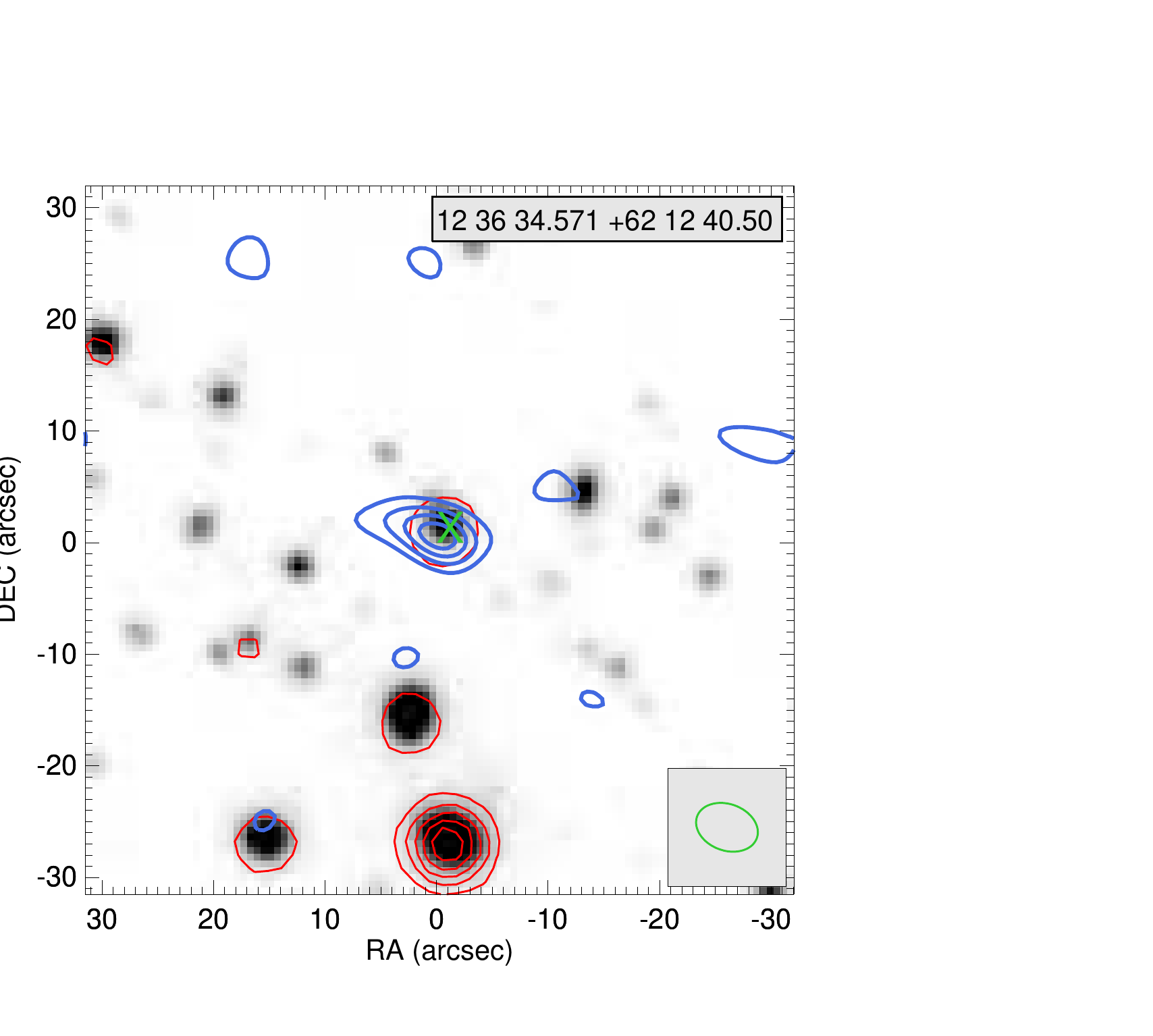}} \hspace{-2cm}
  \subfigure{\includegraphics[width=10cm, clip=true, trim=30 300 70 0]{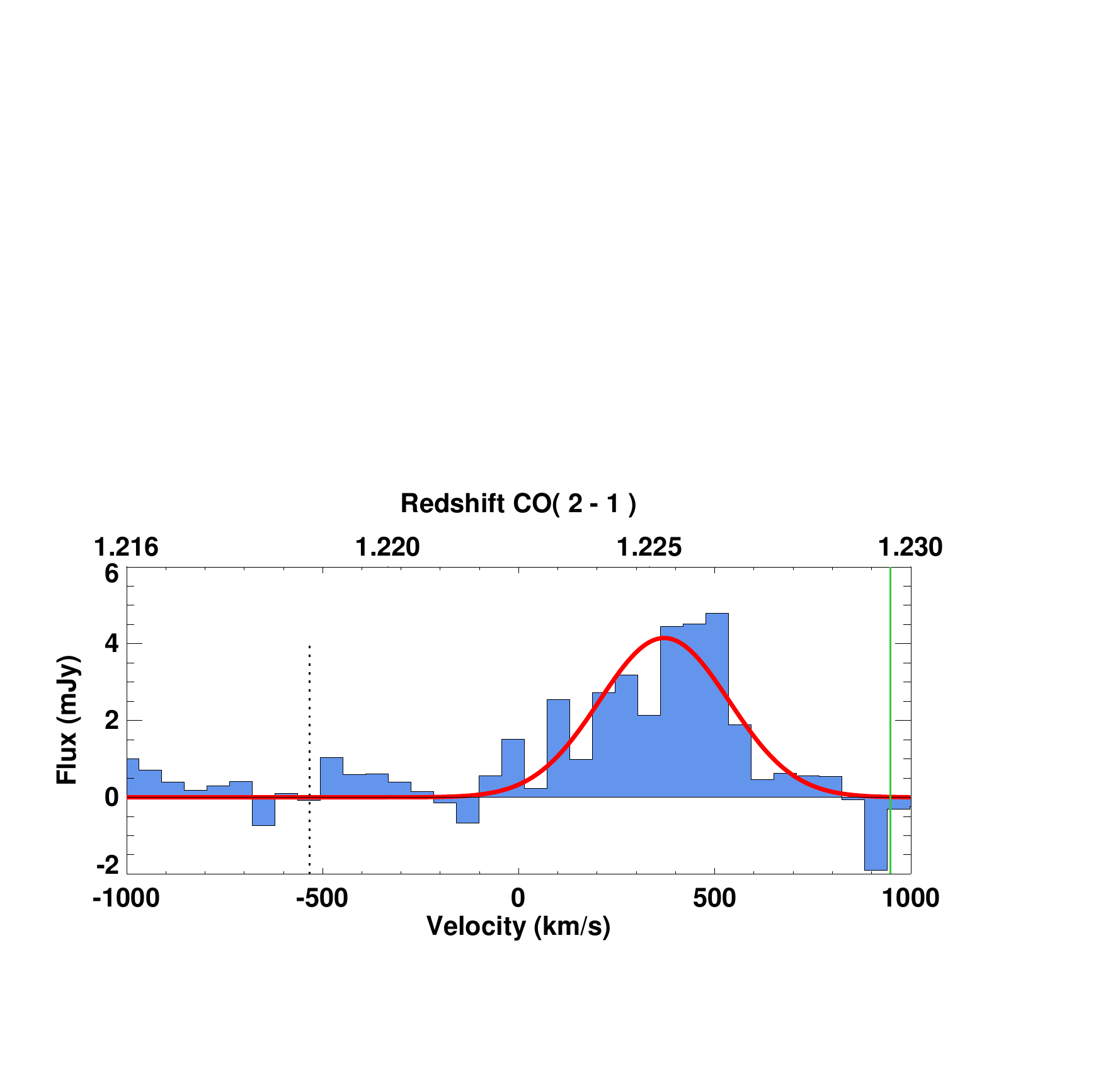}}
}
\caption[SMM123634+6212]{SMM123634+6212. CO contours are double spaced (2$\sigma$, 4$\sigma$, 6$\sigma$, etc). $^{12}$CO is strongly detected at $9.5\sigma$. All emission well aligned with the radio counterpart. This source was also observed in \jtwo\ by \cite{2008ApJ...680L..21F}, who report a $^{12}$CO luminosity and FWHM that matches the values reported here. Higher resolution CO(6-5) data  was also published in Engel et al.\ (2010), where they 
marginally resolve a velocity field within the essentially unresolved source (HWHP 2.2~kpc). }
\label{figure_pe32}
\end{figure*}

%%%%%%%%%%%%%%%%%%%%%%%%%%%%%%%%

\begin{figure*}
\centering
\mbox
{
  \subfigure{\includegraphics[width=8cm, clip=true, trim=50 350 70 0]{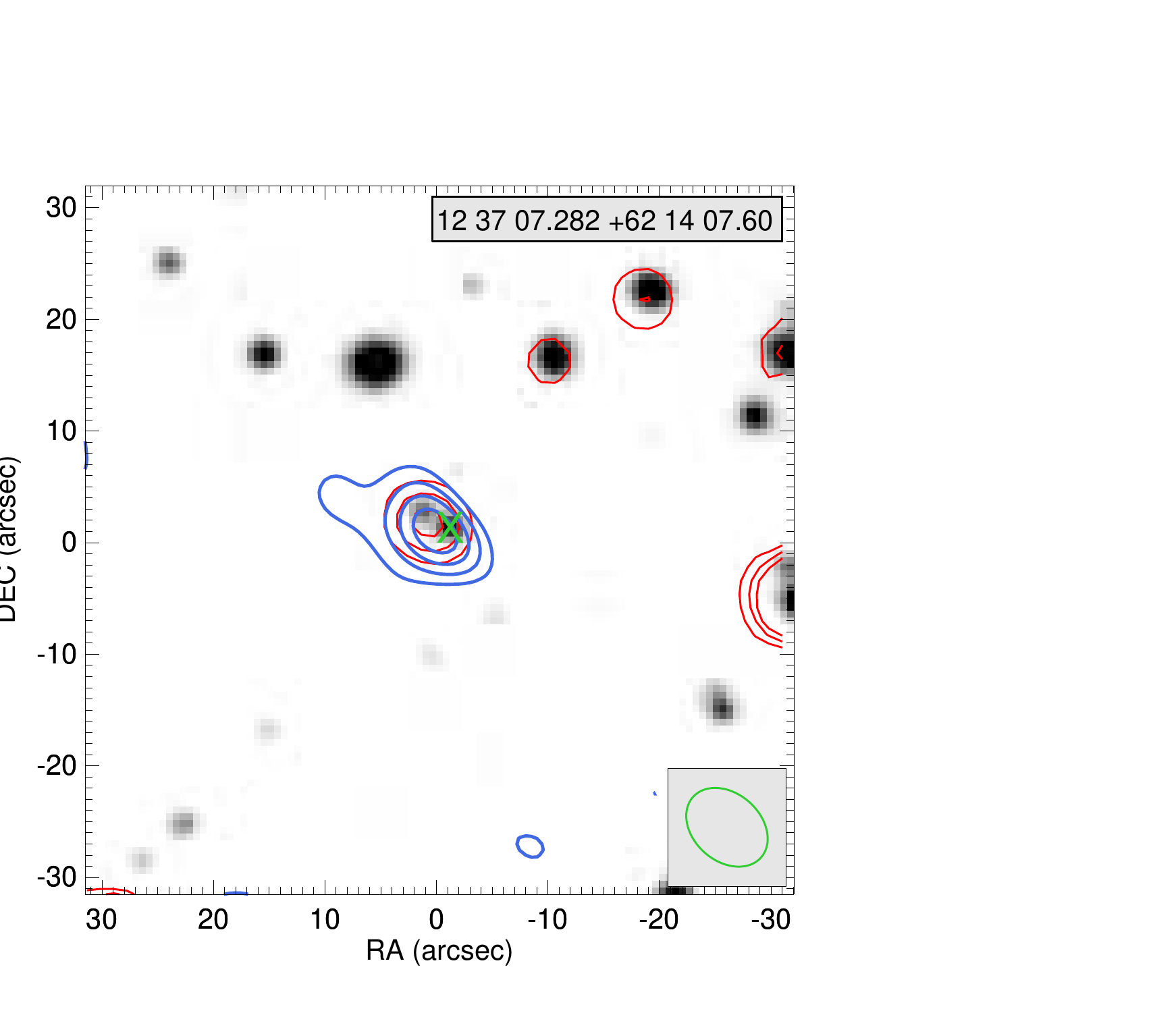}} \hspace{-2cm}
  \subfigure{\includegraphics[width=10cm, clip=true, trim=30 300 70 0]{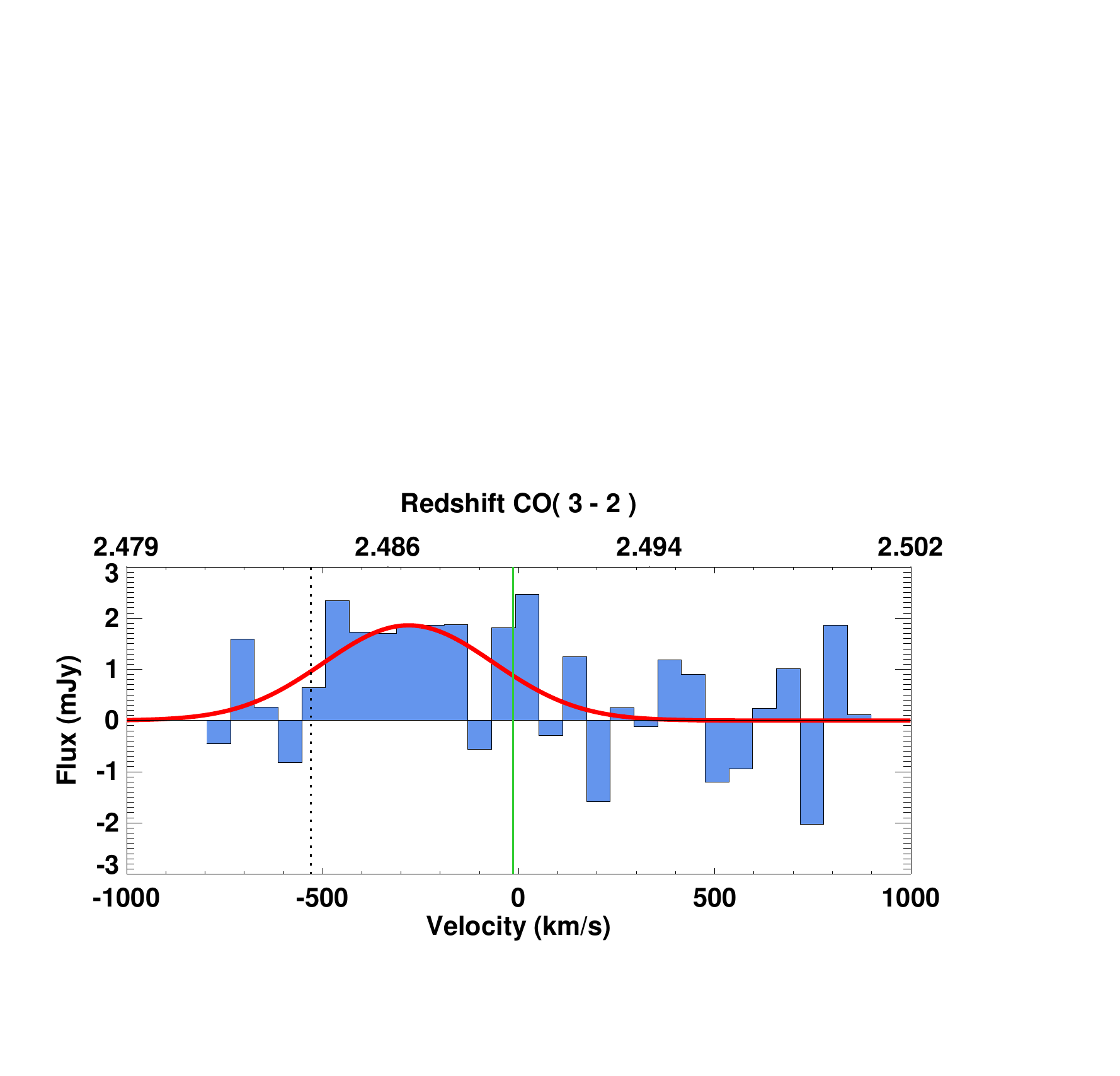}}
}
\caption{SMM123707+6214. $^{12}$CO is well detected at $5.9 \sigma$. All emission components line up well. High resolution imaging (Tacconi et al.\ 2006, 2008) show this source to consist of two $^{12}$CO emission components, separated by $\sim 2.5 ''$, mirrored by the double radio and IRAC sources.}
\label{figure_of2a}
\end{figure*}

%%%%%%%%%%%%%%%%%%%%%%%%%%%%%%%%

\begin{figure*}
\centering
\mbox
{
  \subfigure{\includegraphics[width=8cm, clip=true, trim=50 350 70 0]{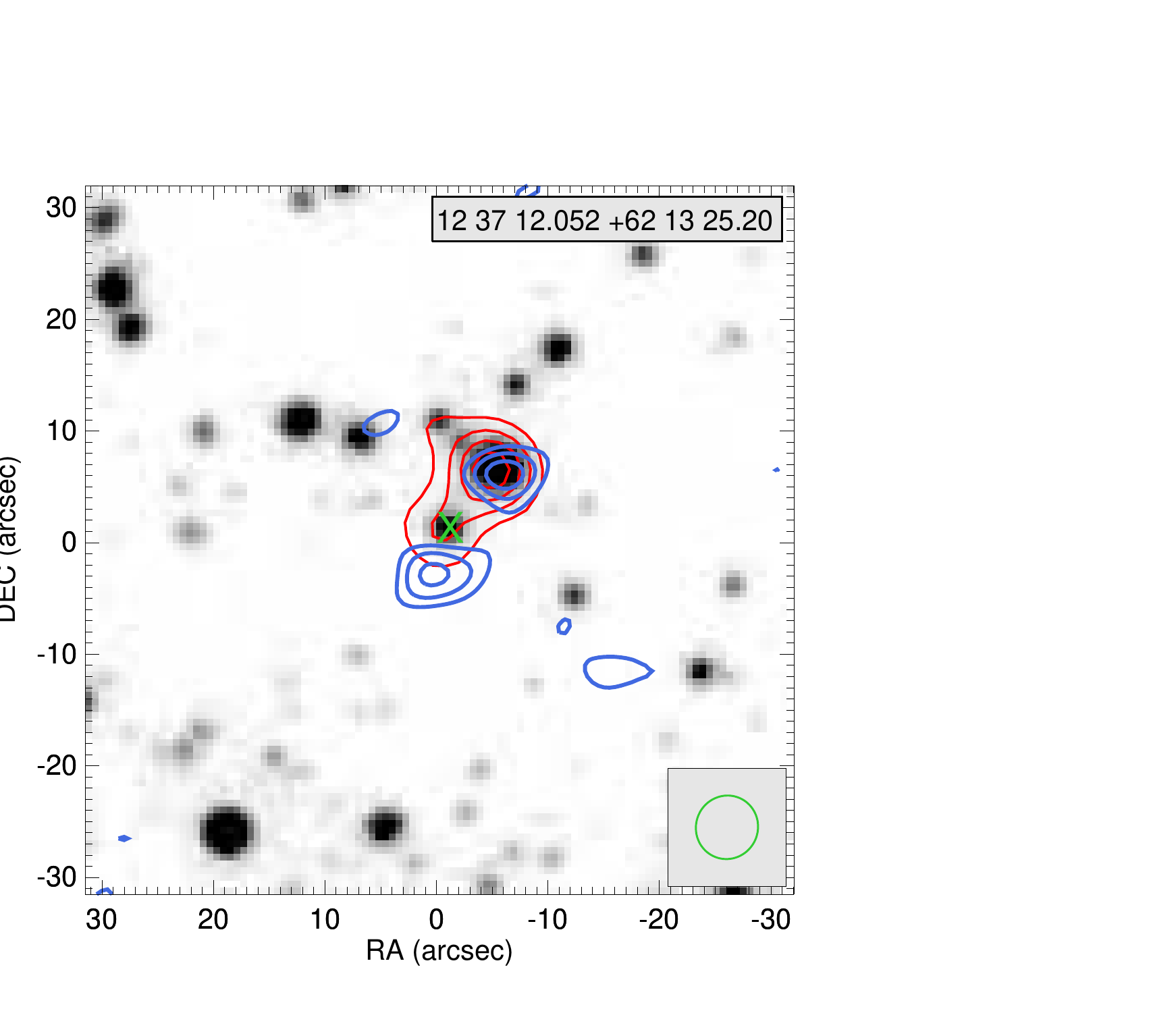}} \hspace{-2cm}
  \subfigure{\includegraphics[width=10cm, clip=true, trim=30 300 70 0]{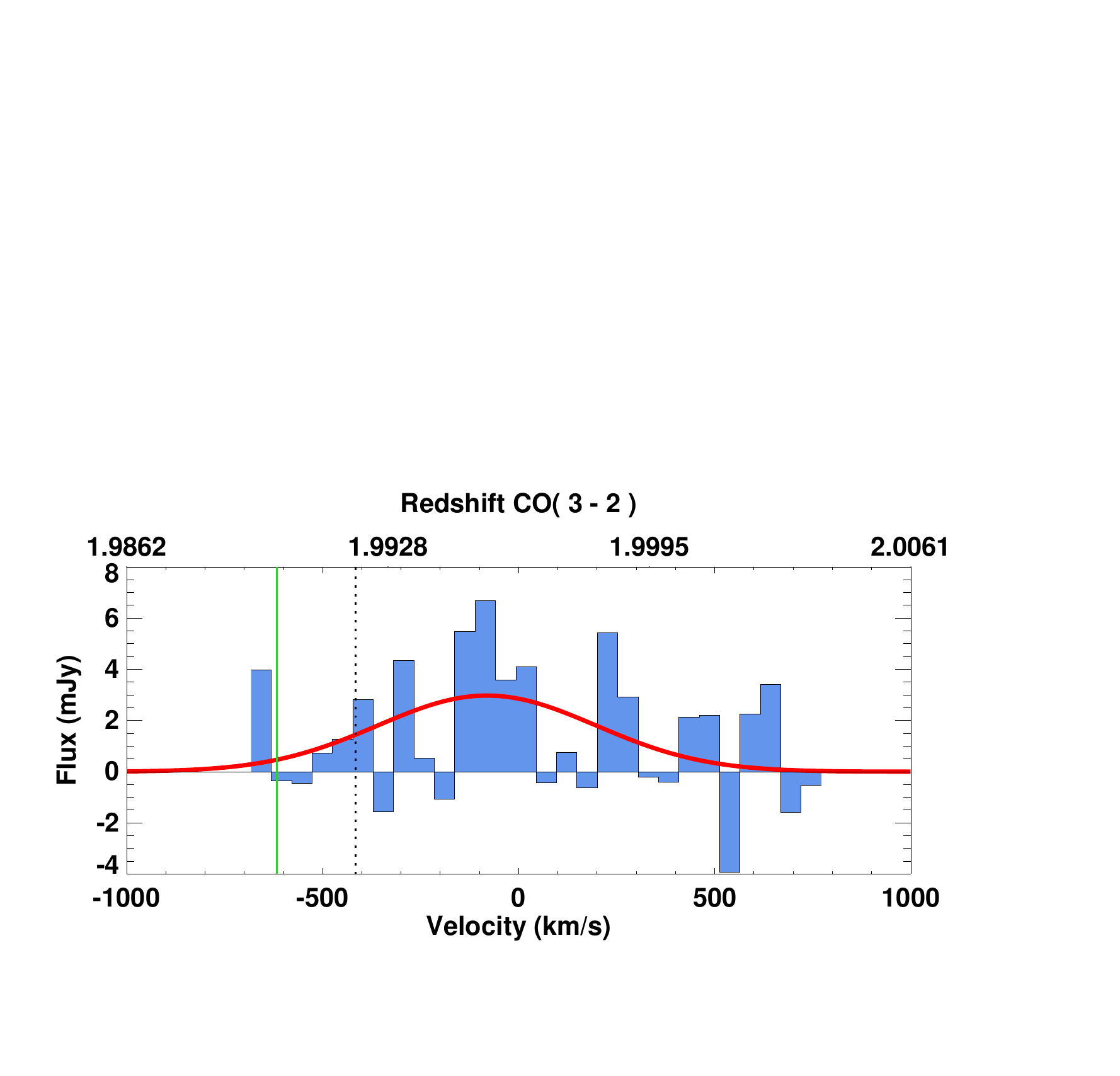}}
}
\caption[SMM123711+6213]{SMM123711+6213. The CO(3-2) detection for this source was originally published by \cite{2008ApJ...689..889C}. The spectrum shown is the NW $^{12}$CO source. $^{12}$CO is detected at $\sim 4.8 \sigma$ --  well aligned with radio, 24$\mu$m and IRAC emission. 
The green cross here indicates the phase centre lying between the two $^{12}$CO sources (see SMM123712+6213 below). 
SMM123711+6213 is also well detected at higher resolution (A-config) in $^{12}$CO$(4-3)$ at $>6 \sigma$, as reported in Bothwell et al.\ (2010), with lower resolution D-config
$^{12}$CO$(4-3)$ data presented in Casey et al.\ (2011). }
\label{figure_ne28_1}
\end{figure*}

%%%%%%%%%%%%%%%%%%%%%%%%%%%%%%%%

\begin{figure*}
\centering
\mbox
{
  \subfigure{\includegraphics[width=8cm, clip=true, trim=50 350 70 0]{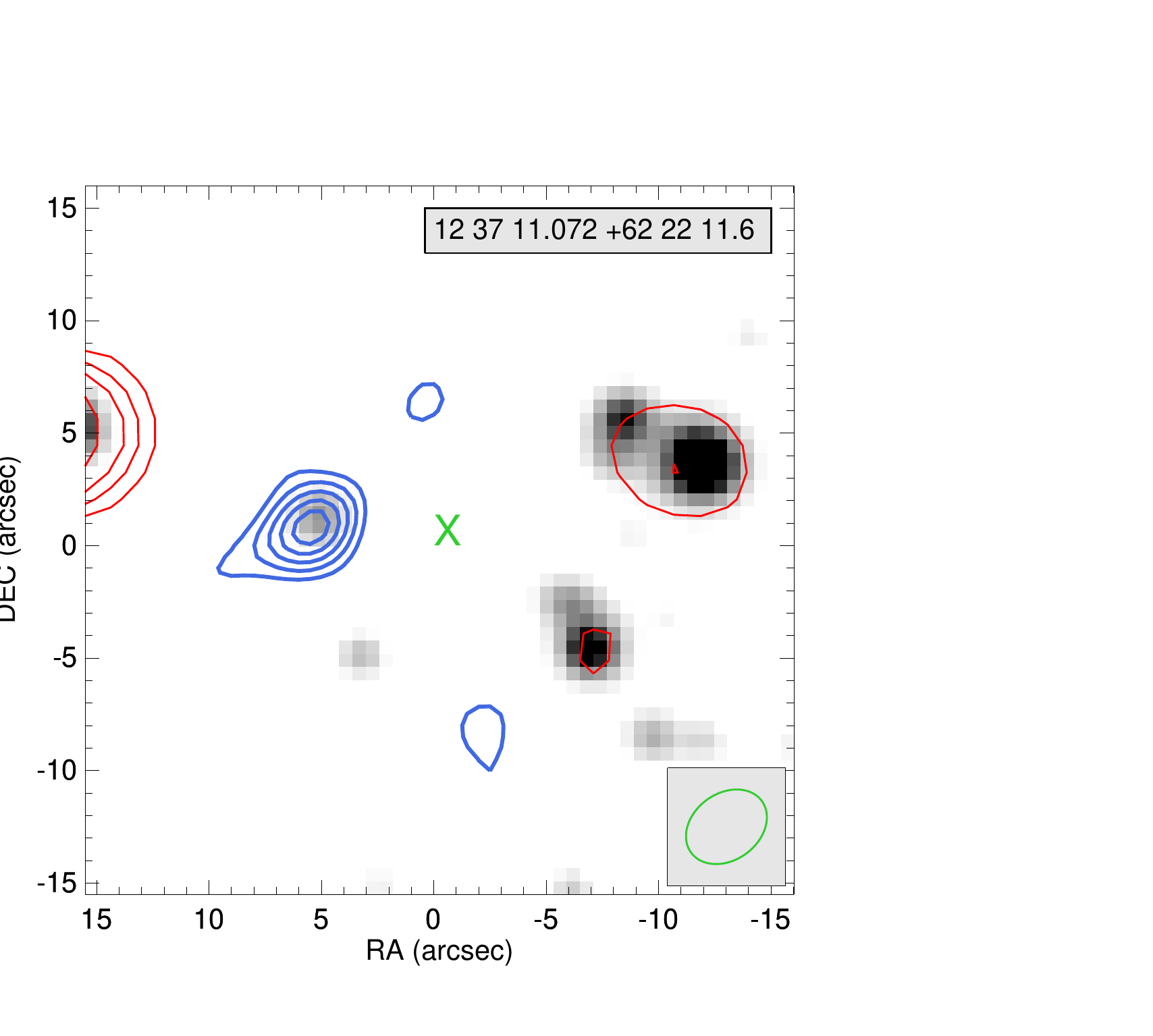}} \hspace{-2cm}
  \subfigure{\includegraphics[width=10cm, clip=true, trim=30 300 70 0]{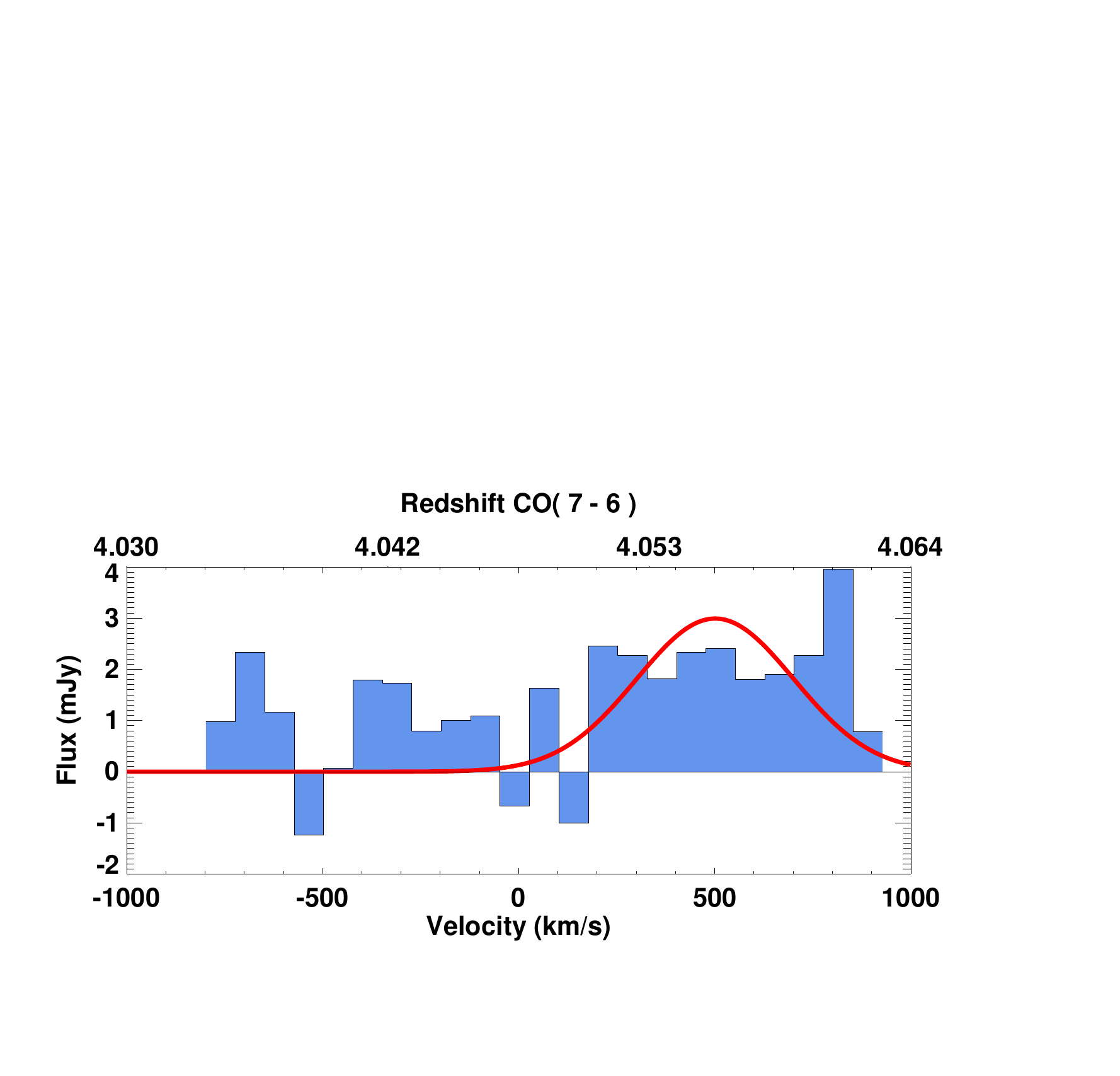}}
}
\caption{SMM123711+6222 = GN20. The phase centre (green cross) was chosen to lie between GN20 (east), and GN20.2 (west; not detected). Both galaxies are detected in lower-J imaging (Daddi et al. 2009). The spectrum consists of two lines; \jseven\ (fitted with a Gaussian function), and {\sc Ci}, approximately 700 km/s bluewards of this.
The data was previously presented in Casey et al.\ (2009), although their overestimated 2mm continuum strength led them to question the detection of the CO(7-6) line.}
\label{figure_sa3b}
\end{figure*}

%%%%%%%%%%%%%%%%%%%%%%%%%%%%%%%%

\begin{figure*}
\centering
\mbox
{
  \subfigure{\includegraphics[width=8cm, clip=true, trim=50 350 70 0]{catalogue_imgs/ne28-CO_spitzer}} \hspace{-2cm}
  \subfigure{\includegraphics[width=10cm, clip=true, trim=30 300 70 0]{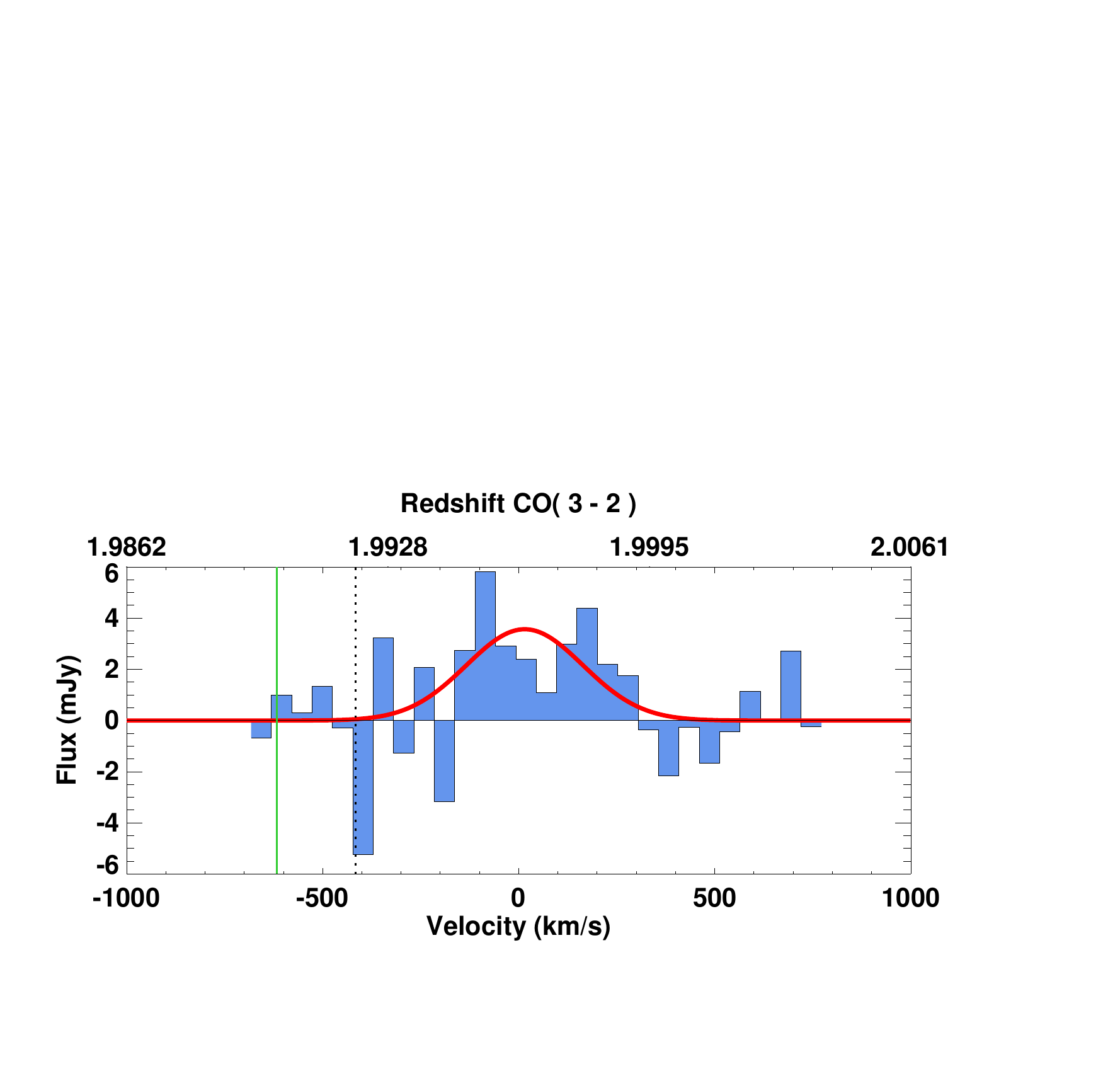}}
}
\caption[SMM123712+6213]{SMM123712+6213. The spectrum shown is the SE $^{12}$CO source. $^{12}$CO is detected at $\sim 4.7 \sigma$, but the source is offset from 5$''$ to the south of the radio position, and the mid-IR IRAC position. However, this source is also detected at 4.5$\sigma$ 
in $^{12}$CO$(4-3)$ at the same southern offset position, as reported in Bothwell et al.\ (2010), lending additional weight to the reality of the 5$''$ offset.}
\label{figure_ne28_2}
\end{figure*}

%%%%%%%%%%%%%%%%%%%%%%%%%%%%%%%%%%%%%%%%%%%%%%%%

\begin{figure*}
\centering
\mbox
{
  \subfigure{\includegraphics[width=7.8cm, clip=true, trim=0 0 0 0]{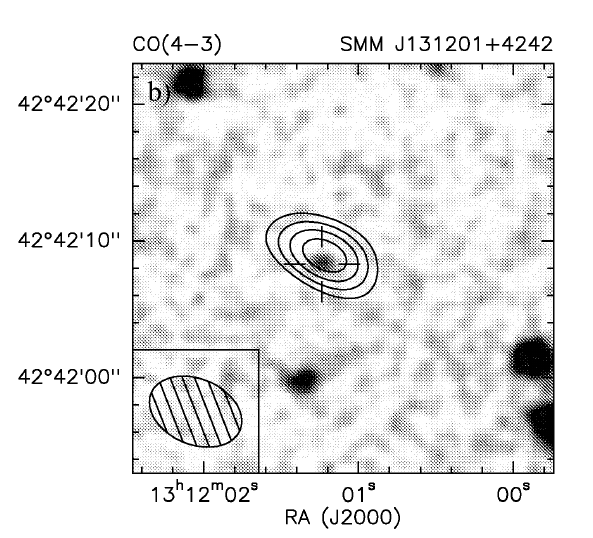}} \hspace{0cm}
  \subfigure{\includegraphics[width=10cm, clip=true, trim=30 300 70 0]{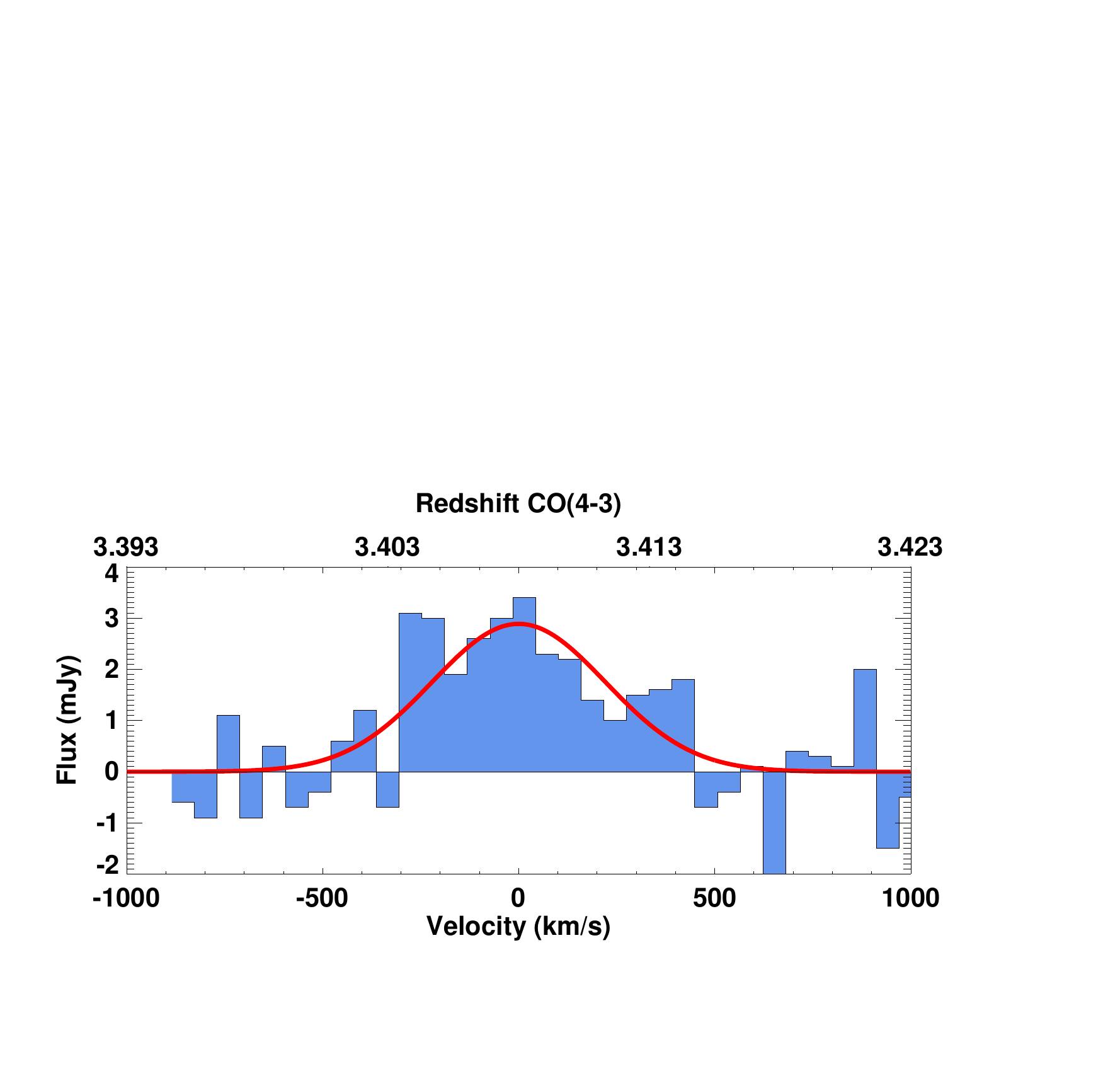}}
}
\caption{SMM131201+4242.  $^{12}$CO is detected at $\sim 5.7 \sigma$. The image is taken from Greve et al.\ (2005). Higher resolution CO(6-5) data is published in Engel et al.\ (2010), resolving the CO into three components.}
\label{figure_sa13}
\end{figure*}

%%%%%%%%%%%%%%%%%%%%%%%%%%%%%%%%

\begin{figure*}
\centering
\mbox
{
  \subfigure{\includegraphics[width=8cm, clip=true, trim=50 350 70 0]{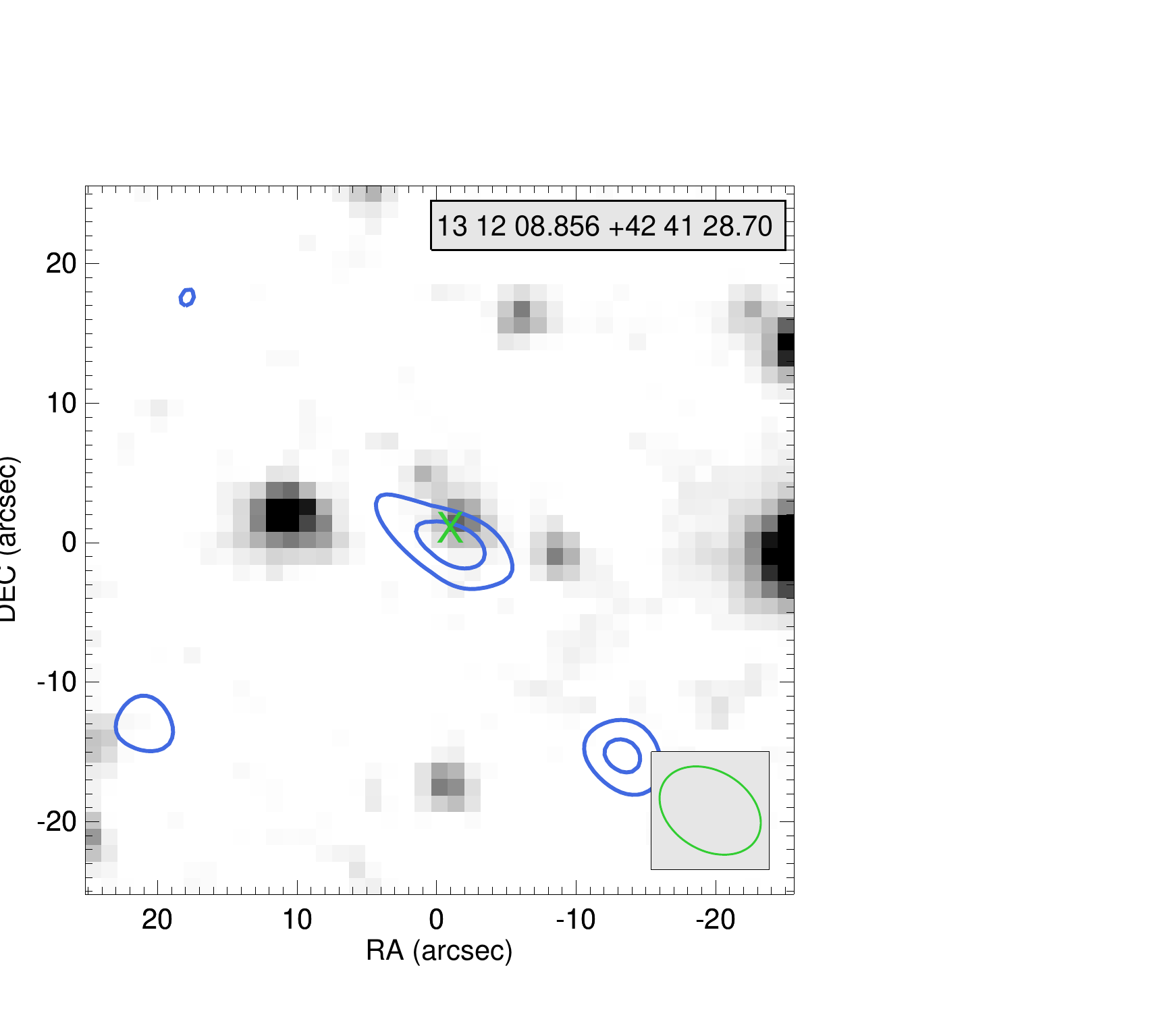}}  \hspace{-2cm}
  \subfigure{\includegraphics[width=10cm, clip=true, trim=30 300 70 0]{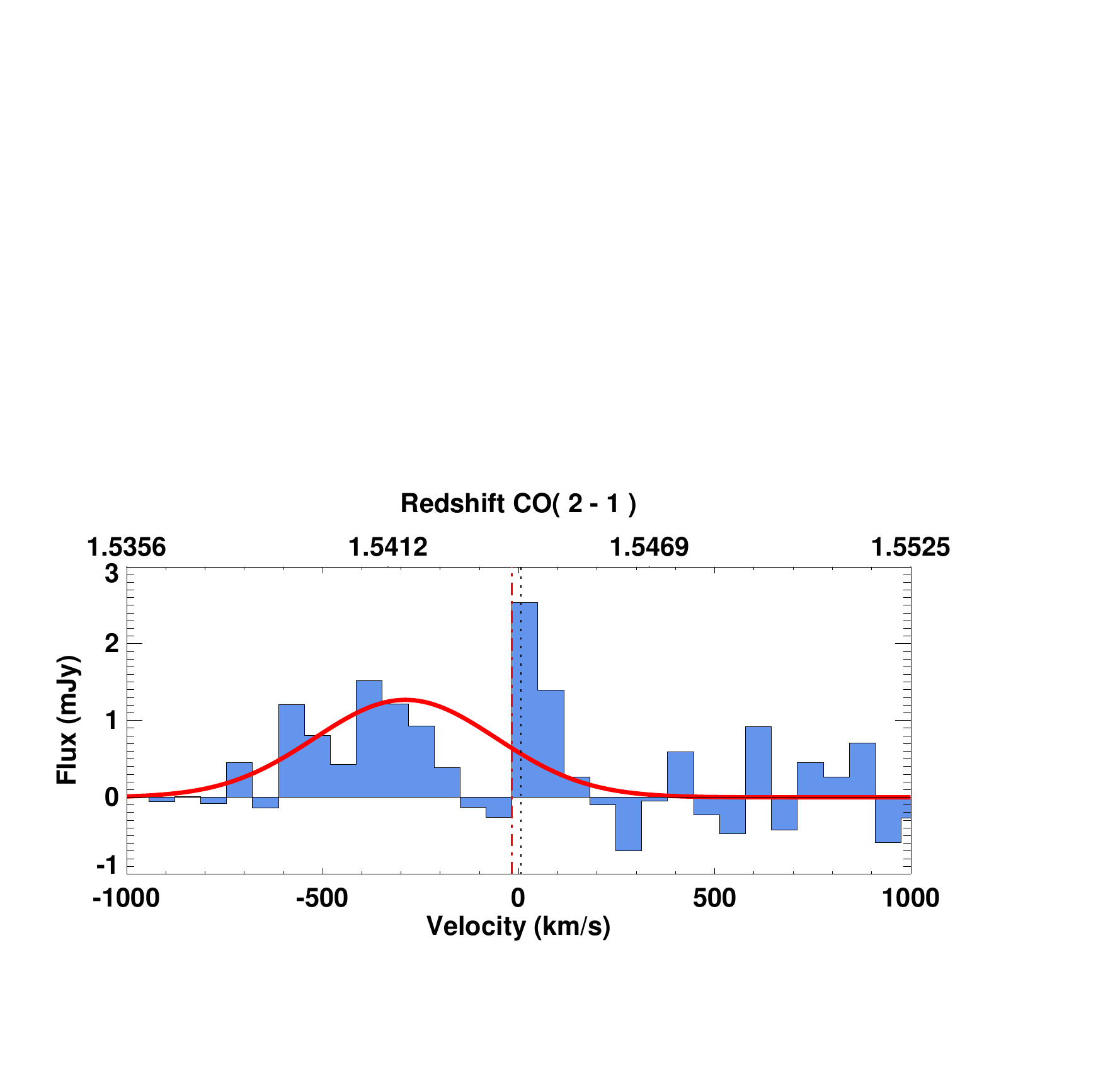}} 
}
\caption{SMM131208+4241. The source is weakly detected in $^{12}$CO at $3.6 \sigma$, but is aligned with the radio counterpart, and with the weak IRAC identification where the optical redshifts were measured. We identify this as a candidate detection mainly based on the weak CO detection.}
\label{figure_nb61}
\end{figure*}

%%%%%%%%%%%%%%%%%%%%%%%%%%%%%%%%

\begin{figure*}
\centering
\mbox
{
  \subfigure{\includegraphics[width=8cm, clip=true, trim=50 350 70 0]{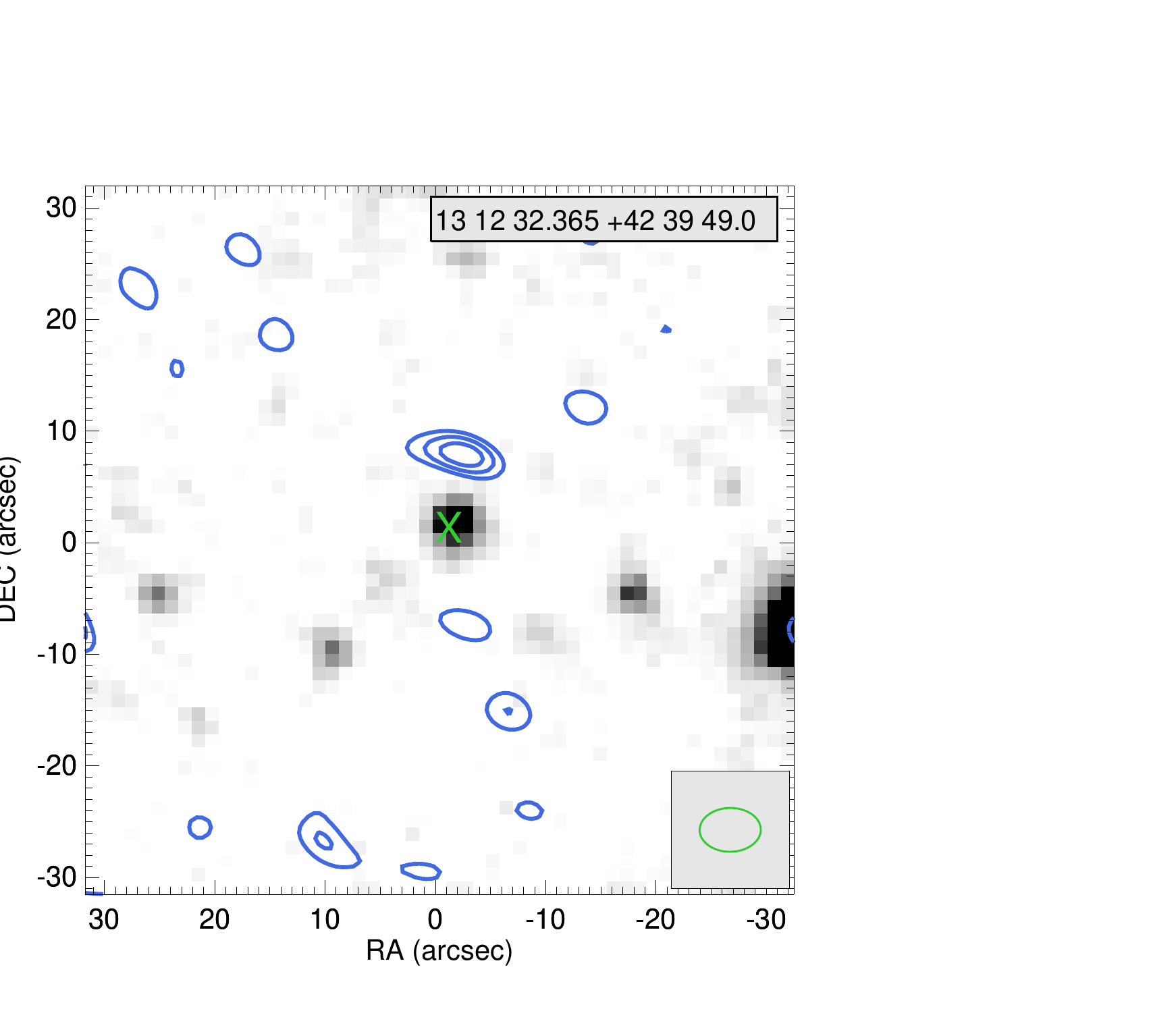}} \hspace{-2cm}
  \subfigure{\includegraphics[width=10cm, clip=true, trim=30 300 70 0]{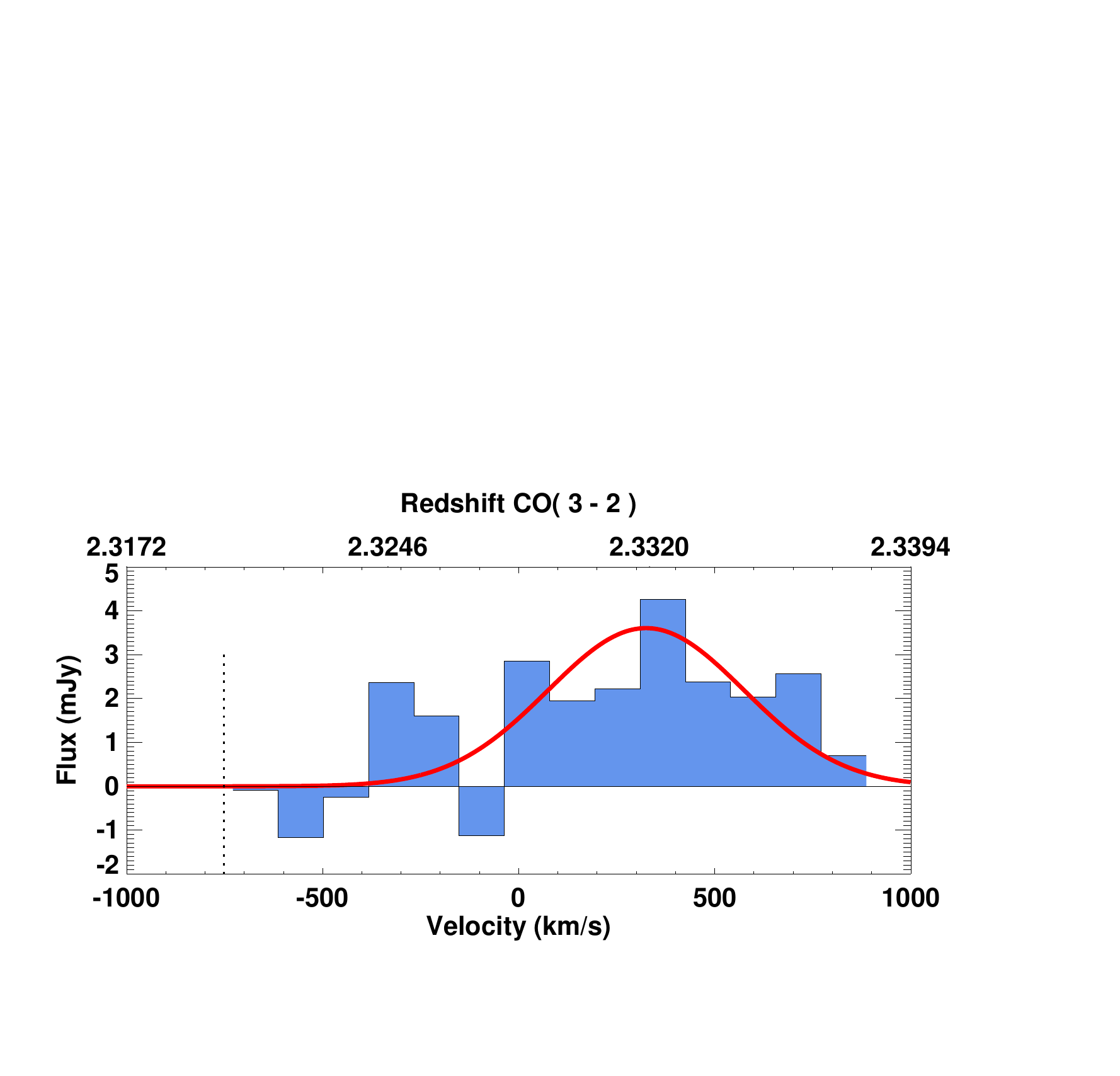}}
}
\caption{SMM131232+4239. $^{12}$CO is detected at $4.9\sigma$, but is offset to the north by 8''. The H$\alpha$ redshift  (obtained for the radio-identified IRAC source shown) is significantly offset from the CO as well  -- ($z = 2.300$ from H$\alpha$, while  $z = 2.320$  is measured from UV Ly$\alpha$ line, and the $^{12}$CO line gives $z=2.332$.). This source could represent a spatially extended merging system, similar to the SMG complex reported by Ivison et al. (2011). We identify this as a candidate detection based on the large offsets spatially and in velocity. }
\label{figure_og2a}
\end{figure*}

%%%%%%%%%%%%%%%%%%%%%%%%%%%%%%%%

\begin{figure*}
\centering
\mbox
{
  \subfigure{\includegraphics[width=8cm, clip=true, trim=50 350 70 0]{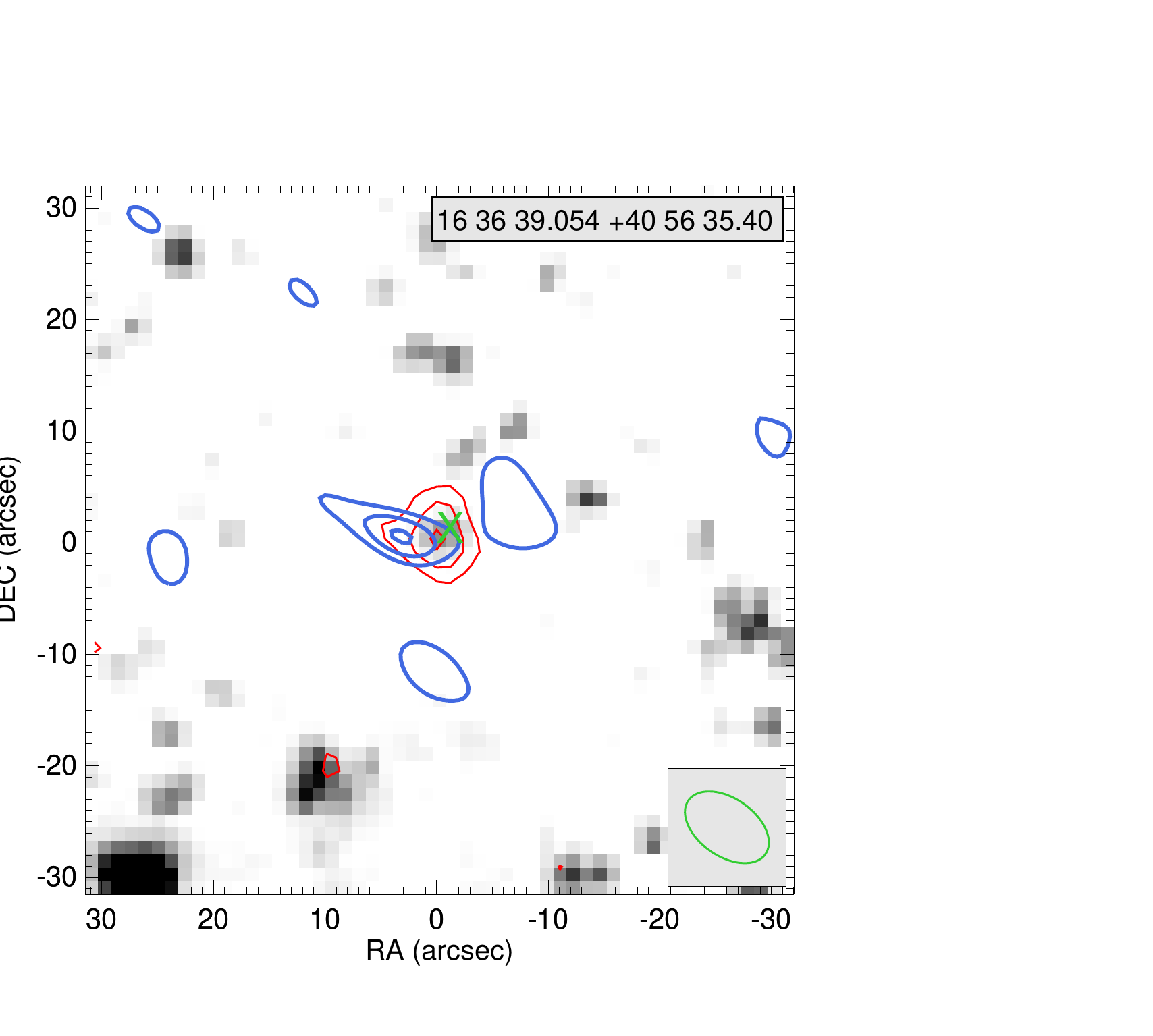}} \hspace{-2cm}
  \subfigure{\includegraphics[width=10cm, clip=true, trim=30 300 70 0]{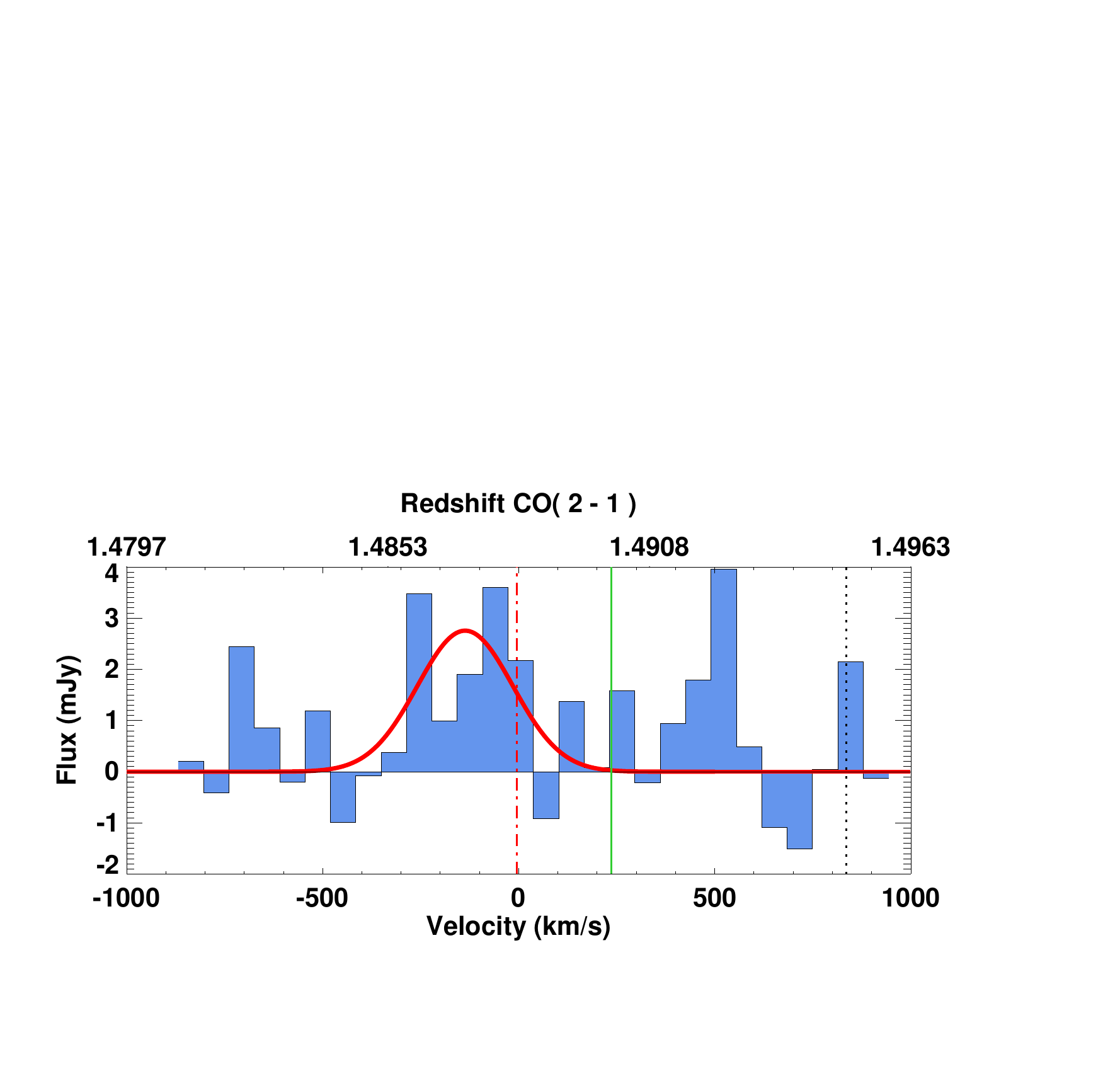}}
}
\caption{SMM163639+4056. $^{12}$CO is detected at $4.5 \sigma$, is offset 6$''$ to the east of the radio position, but is consistent within 2$\sigma$ within the elongated beam shape.
 We identify this as a candidate detection. }
\label{figure_nk28}
\end{figure*}

%%%%%%%%%%%%%%%%%%%%%%%%%%%%%%%%%%%%%%%%%%%%%%%%

\begin{figure*}
\centering
\mbox
{
  \subfigure{\includegraphics[width=8cm, clip=true, trim=50 350 70 0]{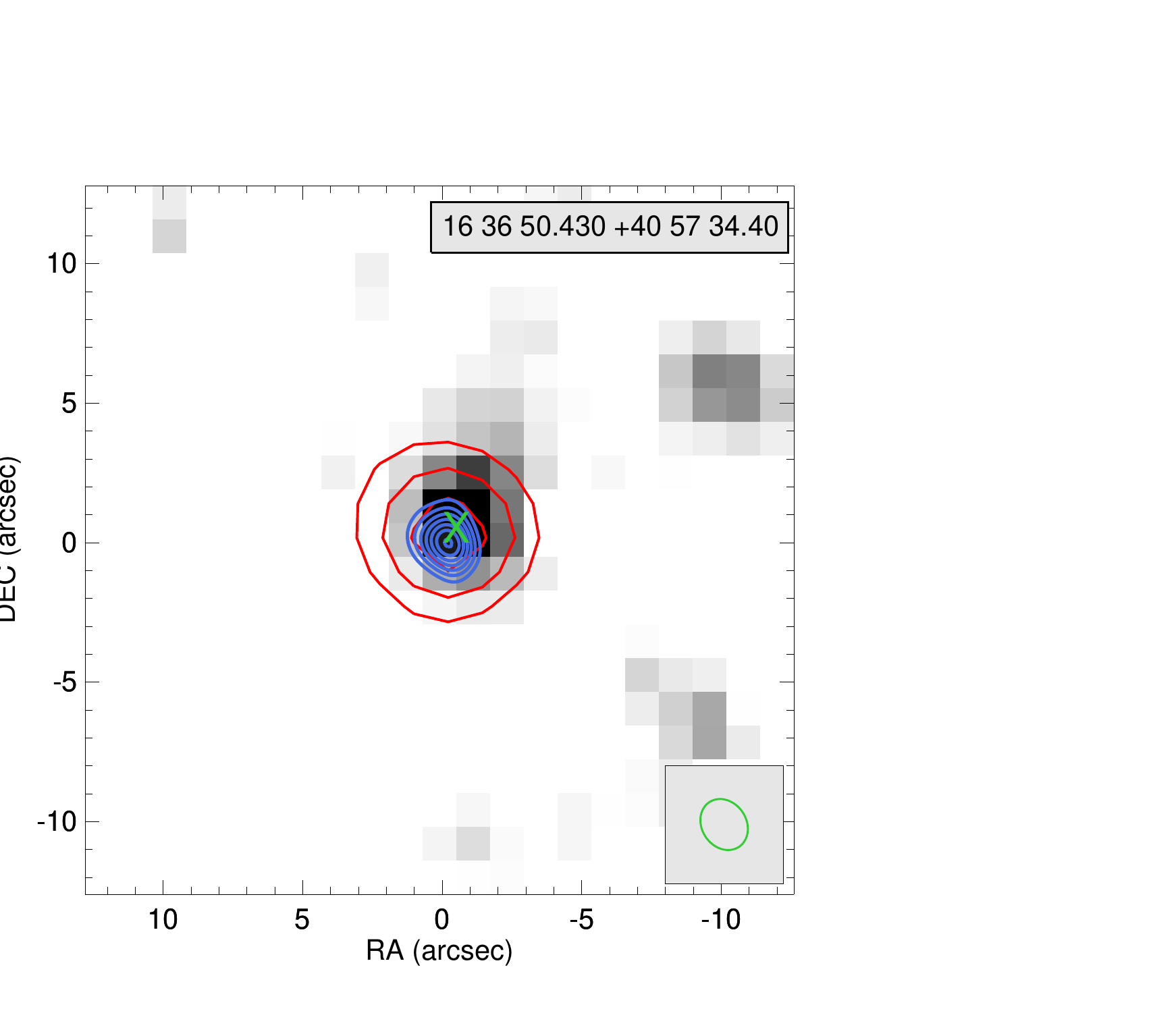}} \hspace{-3cm}
  \subfigure{\includegraphics[width=10cm, clip=true, trim=30 300 70 0]{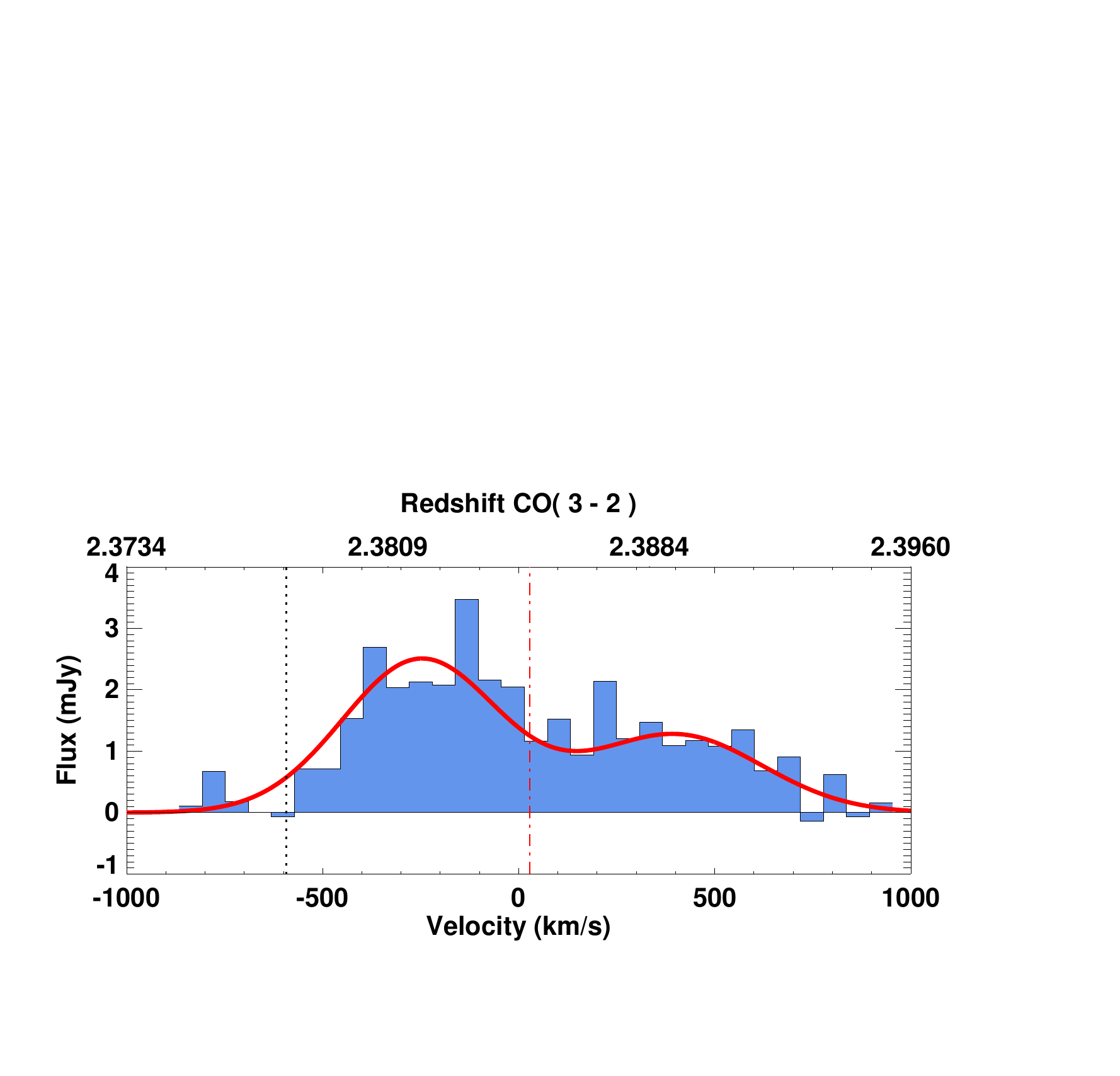}} 
}
\caption[SMM163650+4057]{SMM163650+4057. $^{12}$CO is strongly detected at $8.1 \sigma$. The source is lined up with 24$\mu$m and Near-IR emission at the position of the radio counterpart. This is a higher resolution image, and as a result the map size is shown as 25x25''. The data was previously published by Tacconi et al.\ (2006, 2008) along with higher resolution CO(7-6) data showing an elongated linear feature. }
\label{figure_nb38}
\end{figure*}

%%%%%%%%%%%%%%%%%%%%%%%%%%%%%%%%%%%%%%%%%%%%%%%%

\begin{figure*}
\centering
\mbox
{
  \subfigure{\includegraphics[width=8cm, clip=true, trim=50 350 70 0]{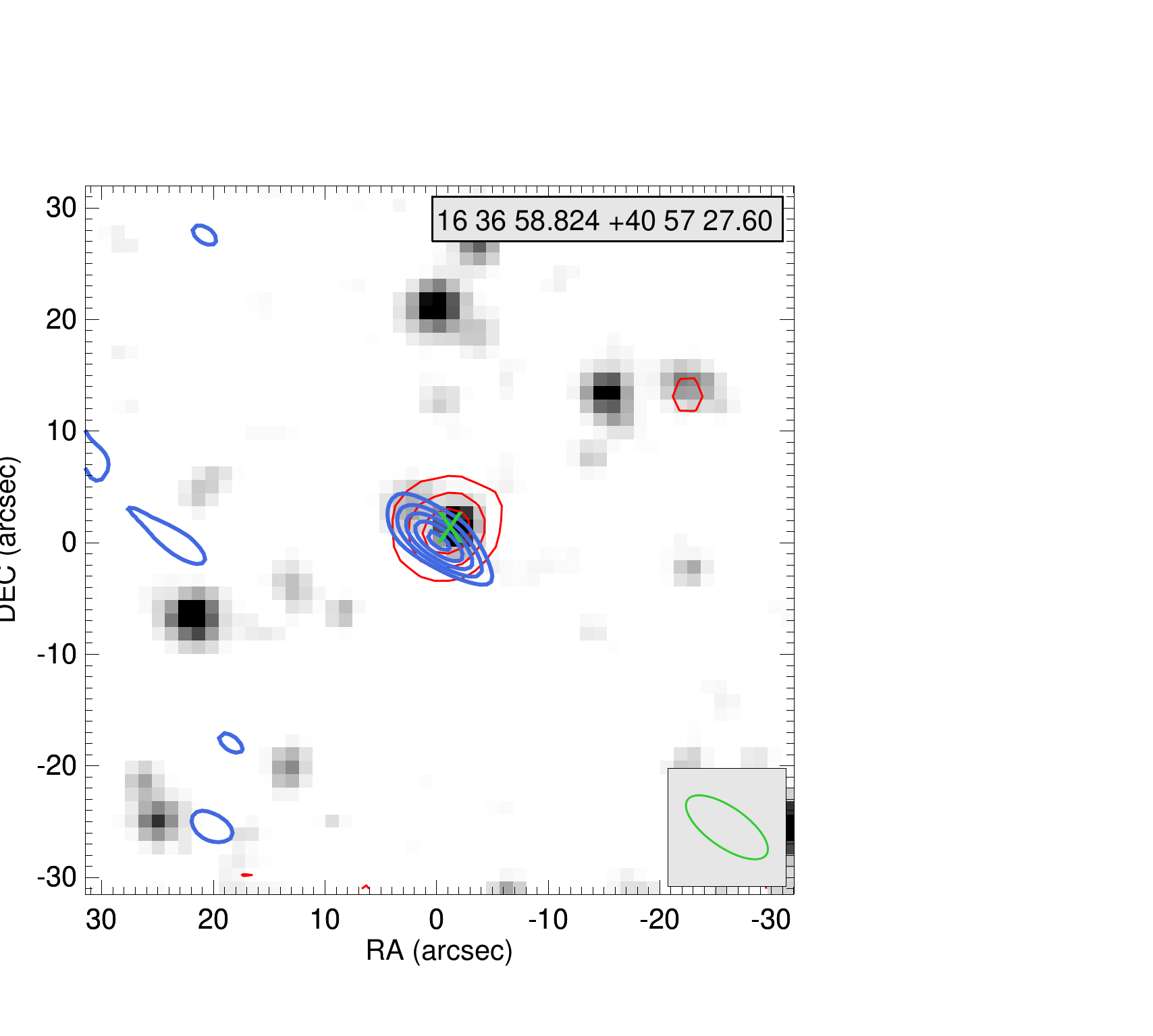}} \hspace{-2cm}
  \subfigure{\includegraphics[width=10cm, clip=true, trim=30 300 70 0]{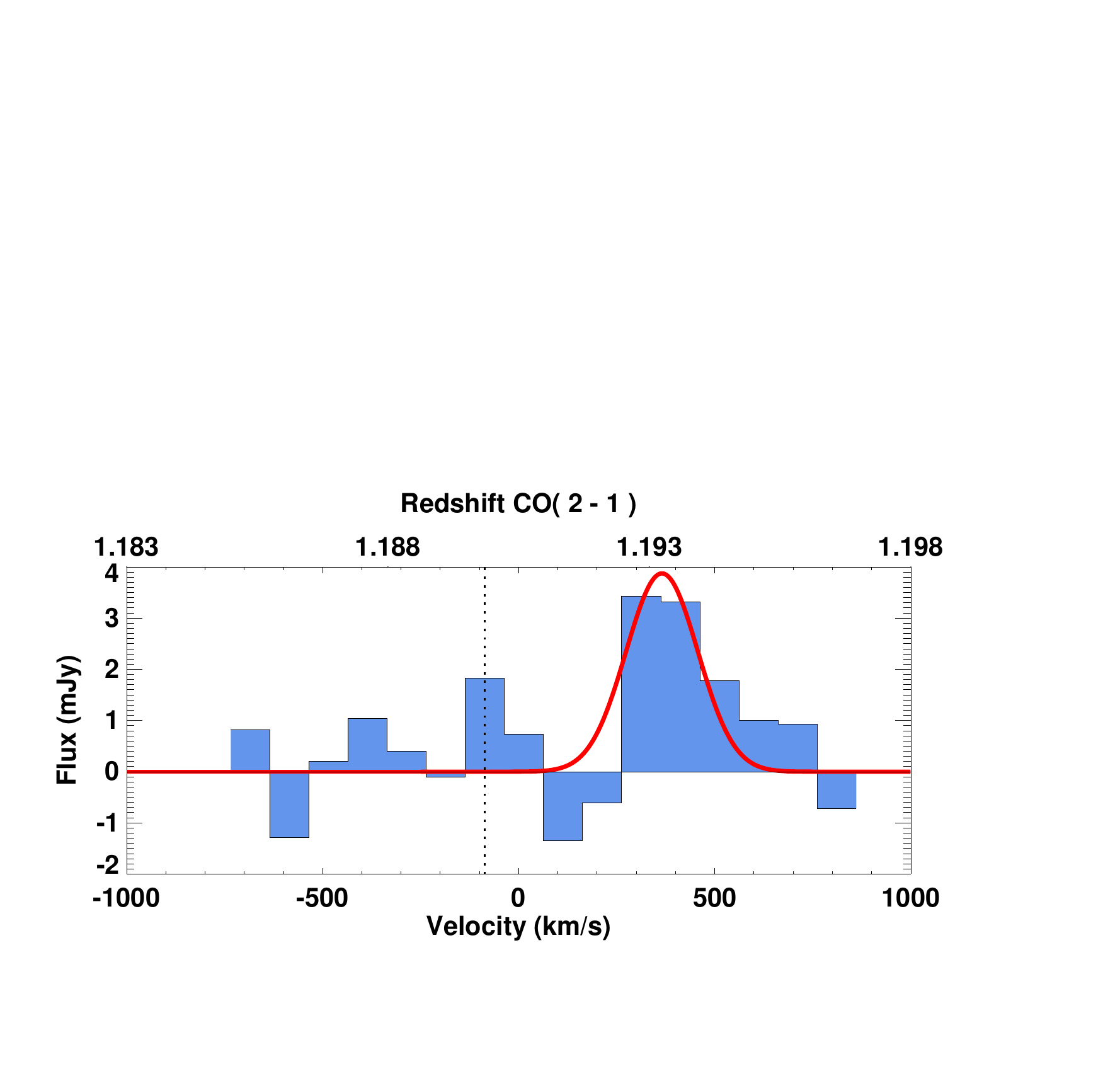}}
}
\caption{SMM163658+4057. $^{12}$CO is detected at $6.3 \sigma$. The PAH redshift is slightly high, at $z=1.20$ (but still consistent, given the typical $dz\sim0.02$ error in fitting PAH features). 
}
\label{figure_oh67}
\end{figure*}

%%%%%%%%%%%%%%%%%%%%%%%%%%%%%%%%%%%%%%%%%%%%%%%%

\begin{figure*}
\centering
\mbox
{
  \subfigure{\includegraphics[width=8cm, clip=true, trim=50 350 70 0]{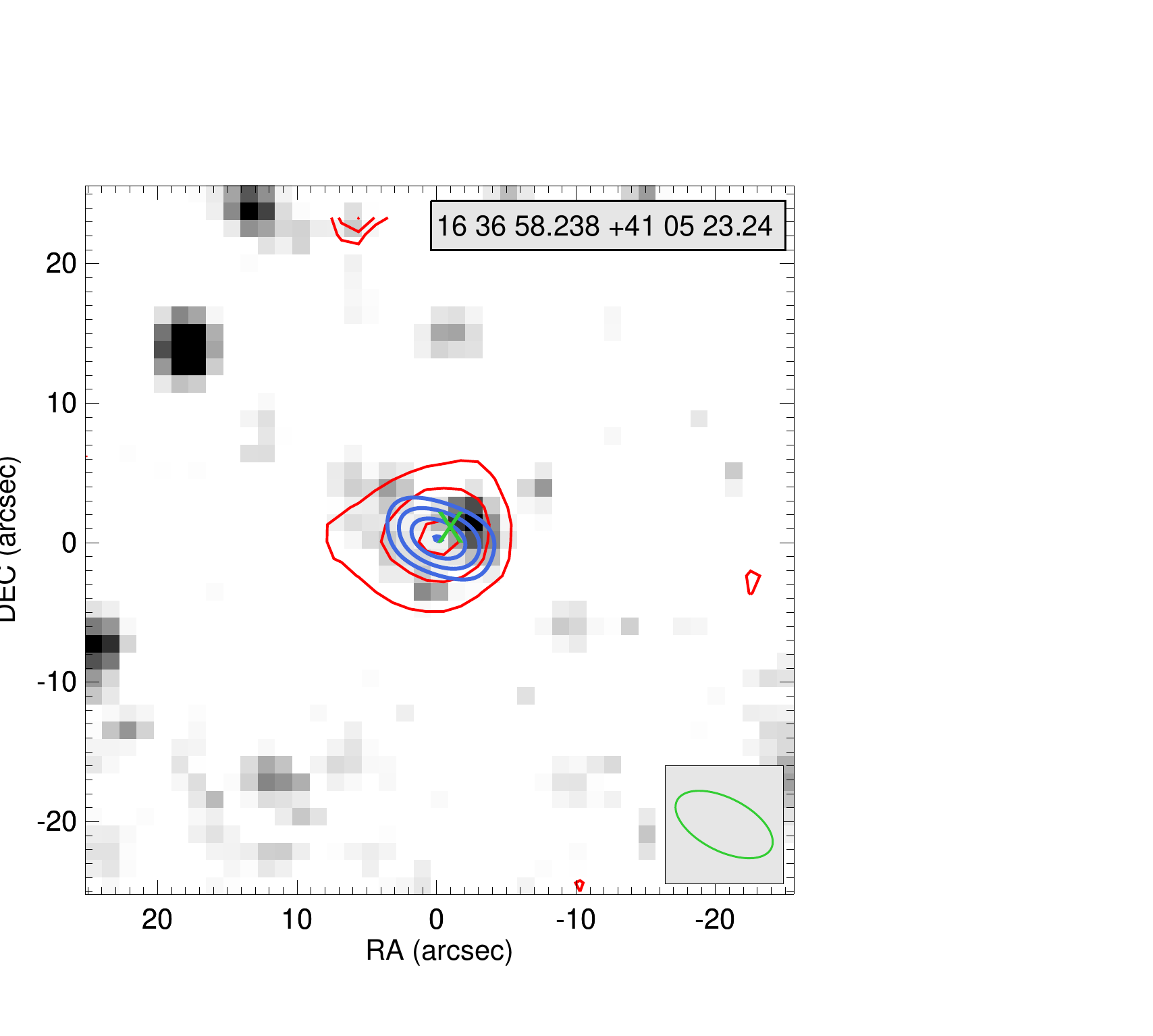}} \hspace{-2cm}
  \subfigure{\includegraphics[width=10cm, clip=true, trim=30 300 70 0]{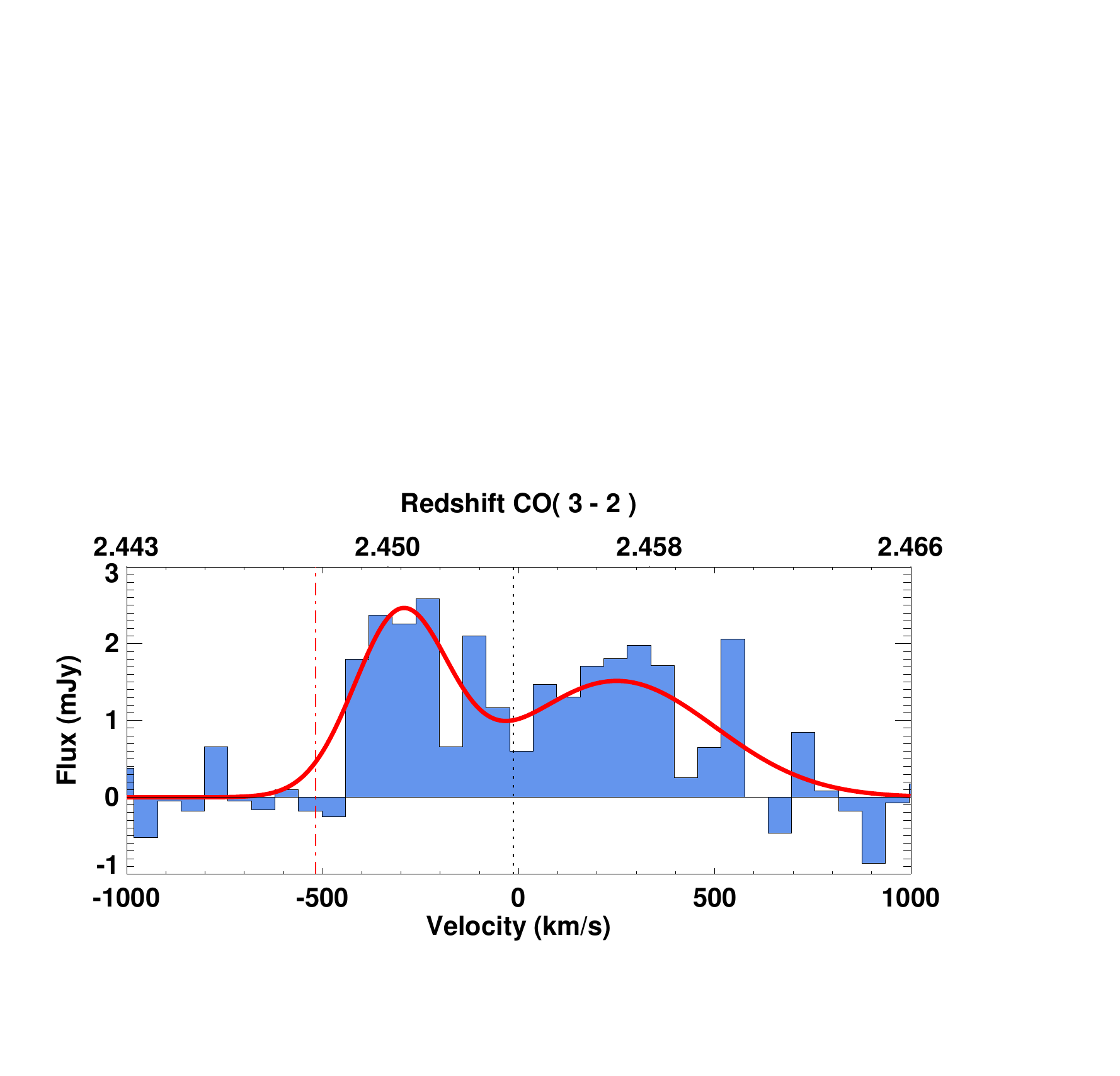}} 
}
\caption{SMM163658+4105. $^{12}$CO is well detected at $5.0 \sigma$. Source is well aligned with 24$\mu$m emission. This source was previously published by Greve et al.\ (2005) and Tacconi et al. (2006, 2008), the higher
resolution, higher J data showing a very compact CO(7-6) source (HWHP=0.8 kpc).  
}
\label{figure_nc38}
\end{figure*}

%%%%%%%%%%%%%%%%%%%%%%%%%%%%%%%%

\begin{figure*}
\centering
\mbox
{
  \subfigure{\includegraphics[width=8cm, clip=true, trim=50 350 70 0]{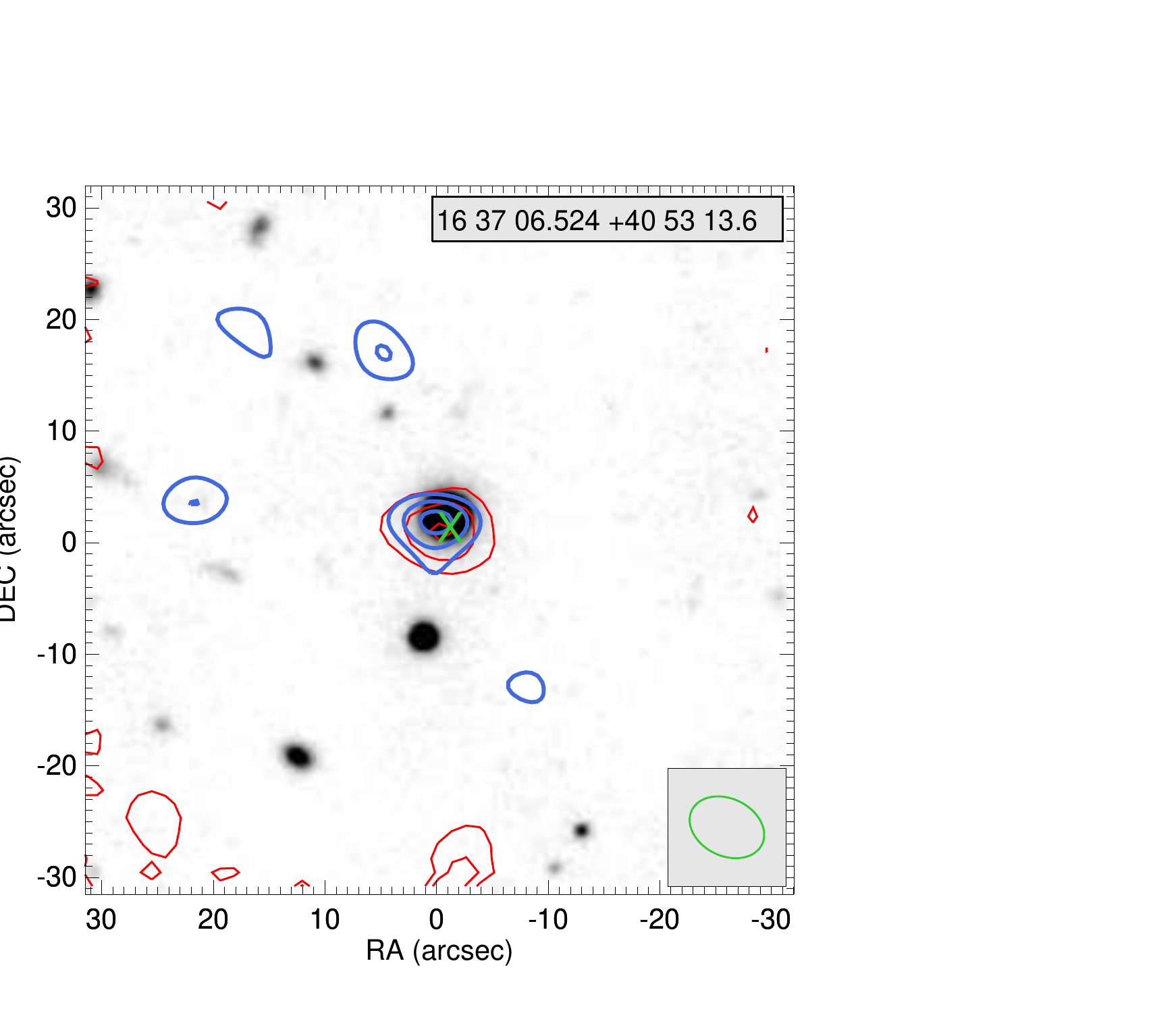}} \hspace{-2cm}
  \subfigure{\includegraphics[width=10cm, clip=true, trim=30 300 70 0]{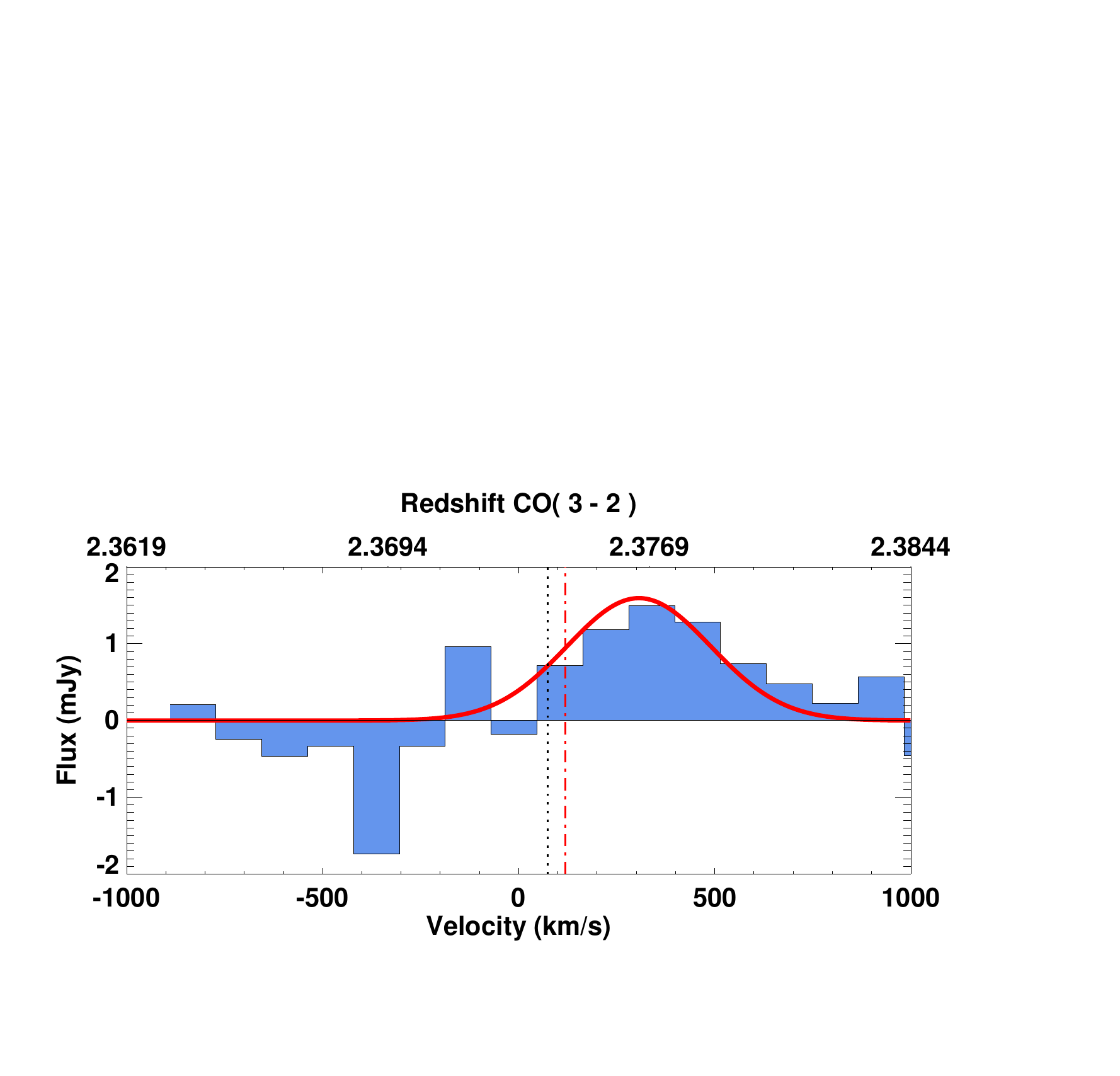}}
}
\caption[SMM163706+4053]{SMM163706+4053. The background image is $I$ band, rather than a composite IRAC image. $^{12}$CO is detected at $4.4\sigma$ and aligned with the radio counterpart.
The source was previously published in Greve et al.\ (2005). }
\label{figure_nj28}
\end{figure*}

%%%%%%%%%%%%%%%%%%%%%%%%%%%%%%%%

\begin{figure*}
\centering
\mbox
{
  \subfigure{\includegraphics[width=8cm, clip=true, trim=50 350 70 0]{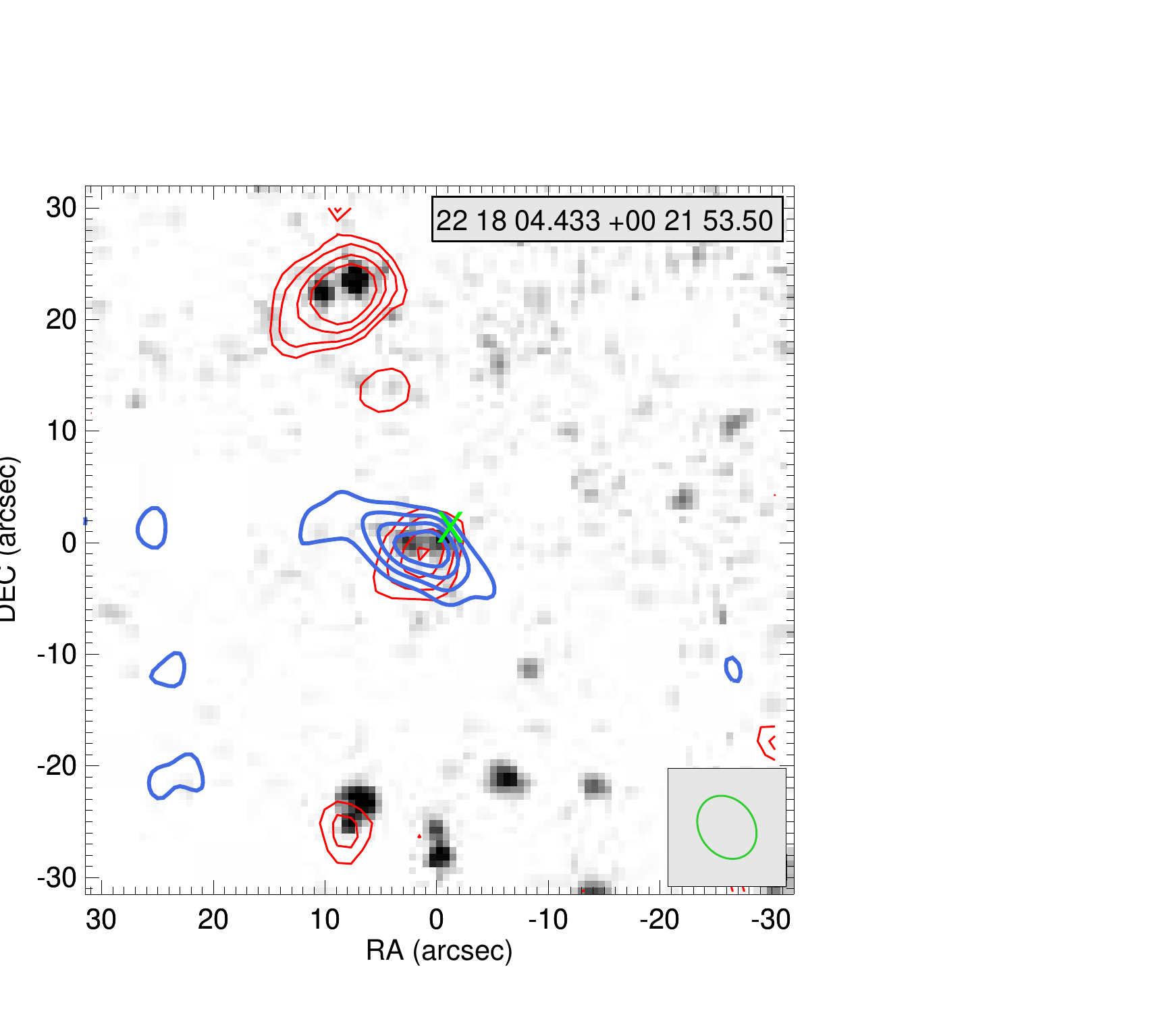}} \hspace{-2cm}
  \subfigure{\includegraphics[width=10cm, clip=true, trim=30 300 70 0]{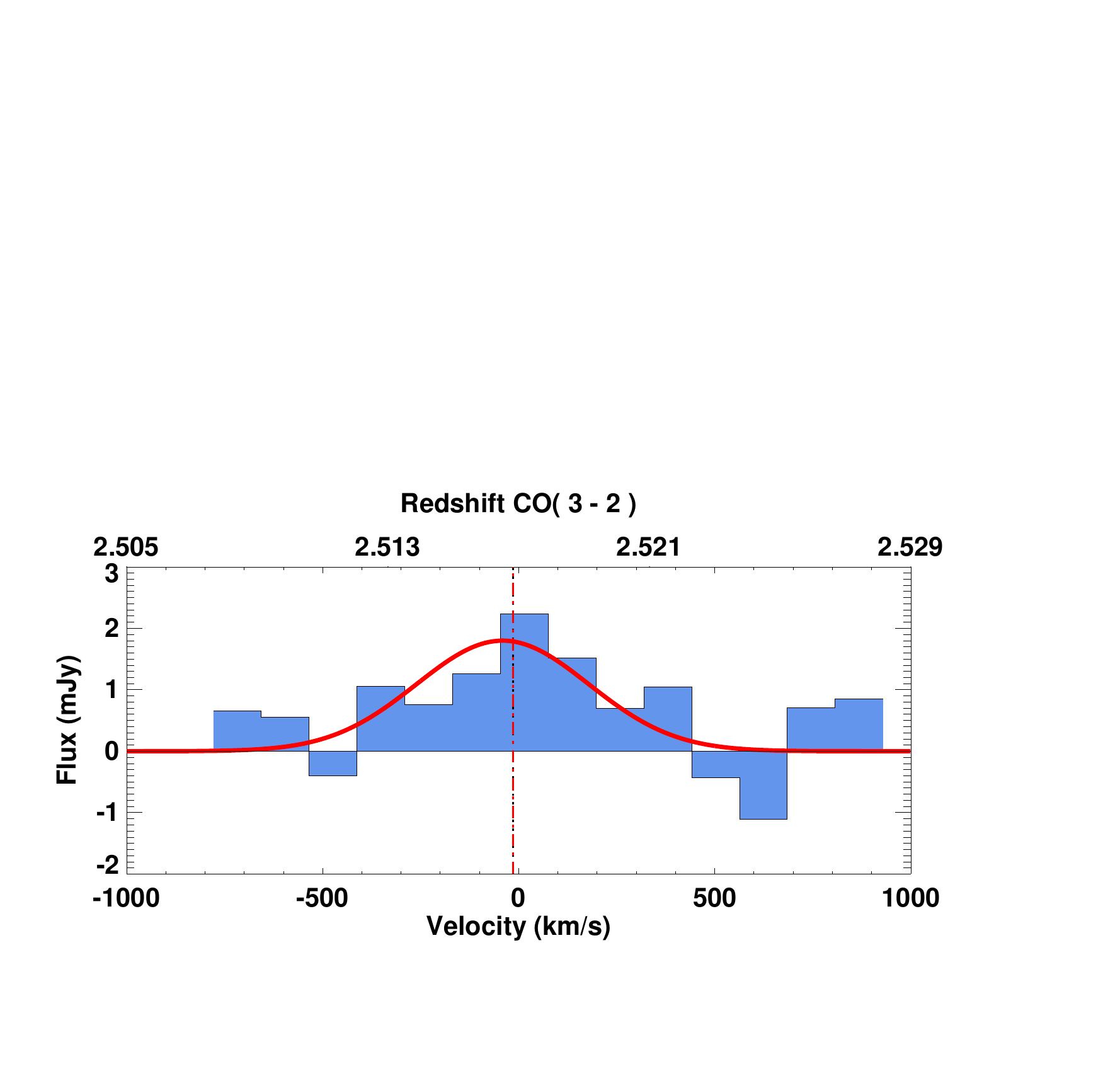}}
}
\caption{SMM221704+0021. $^{12}$CO detected at $6.0 \sigma$, and is lined up with the 24$\mu$m emission, and two IRAC sources, both of which are detected in H$\alpha$. The PAH redshift (z=2.55) is slightly inconsistent (H$\alpha$ and UV give a similar redshift to the CO, $z = 2.517$), but again within the typical PAH fitting errors. This source is detected in \jfour\   and in  [CI](1-0) in Alaghband-Zadeh et al. (in prep).}
\label{figure_nf28}
\end{figure*}

%%%%%%%%%%%%%%%%%%%%%%%%%%%%%%%%

\begin{figure*}
\centering
\mbox
{
  \subfigure{\includegraphics[width=8cm, clip=true, trim=50 350 70 0]{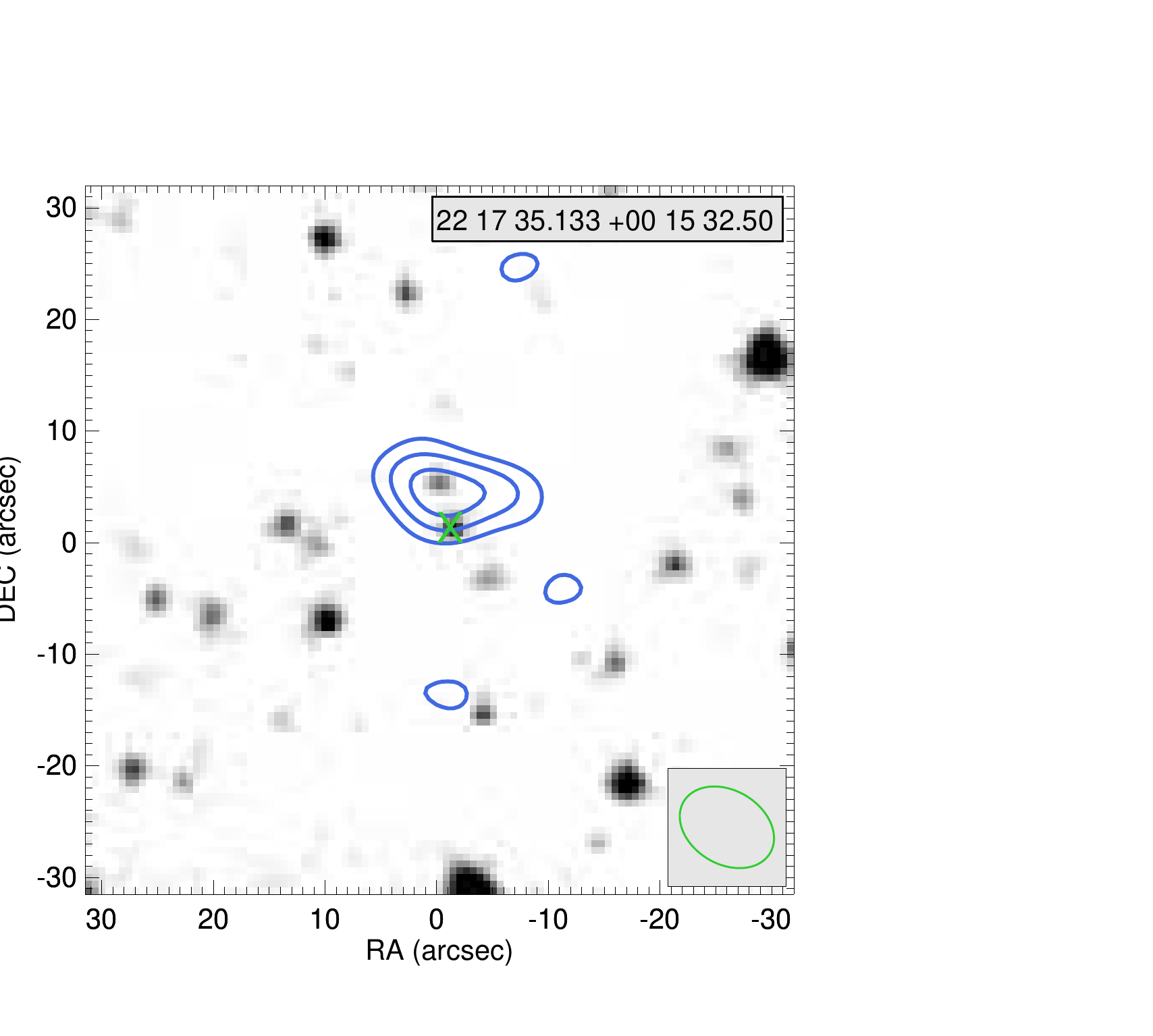}} \hspace{-2cm}
  \subfigure{\includegraphics[width=10cm, clip=true, trim=30 300 70 0]{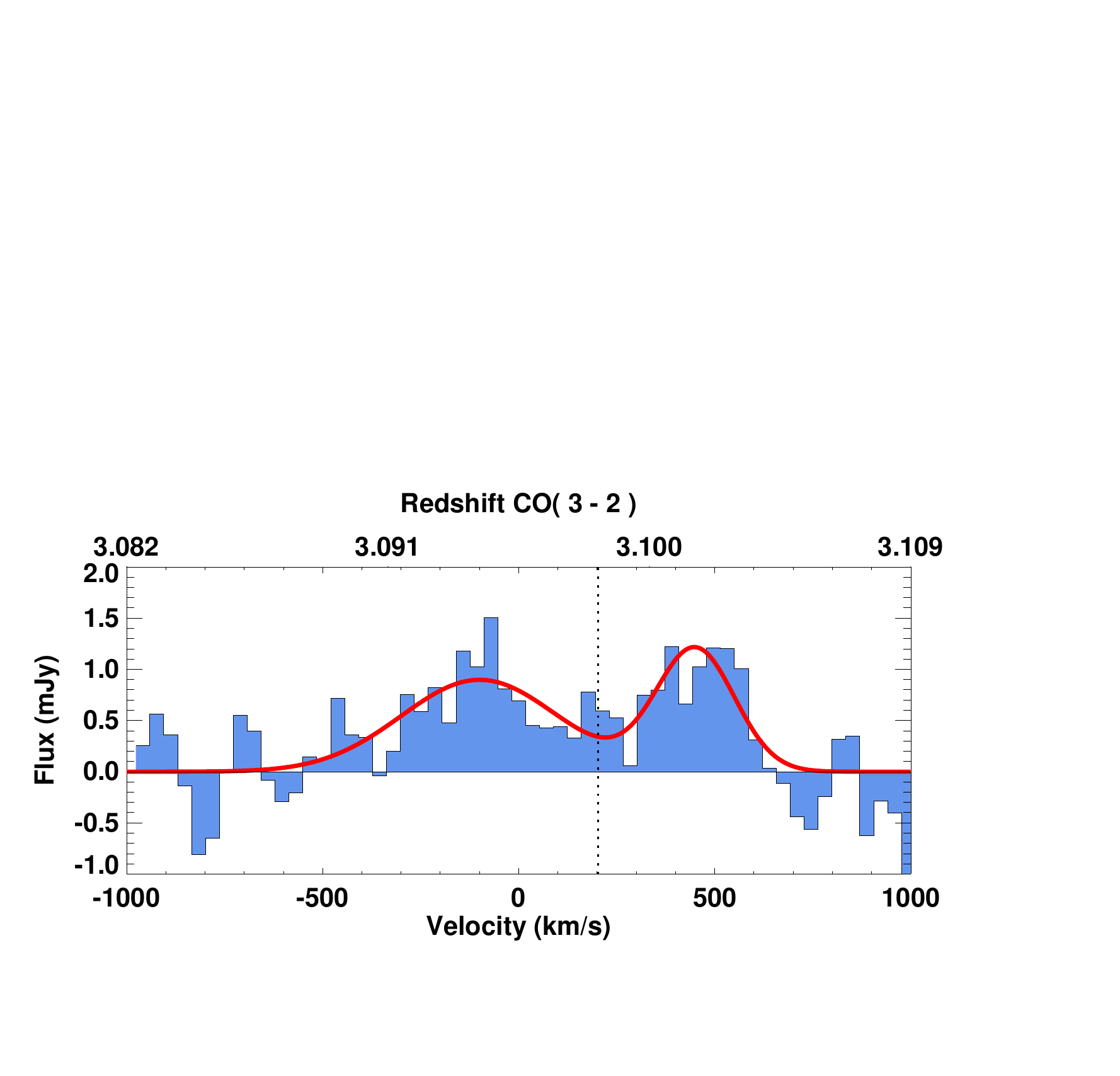}}
}
\caption{SMM221735+0015. $^{12}$CO is detected at $4.7 \sigma$. There is  no 24 $\mu$m emission associated with this source
(Menendez-Delmestre et al.\ 2009). The PAH redshift ($z=3.21$) is inconsistent with the CO data, however the  UV spectrum implies a redshift consistent with the CO, $z = 3.098$.
The source was previously published in Greve et al.\ (2005). }
\label{figure_ng28}
\end{figure*}

%%%%%%%%%%%%%%%%%%%%%%%%%%%%%%%%

\begin{figure*}
\centering
\mbox
{
  \subfigure{\includegraphics[width=8cm, clip=true, trim=50 350 70 0]{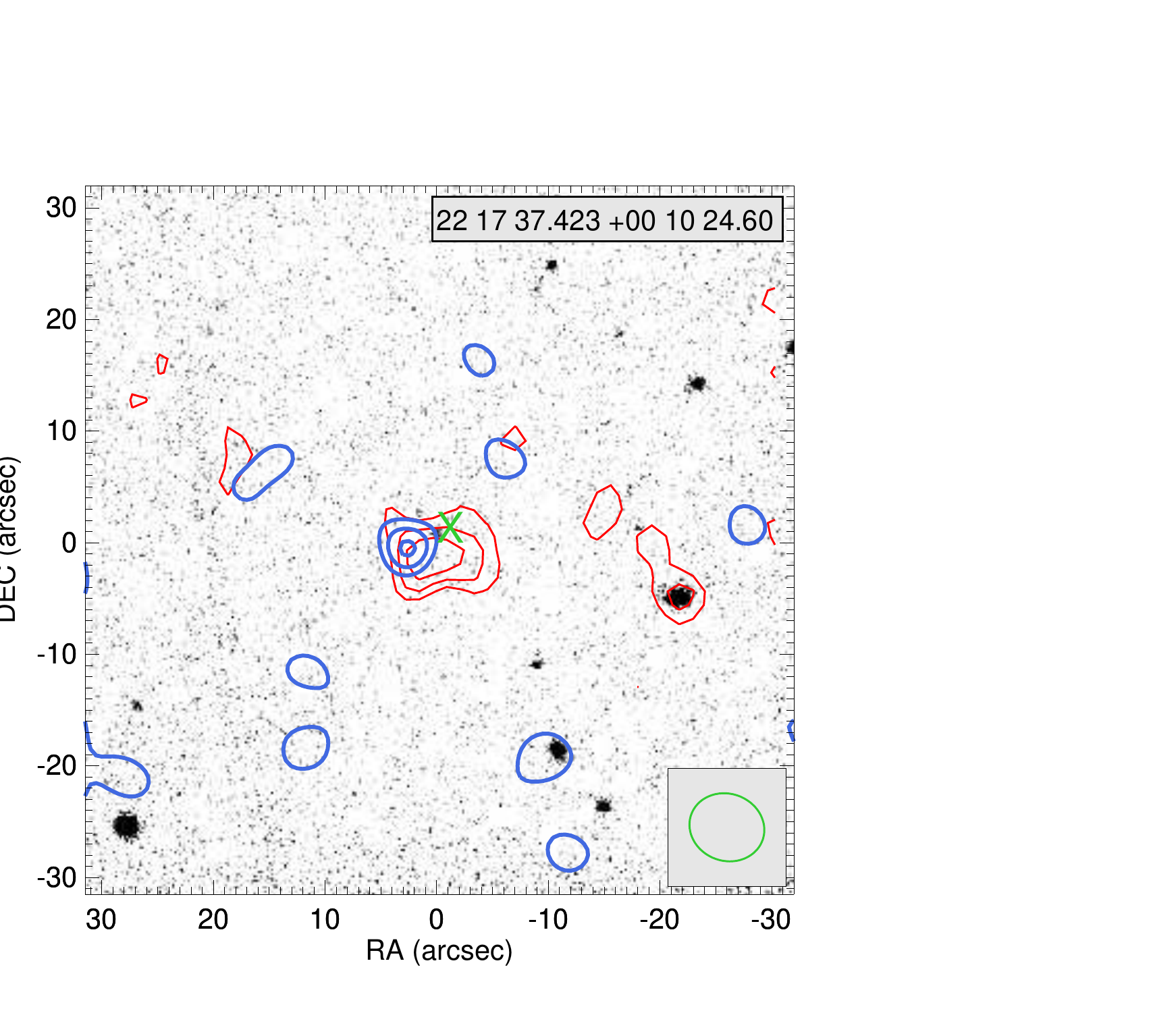}} \hspace{-2cm}
  \subfigure{\includegraphics[width=10cm, clip=true, trim=30 300 70 0]{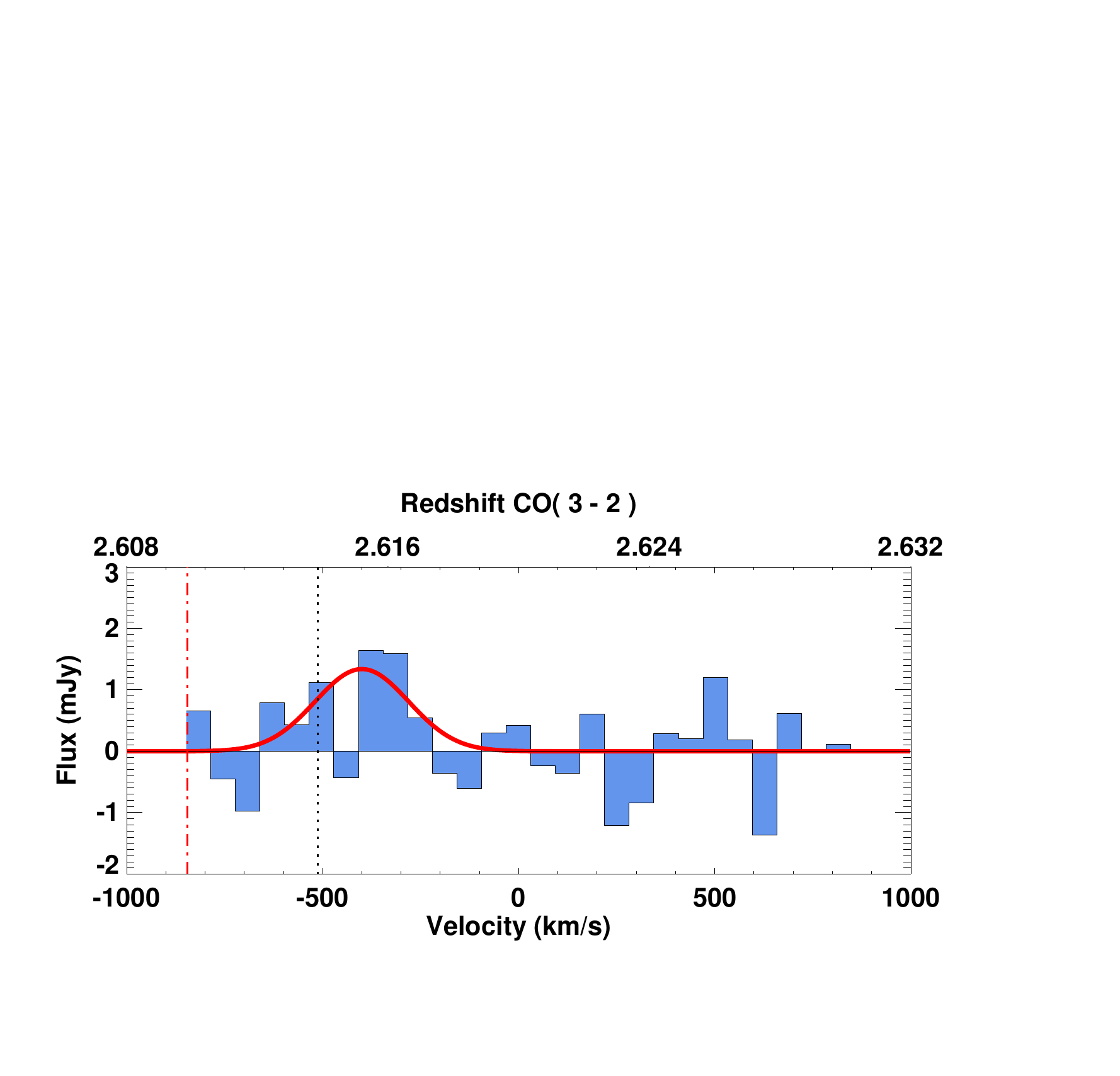}}
}
\caption[SMM221737+0010]{SMM221737+0010. The background image is $K$-band. $^{12}$CO is detected at $4.2 \sigma$, approximately 5'' to the east of the radio centre. There is extended 24um over the CO region, and a $K$-band source 3" to the West of the CO emission, from which the optical and near-IR spectra were obtained. 
We identify this as a candidate detection based on the weak CO detection and the offset from the optical source.
}
\label{figure_nh61}
\end{figure*}

%%%%%%%%%%%%%%%%%%%%%%%%%%%%%%%%

\section{Non-detections}

%%%%%%%%%%%%%%%%%%%%%%%%%%%%%%%%
%%%%%                   NON DETECTIONS                    %%%%
%%%%%%%%%%%%%%%%%%%%%%%%%%%%%%%%

\begin{figure*}
\centering
\mbox
{
  \subfigure{\includegraphics[width=8cm, clip=true, trim=50 350 70 0]{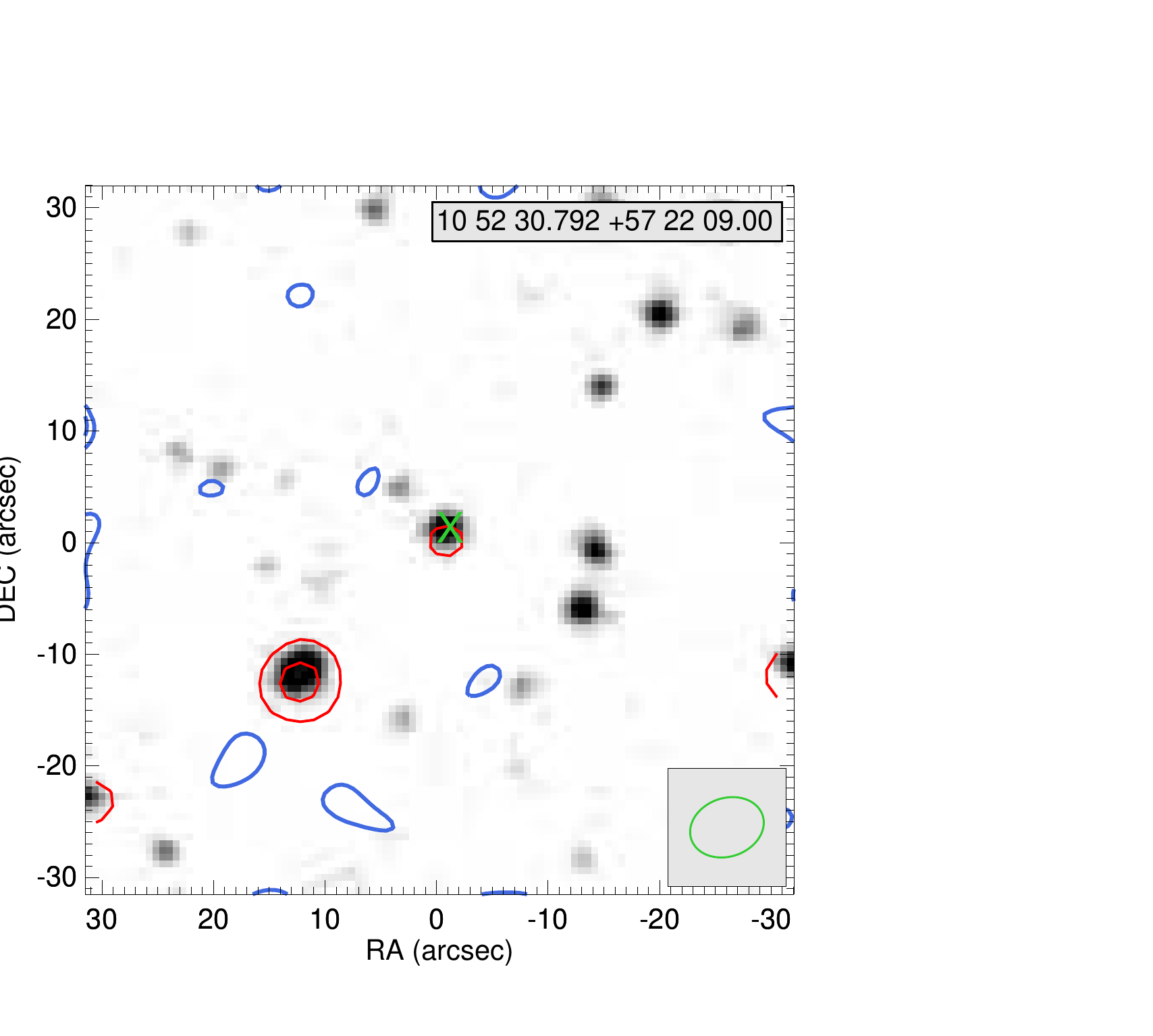}} \hspace{-2cm}
  \subfigure{\includegraphics[width=10cm, clip=true, trim=30 300 70 0]{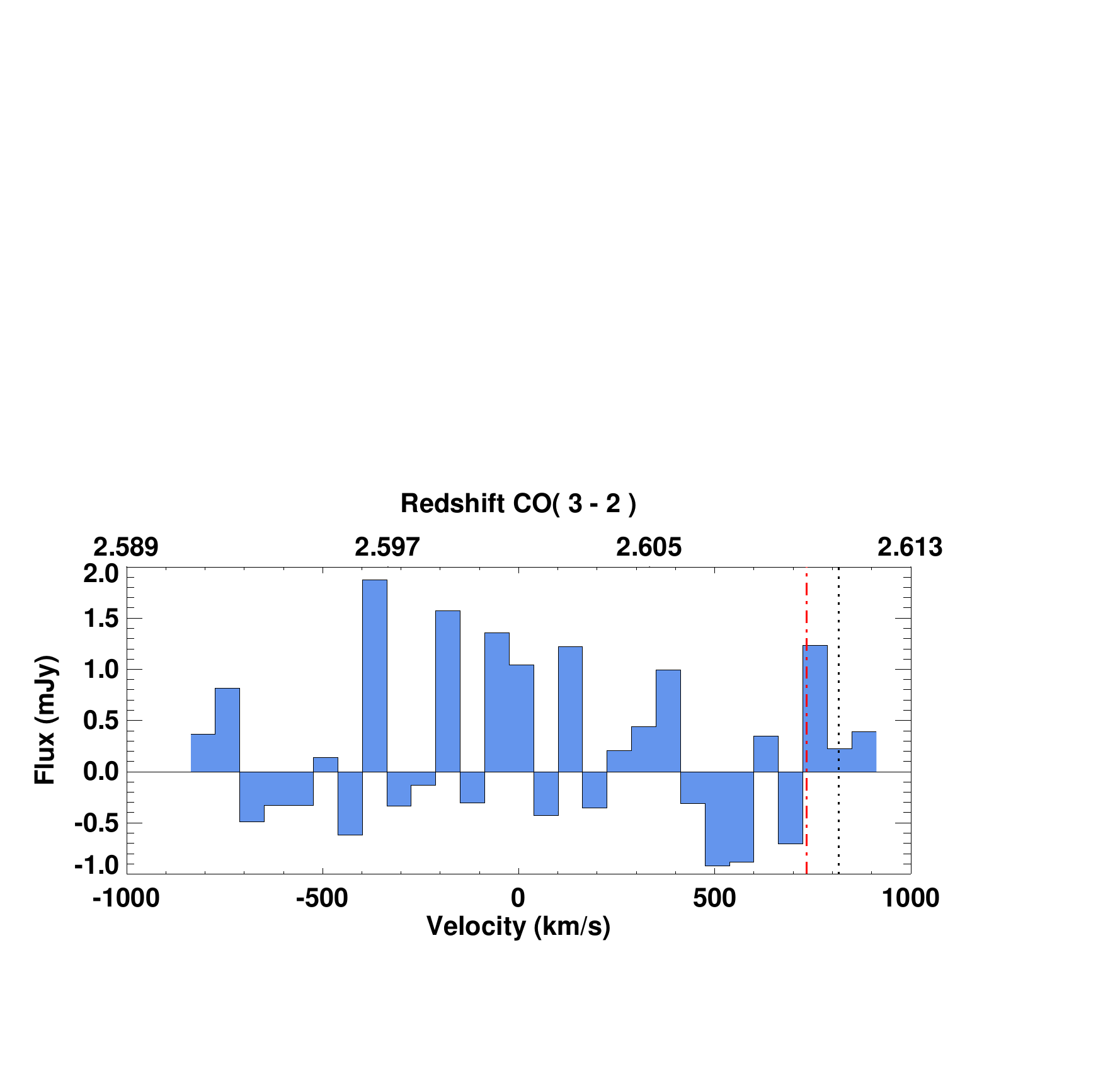}}
}
\caption{SMM105230+5722. Non-detection. All the sources shown in Appendix B are true non-detections, which allow useful constraints to be placed on the CO flux.}
\label{figure_mc5d}
\end{figure*}

%%%%%%%%%%%%%%%%%%%%%%%%%%%%%%%%

\begin{figure*}
\centering
\mbox
{
  \subfigure{\includegraphics[width=8cm, clip=true, trim=50 350 70 0]{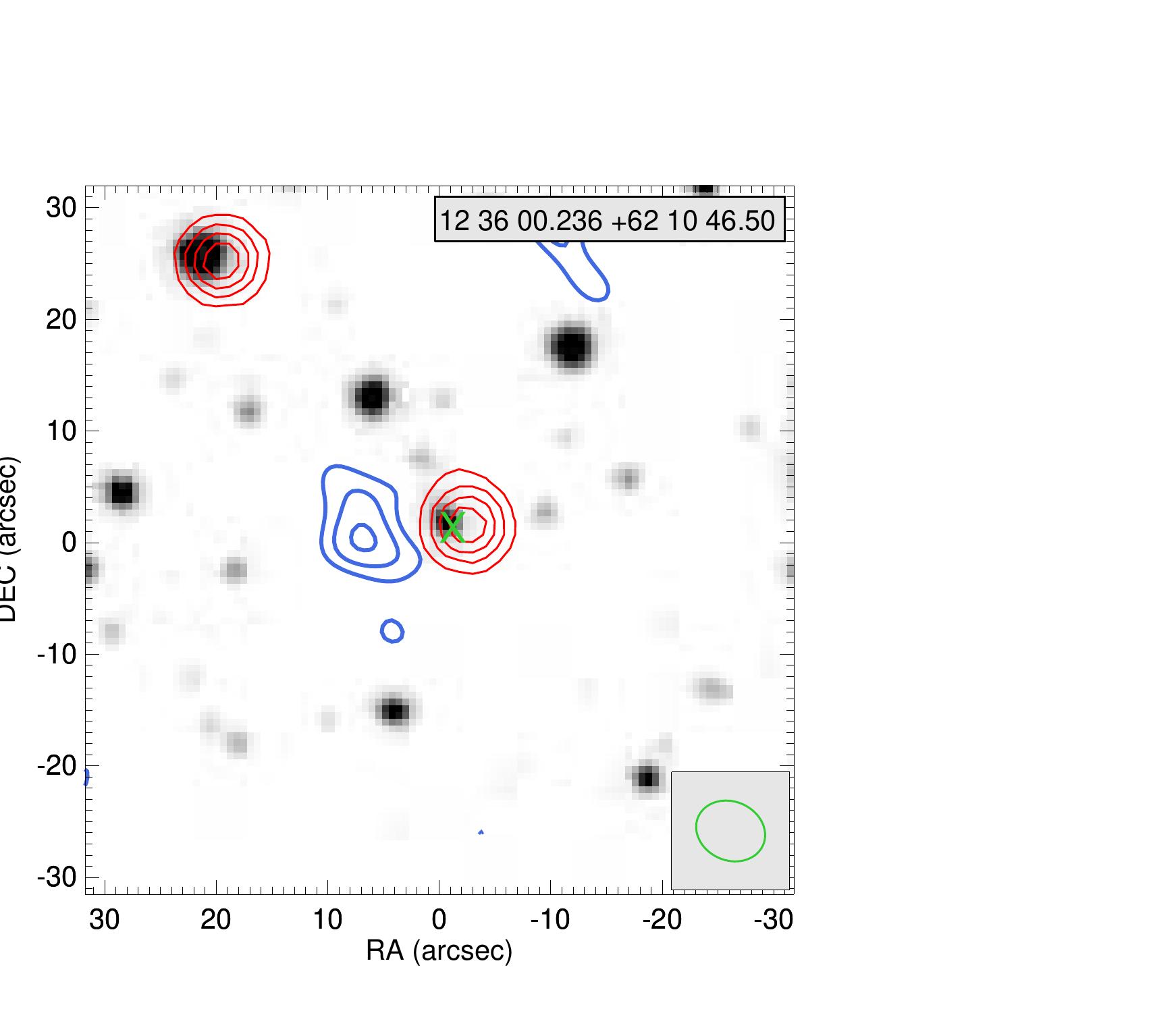}} \hspace{-2cm}
  \subfigure{\includegraphics[width=10cm, clip=true, trim=30 300 70 0]{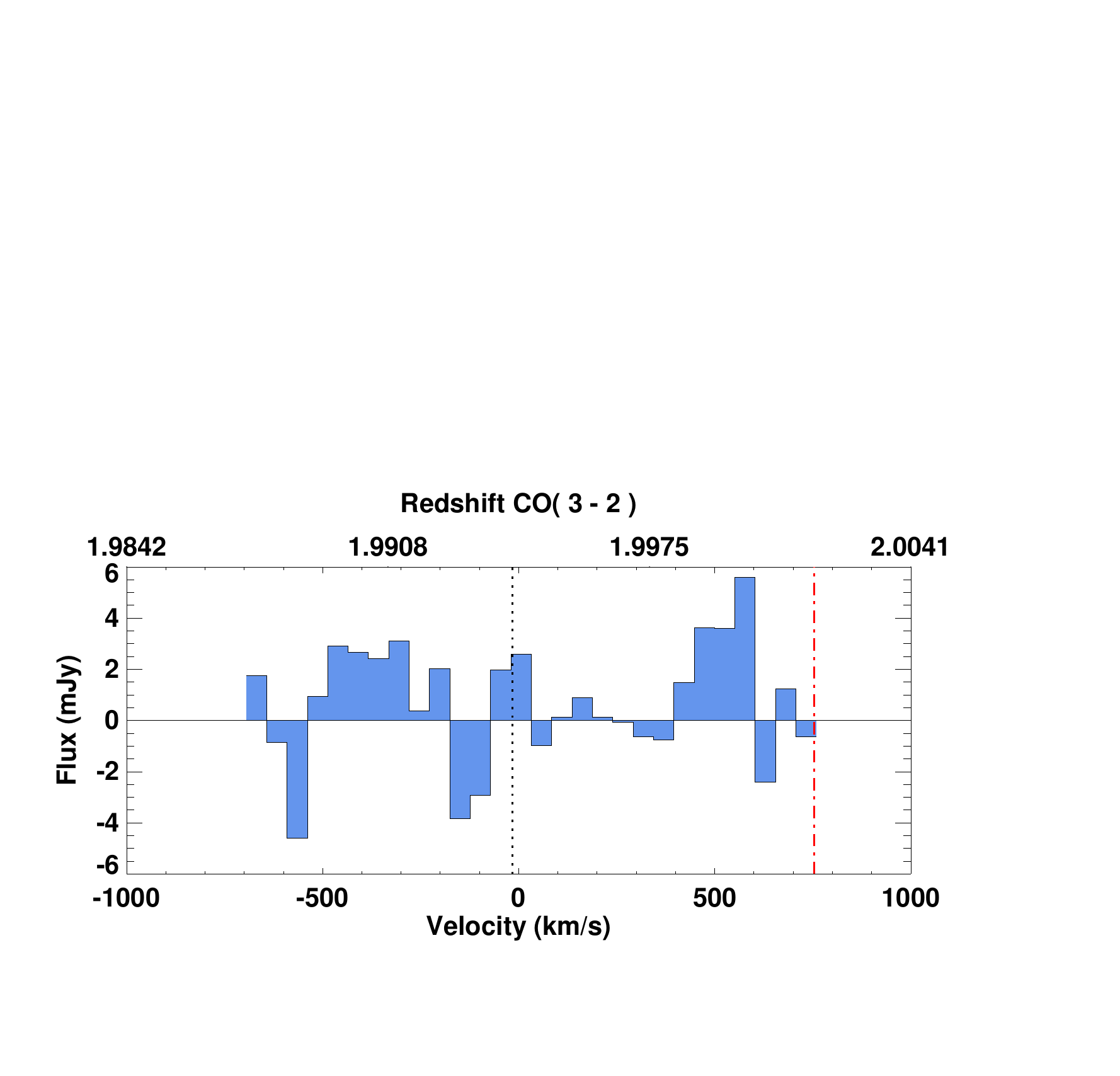}}
}
\caption{SMM123600+6210.  Non-detection.}
\label{figure_ni28}
\end{figure*}

%%%%%%%%%%%%%%%%%%%%%%%%%%%%%%%%

\begin{figure*}
\centering
\mbox
{
  \subfigure{\includegraphics[width=8cm, clip=true, trim=50 350 70 0]{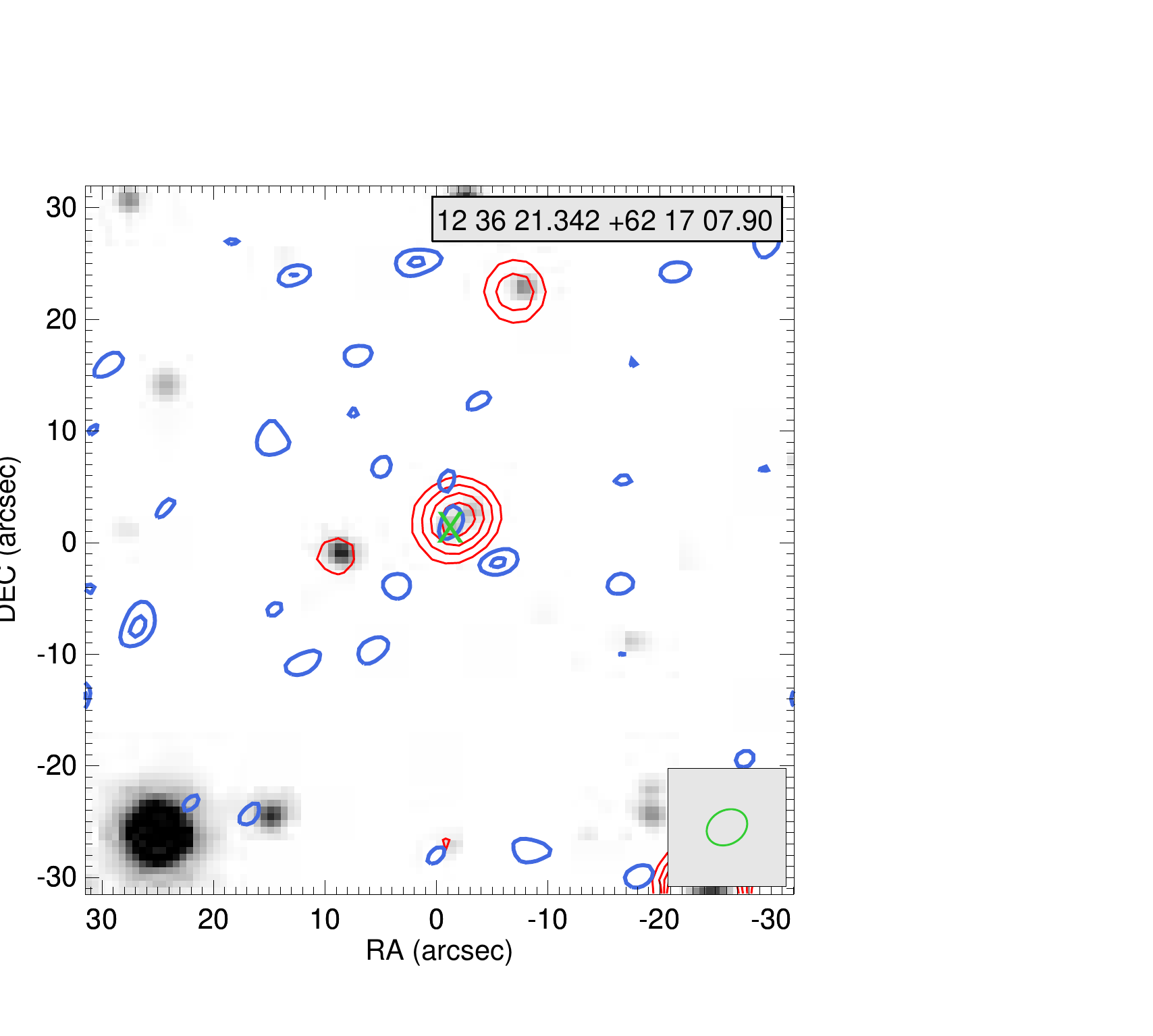}} \hspace{-2cm}
  \subfigure{\includegraphics[width=10cm, clip=true, trim=30 300 70 0]{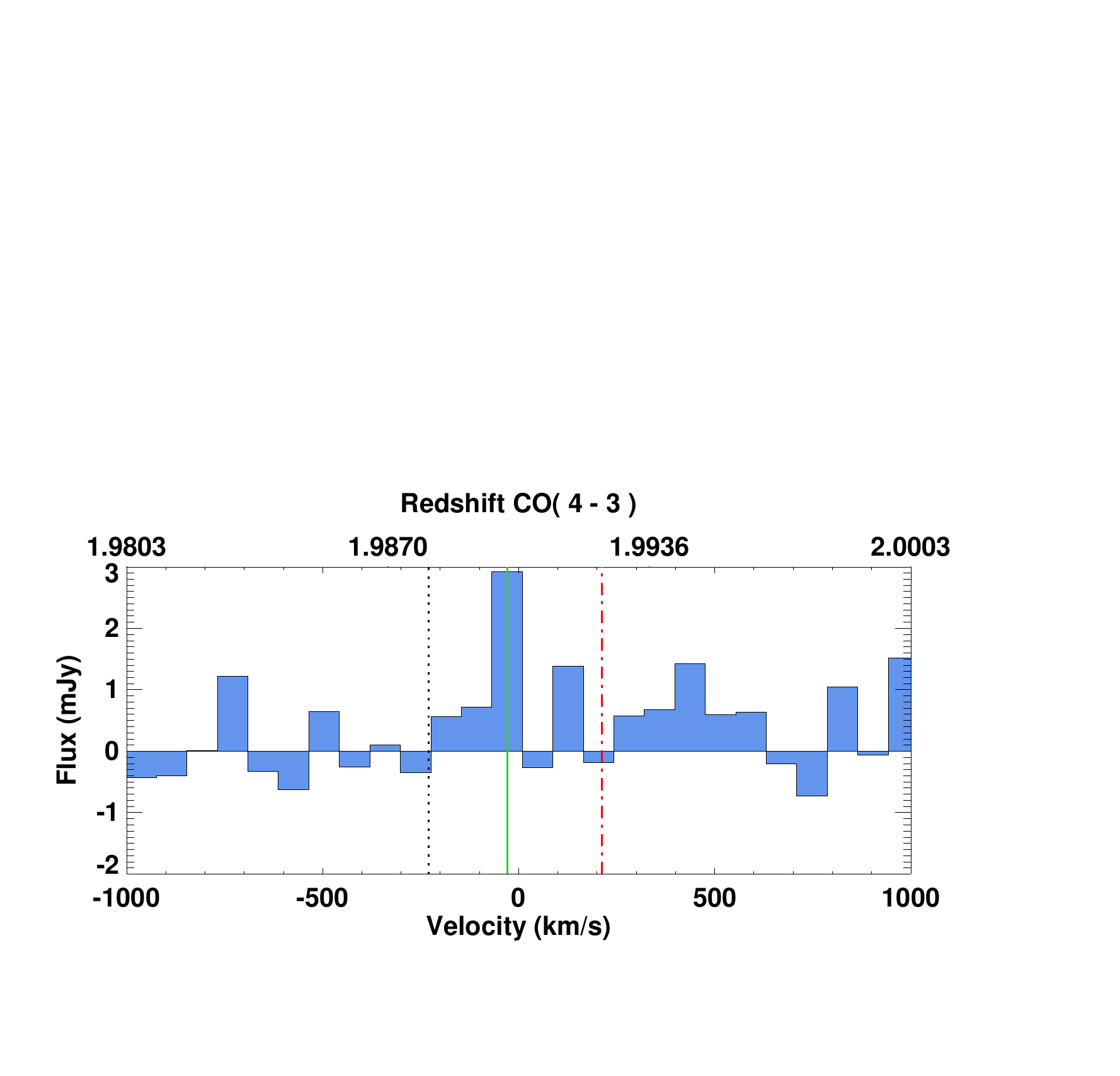}}
}
\caption{SMM123621+6217. Non-detection.}
\label{figure_sd3b}
\end{figure*}

%%%%%%%%%%%%%%%%%%%%%%%%%%%%%%%%

\begin{figure*}
\centering
\mbox
{
  \subfigure{\includegraphics[width=8cm, clip=true, trim=50 350 70 0]{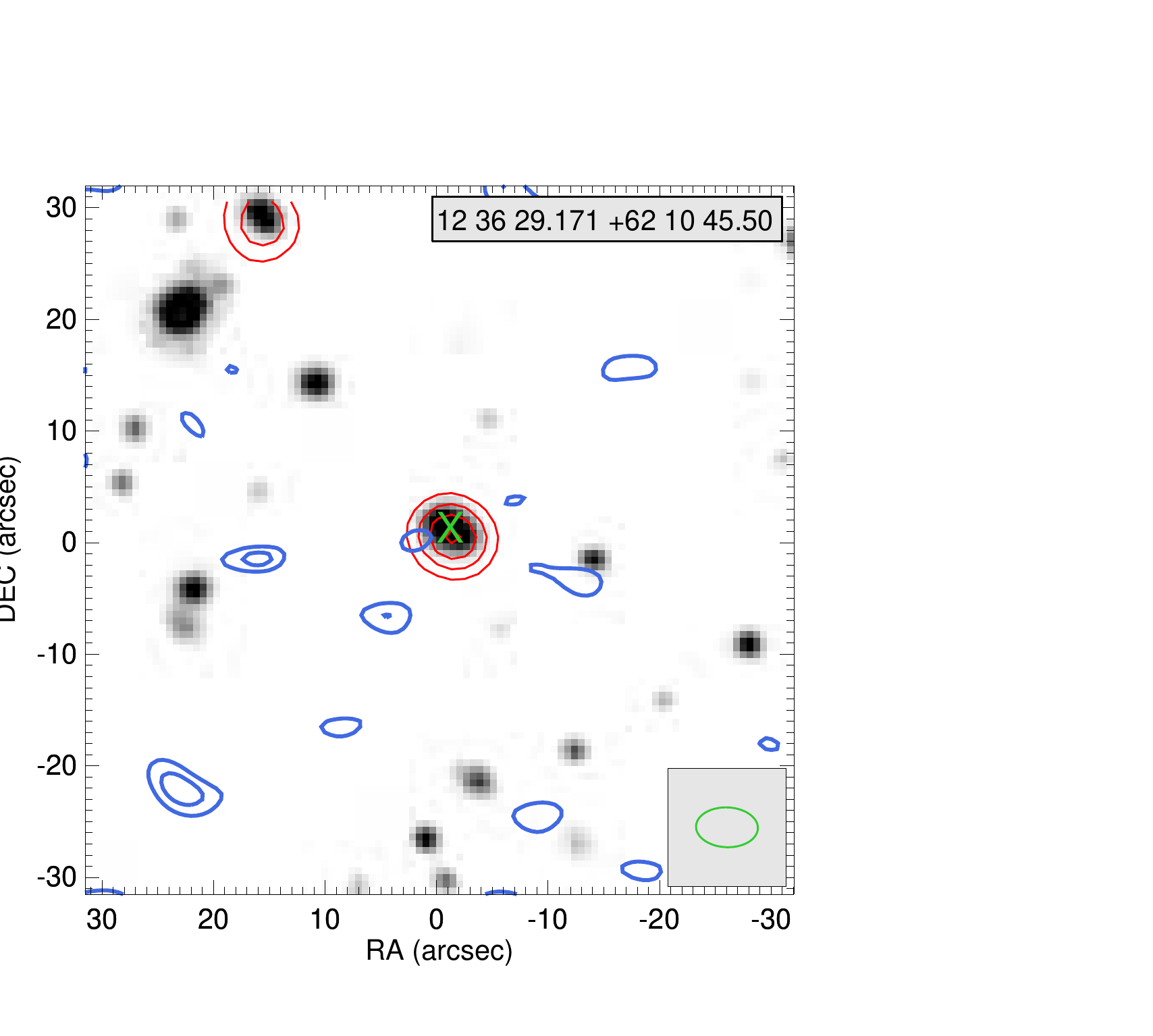}} \hspace{-2cm}
  \subfigure{\includegraphics[width=10cm, clip=true, trim=30 300 70 0]{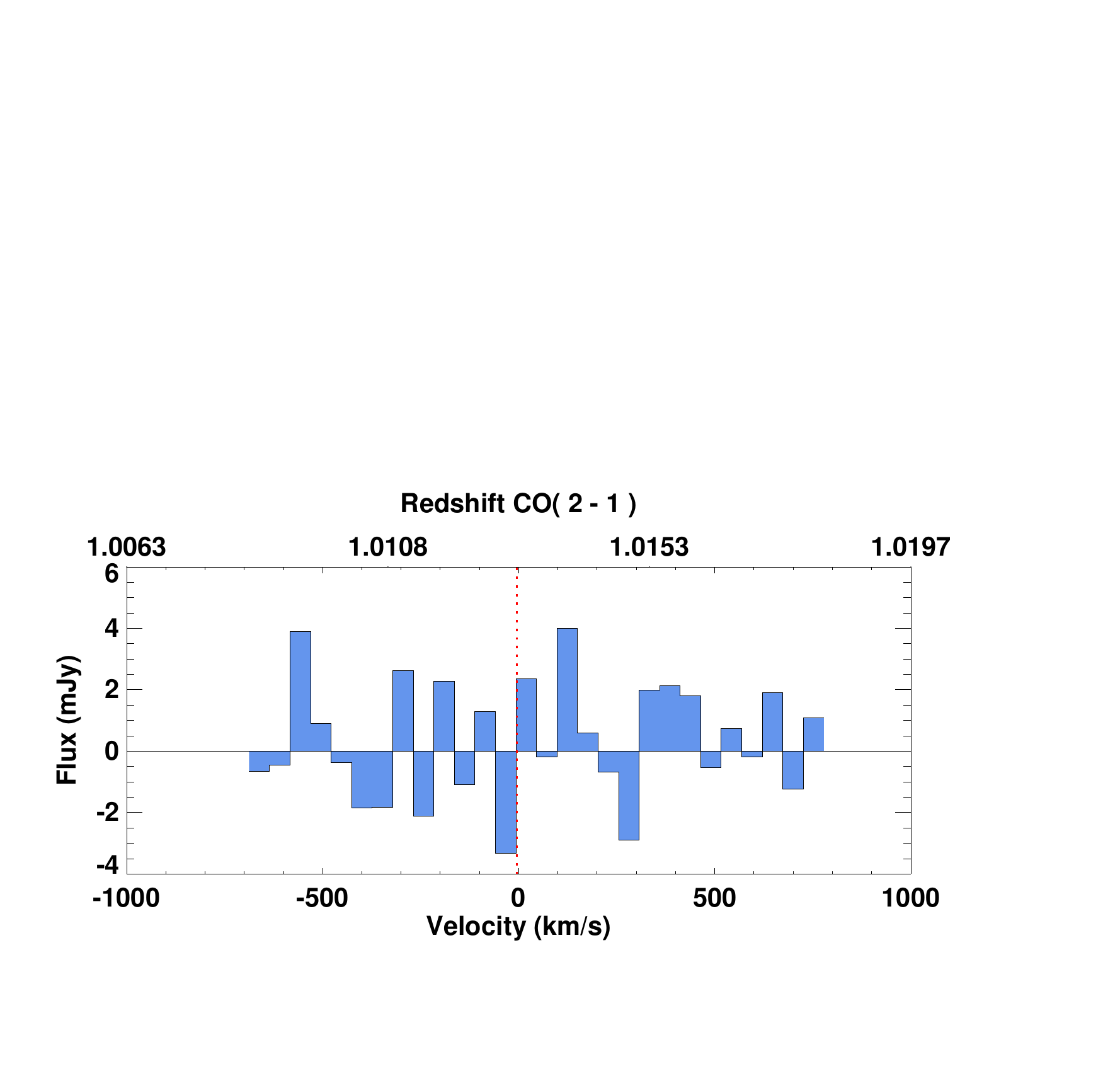}}
}
\caption{SMM123629+6210.  Non-detection.}
\label{figure_pd32}
\end{figure*}

%%%%%%%%%%%%%%%%%%%%%%%%%%%%%%%%

\begin{figure*}
\centering
\mbox
{
  \subfigure{\includegraphics[width=8cm, clip=true, trim=50 350 70 0]{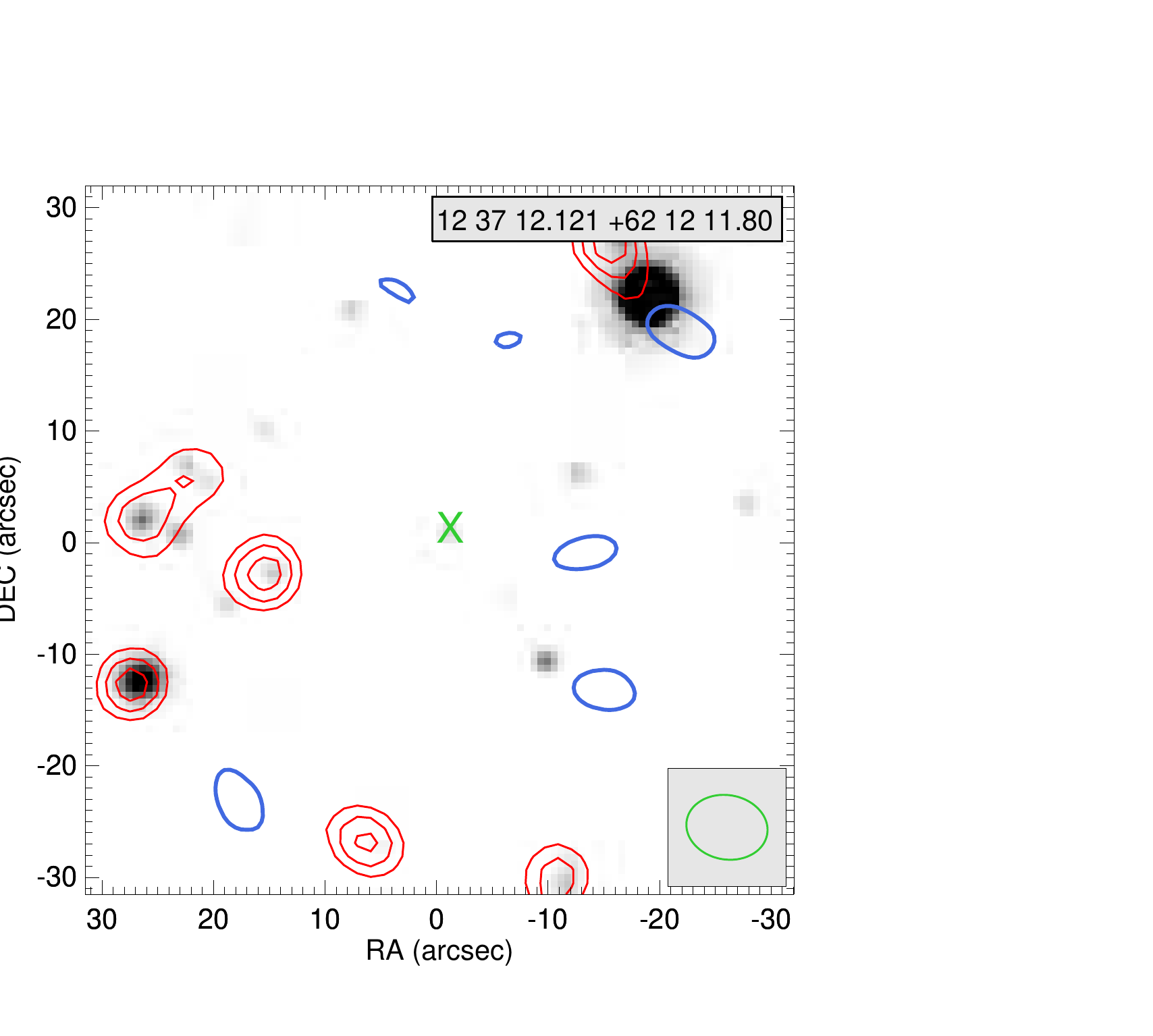}} \hspace{-2cm}
  \subfigure{\includegraphics[width=10cm, clip=true, trim=30 300 70 0]{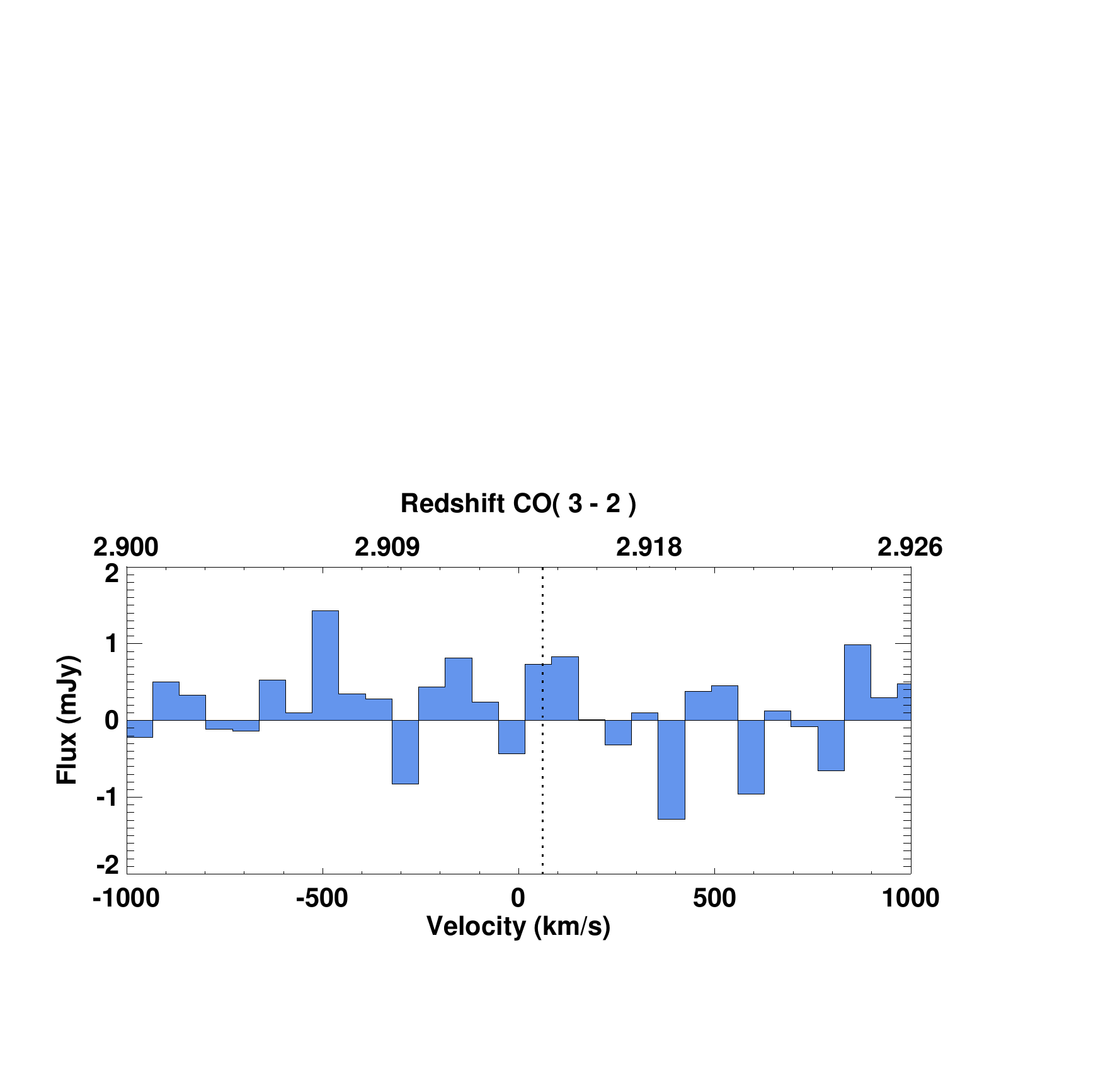}}
}
\caption{SMM123712+6212. Non-detection.}
\label{figure_pf32}
\end{figure*}

%%%%%%%%%%%%%%%%%%%%%%%%%%%%%%%%

\begin{figure*}
\centering
\mbox
{
  \subfigure{\includegraphics[width=8cm, clip=true, trim=50 350 70 0]{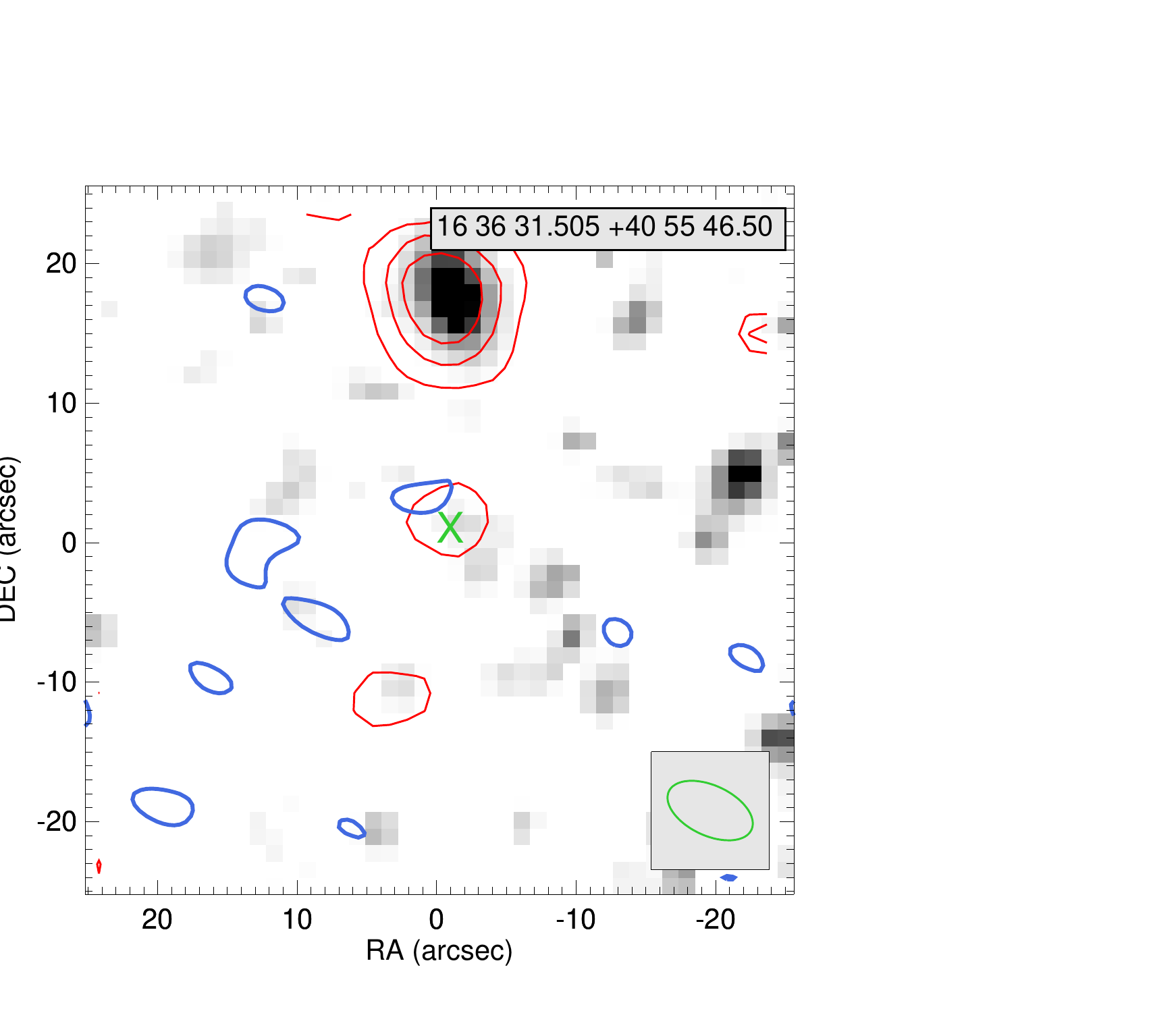}} \hspace{-2cm}
  \subfigure{\includegraphics[width=10cm, clip=true, trim=30 300 70 0]{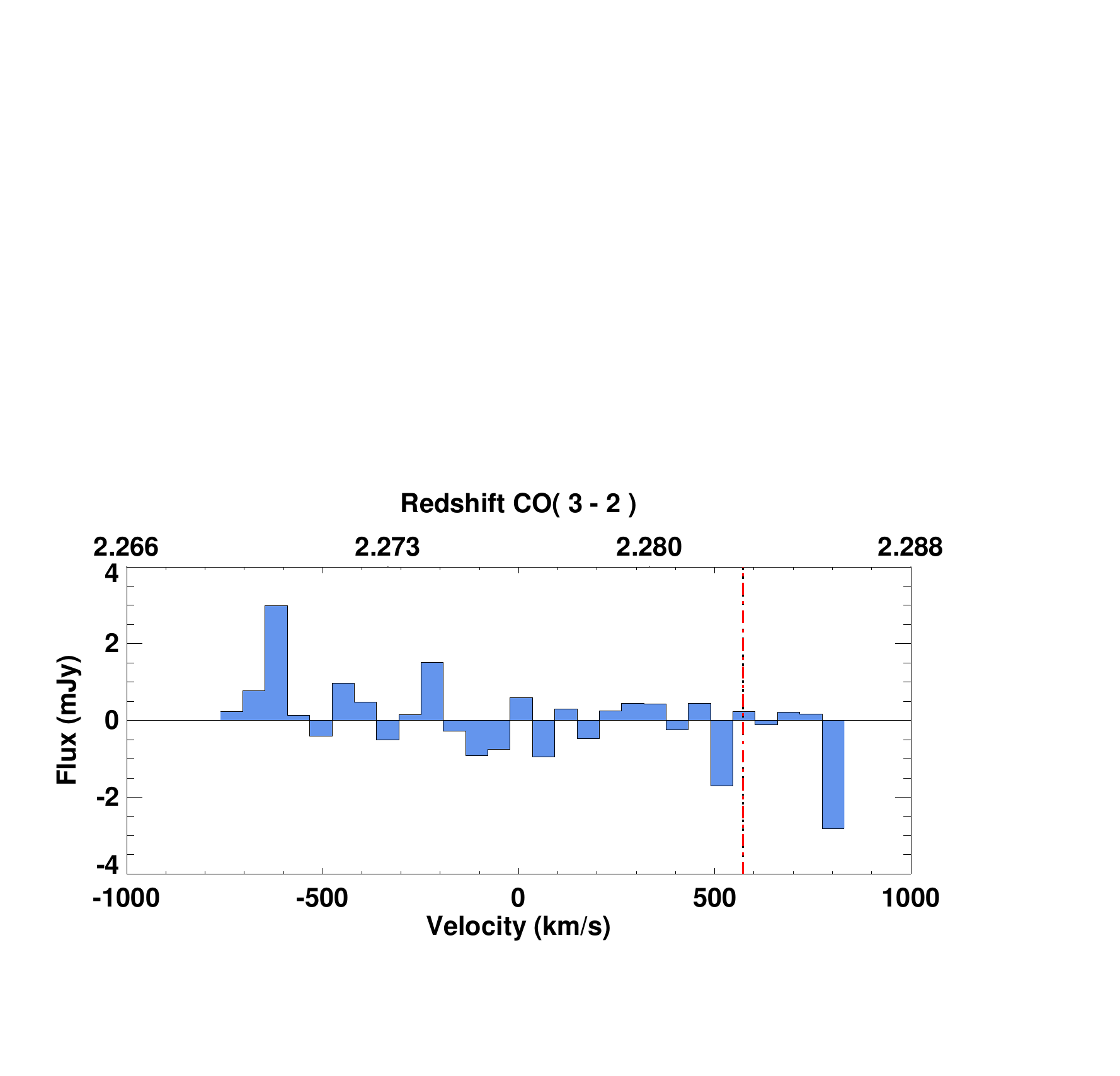}}
}
\caption{SMM163631+4055. Non-detection.}
\label{figure_me5d}
\end{figure*}

%%%%%%%%%%%%%%%%%%%%%%%%%%%%%%%%

\begin{figure*}
\centering
\mbox
{
  \subfigure{\includegraphics[width=8cm, clip=true, trim=50 350 70 0]{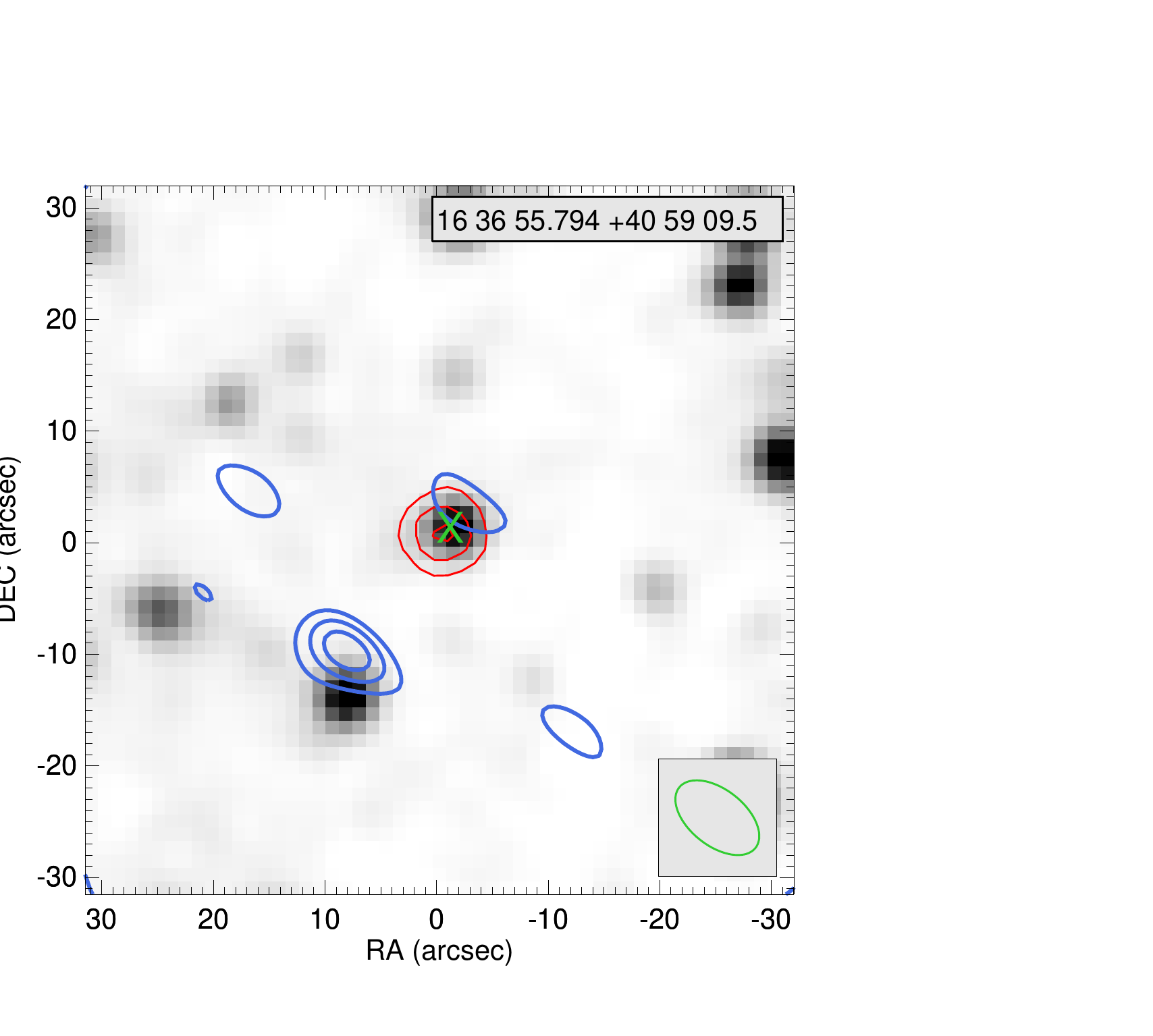}} \hspace{-2cm}
  \subfigure{\includegraphics[width=10cm, clip=true, trim=30 300 70 0]{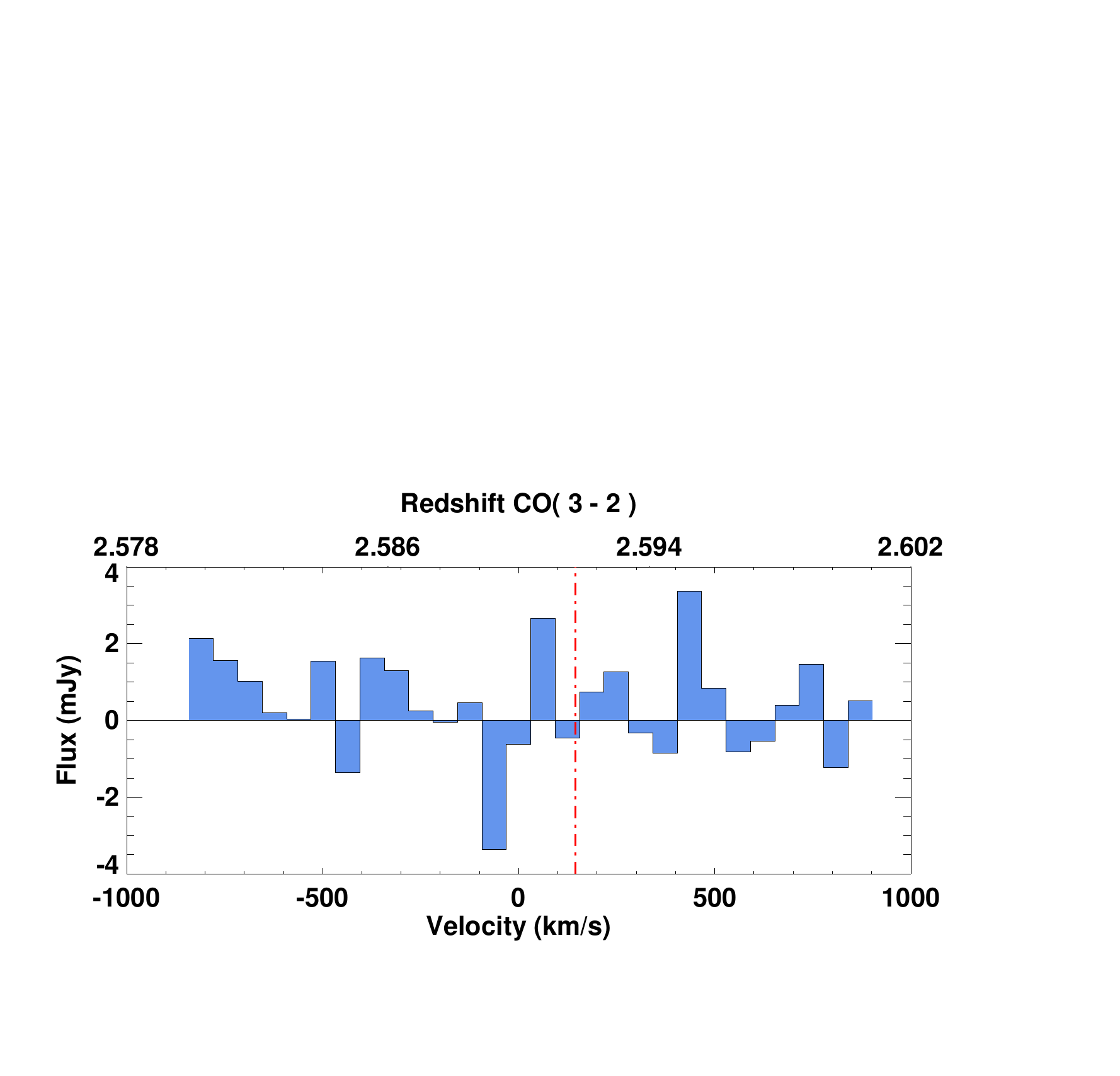}}
}
\caption{SMM163655+4059. Non-detection. The emission to the SE of phase centre is most likely spurious, as indicated by IRAC photometric redshift estimates.}
\label{figure_ng61}
\end{figure*}

%\begin{figure*}
%\centering
%\mbox
%{
%  \subfigure{\includegraphics[width=8cm, clip=true, trim=50 350 70 0]{catalogue_imgs/oe2a-CO_spitzer}} \hspace{-2cm}
%  \subfigure{\includegraphics[width=10cm, clip=true, trim=30 300 70 0]{catalogue_imgs/oe2a-co32-spectrum}}
%}
%\caption{SMM123622+6216 = HDF143. Non-detection.}
%\label{figure_oe2a}
%\end{figure*}

\end{document}